\documentclass[]{aastex63}

\newcommand{\kms}{km~s$^{-{\rm 1}}$}
\newcommand{\msun}{M$_{\odot}$}
\newcommand{\rsun}{R$_{\odot}$}
\newcommand{\Tratio}{$T_{\rm 2}$/$T_{\rm 1}$}
\newcommand{\Rratio}{$R_{\rm 2}$/$R_{\rm 1}$}
\newcommand{\Ronemax}{$R_{\rm 1}$/$R_{\rm 1max}$}
\newcommand{\Rtwomax}{$R_{\rm 2}$/$R_{\rm 2max}$}
\newcommand{\Tapprox}{$T_{\rm 2}$$\approx$$T_{\rm 1}$}
\newcommand{\redchi}{$\chi^2_{\nu}$}
\newcommand{\logq}{$\log q$}
\newcommand{\logp}{$\log P$}
\newcommand{\cosi}{$\cos i$}
\newcommand{\OC}{$O-C$}

\newcommand{\numsample}{ten}
\newcommand{\numksample}{809} % total from Kirk? formerly 452
\newcommand{\nummodeled}{783} % number modeled after removing flat and junky ones
\newcommand{\numcontacts}{178} %  number of probable contacts
\newcommand{\numcontactsplusambiguous}{669} % number contacts & ambiguous
\newcommand{\numambiguous}{491} %  number of ambiguous cases
\newcommand{\numdetached}{114} %  number of probable detached 
\newcommand{\numtextreme}{387} % num with extreme temp ratios
\newcommand{\numbadfit}{230} %  number of bad fits due to spots
\newcommand{\numremoved}{23} % for small variation or junk
\received{Accepted for publication in the Astronomical Journal \today}

\shorttitle{Extreme-Mass-Ratio Binaries}
\shortauthors{Kobulnicky et al.}
\begin{document}

\title{A Bayesian Analysis of Physical Parameters for \nummodeled\ {\it Kepler} Close Binaries: Extreme-Mass-Ratio Systems and a New Mass Ratio versus Period Lower Limit}

\correspondingauthor{Henry A. Kobulnicky}
\email{chipk@uwyo.edu}

\author[0000-0002-4475-4176]{Henry A. Kobulnicky}
\affiliation{Department of Physics \& Astronomy \\
University of Wyoming \\ 1000 E. University \\
Laramie, WY 82070, USA}

\author[0000-0002-4475-4176]{Lawrence A. Molnar}
\affiliation{Department of Physics \& Astronomy \\
Calvin College \\ 
Grand Rapids, MI 49546, USA}

\author[0000-0002-4475-4176]{Evan M. Cook}
\affiliation{Department of Physics \& Astronomy \\
University of Wyoming \\ 1000 E. University \\
Laramie, WY 82070, USA}

\author[0000-0001-6207-4388]{Lauren E. Henderson}
\affiliation{Department of Physics \& Astronomy \\
Calvin University \\ 
Grand Rapids, MI 49546, USA}

\begin{abstract}
Contact binary star systems represent the long-lived penultimate phase of binary evolution.  Population statistics of their physical parameters inform understanding of binary evolutionary pathways and end products.  We use light curves and new optical spectroscopy to conduct a pilot study of ten (near-)contact systems in the long-period ($P$$>$0.5~d) tail of close binaries in the {\it Kepler} field.   We use PHOEBE light curve models to compute Bayesian probabilities on five principal system parameters. Mass ratios and third-light contributions measured from spectra agree well with those inferred from the light curves.  Pilot study systems have extreme mass ratios $q$$<$0.32.  Most are triples.  Analysis of the unbiased sample of \nummodeled\ $0.15$~d$<$$P$$<$2 d (near-)contact binaries results in \numcontacts\ probable contact systems, \numdetached\ probable detached systems, and \numambiguous\ ambiguous systems for which we report best-fitting and 16th/50th/84th percentile parameters.  Contact systems are rare at periods $P$$>$0.5 d, as are systems with $q$$>$0.8.  There exists an empirical mass ratio lower limit $q_{\rm min}(P)$$\approx$0.05--0.15 below which contact systems are absent, supporting a new set of theoretical predictions  obtained by modeling the evolution of contact systems under the constraints of mass and angular momentum conservation.  Pre-merger systems should lie at long periods and near this mass ratio lower limit, which rises from $q$=0.044 for $P$=0.74 d to $q$=0.15 at $P$=2.0 d.  These findings support a scenario whereby  nuclear evolution of the primary (more massive) star drives mass transfer to the primary, thus moving systems toward extreme $q$ and larger $P$ until the onset of the Darwin instability at $q_{\rm min}$ precipitates a merger.    
\end{abstract}

\section{Introduction} \label{sec:intro}
\subsection{Contact Binary Stars}

The evolutionary paths of many stars culminate in a merger with a close stellar companion.  As currently envisioned, these cataclysmic events occur after the stars undergo long geriatric episodes, exchanging mass as members of the ubiquitous population of contact binary systems.  V1309 Sco became the prototype for stellar merger events when this 1.4~d contact binary exhibited an exponentially decreasing period, brightened, underwent a rapid evolution in light curve morphology, and erupted in a ``luminous red nova'' \citep{Kulkarni2007} event similar to the 2002 V838 Mon eruption \citep{Munari2002}, leaving only a single cool inflated star \citep{Tylenda2011}.  Stellar mergers may be as frequent as 0.2 yr$^{-1}$ in the Milky Way \citep{Kochanek2014, Howitt2020}. The number of observed red novae in nearby galaxies is currently small, but the advent of wide-area deep sky surveys may precipitate detection of hundreds per year in the local universe.  A comprehensive picture of the evolutionary sequence(s) yielding contact binaries and subsequent mergers is still elusive, but rapid advances are possible in this nascent era of all-sky time-domain datasets.  

\citet{Eggleton2012} summarized a working theoretical sequence for the evolution of contact binaries terminating in a merger \citep[see also][]{Lucy1976, Webbink1976, Hilditch1989, Stepien1995, Yakut2005}.  The sequence commences with a wide (period $P$=months--1,000 yr) binary with main-sequence components orbited by a distant tertiary that induces Kozai-Lidov cycles \citep{Kozai1962, Lidov1962} with tidal friction that shrinks the inner orbit to a few days, a point where magnetic braking can continue to rob the orbit of angular momentum.  When the semi-major axis becomes small enough that the more massive primary star fills its Roche lobe, mass transfer to the secondary commences.  Conservative mass transfer to the less massive secondary component drives the orbit toward smaller separations and heats the secondary which fills its Roche lobe, eventually inducing mass transfer the other direction.  A series of ``thermal relaxation oscillations'' ensue with mass transfer alternating direction on a thermal time scale. This complex interaction makes it impossible to follow the evolution of either star in detail, but it is presumed that the long-term average transfer is to the primary star and hence toward small mass ratios ($q$=$M_{\rm 2}$/$M_{\rm 1}$).  Once $q$ reaches a critical threshold in the vicinity of 0.15, dictated by the point at which the tidally locked components' rotational angular momentum exceeds 1/3 their orbital momentum, the Darwin instability \citep{Darwin1893, Counselman1973} instigates a rapid loss of angular momentum driven by tidal dissipation and non-conservative mass loss through the second Lagrange point. The runaway angular momentum loss culminates in a phase of dynamic friction in a common envelope as the primary subsumes the secondary. 

\citet{Molnar2019} and \citet{Molnar2022} show that contact binary systems with steady mass transfer to the primary (i.e. without oscillations) is possible for all but the largest mass ratios. They computed evolutionary models driven by the nuclear evolution of the primary (i.e., mass-receiving star) and so derived the lifetimes for a grid of initial mass values and the evolution of the moment of inertia of the primary. These showed the mass ratio for the onset of the Darwin instability depends modestly on the initial mass ratio and total mass, with the limit being reached for final period $P>$0.75 d at mass ratios increasing from 0.05 to 0.15 with increasing $P$. This scenario predicts a paucity of systems with mass ratios more extreme than 0.10 and entails the requirement that contact binaries evolve from short-period (0.3--0.5 d) orbits toward $\approx$1 d orbits prior to coalescence.  However, even if this sequence describes one dominant paradigm for contact binary evolution, alternate channels (e.g., systems that first come into contact after significant individual evolution) are likely to operate as well. 

Population studies of contact binaries lend some support to this ``standard'' scenario but are also consistent with the \citet{Molnar2022} scenario.  W-type \citep[][]{Binnendijk1970} contact binaries (those where the less massive secondary star appears to be hotter and produces the deeper minimum when eclipsed) are the most numerous and preferentially have shorter orbital periods in the 0.3--0.5 d range.  By contrast, A-type contact binaries appear to have hotter primaries and tend to have longer orbital periods, lower mass ratios \citep{Yakut2005} and larger component radii, indicating a more evolved state \citep{Mochnacki1981}.  These observations are consistent with the grid of evolutionary tracks computed in \citet{Molnar2019} and refined in \citet{Molnar2022}. \citet{Molnar2019} also identified a small (N$=$7) class of long-period $P\approx$1 d contact binaries from among 22,400 close binaries in the 14-year OGLE photometric survey \citep{Pietrukowicz2017} that exhibit large ($\vert dP/dt\vert > 1.4\times10^{-8}$) negative period derivatives, consistent with being pre-merger candidates. \citet{Molnar2020} analyzed the light curves of 184,000 OGLE contact binaries to demonstrate an anti-correlation between mass ratio and period, supporting the evolution from W-type short-period toward A-type longer-period systems, a trend that we re-examine here using a sample of \nummodeled\ short-period binaries having high-precision {\it Kepler} photometric data.    

\begin{figure}[ht!]
%\fig{fillouts.pdf}{7in}{}
\fig{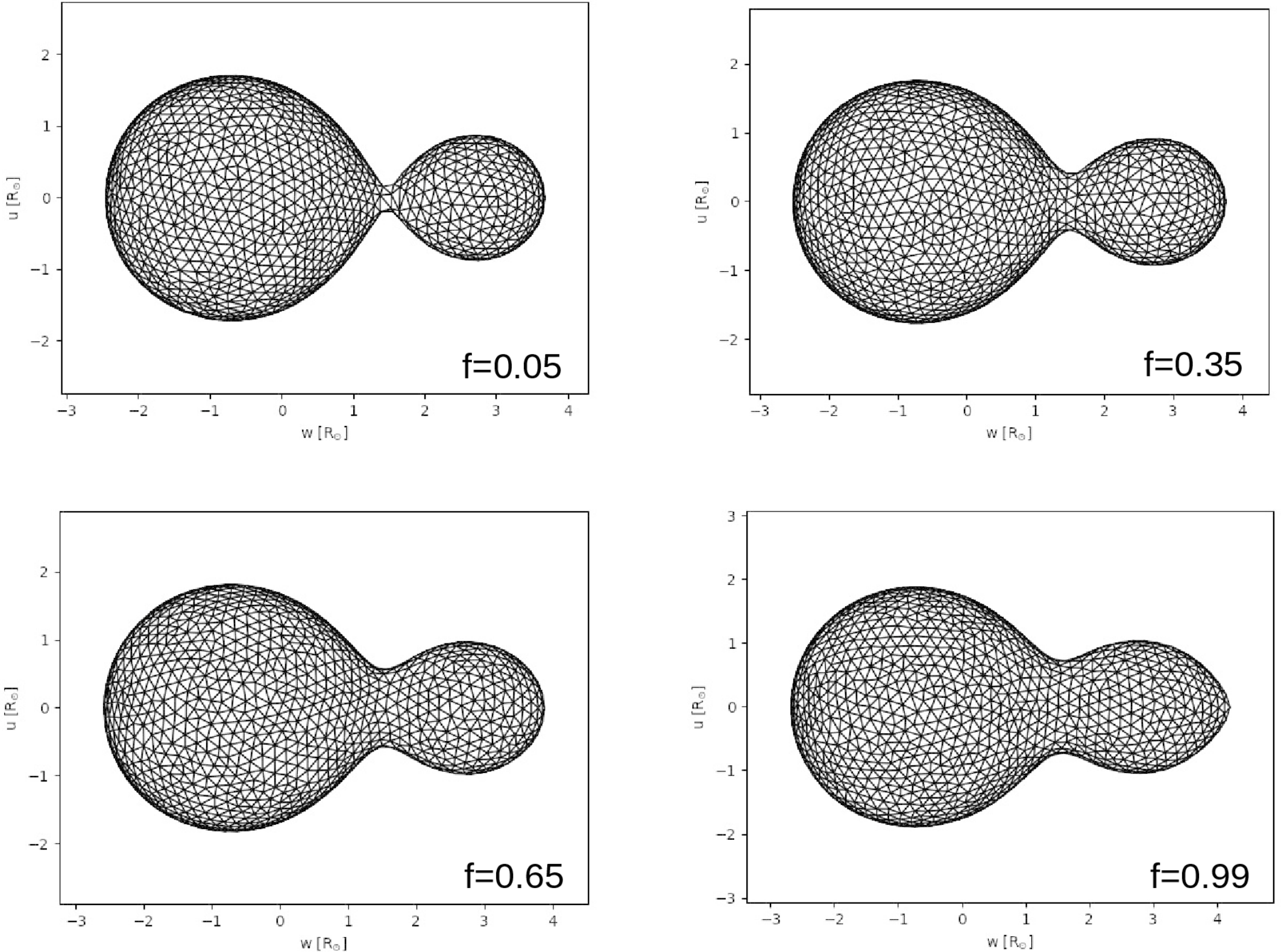}{5in}{}
\caption{Depiction of the Roche surfaces for four $q$=0.25 systems having fillout factors $f$=[0.05, 0.35, 0.65, 0.99]. \label{fig:fillouts}}
\end{figure}

High-quality photometry can provide excellent constraints on most system parameters in contact binaries and even in some detached binaries having large ellipsoidal light curve modulations.  Fundamental parameters include periods, period derivatives, orbital inclination $i$, mass ratio $q$, fillout factor $f$, and ratio of stellar temperatures \Tratio.  The fillout factor is a measure of the Roche lobe volume occupied and is defined in terms of the potential at the poles, $\Omega$, and the critical potentials at the first and second Lagrange points, e.g., \citet[][eqn. 3.67 and discussion therein]{Prsa2018}. 
\begin{equation} 
f=(\Omega - \Omega_{\rm L1})/(\Omega_{\rm L2}-\Omega_{\rm L1}). 
\end{equation}
\noindent Fillout factors $f$$<$0 correspond to detached systems while $f$$>$0--1  correspond to contact systems. Figure~\ref{fig:fillouts} presents a graphical depiction of the Roche surfaces for fillout factors f=[0.05, 0.35, 0.65, 0.99] generated with PHOEBE \citep{Prsa2016} for a $q$=0.25 system with a 1 \msun\ primary star.  It illustrates the transition from nearly detached systems at $f$=0.05 (upper left panel) to the highly distorted secondary (nearly overflowing at the $L_{\rm 2}$ point) in the $f$=0.99 case (lower right panel).  

\begin{figure}[ht!]
%\fig{fourbyfourmodels.pdf}{7in}{}
\fig{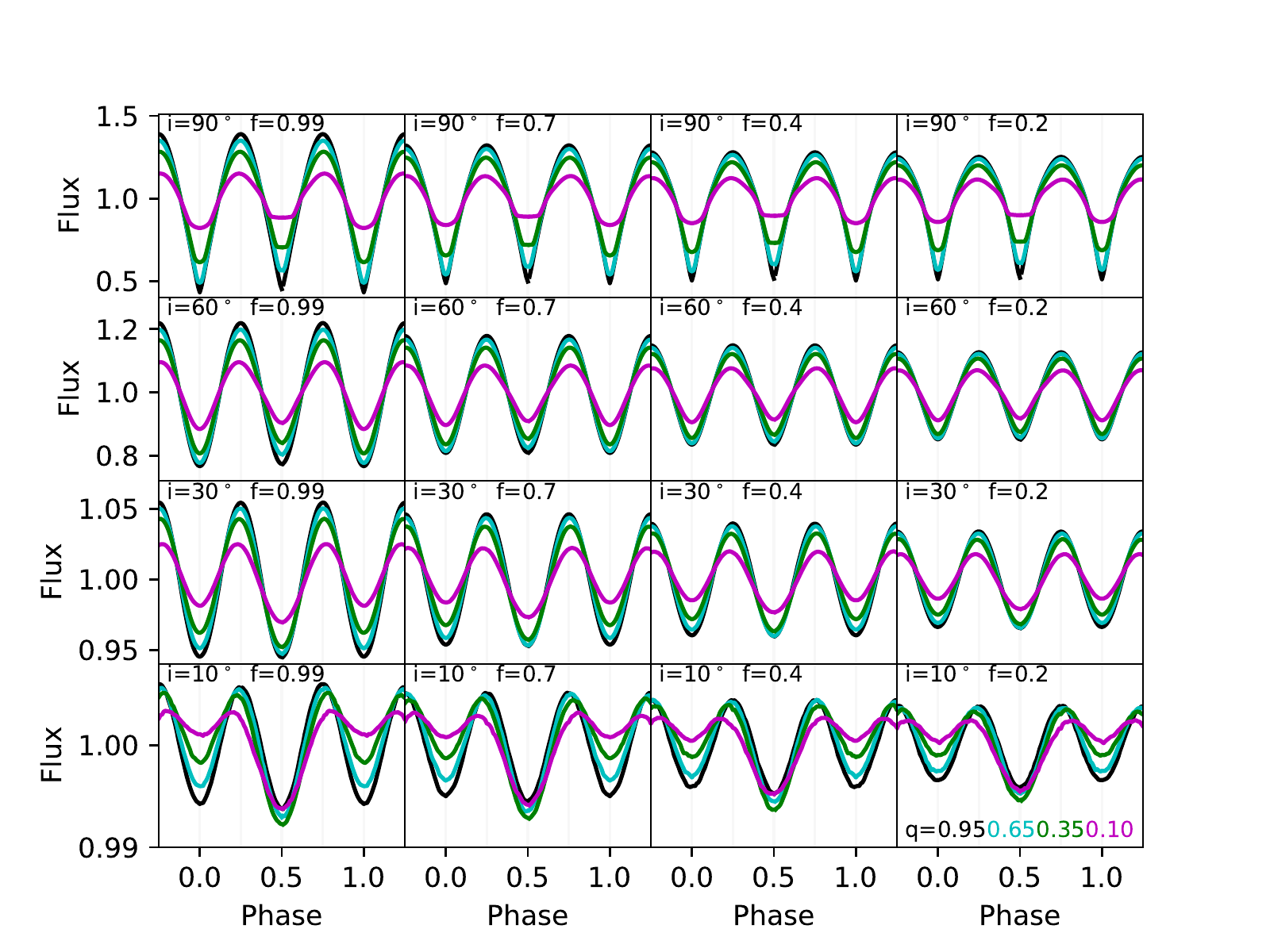}{7in}{}
\caption{Grid of model light curves for four selected values of $i$ (rows from top to bottom), $f$ (columns from left to right), and $q$ (colors) as indicated by labels within the panels.  In all cases the third-light fraction $l3$=0 and $T_{\rm 1}$=$T_{\rm 2}$.  \label{fig:fourbyfourmodels}}
\end{figure}

Figure~\ref{fig:fourbyfourmodels} shows a grid of model contact binary light curves produced using PHOEBE for inclination angles $i$=[90\degr, 60\degr, 30\degr, 10\degr] in rows top-to-bottom, fillout factors $f$=[0.99, 0.70, 0.40, 0.20] in columns left to right, and mass ratios $q$=[0.95, 0.70, 0.35, 0.10] coded by color as indicated in the legend.  Here we adopt equal-temperature stars, $T_{\rm 1}$=$T_{\rm 2}$, since contact binaries have atmospheres in thermal contact, but actual systems can exhibit slightly unequal component temperatures \citep{Hilditch1988, Yakut2005}, sometimes as a consequence of spots \citep{Barnes2004, Nelson2014}.  Phase $\phi$=0 is defined, for purposes of this plot, when the more massive star is at superior conjunction.  At large inclinations (top row, $i$=90\degr) the distinctive v-shaped or flat minima provide information on the inclination and mass ratio, while the overall amplitude of modulation, which decreases from left to right, provides constraints on $f$. Secondary minima are flatter and less deep as mass ratios becomes more extreme from $q$=0.95 toward $q$=0.10.  At intermediate inclinations (second row, $i$=60\degr) the light curves become quasi-sinusoidal and more uniform, with the overall amplitude of modulation decreasing with decreasing $q$ and $f$.  At this inclination the more massive component still produces a deeper eclipse at $\phi$=0.  At still lower inclinations (third row, $i$=30\degr) the secondary minima (phase $\phi$=0.5 when the less massive star is at superior conjunction) become deeper than the one at $\phi$=0, more dramatically so as $q$ becomes more extreme. The full  amplitude of modulation is now $\lesssim$10\%.  At the lowest inclinations (bottom row) the minima at $\phi$=0.5 become much deeper and wider than those at $\phi$=0.0, most dramatically so at extreme $q$.  The amplitude of modulation is also very small, on the order of 1\%. {\it One consequence of this reversal in timing of the deeper minimum with decreasing inclination is that the standard observational practice of locating the deeper minimum at $\phi$=0 results in mass ratios $M_{\rm 2}$/$M_{\rm 1}$$\equiv$$q$$<$1 being observed at $i$$\gtrsim$50\degr\ and $q$$>$1 for $i$$\lesssim$50\degr---i.e., an eclipse of the more massive star always produces the deeper minimum (for equal temperatures).}

\begin{figure}[ht!]
%\fig{showl3effect.pdf}{5in}{}
\fig{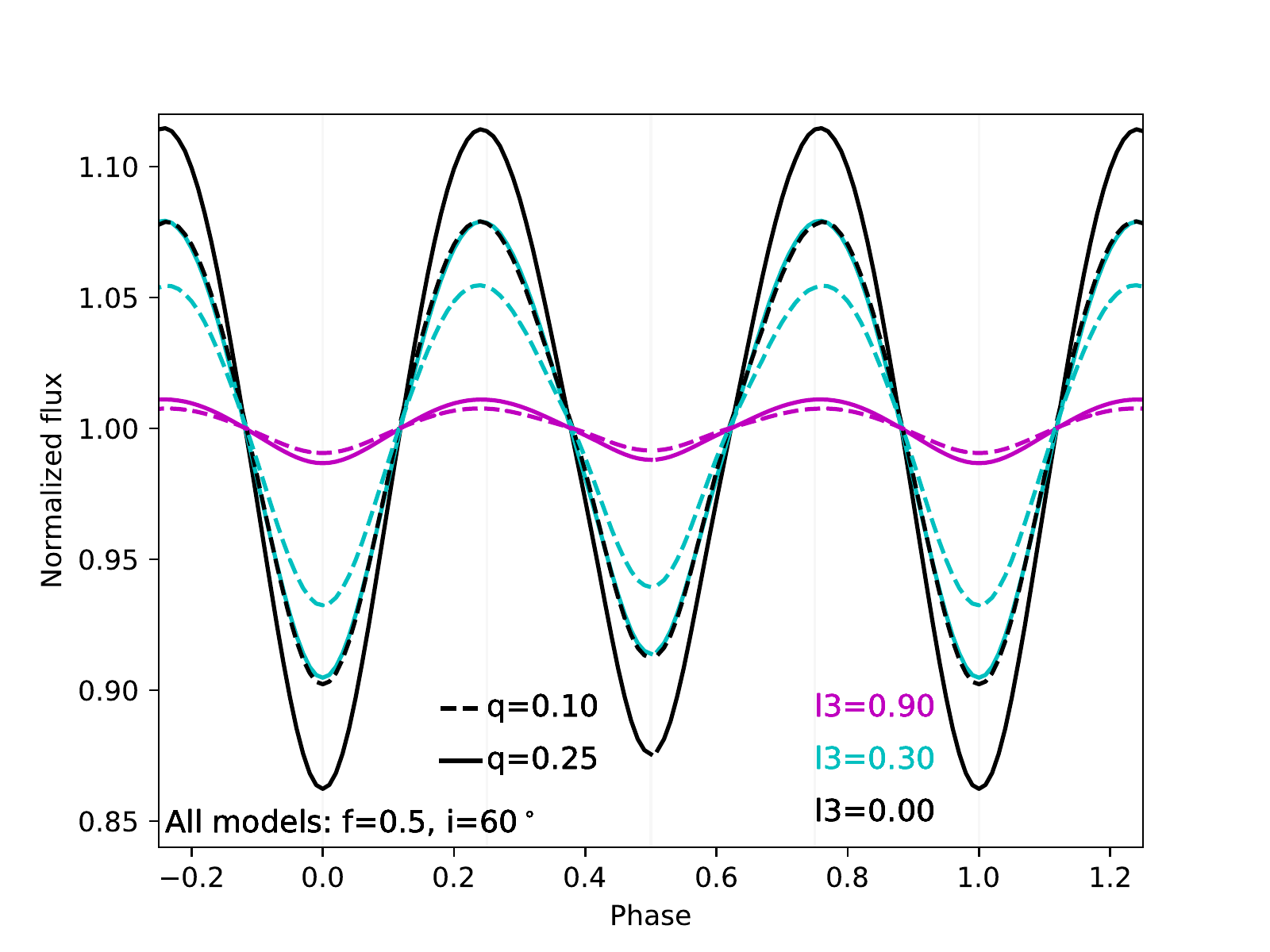}{5in}{}
\caption{PHOEBE model light curves for contact binaries with $f$=0.5, $i$=60\degr, two different mass ratios (indicated by line style) and three different third-light fractions (indicated by color).  Modest mass ratios with substantial third light closely resemble more extreme mass ratios with minimal third light.  \label{fig:l3}}
\end{figure}

A fourth fundamental parameter, $l3$, the fraction of light from an unresolved third stellar component (either physically related or projected along the line of sight; not illustrated in Figure~\ref{fig:fourbyfourmodels}) serves to dilute the light curve modulation at all phases and alter the ratios of minima to maxima in a way that is partially degenerate with other parameters,  including the temperature ratio, $T_{\rm 1}$/$T_{\rm 2}$.   Figure~\ref{fig:l3} shows model light curves for contact binaries having $i$=60\degr and $f$=0.5 for two mass ratios ($q$=[0.10, 0.25]) and three third-light fractions ($l3$=[0.0, 0.30, 0.90]), as coded by color and line style in the legend.  Third-light contributions dilute the amplitude of modulation such that modest mass ratios with substantial third light become almost indistinguishable from more extreme mass ratios having small third light.  For example, the $q$=0.10, $l3$=0.00 model (dashed black curve) is nearly (but not {\it exactly}) identical to the  $q$=0.25, $l3$=0.30 model (solid cyan curve).  Failing to diagnose third light correctly can lead to an erroneous mass ratio.  The degeneracy is less severe at high inclinations and large fillout factors where light curve shapes are less ambiguous.   With high-quality light curves, such as those afforded by {\it Kepler}, it is often possible to  model and recover several or all of the principal system parameters. However, Figure~\ref{fig:l3}  serves to illustrate danger in attempting to recover mass ratios solely from low-quality light curves when third-light contributions are unconstrained.  If the third component has a spectral energy distribution substantially different from the contact binary, multi-color photometry can help break this degeneracy. However, in this work, we focus on the constraints afforded by high-quality single-band light curves.   

Additional free parameters such, as the ratio of stellar temperatures, $T_{\rm 2}$/$T_{\rm 1}$ ($\simeq$1 for contact systems), and the possibility of inhomogeneities (e.g., hot or cool spots on the stellar surfaces) introduce additional signatures in the light curve that may be partially degenerate with other parameters but can nevertheless be modeled statistically, especially if multi-color light curves are available.  Although detailed modeling of high-quality light curves can constrain system parameters of close binaries in many cases, phase-resolved kinematic measurements from spectroscopic data are required to measure individual component masses, total system masses, and constrain the structure and location of asymmetries in the systems such as hot or cool spots \citep[e.g., techniques usually known as Doppler or Roche imaging; ][]{Vogt1983, Rutten1994}. Spectroscopic data also serve to validate the results obtained from light curve modeling or provide mass ratios for detached binaries which can then be modeled to recover remaining parameters on the basis of light curves.   

\subsection{Goals of this Investigation}

Our goals in this contribution are 1) to critically assess the prospects for identifying extreme-mass-ratio contact binary systems from high-quality photometry in conjunction with state-of-the-art stellar binary model light curves, and 2) to test the \citet{Molnar2019,Molnar2022} hypothesis that extreme-mass-ratio contact binaries are rare at periods exceeding $\approx$0.5 d and non-existent below a critical $q$ threshold demarcating the onset of rapid stellar mergers.  Section~\ref{sec:data} describes the acquisition and reduction of new optical spectroscopic data of ten (near-)contact  binaries obtained near quadrature phases.  The spectra directly provide the mass ratios and velocity amplitudes, informing light curve models that subsequently constrain the individual component masses and total system mass once the inclination is known.  Furthermore, the spectral line velocity profile of the system serves to validate the remaining system parameters retrieved from light curve analysis alone ($f$, $q$, $l3$).  Section~\ref{sec:analysis} describes our application of the spectroscopic Broadening Function \citep{Rucinski1992} analysis techniques to determine the velocity profile of each of the \numsample\ contact systems in the pilot spectroscopic study. Section~\ref{sec:analysis} also introduces our application of PHOEBE light curve models in conjunction with {\tt emcee} \citep{Foreman-Mackey2013} Markov-Chain Monte Carlo (MCMC) software to retrieve the Bayesian (posterior) probability distribution of system parameters.  Section~\ref{sec:sample} employs {\it Kepler} spacecraft light curves in tandem with our spectroscopic data to measure the full set of system parameters for \numsample\ (near-)contact systems.  We demonstrate the power of these combined datasets to vet state-of-the-art binary models while identifying extreme-mass-ratio systems possibly in the penultimate phase of evolution.  We also use PHOEBE in conjunction with an MCMC analysis to sample the posterior probability distributions of system parameters as constrained by the data and obtain rigorous uncertainties on each. Section~\ref{sec:MCMC} extends use of these PHOEBE+MCMC retrieval tools to the entire set of $>$800 close {\it Kepler} binaries to obtain best-fitting and probabilistic system parameters in a uniform manner for an unbiased sample of unprecedented size and photometric precision.  We investigate both contact and detached configurations for each system and identify the best-fitting configuration, where possible.  Section~\ref{sec:MCMC} also investigates interesting statistical correlations among system parameters, providing insight regarding the evolutionary paths of close binaries.  Section~\ref{sec:qp} provides a synopsis of the \cite{Molnar2019} and \citet{Molnar2022} evolutionary scenario for contact binaries and tests key predictions against the distribution of $q$ and $P$ derived from the {\it Kepler} sample.  Section~\ref{sec:conclusions} provides a summary of substantial successes of the joint PHOEBE+MCMC approach and reviews the insights gleaned from the pilot study and the large-scale analysis of {\it Kepler} binaries that inform evolutionary scenarios for contact systems.  

Our intention throughout is to be pedagogical regarding some aspects of contact binary light curve analysis where we feel the modern literature is lacking and to be prescriptive in ways that help pave a path for large-scale analyses of binary systems in the age of vast time-domain datasets. We adopt the classical observational definition for eclipsing binaries that the primary star (mass $M_{\rm 1}$, radius $R_{\rm 1}$, effective temperature $T_{\rm 1}$) is the component that is eclipsed at the time of superior conjunction ($t_0$), producing the deeper eclipse.  By this definition, the primary star is not necessarily the most massive or largest or hottest.  This may lead to reported mass ratios, $q$=$M_{\rm 2}$/$M_{\rm 1}$, either greater than unity (corresponding to the W-type class of contact binaries in which the less massive component appears to be hotter) or $q$ less than unity (corresponding to the A-type class of contact binaries in which the more massive component appears to be hotter).  In some cases the primary minimum (orbital phase $\phi$=0.0) and secondary minimum (orbital phase $\phi$=0.50) have essentially the same depth, so the distinction between primary and secondary becomes ambiguous and arbitrary. During the analyses we are often interested only in the {\it magnitude} of the mass ratio without regard to which component is more massive.  In such cases we invert mass ratios $>$1 to obtain what can be regarded as an absolute mass ratio, $q_{\rm a}$$\equiv$[0--1].

\section{Photometric and Spectroscopic Data \label{sec:data}}
\subsection{Target Selection}

\begin{figure}[ht!]
%\fig{PvsMorph+depth.pdf}{7in}{}
\fig{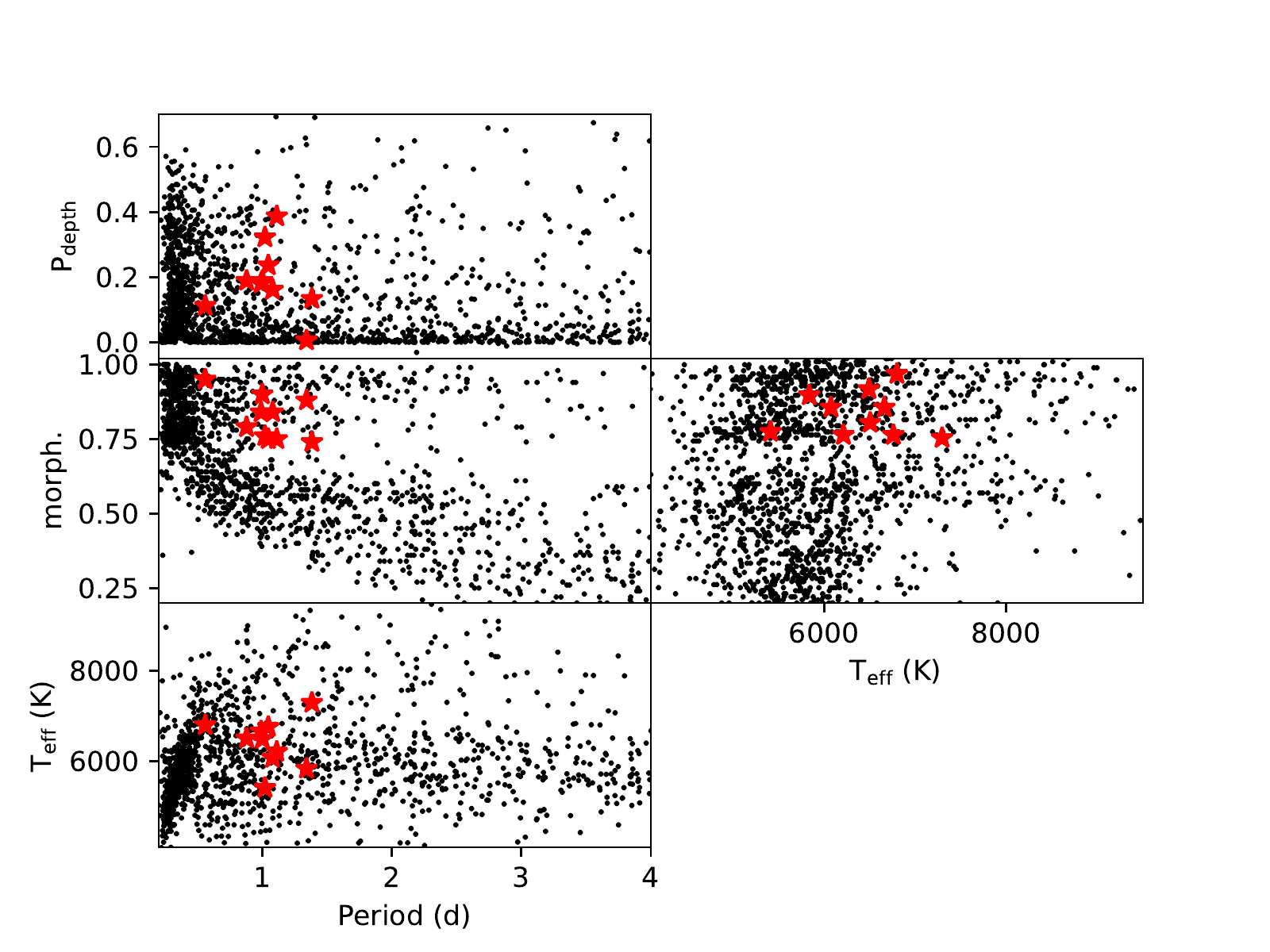}{7in}{}
\caption{Distribution of orbital period, effective temperature, light curve shape morphology parameter, and primary eclipse depth for binary stars in the {\it Kepler} field \citep[][black points]{Kirk2016} and our \numsample\ targets (red stars). \label{fig:selection}}
\end{figure}

A small sample of ten long-period ($P$=0.55--1.38 d) contact binary stars were selected from the compilation of 2878 {\it Kepler} binary stars \citep{Kirk2016} for spectroscopic observation.  The targets were selected for having light curves consistent with contact or near-contact systems and longer-than-average periods.  Figure~\ref{fig:selection} shows the distribution of period ($P$) in days, effective temperature ($T_{\rm eff}$), primary eclipse depth ($p_{\rm depth}$), and light curve morphology parameter ($morph$)\footnote{This morphology parameter represents an attempt to represent the menagerie of binary light curves using a higher order manifold  down-projected onto one dimension, introduced by \citet{Matijevic2012} and described further in \citet{Prsa2018}.}   for all \citet{Kirk2016} binaries with periods between 0.2 d and 4 d (black points) and our targets ({\it red star symbols}).   \citet{Kirk2016} note that morphology parameters between 0.5--0.7 correspond to semi-detached systems while 0.70 and higher belong to contact systems and ellipsoidal variables (i.e., tidally deformed detached systems). The dense locus of points in the lower left panel forming a linear trend of increasing temperature with period marks the population of close binaries with main-sequence components.  The correlation reflects the relation between temperature and radius for main-sequence stars. This trend becomes less pronounced above about 7000~K, reflecting the small number of high-mass ($M\gtrsim$2 \msun) hot stars observed by {\it Kepler}, a consequence of both the stellar initial mass function and possible selection biases imposed by the {\it Kepler Input Catalog} \citep[][KIC]{Brown2011, Batalha2010}.   These hotter stars preferentially have morphology parameters greater than about 0.75 (appropriate to contact and ellipsoidal variables) and primary eclipse depths reflecting the full range of values, as expected from a random distribution of inclination angles.  At periods longer than about 0.5 d the distribution of morphologies for {\it Kepler} binaries bifurcates into a lower branch reflecting detached morphologies and an upper branch indicating contact and ellipsoidal variables.  Another narrow locus of points in the upper left panel near $p_{\rm depth}\simeq0$ stretching from short  toward long periods reflects the population of statistically numerous grazing-eclipse systems.  Our target sample ({\it red star symbols}) has morphology parameters ranging from 0.74 to 0.95, relatively large primary eclipse depths, and periods $\approx$ 1 day, placing them on the long-period tail of the distribution.  

Table~\ref{tab:sample} lists the ten objects in the pilot study by Kepler Input Catalog identifier (column 1),  mean {\it Kepler} band magnitude (column 2), stellar effective temperature from the KIC (column 3), light curve morphology parameter (column 4), orbital period (column 5), and reference time ($t_{\rm 0}$) of superior conjunction from \citet{Kirk2016} with updates in this work  (column 6).

\subsection{Photometric Data}

We assembled calibrated {\it Kepler} photometry on the targets available from the the public MAST\footnote{Mukulski Archive for Space Telescopes; https://archive.stsci.edu/} archive as cleaned and detrended by \citet{Kirk2016}\footnote{Light curve data were obtained 2021 August from http://keplerebs.villanova.edu/data/.}.  Data were generally available from a majority of the {\it Kepler} operational quarters, from as few as seven to as many as 17, yielding tens of thousands of measurements in the broad {\it Kepler} bandpass spanning four years, from 2009 May through 2013 May.  We determined mean periods for each system using a Lomb-Scargle periodigram analysis\footnote{As implemented in Python 2.7, {\tt astropy.stats}} which were found to be in good agreement with those tabulated by \citet{Kirk2016}.  These periods were then used to fold the light curve and calculate a time of superior conjunction, $t_{\rm 0}$---the reference time centered on the deeper minimum in the light curve.  In the manner of \citet{Molnar2017} we fit the full light curve using a sum of three--six Fourier components to define an analytic function characterizing the mean light curve.  This function was then used to fit subsets of the light curve, one {\it Kepler} quarter at a time, to measure time-dependent phase shifts that indicate variations in the times of minima or maxima.  Such variations, termed ``eclipse timing variations'', may indicate a changing orbital period or a light travel time delay that results from an orbit about a third body.   In a few cases where suitable data were available and helpful, we added recent 2018--2019 photometric measurements from the Zwicky Transient Factory \citep[ZTF,][]{ZTF2014} as a way of extending the time baseline.          

\subsection{Spectroscopic Data}

We obtained optical spectra on the targets near each of the two quadrature orbital phases ($\phi$=[0.25,0.75]) with the longslit spectrographs at the Wyoming Infrared Observatory (WIRO) 2.3 meter telescope and/or the Apache Point Observatory (APO) 3.5 meter telescope.  At APO we used the Double Imaging Spectrograph (DIS) red arm with a 1200 line mm$^{-1}$ grating in first order to acquire spectra over the range 5800--6900 \AA.  The spectral resolution was $\approx$3000 using slit widths of 0\farcs{9} or 1\farcs{2} at a reciprocal dispersion of 0.62 \AA\ pix$^{-1}$.  The wavelength calibration was performed with the aid of HeNeAr lamp spectra obtained close in time to the target exposures and has an RMS of 0.06 \AA, or about 3 \kms\ at 6500 \AA. Spectra acquired at WIRO used the Longslit spectrograph with a 1\farcs{2} slit and a 2000 line mm$^{-1}$ grating in first order to cover 5400--6700 \AA\ at a reciprocal dispersion of 0.61 \AA\ pix$^{-1}$ and resolution R$\simeq$4000.  The CuAr lamp exposures acquired after each science exposure provided wavelength calibration to an RMS of 0.019 \AA\ ($\sim$1 \kms).  At both observatories, target exposures times varied from two$\times$600 s to four$\times$600, depending on seeing and source brightness, yielding spectra with continuum signal-to-noise ratios (SNRs) of 40:1--100:1 near 6500 \AA\ in the final combined spectra.  Averaging spectra over $\leq$40 minutes introduces a small amount of phase smearing which is small compared to the rotational broadening of the tidally-locked short-period stellar components.  Spectra were reduced using standard techniques in IRAF \citep{Tody1986}, including 1D spectral extraction and local sky background subtraction, flat fielding with quartz continuum lamps, and transformation to the Heliocentric velocity frame of reference.  Observations of radial velocity standard stars confirm that the velocity calibration is precise to $\pm$6 \kms\ between observatories and epochs, with deviations primarily attributed to variable placement of a target within the slit---an inevitable limitation of slit spectrographs.      

\section{Analysis Techniques \label{sec:analysis}}
\subsection{Broadening Function Analysis}

We analyzed the optical spectra using a custom {\tt python} version of the broadening function algorithm (BF) described by \citet{Rucinski1992, Rucinski2002} to recover the velocity profile of each contact binary at each quadrature phase.  The BF code performs a true linear deconvolution of a broadened stellar spectrum given a narrow-lined spectral template of the appropriate effective temperature.  It is superior to cross-correlation methods when used to separate the blended components of close binary systems having similar temperature \citep{Rucinski1999}.  The resulting BF for a contact binary is the light-weighted velocity profile of the combined system at the time of observation, as broadened by the instrumental profile (which is $\simeq$75 \kms\ FWHM, smaller than the $\approx$200 \kms\ rotational profiles) and any temporal broadening from finite exposure durations.  The BF can also reveal the signatures of third components in the spectra of rotationally broadened binary systems \citep[e.g.,][]{Dangelo2006}.  

Our BF analysis used as a template a high-resolution high-SNR spectrum of the appropriate $T_{\rm eff}$ (we adopt solar metallicity and $\log~g$=3 models for inflated stars; the exact choice in inconsequential for our purposes) from the PHOENIX model atmospheres  \citep{Husser2013}.  However, gross mismatches in $T_{\rm eff}$ between the spectra and the template of more than about 1000~K produce negative ``bowls'' on both sides of the BF peak.  Good matches between the data and the template produce the largest BF amplitudes, serving as a check on the suitability of the effective temperature listed in the {\it Kepler Input Catalog}.  We compared BFs resulting from the full spectral coverage (usually $\approx$5450--6650 \AA) and from a spectral subregion that excludes H$\alpha$, the strongest single spectral feature.  We found that BFs are consistent with each other regardless of spectral regime, but they have larger SNRs when H$\alpha$ is included.  In principle, low levels of H$\alpha$ emission in the core of the line could lead to a skewed BF because of the large weight of this feature in the spectrum, but we found no evidence for such effects.

\subsection{PHOEBE Bayesian Modeling}

For the ten systems studied spectroscopically (Section \ref{sec:sample}) and the entire ensemble of nearly 800  candidate contact binaries from the compilation of \citet{Kirk2016} (Section~\ref{sec:MCMC}) we modeled the light curves and velocity curves using the binary modeling code PHOEBE 2.2\footnote{As this work was being completed PHOEBE 2.3 \citep{Conroy2020} which includes support for MCMC analysis was released in late 2020. Our analysis essentially followed the methods described in \citet{Conroy2020} in regards to the merit function and MCMC techniques which we implemented separately.  Tests on a small subset of systems revealed no differences between the light curves produced by the two PHOEBE releases.} \citep{Prsa2016} in conjunction with the Markov-Chain Monte Carlo code {\tt emcee} \citep{Foreman-Mackey2013} to explore the posterior probability distributions of system parameters.  We adopted the period ($P$) and time of superior conjunction ($t_{\rm 0}$) from \citet[][]{Kirk2016}, except in a few cases where we recomputed slightly different values using a subset of the {\it Kepler} data. We adopted $T_{\rm 1}$  from the {\it Kepler input Catalog} \citep{Brown2011}, using 6200~K if no value was listed.  The models are insensitive to the adopted $T_{\rm 1}$ because the limb darkening coefficients vary only modestly over the 4500~K$<T_{\rm eff}<$7500~K range of contact binaries of concern here \citep[e.g., limb darkening tables of][]{VanHamme1993, Claret2004}.  Given the short periods of these systems, we set zero orbital eccentricity ($e$=0) as a fixed parameter.  We used the mean {\it Kepler} passband in the model light curve and the default \citet{Castelli2003} stellar atmosphere models with limb darkening coefficients interpolated from these models as implemented in PHOEBE 2.2.    All systems were modeled  using four different computational approaches.

\begin{itemize}
\item{Fixed-temperature-ratio ($T_{\rm 1}$=$T_{\rm 2}$) model---a contact binary geometry with equal temperature components (\Tratio=1) and four free parameters: inclination $i$, fillout factor $f$, mass ratio $q$, and third-light fraction $l3$.  We ultimately concluded that this model is too restrictive, leading to incorrect solutions as the fitting process is forced to alter other model parameters to compensate for lack of flexibility in \Tratio.}
\item{Variable-temperature-ratio model---a contact binary geometry with five free parameters: inclination $i$, fillout factor $f$, mass ratio $q$, third-light fraction $l3$, and temperature ratio  0.7$<$\Tratio$<$1.4.  We concluded that this model is  too flexible, permitting demonstrably wrong solutions as a consequence of degeneracy between model parameters---\Tratio\ and $q$, in particular.  }
\item{\Tapprox\ model---a contact binary geometry with nearly equal temperature components (0.95$<$\Tratio$<$1.05) and five free parameters: inclination $i$, fillout factor $f$, mass ratio $q$, third-light fraction $l3$, and \Tratio.  We ultimately adopted this model as the best general approach to solving the inverse problem and retrieving system parameters for (near-)contact binaries.  }
\item{Detached model---a detached geometry with six free parameters: inclination $i$, mass ratio $q$, third-light fraction $l3$, temperature ratio 0.7$<$\Tratio$<$1.4, primary star radius $R_{\rm 1}$, and ratio of component effective radii \Rratio.  We show that this model correctly identifies detached systems in cases where the models provide a superior fit to the light curve over the contact models.  }
\end{itemize}

We assigned broad flat priors: 0.0$<$\cosi$<$1.0, 0.03\footnote{This fillout factor lower limit of 0.03 was adopted primarily to avoid numerical difficulties that occur when the Roche Lobes are only tenuously in contact at low $f$.}$<$$f$$<$0.99, $-$1.4$<\log(q)<$1.4, 0$<$$l3$$<$0.99, $T_{\rm min}$$<$\Tratio$<$$T_{\rm max}$, 0.1 \rsun$<$$R_{\rm 1}$$<$4 \rsun, and 0.15$<$\Rratio$<$5  (i.e., no $a priori$ preference for parameters within the given broad physically plausible limits).  Using logarithmic intervals in mass ratio is necessary to capture the large dynamic range in that parameter.   

The distinction between a contact and a detached system is, in actuality, an artificial one necessitated by modeling limitations.  That is, a contact system where only a small fraction of the Roche lobes are filled will look very much like a detached system where one or both components fill their Roche lobes nearly to overflowing.  A contact system with an extreme mass ratio and a small fillout factor will look very much like a ``semi-detached'' system where only one component (nearly) fills its Roche lobe.  Nevertheless, the modeling approaches described here will serve to constrain the probable system parameters across this physical continuum.  

We assessed the goodness of fit for each forward PHOEBE model using the $\chi^2$ statistic computed using the phased model and observed light curves.   The phased light curve was divided into 100 phase bins, so that one bin represents $\approx$2 minutes for the shortest period (0.16 d) systems and 14 minutes for the $\approx$1 d systems, thereby oversampling the 30-minute {\it Kepler} download cadence.   The mean of all data points in a bin defines the average measurement.  

To quantify the uncertainty on each bin in the phased light curve we considered several approaches.  Adopting the RMS deviation of the data points in each bin often led to very small \redchi\ values $\ll$1 because the dispersion in the data is dominated by {\it real variations} in the system (e.g., spots/pulsations/flares) that are much larger than the {\it Kepler} photometric uncertainty of $\approx$0.01\%.  By this measure, best-fitting models are often very good---{\it too good}.  One consequence of this choice is that the posterior range of acceptable model parameters is overly large and that the envelope of Bayesian posteriors contains models that are demonstrably inconsistent with the data.  Nevertheless, this served as a useful indicator of the goodness of fit for a first round of Monte Carlo iterations that identified the global locus of best-fitting parameters.  We also experimented with using an error-of-the-mean (RMS/$\sqrt{N_{\rm data}}$) as the uncertainty on each phase bin.  This approach---ultimately abandoned---yielded \redchi$\gg$1 in all cases because the nominal models do not include the real physical features (spots/pulsations/flares) that are apparently present in essentially {\it all} systems  at levels greatly exceeding the measurement precision.  As a compromise that lies between these two extremes, we chose an approach that marginalizes over the (unknown) additional physical phenomena shaping each light curve by performing a second round of Monte Carlo iterations. Here, we multiplied the standard deviation of measurements within each phase bin $\sigma$ by a scale factor\footnote{This scale factor is the square root of the \redchi\ from the best-fitting model found in the initial MCMC runs.} to obtain an effective uncertainty $\sigma_{\rm e}$ that forced the \redchi\ of the best-fitting model to lie near unity.  This scaling allows the relative probabilities of competing models to be adjudicated\footnote{We use Python's {\tt scipy.stats.chi2.logpdf(chisq,dof)}.} and provides appropriate statistical error estimates on model parameters but does not, by itself, tell whether the model provides a good fit to the light curve. Good and poor models are decided later on the basis of the RMS of the best-fitting model. 

We employed two rounds of MCMC analysis on each system, the first to localize the global minimum in parameter space and the second to rigorously define its shape and extent.  In the first round we distributed 40 walkers randomly across the allowed parameter space and used a combination of  walker movement algorithms {\tt DEMove} (80\%) and {\tt DESnookerMove} (20\%) as implemented in {\tt emcee 3.0} to rapidly explore parameter space.\footnote{Such a mixture of move strategies is suggested within the {\tt emcee} documentation for a complex multi-modal parameter space. A rigorous exploration of optimal Monte Carlo strategy is beyond the scope of this paper.}  In PHOEBE we used 8000 triangles to comprise the mesh surface of the stars and {\tt ``irradiation method''=None}, at least initially, to disable reflection, absorption, and re-radiation effects and achieve shorter computation times.  Experiments with differing numbers of triangles showed that as few as 3000 are often adequate for typical geometries, but 10,000 (or more!) were required to obtain consistent results for systems having exquisite photometric precision and/or systems where one or more parameter is extreme.\footnote{We did not explore meshing optimization. We set a conservative $N_{\rm triangles}$=8,000 triangles in the mesh that generally achieved convergence.}  For the initial round of models we used 1500 steps per MC walker.  For the log of the probability for each model we use the negative of the chi-squared value to force walkers toward the global minimum.   Walkers often spent a disproportionate number of steps exploring local minima before finding the global minimum, and often some walkers remained trapped in a local minimum.  Nevertheless, the global minimum was always singular and unimodal (i.e., no local minima comparable to the global minimum), although sometimes the global minimum was quite broad in one or more parameters.  Subsequently, we conducted a second round of PHOEBE simulations using 10,000 triangles and MCMC to perform 2,000 steps with 10 or 12 walkers (twice the number of free parameters). Trials using 6,000 MC steps changed results negligibly.  Each model included the more computationally expensive option for a detailed treatment of irradiated and reflected light as described in \citet{Horvat2019}.  The probability of each model in this round was computed using {\tt scipy.stats.chi2.logpdf(chisq,dof)}, the standard probability of $\chi^2$ for a given number of degrees of freedom $\nu$.   We discarded the first 300 steps of each walker as ``burn-in'' iterations.  The initial walker positions were clustered in a small Gaussian blob near the best-fitting parameters from the first round of models.  This second round served to define the shape of the global minimum and compute Bayesian 16th, 50th, and 84th percentile values for each free parameter, which we tabulate as a 1$\sigma$ uncertainty.\footnote{Our approach using two rounds of MCMC may not be the most efficient, given the availability of other optimizing algorithms in more recent PHOEBE releases.}  We caution that the posterior distributions of parameters are often non-Gaussian and asymmetric, as illustrated with specific examples in the ensuing sections.

\section{Pilot Sample of Ten (near-)Contact Binaries \label{sec:sample} }
For the ten systems having spectroscopic data obtained at quadrature phases, we present a detailed comparison between the broadening functions, mass ratios, and component velocities obtained from the spectra to the parameters determined from best-fitting PHOEBE models to illustrate the reliability of light-curve-based solutions for (near-)contact systems. 

\subsection{KIC04853067}

\begin{figure}[ht!]
%\fig{Spectra/KIC04853067/KIC04853067_phased_all_smaller.pdf}{4in}{}
\fig{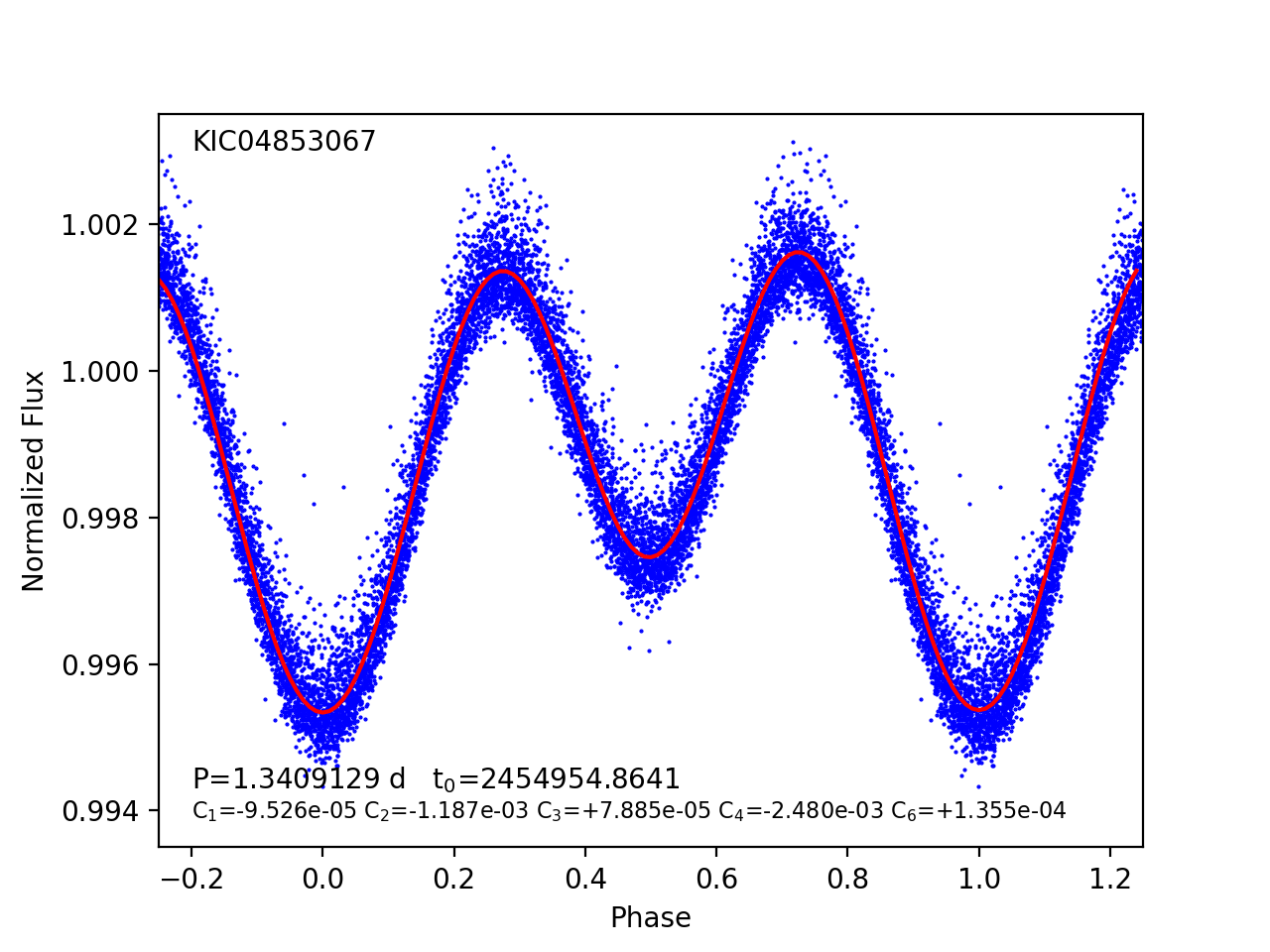}{5in}{}
\caption{Folded {\it Kepler} light curve of KIC04853067 and mean light curve shape (red curve) generated from the sum of low-order Fourier coefficients, as labeled.  Only a fraction of the {\it Kepler} data are plotted, here and subsequently, to reduce figure file size and improve clarity.  \label{fig:067LC}}
\end{figure}

KIC04853067 is the longest period system in our spectroscopic sample at $P$=1.34 d. The light curve exhibits the classical shape of a (near-)contact binary system and has slightly different depths at the two minima.  The modulation semi-amplitude is quite low, 0.2\%, suggesting a low angle of inclination and/or a large third-light contribution.  Figure~\ref{fig:067LC} displays the mean folded light curve over all {\it Kepler} quarters spanning 1460 days (blue dots) fit with a Fourier series (red curve) consisting of five\footnote{ We use only the minimum number of Fourier components needed for a functional  approximation to the light curve so the $O-C$ eclipse timing residuals, computed subsequently, converge and are well-characterized.  Although additional components achieve a better fit, there is no physical or utilitarian basis for higher orders.} components plus a zero point offset, $C_0$,
\begin{equation}
I(\phi) = C_0 +  C_1 \cos \left[ 1\pi (t-t_0)/P \right] + C_2 \cos \left[ 2\pi (t-t_0)/P \right] + C_3 \cos \left[ 3\pi (t-t_0)/P \right] +  C_4 \cos \left[ 4\pi (t-t_0)/P \right] + ...
\end{equation}
\noindent A doubly periodic contact binary light curve with equally deep minima will be dominated by the $C_4$ term, while an increasing contribution from the $C_2$ term is required to produce a secondary minimum less deep than the primary minimum.  The odd terms ($C_1$, $C_3$, ...) will be negligible unless there is an appreciable asymmetry in the light curve, such as when one maximum is brighter than the other.  Labels in Figure~\ref{fig:067LC} list $P$, $t_0$, and coefficients of the Fourier components used to fit to the mean light curve.  In the case of KIC04853067, $C_2/C_4$=0.47, indicating a secondary minimum substantially less deep than the primary minimum.  The model light curve sequences plotted in Figure~\ref{fig:fourbyfourmodels} show that this ratio of secondary to primary minimum is best reproduced by a system with a low $i$ and a mass ratio significantly different than 1.0.     

\begin{figure}[ht!]
%\fig{Spectra/KIC04853067/KIC04853067_O-C.pdf}{4in}{}
\fig{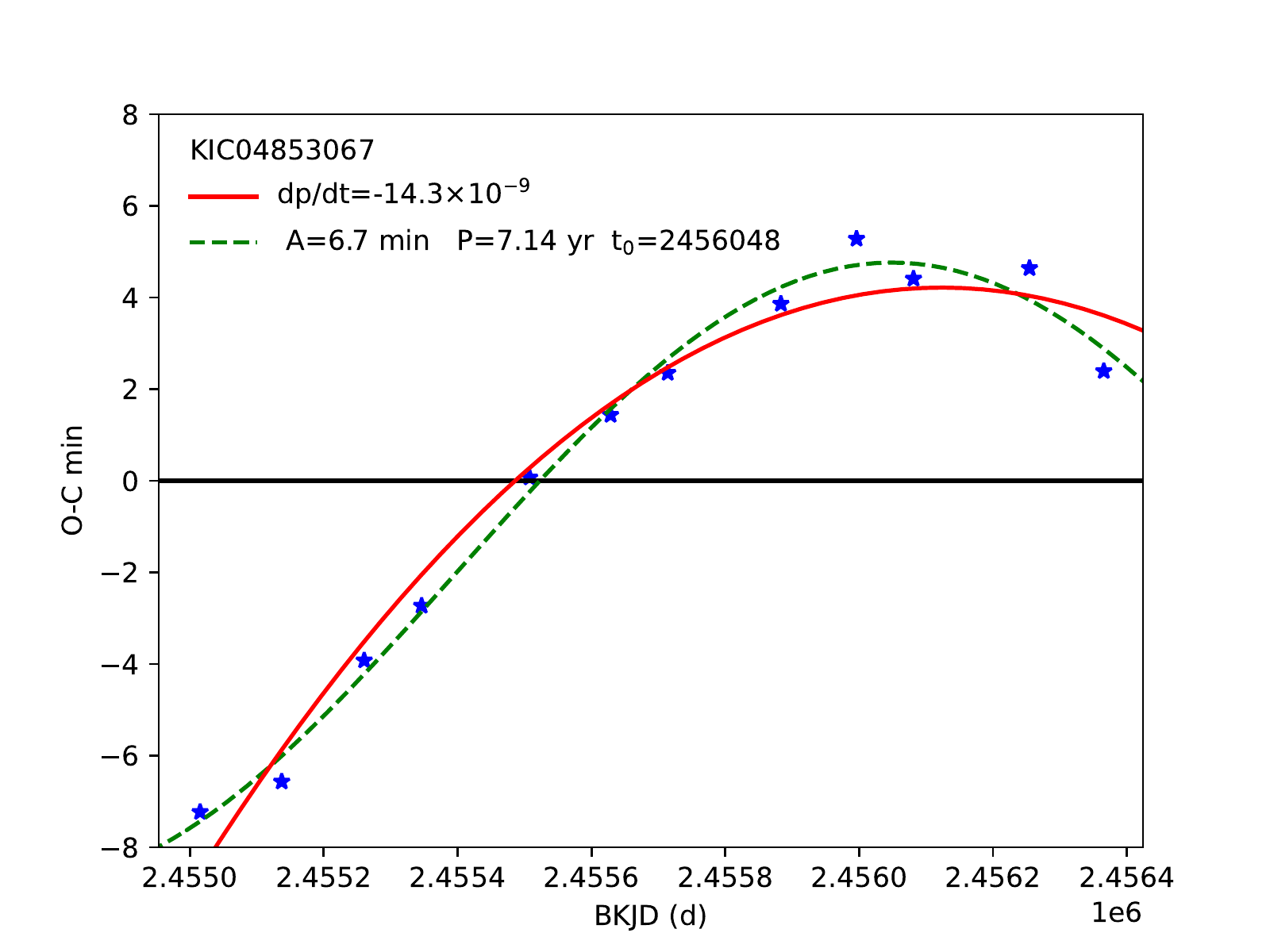}{5in}{}
\caption{\OC\ eclipse timing residuals of KIC04853067. The red curve is the best parabolic (constant period derivative) fit to the data, while the green dashed curve is the best sine function (periodic) fit.  \label{fig:067OC}}
\end{figure}

There is evidence for deviation from the mean period.  Figure~\ref{fig:067OC} shows the observed minus computed ($O-C$) time of primary eclipse versus Barycentric {\it Kepler} Julian Date (BKJD), calculated by fitting the mean light curve shape to the data divided into twelve equal time intervals. A positive \OC\ corresponds to a later-than-expected eclipse.  The red curve shows a parabolic function fit to the \OC\ data, consistent with a shortening of the orbital period and period derivative $dP/dt$=$-$14.3$\times10^{-9}$.   This signature may be caused either by true changes in the orbital period or the presence of a third body that introduces light travel time effects mimicking a period change.  A sine function fit to the \OC\ data (green dashed curve) provides a superior fit and yields a semi-amplitude of $A$=6.7 minutes, $P$=7.14 yr, and $t_0$=2456048 (the time when the eclipsing binary is at superior conjunction relative to a third body).  The semi-amplitude measures the light travel time delay of the contact binary as it orbits the barycenter of the (probable triple) system.  This light travel time leads to a {\it projected} semi-major axis of $a$ sin $i$=0.85 AU (assuming an $e$=0 orbit).  For an estimated contact binary total mass of 1 \msun, the period and projected semi-major axis imply a minimum tertiary mass of 0.27 \msun.  The analysis of the light curve that follows suggests a substantial third-light contribution in this system.  Given the probability of a third body creating the observed \OC\ variations rather than a  decreasing orbital period, we adopt a linear ephemeris.  

\begin{figure}[ht!]
%\fig{Spectra/KIC04853067/KIC04853067_twocomp.pdf}{4in}{}
\fig{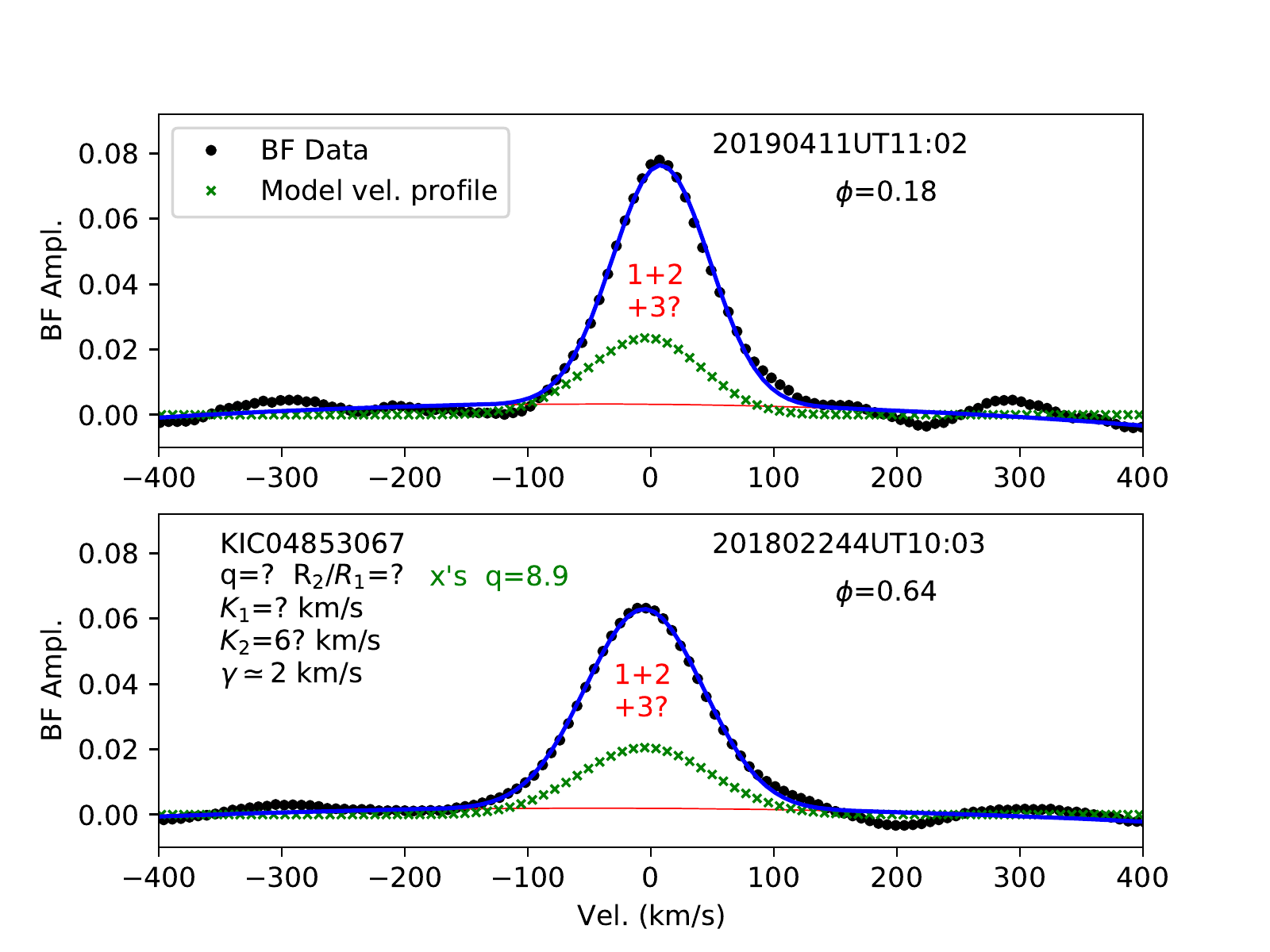}{5in}{}
\caption{Broadening function of KIC04853067 (black dots) showing a single peak with a small velocity difference between the two epochs. Red curves depict two Gaussian components fit to the data, while the blue curve is the sum of those components.  Green x's show the theoretical line profile generated by PHOEBE for the best-fitting contact binary system parameters as convolved with the instrumental spectral profile. The normalization of both curves is arbitrary.  The stellar components are viewed at low $i$ and are unresolved.  The BF may be dominated by a third component that is not represented in the model line profile function.  This system may also be a detached configuration.   \label{fig:067BF}}
\end{figure}

The BF of this $T_{\rm eff}$=5800~K \citep{Brown2011} $P$=1.34 d system in Figure~\ref{fig:067BF} (black dots) shows a single strong peak at both epochs with $V_\odot$=8 \kms\ at $\phi$=0.18 and $V_\odot$=$-$5 \kms\ at $\phi$=0.64.\footnote{A third spectrum obtained 20200420UT10:16 at $\phi$=0.81 shows a very similar broadening function.}  The larger slit width used in the $\phi$=0.64 epoch results in a slightly broader instrumental profile than the 0\farcs{9} slit used for the $\phi$=0.18 epoch observation. Red curves in the Figure depict two Gaussian functions fit to the BF---in this case one is dominant and one is negligible.  The blue curve is the sum of the two Gaussian components, which provides an excellent match to the BF.  Owing to the blended profiles, we cannot use the BF to measure component velocities or a mass ratio.  The estimated systemic velocity is $\gamma$=2 \kms.   The 13 \kms\ radial velocity difference between the two epochs is attributable partly to uncertainties in the velocity calibration from epoch to epoch ($\sim$3 \kms), placement of the target within the spectrograph slit  ($\sim$4 \kms) and the possibility of real radial velocity variations owing to an orbit about a third body ($\sim$few \kms\ for plausible ranges of third body masses).  

\begin{figure}[ht!]
%\fig{IndivSpectra/KIC04853067/KIC04853067LC+VCnew.pdf}{4in}{}
\fig{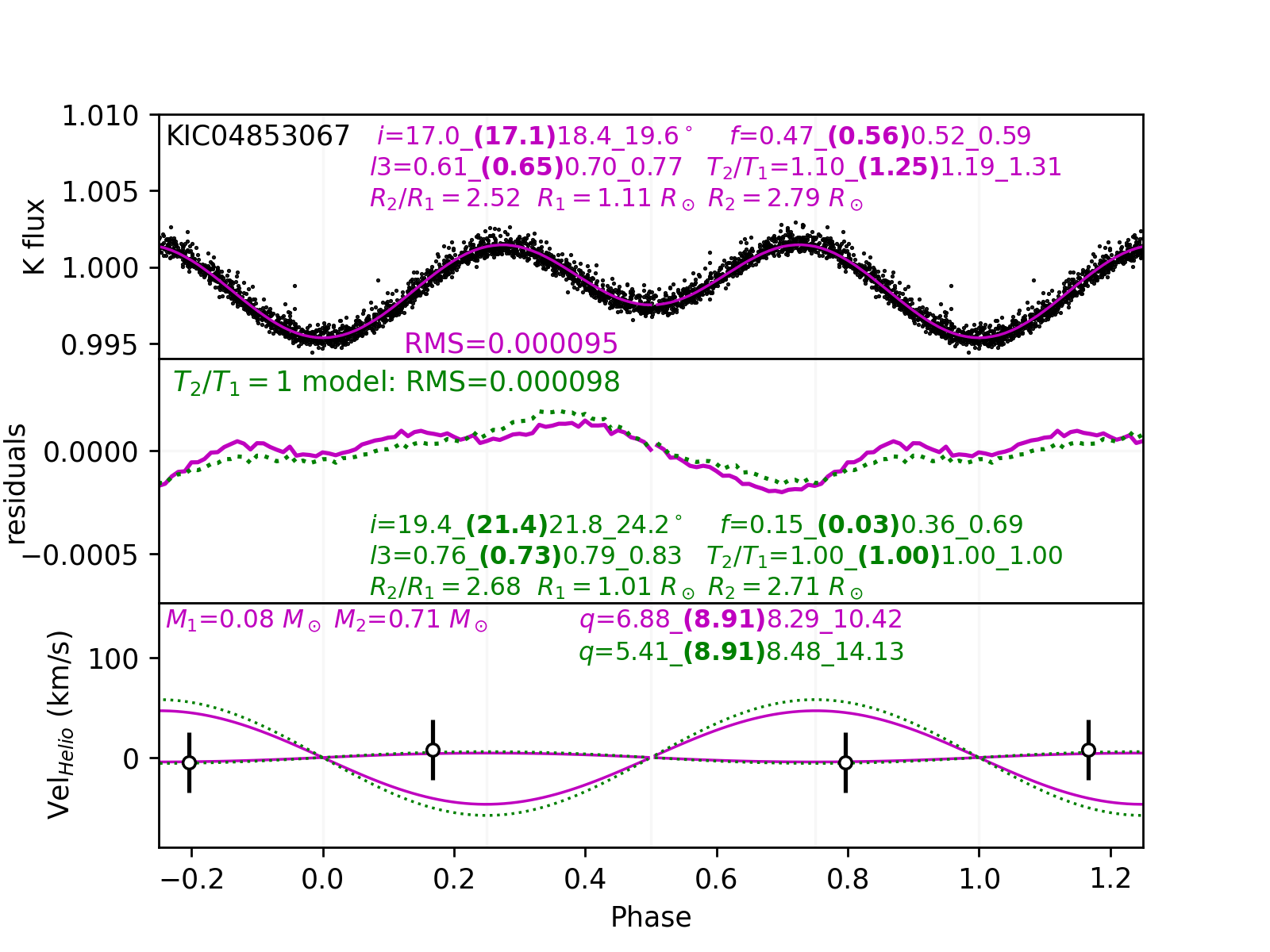}{5in}{}
\caption{Light curve and velocity curve of KIC04853067, along with best-fitting contact binary models (magenta: variable-temperature-ratio; green: fixed-temperature-ratio). \label{fig:067LCVC}  Individual points represent {\it Kepler} photometric data and residuals (upper and middle panels) or spectroscopic measurements (lower panel).  Labels denote 16th\_{\bf (best)}50th\_84th percentile values for the principal model parameters.   The components' line profiles are likely blended and contaminated by third light so that individual velocity amplitudes cannot be measured for this low-inclination extreme-mass-ratio binary.  }
\end{figure}

The very shallow depth of modulation in the light curve is best modeled using a  small inclination angle and a substantial third-light contribution.   Variable-temperature-ratio models  provide a slightly better match to the data than the fixed-temperature-ratio models. Figure~\ref{fig:067LCVC} shows the best-fitting contact configuration light curves (upper panel), residuals (middle panel), and velocity curves (lower panel).  The best-fitting variable-temperature-ratio (magenta curve) contact model\footnote{As a comparison, we ran the same series of MCMC simulations using the \citet{Horvat2019} and \citet{Wilson1990} approach including of irradiation and re-radiation effects, and we found that they both yielded a similar \redchi\ and range of parameters.} produces a very small RMS=0.000095 for parameters $i$=17.1\degr, $f$=0.56, $q$=8.9 ($q_{\rm a}$=1/$q$=0.11),  $l3$=0.65, and \Tratio=1.25.  The fixed-temperature-ratio models yield a nearly identical  RMS and $i$=21.4\degr, $f$=0.03, $q$=8.9,  $l3$=0.73.   The best-fitting detached models produce a similar RMS and suggest that both of the components are close to filling their Roche lobes: \Ronemax$\geq$0.9 and \Rtwomax$\approx$0.9 in all of the best-fitting parameter sets.  On account of the extreme mass ratio and the small dispersion in the data, the exact RMS of the best-fitting model is sensitive to the number of triangles used in the model stellar surface mesh.  We found that as many as $\geq$50,000 triangles are required to produce consistent results on this system.

Without knowing the mass of either component, neither the individual masses or radii can be known.  If we were to adopt, for illustration, $M_{\rm 1}$=0.08 \msun, then $M_{\rm 2}$=0.71 \msun,  $R_{\rm 1}$=1.1 \rsun, $R_{\rm 2}$=2.8 \rsun, and radii ratio $R_{\rm 2}$/$R_{\rm 1}$=2.8.  However, a range of parameters are possible, as indicated by Monte Carlo simulations to follow.  The less massive but hotter component produces the deeper minimum at $\phi$=0, a feature consistent with the low implied inclination and extreme mass ratio with $M_{\rm 2}>M_{\rm 1}$.  The model velocity semi-amplitude of the more massive and larger contact binary component is 4.2 \kms, consistent with the single peak at nearly constant velocity in the BF.  The extreme mass ratio and inclination implies that the less massive component is faint and would be hard to detect (indeed, it is not detected) in the spectral profile despite its large velocity amplitude, especially if the third-light contribution to the BF is substantial.\footnote{At present PHOEBE cannot include a third component in the line profile function.}
 
The green x's in Figure~\ref{fig:067BF} show the theoretical line profile generated by the best-fitting PHOEBE model at the observed orbital phases after convolution with the instrumental spectral profile.\footnote{We approximate the instrumental profiles by a Gaussian function of FWHM$\simeq$1.26--1.50 \AA, depending on the slit width used.}   The normalization of both the BF and the theoretical line profile is arbitrary, so they have been scaled to approximate the suggested $l3$$\approx$0.7 third-light contribution which dominates the BF.  It is not possible to draw conclusions about the luminosity of a third body on the basis of the BF since it is expected to be blended with the profile of the contact binary.  In this system, only one component is apparent in the BF.  We are unable to distinguish whether this is the brighter component of the contact binary or a third component.\footnote{Resolution R=22,000 echelle spectra to be presented in \citet{Cook2022} show an unresolved velocity profile consistent with a low inclination and some radial velocity variability of the strongest peak.}  The small offset of about 10 \kms\ between the theoretical profile and the BF at $\phi$=0.18 is consistent with epoch-to-epoch wavelength calibration uncertainties, but it could also include a contribution from orbital motion of the contact binary about a putative third star implied both by the \OC\ analysis and the ensuing third-light analysis.  The inclination of the third body's orbit is unconstrained.  

\begin{figure}[ht!]
%\fig{/d/zem1/hak/chip/research/Larry/KB/phoebe/KB/Tfixfudgechi2/04853067/04853067_triangle_plot_2k_9th_Tvarfudgechi2alt.pdf}{5in}{}
\plottwo{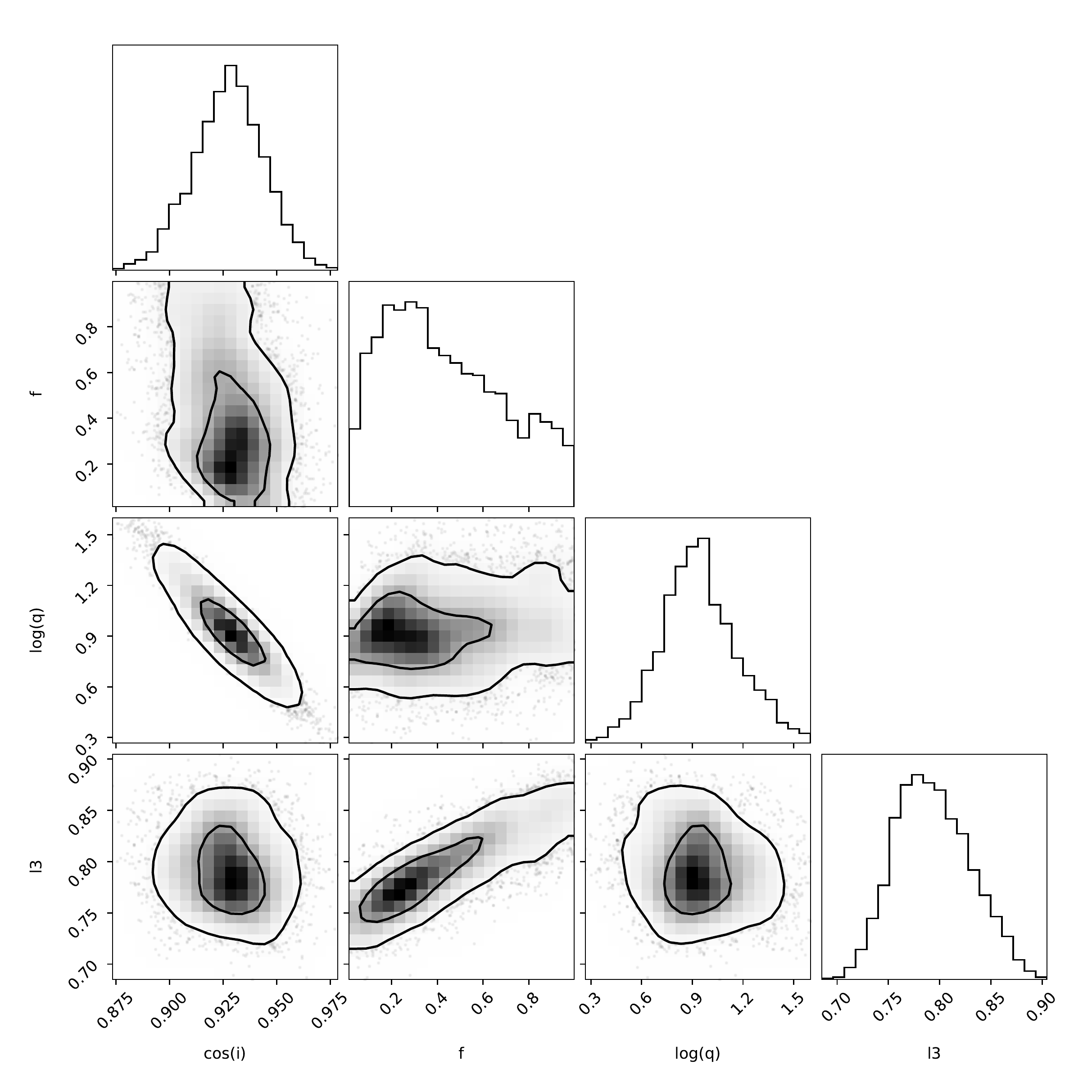}{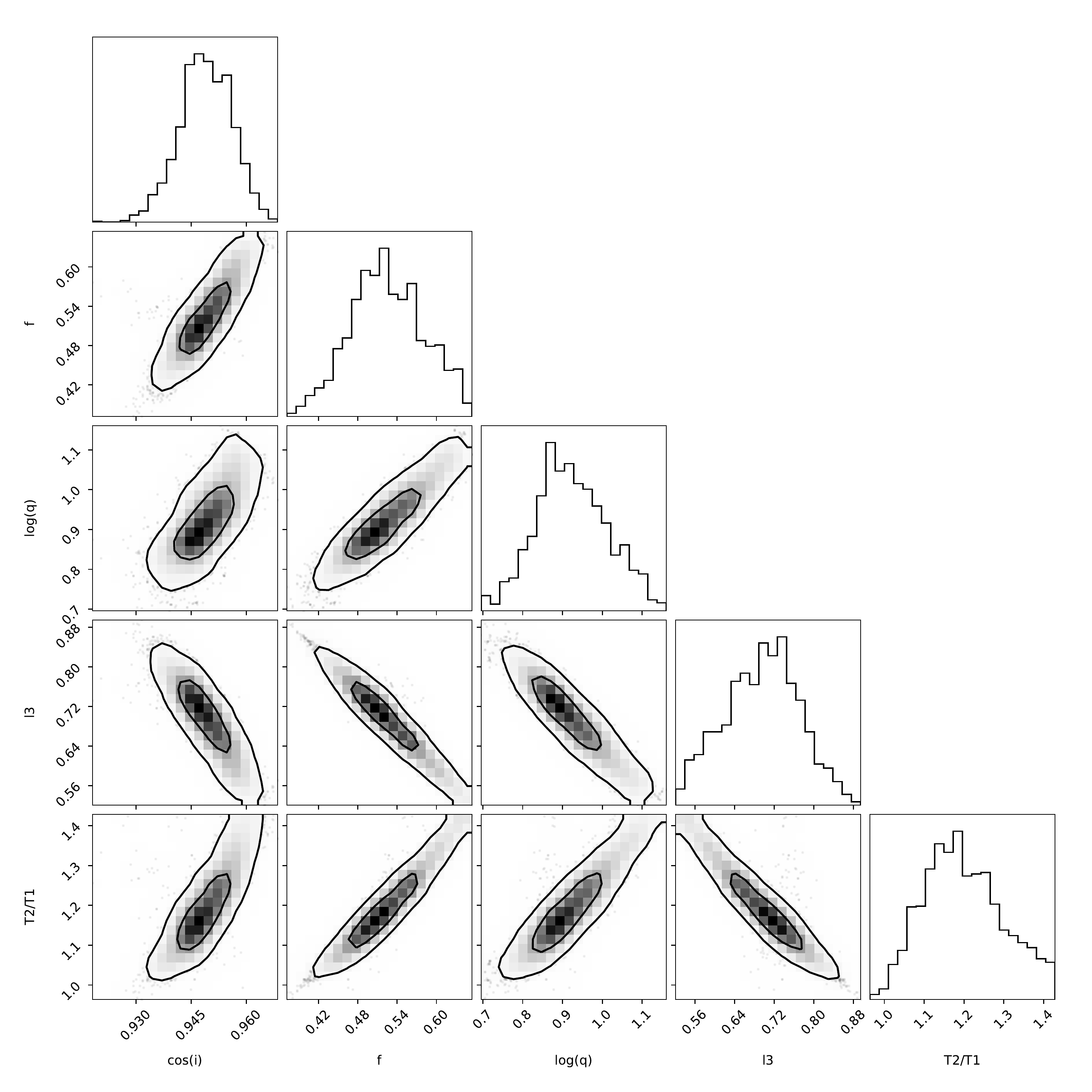}
\caption{Two dimensional contours and one-dimensional histograms depicting the posterior probability distribution for combinations of free parameters for the fixed-temperature-ratio models (left) and variable-temperature-ratio models (right) used to model the KIC04853067 light curve.  The inner and outer contours enclose 39\% and 86\% of the samples, respectively, corresponding to 1$\sigma$ and 2$\sigma$ uncertainties for normal distributions, but note the significantly non-Gaussian histograms in each parameter.  Degeneracies between temperature ratio and other parameters are prominently visible by the elongated contours in the right panel.   \label{fig:067MCMC}}
\end{figure}

To investigate the power of the light curve to constrain the system parameters in the absence of kinematic measurements, we performed a PHOEBE+MCMC analysis for each of the three competing models. Figure~\ref{fig:067MCMC} shows contour plots (2D) and histograms (1D) of the relative probabilities of the four free parameters used in the fixed-temperature-ratio contact binary models (left panel) and the five free parameters of the variable-temperature-ratio models (right panel).  The inner and outer contours enclose 39\% and 86\% of the samples, respectively, corresponding to 1$\sigma$ and 2$\sigma$ levels for a two-dimensional normal distribution. The simulations show that all parameters  are unimodal and well-constrained.  The one-dimensional histograms are notably non-Gaussian in all parameters.  The most probable parameters and 1$\sigma$ one-dimensional uncertainties of the fixed-temperature-ratio models---as represented by the 16th/50th/84th percentiles on each parameter---are \cosi=0.912/0.928/0.943  ($i$$\simeq$18\degr), $f$=0.15/0.36/0.69, \logq=0.733/0.928/0.949, $l3$=0.75/0.79/0.83.  The parameters of the best-fitting variable-temperature-ratio model  are consistent with these ranges, with the exception of fillout factor which is much larger than the fixed-temperature-ratio models. Visualizations of 100 parameter sets randomly selected from the MCMC ensemble closely follow the best-fitting model depicted in Figure~\ref{fig:067LCVC}, providing assurance that the walkers sampled the allowed parameter space in a suitably probabilistic manner. The best-fitting and the most probable solutions require low inclinations, highly uncertain fillout factors, extreme mass ratios, and significant third-light contributions, consistent with the eclipse timing variations depicted in Figure~\ref{fig:067OC}.  

\begin{figure}[ht!]
%\fig{/d/zem1/hak/chip/research/Larry/KB/phoebe/KB/DetM1var/04853067/04853067_triangle_plot_4k_9th_DetM1varrev.pdf}{5in}{}
\fig{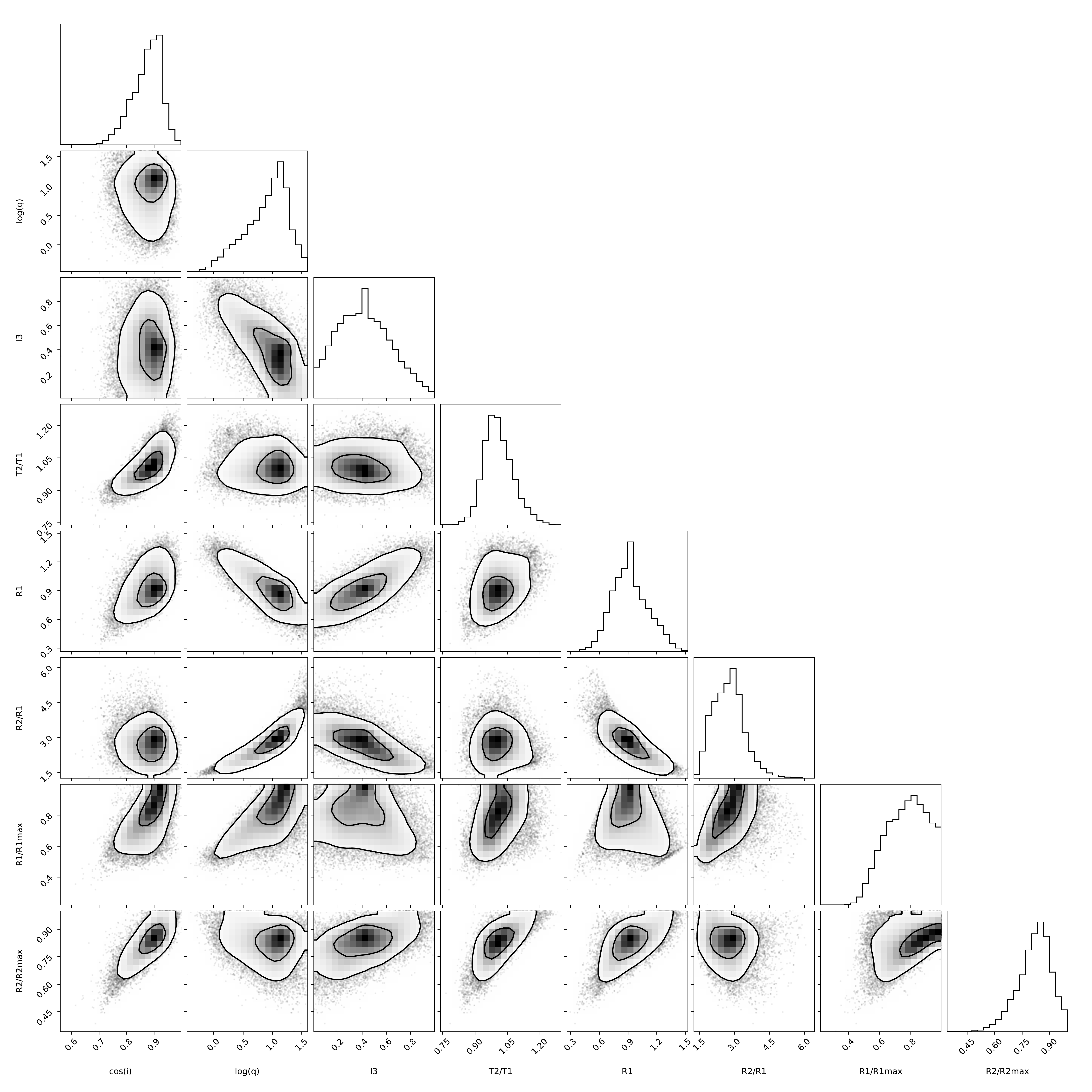}{6in}{}
\caption{Two dimensional contours and one-dimensional histograms depicting the posterior probability distribution for combinations of six free parameters (plus two ancillary derived  parameters \Ronemax\ and \Rtwomax) for the detached configuration models of KIC04853067. Model parameters remain well-constrained and most probable values overlap with the fixed-temperature-ratio models indicating low inclination, extreme $q$, and significant third light.     \label{fig:067MCMCdetached}}
\end{figure}

Figure~\ref{fig:067MCMCdetached} presents two-dimensional contour  and one-dimensional histogram distributions for each of the six free parameters resulting from the detached model MCMC analyses.  The permitted parameter range is large for many of the parameters except inclination. Nevertheless, KIC04853067's light curve provides  meaningful constraints on the majority of key system parameters.  The RMS of the best-fitting solutions is nearly identical to that in Figure~\ref{fig:067LCVC} for the contact configurations.  The most probable \Tratio\ here is near unity, consistent with a contact configuration wherein the stellar components share an atmosphere.  The most probable mass ratio near $q$=10 ($q_{\rm a}$=0.1) is extreme and both components are close to overflowing  with \Ronemax$>$0.9 and \Rtwomax$\approx$0.9.  In such configurations, contact and detached systems become indistinguishable as the larger overflowing or nearly overflowing component dominates the light curve modulation.     

In conclusion, KIC04853067 is a $P$$\approx$1.3 d extreme-$q$ candidate system exhibiting a well-measured low-amplitude light curve that can be modeled equally well as a contact or detached configuration.  This is consistent with its large light curve morphology parameter of 0.88, approaching the regime of ellipsoidal variables.  Either configuration requires a low inclination angle and both stars near overflowing.  The system's velocity profile is kinematically unresolved, consistent with low $i$ and large third light. Systematic \OC\ variations and a preference for significant $l3$ in the Bayesian analyses of the light curve indicate that KIC04853067 is a triple system in both the contact and detached models.   The case of KIC0485306 is a cautionary tale of how even high-precision single-band light curves can fail to discriminate between contact and detached geometries when mass ratios are extreme or inclinations are small. 

\clearpage

\subsection{KIC04999357}

\begin{figure}[ht!]
%\fig{KB/IndivSpectra/KIC04999357/KIC04999357_phased_all.pdf}{4in}{}
\fig{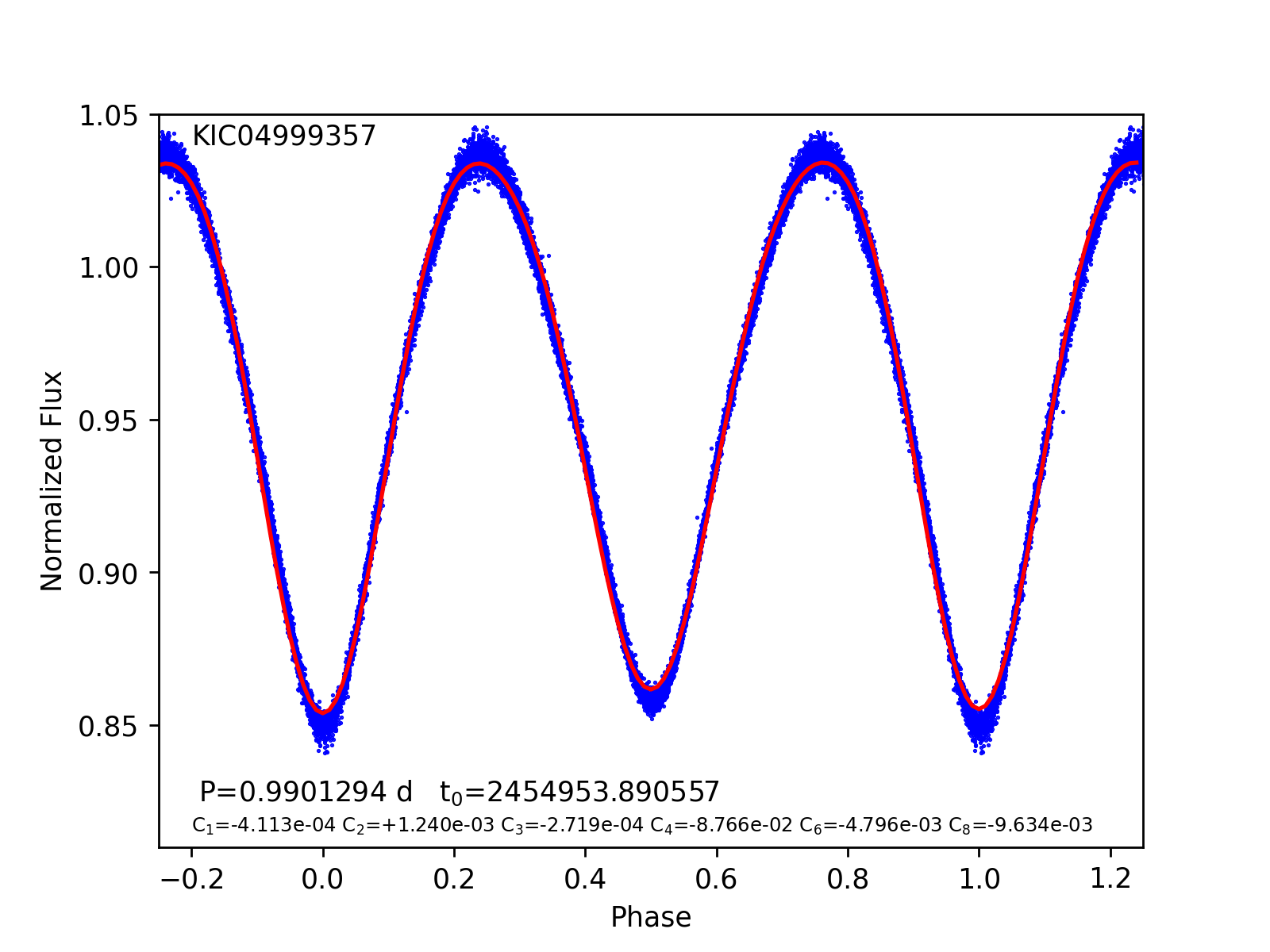}{5in}{}
\caption{Folded {\it Kepler} light curve of KIC04999357 and mean light curve shape (red curve) generated from the sum of low-order Fourier coefficients.  \label{fig:357LC}}
\end{figure}

Figure~\ref{fig:357LC} shows the folded {\it Kepler} light curve of the $P$=0.99 d system KIC04999357.   The Fourier components used to approximate the mean light curve (red curve) are given in the Figure. Primary and secondary eclipse depths are very nearly equal, as are the maxima between eclipses.  The continuous modulation is that of a classical contact binary.  KIC04999357 exhibits no evidence for \OC\ variations over the duration of the {\it Kepler} mission, with an RMS variation of $<$0.2 min.  Accordingly, we adopt a linear ephemeris.  

\begin{figure}[ht!]
%\fig{IndivSpectra/KIC04999357/KIC04999357_twocomp.pdf}{4in}{}
\fig{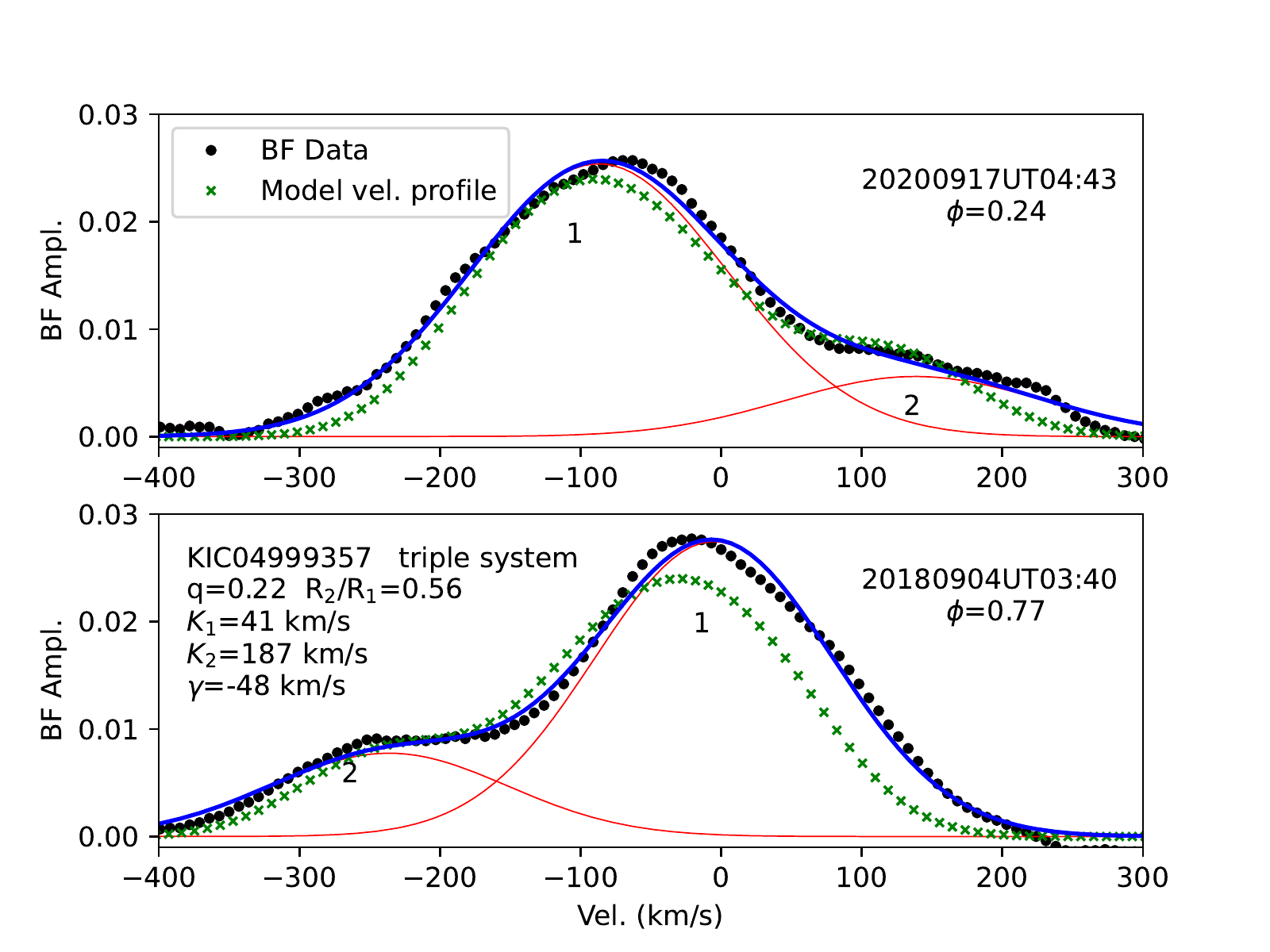}{5in}{}
\caption{Broadening Function of KIC04999357 showing two  velocity components.  The black dotted curve is the BF, the red curves are two Gaussian components fitted to the BF, the blue curve is the sum of the Gaussian components, and the green x's represent the model line profile of the best-fitting PHOEBE model. The normalization of both the data and the model is arbitrary. \label{fig:357BF}}
\end{figure}

Figure~\ref{fig:357BF} shows the BF (black dots) for two epochs of spectroscopy at orbital phases $\phi$=0.24 and $\phi$=0.77.  There are two clear components.  We fit the BF using a two-component Gaussian\footnote{A Gaussian function is not a physically realistic representation of the asymmetric velocity profile of a tidally distorted star, but it serves here to measure the component velocities in a model-independent way. } function (red curves) which provide an estimate of the component velocities at each quadrature phase, and thereby, a mass ratio and a systemic velocity.  The sum of the two Gaussian components (blue curve) provides a good match to the BF data.  The ratio of areas of the two Gaussian components implies a luminosity ratio $L_{\rm 2}$/$L_{\rm 1}$ of about 0.28 (1:3.5) and a radius ratio $\sqrt{ L_{\rm 2}/L_{\rm 1}}$ = $R_{\rm 2}$/$R_{\rm 1}$=0.56.  The radial velocities of the  components yield velocity semi-amplitudes of $K_{\rm 1}$=41 \kms\ and $K_{\rm 2}$=187 \kms,  implying a mass ratio near $q$=0.22 and a systemic velocity of $\gamma=-$60 \kms.  However the velocity of the secondary at $\phi$=0.24 is poorly measured, as it appears considerably fainter and less defined than at $\phi$=0.77.   A second spectrum obtained at $\phi$=0.27 on 20170711 shows the same deficit near the expected peak of the secondary's BF.  This deficit is present regardless of spectral range used in the BF analysis, effectively ruling out the possibility of emission in any one spectral feature or uncorrected Telluric absorption affecting the spectra. We find that introducing a cool spot on the trailing face of the secondary star in the PHOEBE models can roughly reproduce this deficit in the BF. The deficit here, and in some subsequent examples, bears some similarities to that of the secondary in extreme-$q$ system AW UMa which \citet{Rucinski2015} interprets as indicating the presence of an accretion disk or flow of matter to/from the secondary.     

KIC04999357 is the only one among our pilot sample to have large astrometric uncertainties, as judged by the Renormalized Unit Weight Error\footnote{ https://gea.esac.esa.int/archive/documentation/GDR2/Gaia\_archive/chap\_datamodel/} (RUWE=3.00) in the $Gaia$ DR2 dataset \citep{Gaia}. RUWE is a metric that designates a poor single-star astrometric solution when RUWE$>$1.4 and is often used to flag candidate multiple-star systems \citep[e.g.,][]{Kervella2019}. Contact binaries having semi-major axes of several solar radii would not yield poor $Gaia$ astrometric solutions on their own, but tertiaries orbiting contact binaries at distances $\gg$few AU could produce large RUWE values. Hence, there is evidence for a third component in this system or along the line of sight.  

\begin{figure}[ht!]
%\fig{IndivSpectra/KIC04999357/KIC04999357LC+VCnew.png}{}
\fig{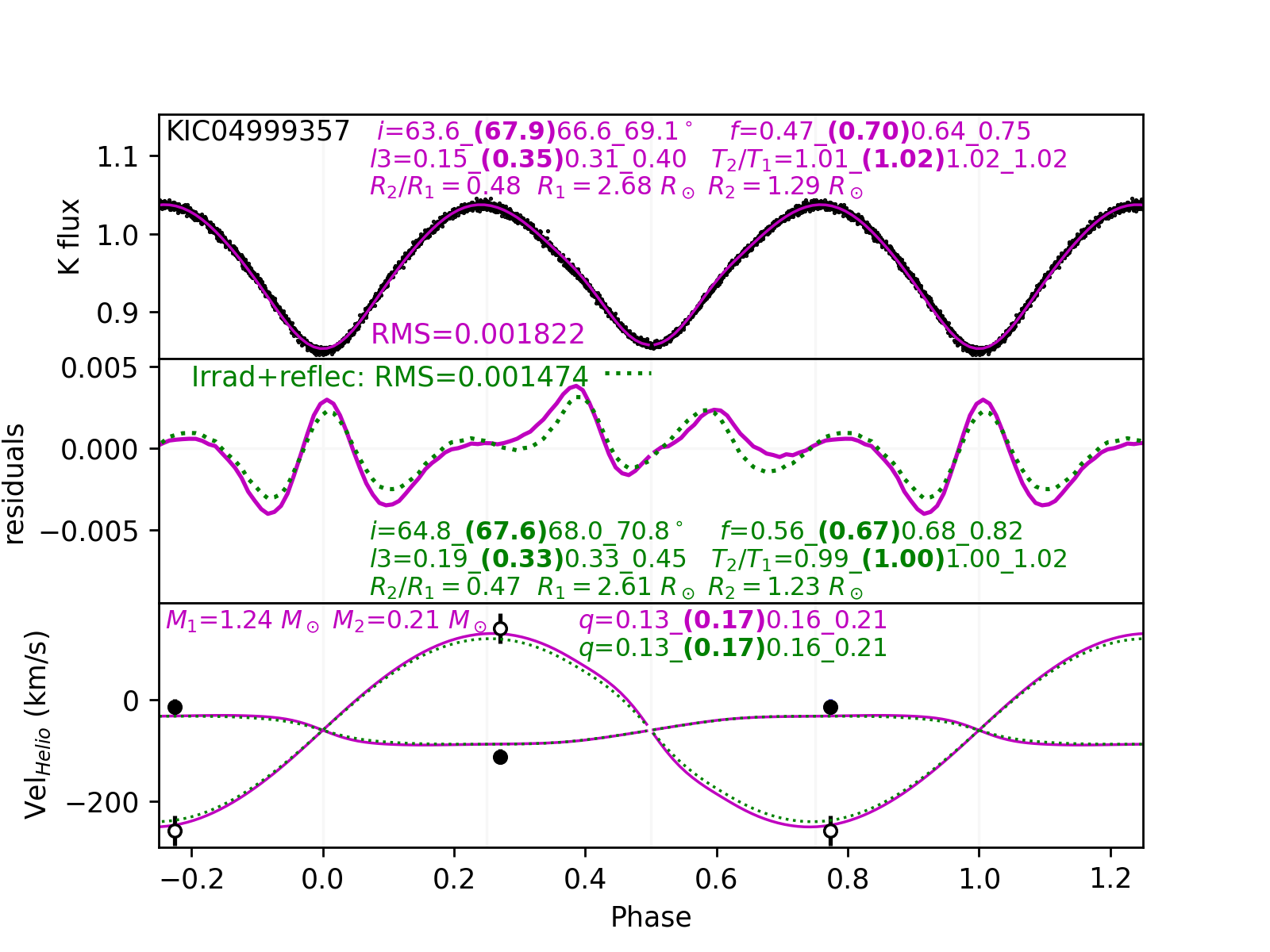}{5in}{}
\caption{Folded light curve and velocity measurements of KIC04999357, along with best-fitting PHOEBE model light curves, residuals (model minus data), and velocity curves without irradiation effects (magenta curves) and with irradiation  (green curves). \label{fig:357LCVC}  }
\end{figure}

Figure~\ref{fig:357LCVC} shows the folded photometric data (upper panel) and the spectroscopic radial velocity data (lower panel), with the best-fitting variable-temperature-ratio PHOEBE model (without irradiation effects) overplotted using solid magenta curves.  Text within the panels also gives the 16th/{\bf(best)}50th/84th percentile ranges from MCMC simulations. The best-fitting model (RMS=0.0018) requires  $i$=67.9\degr, $f$=0.70, $q$=0.17 (in reasonable agreement with the $q$=0.22 from the broadening function), $l3$=0.35, and \Tratio=1.02. This model is superior to the best-fitting detached models which all have larger RMS and require \Ronemax$>$0.99 and \Rtwomax$>$0.99, indicating a contact configuration.  The contact model yields a ratio of component radii $R_{\rm 2}$/$R_{\rm 1}$=0.48 (close to the value inferred above from the BF), $R_{\rm 1}$=2.68 \rsun, and $R_{\rm 2}$=1.29 \rsun.  Adopting this inclination allows the computation of component masses $M_{\rm 1}$=1.24 \msun\ and $M_{\rm 2}$=0.21 \msun.  The line profile function produced by this model (green x's in Figure~\ref{fig:357BF}) shows a reasonable match to the BF when normalized to the fainter component.  However, the amplitude of the brighter component in the model is smaller than the peak in BF.  The difference between the data (black dotted curve) and model (green x's) can readily be explained by the light of a third component amounting to $\approx$10\% of the total system light, {\it assuming the third component has a similar temperature to the contact binary.} As the present release of PHOEBE does not support a full inclusion of tertiary components in the line profile, we cannot model the contribution of the putative third star to the BF here or in subsequent examples where third components are probable. The presence of a third star is likely to skew the primary peak in the BF, making measurement of the primary star's velocity uncertain.        

The middle panel of Figure~\ref{fig:357LCVC} plots the model residuals (model minus data) of the best fitting variable-temperature-ratio light curve (magenta curve) and the best-fitting light curve when irradiation and reflection effects are included in the model (green dashed curve).  The best-fitting model with irradiation has a modestly smaller RMS (0.0014 versus 0.0018), but the 16th/50th/84th percentile ranges are very similar to those listed in the top panel and the best-fitting parameters are all nearly identical.   Residuals are smaller, primarily in the region centered on the primary eclipse ($\phi$=0), as expected based on the comparisons presented in \citet{Horvat2019}. While an attempt at proper treatment of irradiation effects improves the best-fitting models in this well-measured system having small intrinsic dispersion in the data, the impact on the Bayesian constraints on each parameter is minimal.    

\begin{figure}[ht!]
%\fig{Ksample_V2/KIC04999357/triangle_plot_6k_6th.pdf}{5in}{}
\fig{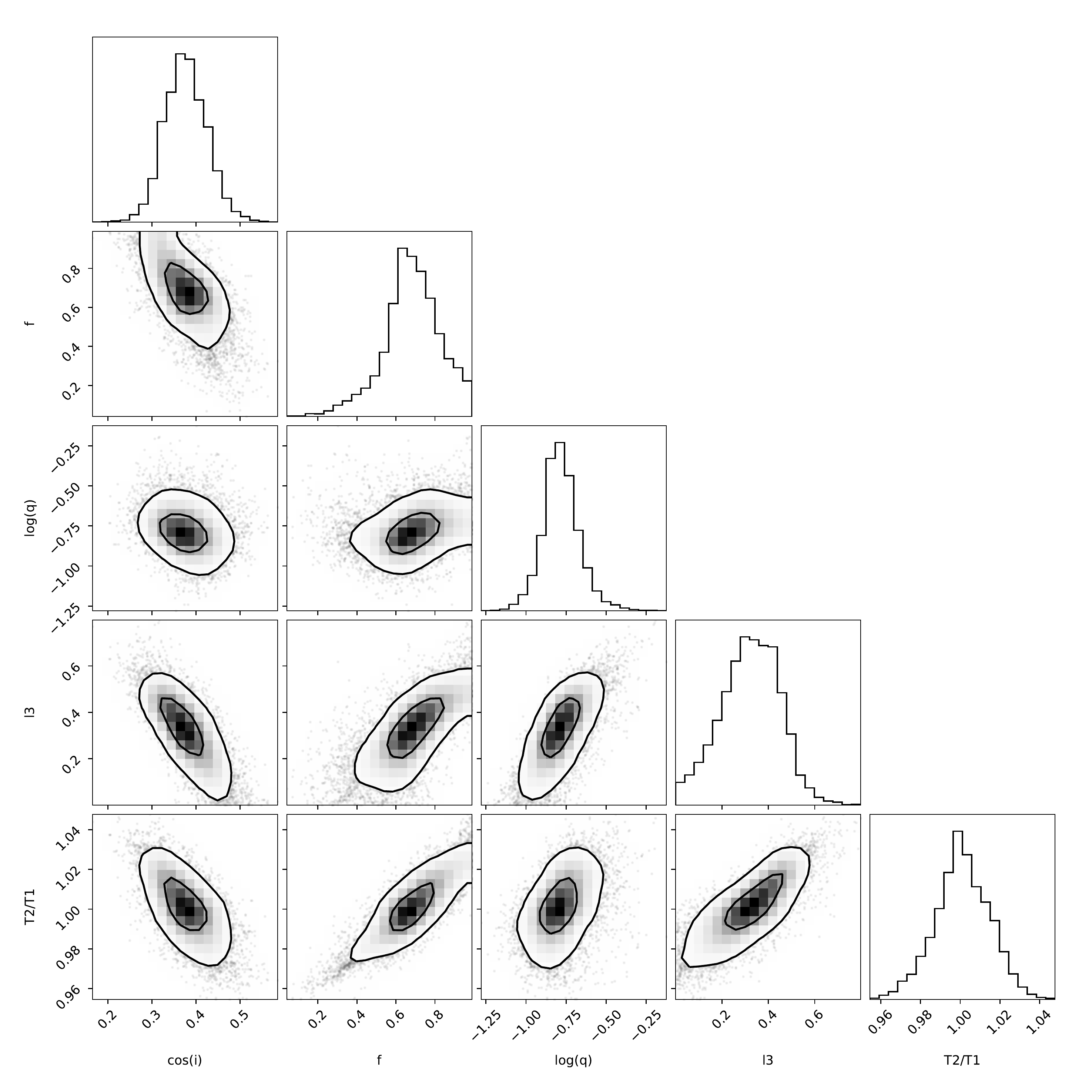}{5in}{}
\caption{Posterior probability distributions for combinations of five free system parameters in the KIC04999357 system from the nominal \Tapprox\ models. \label{fig:357MCMC}}
\end{figure}

Figure~\ref{fig:357MCMC} shows the results of the MCMC analysis for KIC04999357 stemming from the nominal \Tapprox\ models.  The contours and peak density of points indicates that the most 16th/50th/84th percentile system parameters are \cosi=0.34/0.38/0.43, $f$=0.52/0.67/0.81, \logq=$-$0.90/$-$0.79/$-$0.67 ($q$=0.18, in general agreement with the  BF), $l3$=0.16/0.31/0.43, and \Tratio=0.99/1.00/1.01, consistent with the best-fitting parameters given above. All five of the model parameters are well-constrained. The Monte Carlo simulations allow for a modest $l3$$\approx$0.31, not inconsistent with the contribution inferred from the BF discussed in connection with Figure~\ref{fig:357BF} previously.  

In summary, KIC04999357 ($morph$=0.84) is a possible triple system where the inner contact binary contributes most of the luminosity and has a small mass ratio $q$=0.17--0.22.  PHOEBE models with and without irradiation effects produce convincing fits to the data with a component temperature ratio near unity.  KIC04999357 may be an evolved binary that has exchanged mass, on its way toward the Darwin instability limit.  The evidence for a third component---from the large GDR2 RUWE values and the excess in the BF near the systemic velocity---is consistent with the idea that Kozai-Lidov cycles initially play a role in bringing the inner components into contact.  The putative third body must lie at a large separation from the contact binary in order that it produce a large astrometric RUWE and not produce detectable \OC\ variations.      

\clearpage

\subsection{KIC06844489}

\begin{figure}[ht!]
%\fig{Spectra/KIC06844489/KIC06844489_phased_all_smaller.pdf}{4in}{}
\fig{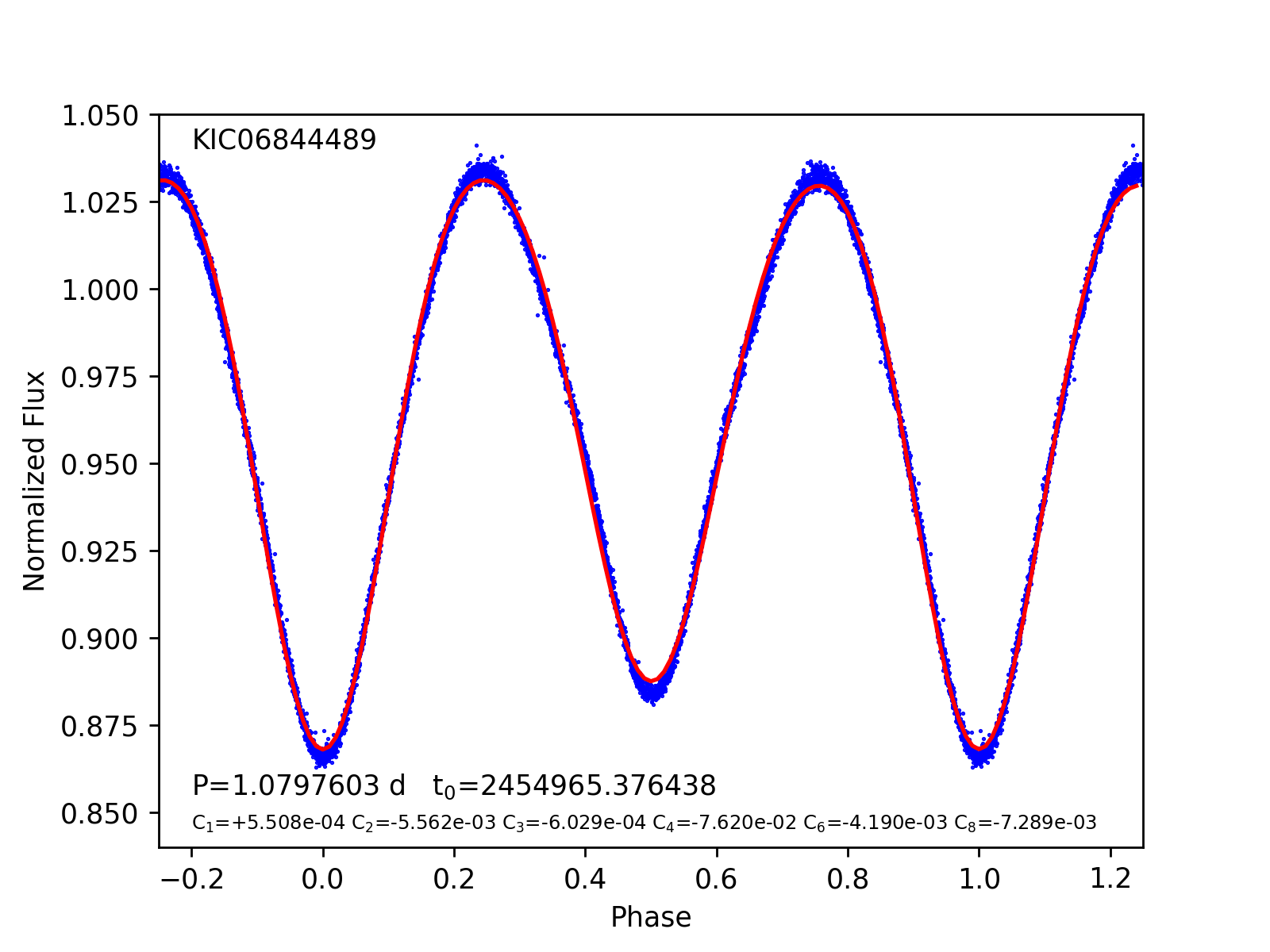}{5in}{}
\caption{Folded {\it Kepler} light curve of KIC06844489 and mean light curve shape (red curve) generated from the sum of low-order Fourier coefficients. \label{fig:489LC}}
\end{figure}

Figure~\ref{fig:489LC} displays the folded {\it Kepler} light curve of the $P$=1.08~d system KIC06844489.  
The mean light curve displays a $\simeq$10\% semi-amplitude, very little dispersion, and is well characterized by the Fourier components labeled in the Figure.  Only seven quarters of $Kepler$ data are available.  The secondary minimum is less deep than the primary minimum.  

\begin{figure}[ht!]
%\fig{Spectra/KIC06844489/KIC06844489_O-C+ztf.pdf}{4in}{}
\fig{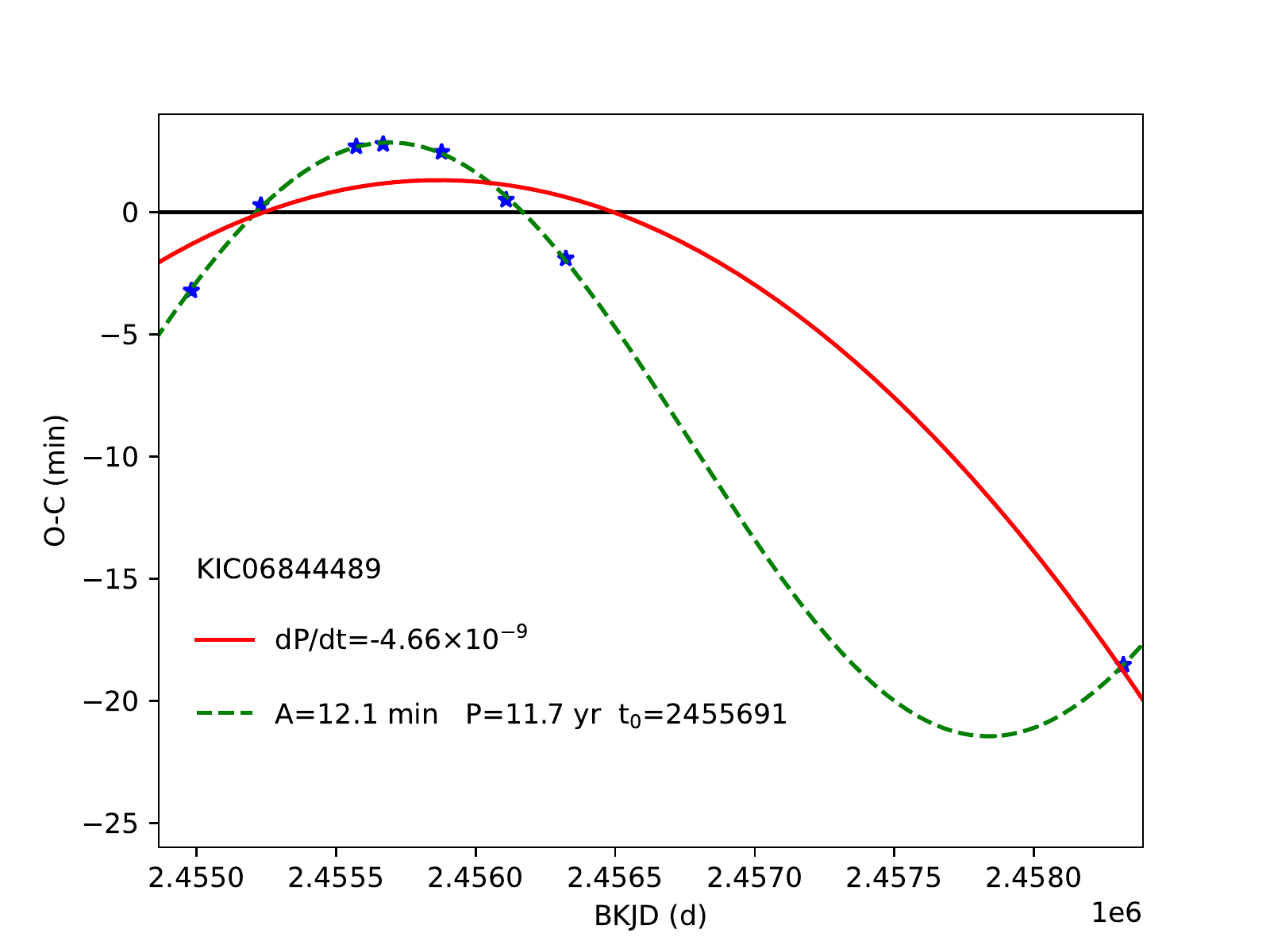}{4in}{}
\caption{\OC\ eclipse timing residuals of KIC06844489.  The red curve shows a parabolic fit corresponding to a constant negative period derivative.  The green dashed curve shows a sinusoidal fit corresponding to light travel time effects arising from an orbit about a third body.  \label{fig:489OC}}
\end{figure}

Figure~\ref{fig:489OC} presents the eclipse timing residuals from a linear ephemeris with $P$=1.0797603 d.  The first six data points are from the 2009--2013 {\it Kepler} mission while the last data point uses 2018--2019 measurements from the Zwicky Transient Factory \citep[ZTF][]{ZTF2014}.   The red solid curve shows a best-fit parabola, representing a quadratic ephemeris with a period derivative of $dP/dT=-$4.66$\times10^{-9}$---a constantly decreasing period.  This model is a poor match to the \OC\ data.  The dashed green curve shows a sinusoidal model with amplitude $A$=12.1 min, $P$=11.7 yr, and $t_0$=2455691 (the peak of the sine curve---the time when the contact binary is at superior conjunction relative to the foreground third component).  This model, predicated on the hypothesis of a third body in orbit with the contact binary, provides an excellent match to the data.  The semi-amplitude of this curve gives the light crossing time across the projected semi-major axis of the contact binary about the system Barycenter, $t$=$a$sin($i_{\rm outer}$)/$c$.  The implied minimum semi-major axis of the contact binary's orbit is 1.45 AU.  A hypothetical 0.34/sin($i_{\rm outer}$) \msun\ tertiary in an 11.7 yr orbit can explain the \OC\ data.

\begin{figure}[ht!]
%\fig{Spectra/KIC06844489/KIC06844489_threecomp.pdf}{4in}{}
\fig{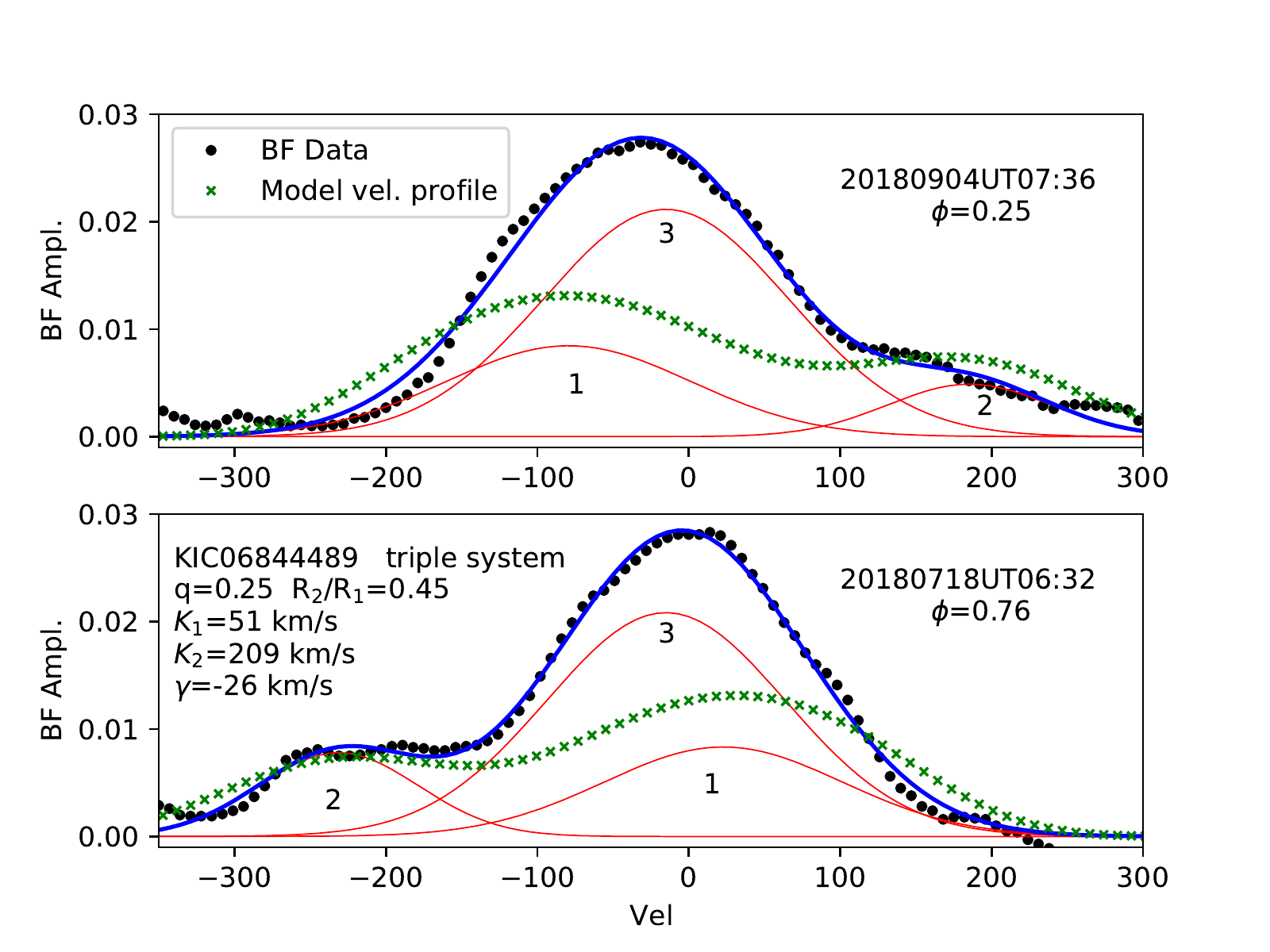}{5in}{}
\caption{Broadening Function of KIC06844489 showing three probable spectral components. The black dotted curve is the BF, the red curves are Gaussian components, the blue curve is the sum of the Gaussian components, and the green x's represent the model velocity profile of the best-fitting PHOEBE model (which does not include a tertiary).  The normalization of both the BF and model velocity profile is arbitrary.   \label{fig:489BF}}
\end{figure}

The broadening function of KIC06844489 in Figure~\ref{fig:489BF} shows two components, one much smaller than the other. The spectroscopic data at phases $\phi$=0.25 and $\phi$=0.76 (assuming a linear ephemeris) were obtained 1.5 months apart.   The primary BF component has a distinctly non-Gaussian line profile, as if it may be a blend of unresolved subcomponents.  The ratio of component velocities in a two-component fit indicates an extreme mass ratio near $q$=0.10.  However the presence of a bright third component spectrally blended with the primary would result in the $K_{\rm 1}$ amplitude and $q$ being underestimated. Given the evidence for a third body from the large \OC\ variations, we fit the BF using three components, fixing the velocity of the tertiary near $-$25 \kms.  The sum of the three components (blue curve) shown by the three red curves in Figure~\ref{fig:489BF} provides a good overall fit to the BF, with the third component contributing 60\% of the total light (under the assumption of a similar temperature to the contact binary).  The implied mass ratio is $q$$\simeq$0.25, but the velocity of the fainter component is rather uncertain. There is considerable degeneracy between the strengths of the primary and tertiary, rendering the velocity of the primary highly uncertain.  As in KIC04999357, the secondary's peak near $\phi$=0.25 is smaller than at the other quadrature phase. 

\begin{figure}[ht!]
%\fig{Spectra/KIC06844489/KIC06844489LC+VCnew}{4in}{}
\fig{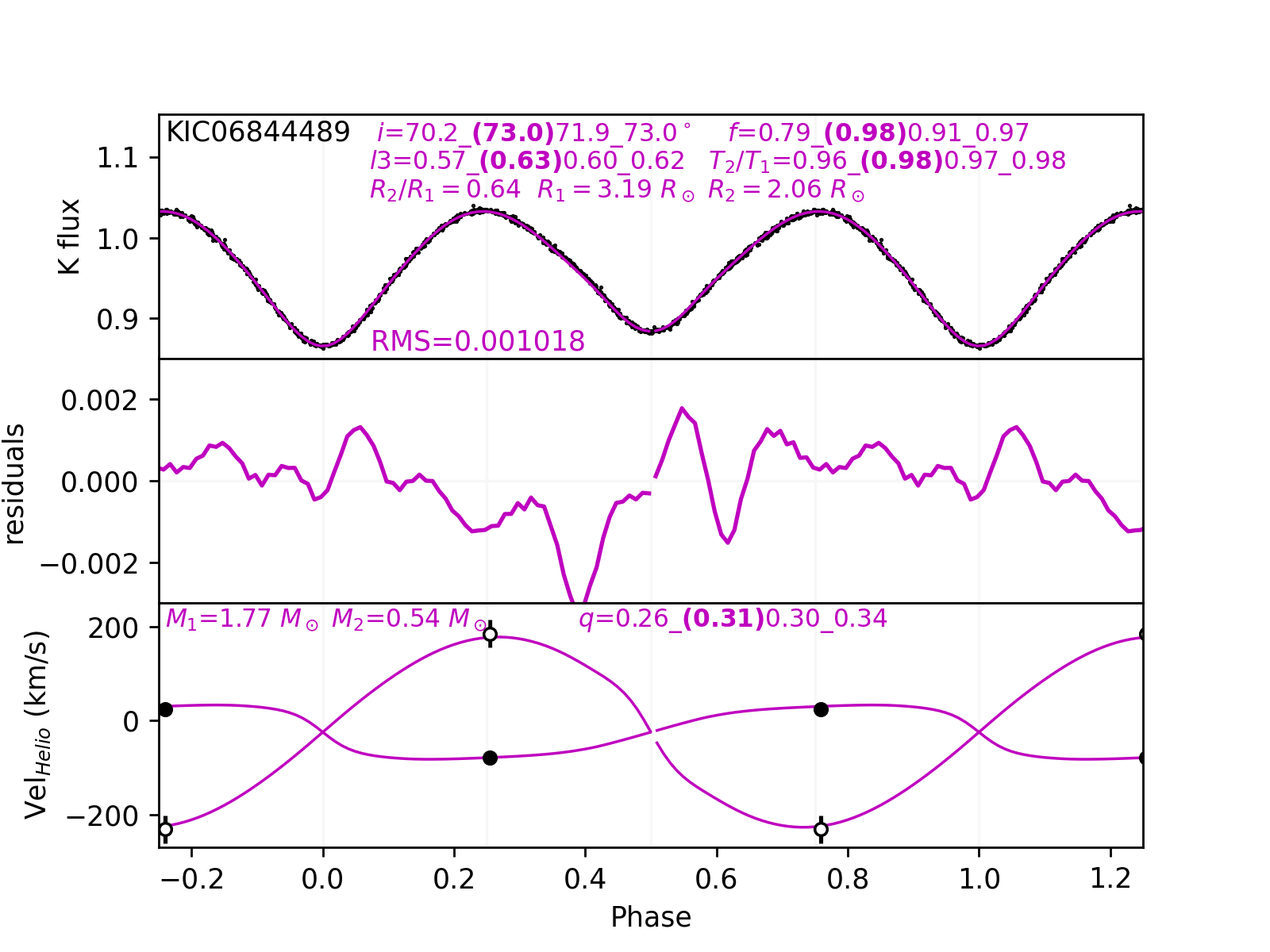}{5in}{}
\caption{Folded light curve and velocity measurements of KIC06844489, along with best-fitting PHOEBE model light curve, light curve residuals, and velocity curve.  \label{fig:489LCVC}}
\end{figure}

Figure~\ref{fig:489LCVC} shows a light curve and velocity curve of a best-fitting variable-temperature-ratio PHOEBE model. A model with $i$=73.0\degr, $f$=0.98, $q$=0.31 (in good agreement with the BF), $l3$=0.63 (consistent with the deficit in the primary's BF peak in Figure~\ref{fig:489BF}), and \Tratio=0.98 produces an excellent fit (RMS=0.0010).  Models implementing irradiation effects require nearly identical system parameters and do not improve the fit. The residuals in the middle panel are small and show no systematic offsets near phases 0 or 0.5 where irradiation effects would be expected to create the largest signatures. At this inclination the measured component velocities imply masses $M_{\rm 1}$=1.77 \msun\ and $M_{\rm 2}$=0.51 \msun.  These masses dictate component equivalent radii $R_{\rm 2}$/$R_{\rm 1}$=0.64,  $R_{\rm 1}$=3.19 \rsun, and $R_{\rm 2}$=2.06 \rsun. The best contact models fit the light curve well but produces a line profile (green x's in Figure~\ref{fig:489BF}) having primary star amplitude smaller than the primary BF peak, suggesting a need for a third component.  By contrast, the best-fitting detached model has a much larger RMS and indicates the primary is nearly overflowing with \Ronemax$\geq$0.99.  Accordingly, we consider this to be a {\it bona fide} contact system with nearly equal temperature components and a large fillout factor.     

\begin{figure}[ht!]
%\fig{Ksample_V2/KIC06844489/triangle_plot_6k_6th.pdf}{5in}{}
\fig{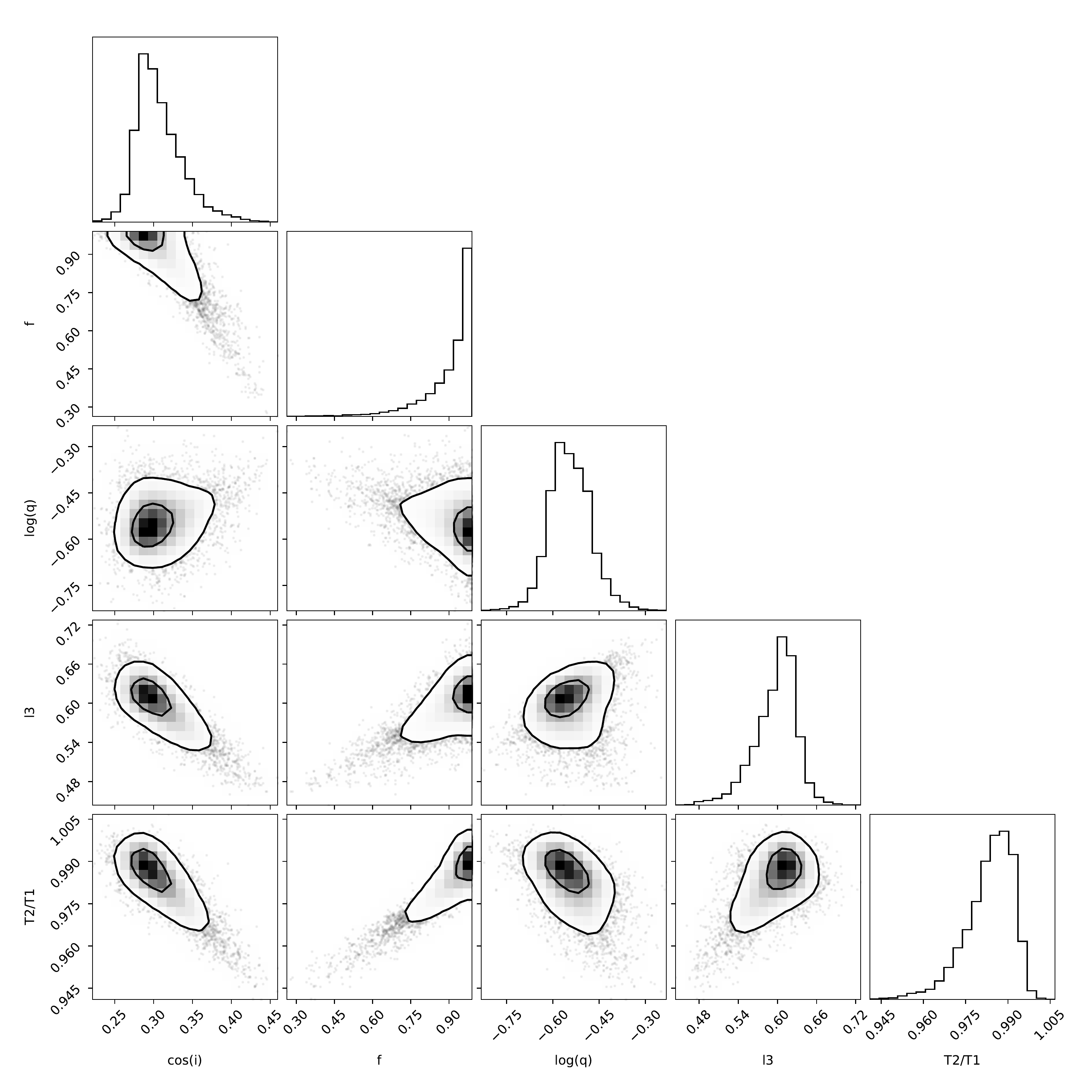}{5in}{}
\caption{Posterior probability distributions for combinations of five free system parameters from the \Tapprox\ models for the KIC06844489 system. \label{fig:489MCMC}}
\end{figure}

Monte Carlo simulations of the KIC06844489 system using the \Tapprox\ model yield the probability distributions in Figure~\ref{fig:489MCMC}.  The most probable values are all well-constrained at \cosi=0.292/0.311/0.338 ($i$$\simeq$73\degr), $f$=0.79/0.91/0.97, \logq=$-$0.58/$-$0.52/$-$0.46, $l3$=0.57/0.60/0.62, and \Tratio=0.96/0.97/0.98. The Figure shows some degeneracy between \cosi\ and $f$ in the sense that larger \cosi\ (smaller inclination) requires smaller fillout factors.  There is also degeneracy between $f$ and \Tratio, where larger $f$ demands larger \Tratio.  Evidence for a substantial third-light component is supported by the MCMC analysis, consistent with the \OC\ residuals in Figure~\ref{fig:489OC} and the excess in the BF near the systemic velocity. The minimum ($i$=90\degr) third-body mass implied by the \OC\ analysis of $M_3$=0.33 \msun\ is not capable of producing 63\% of the system light if it is a main-sequence star.  Orbital inclinations of the $M_3$--($M_1$+$M_2$) system of $i_{\rm outer}<$20\degr\ would result in $M_3\gtrsim$1.5 \msun, allowing a main-sequence F star tertiary to contribute substantially to the system light in the amount suggested by the Monte Carlo simulations.   

In summary, the evidence indicates KIC06844489 is a contact system (consistent with its $morph$=0.84) with a luminous third component that creates large \OC\ systematics well-modeled by a sine function.  The most probable mass ratio of the contact binary system is small at $q$$\simeq$0.30.  It is a somewhat massive ($M_{\rm tot}$=2.28 \msun) long-period contact binary system that may be nearing the putative Darwin instability regime.

\clearpage

\subsection{KIC08913061}

\begin{figure}[ht!]
%\fig{Spectra/KIC08913061/KIC08913061_phased_all.png}{4in}{}
\fig{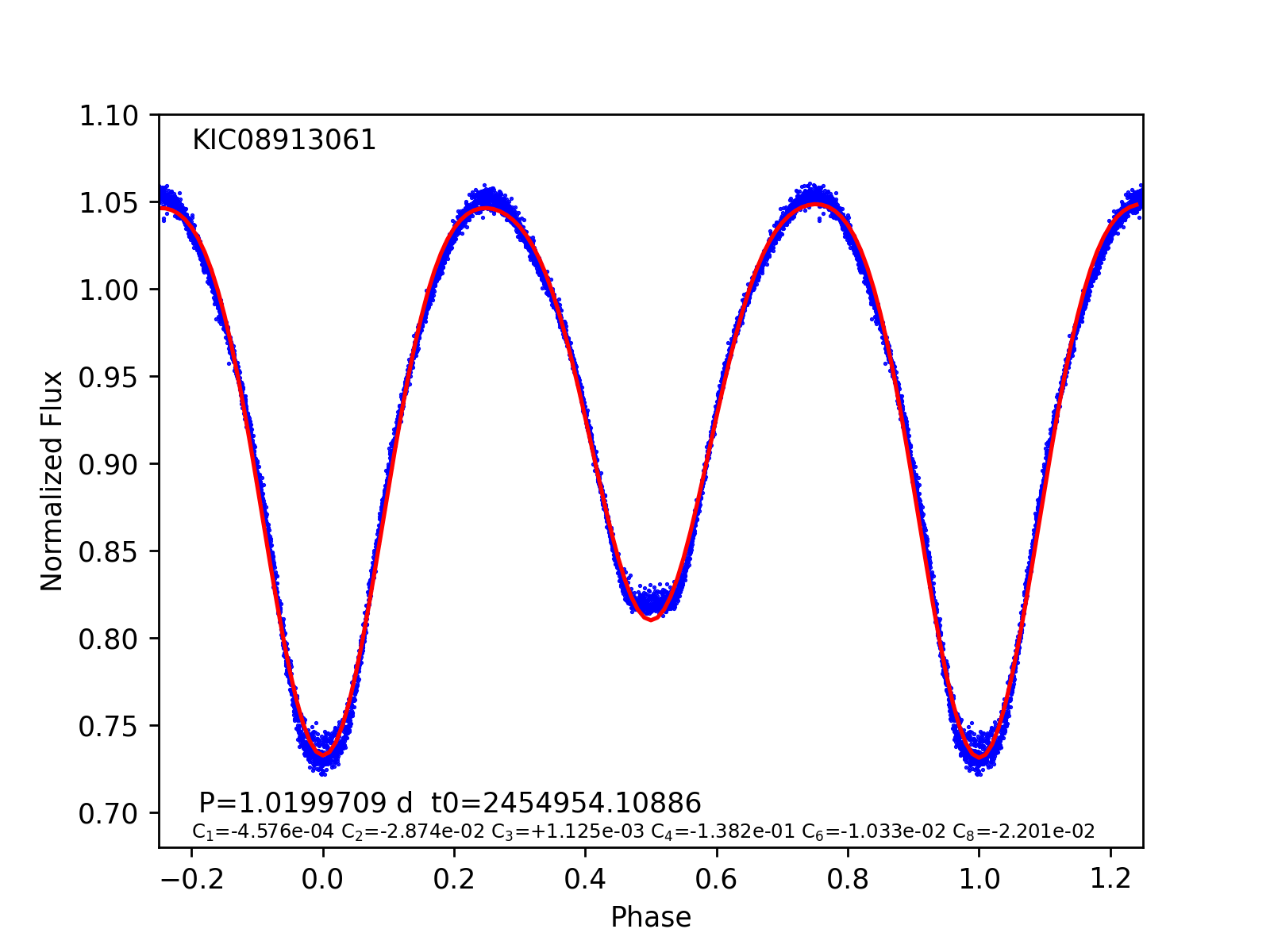}{5in}{}
\caption{Folded {\it Kepler} light curve of KIC08913061 and mean light curve shape (red curve) generated from the sum of low-order Fourier coefficients.  \label{fig:061LC}}
\end{figure}

Figure~\ref{fig:061LC} displays the folded {\it Kepler} light curve of the $P$=1.02~d system KIC08913061. Primary minimum is considerably deeper than the secondary minimum.  The secondary minimum has a flat bottom, suggesting both a high inclination and a low mass ratio (cf. the top row of Figure~\ref{fig:fourbyfourmodels}). The mean light curve displays a $\simeq$15\% semi-amplitude and is well characterized by the Fourier components labeled in the Figure, excepting the flat bottom which would require higher order coefficients.   The eclipse timing residuals show no systematic variation over the time baseline of the {\it Kepler} mission.  

\begin{figure}[ht!]
%\fig{Spectra/KIC08913061/KIC08913061_twocomp.pdf}{4in}{}
\fig{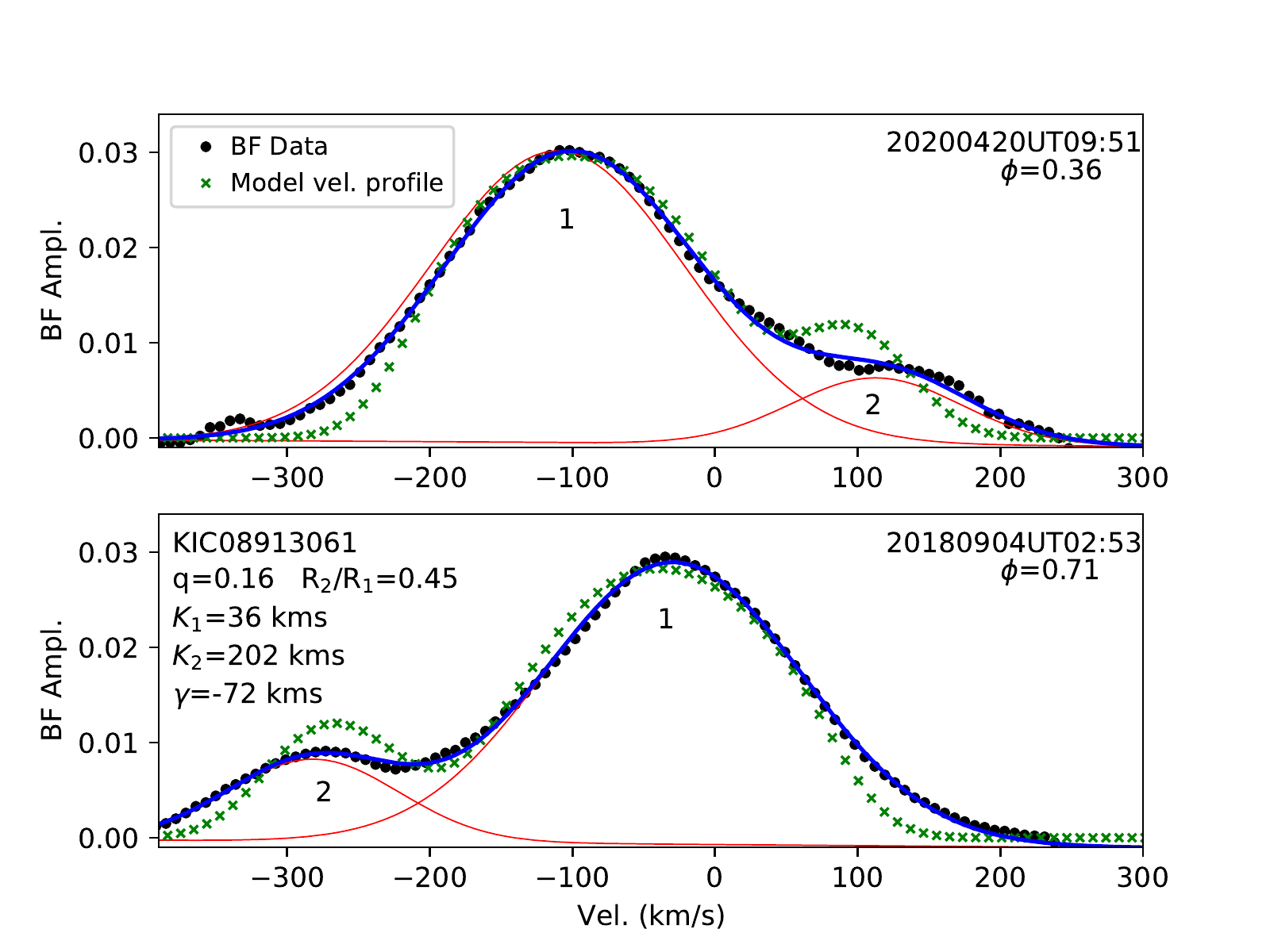}{4in}{}
\caption{Broadening Function of KIC08913061 showing two distinct components with very disparate luminosities. \label{fig:061BF}}
\end{figure}

Figure~\ref{fig:061BF} shows the broadening function of KIC08913061 at phases $\phi$=0.36 and $\phi$=0.71.  There are two distinct peaks in the BF with the small peak at positive velocities at $\phi$=0.36, indicating a mass ratio $q$$<$1.  The component velocities yield $K_{\rm 1}$=36 \kms, $K_{\rm 2}$=202 \kms, $q$=0.18, and systemic velocity $\gamma$=$-$72 \kms.   As in the previous two examples, the secondary's BF peak is again smaller at $\phi$=0.36 than at the opposite quadrature phase.        

\begin{figure}[ht!]
%\fig{Spectra/KIC08913061/KIC08913061LC+VC_smaller.pdf}{6in}{}
\fig{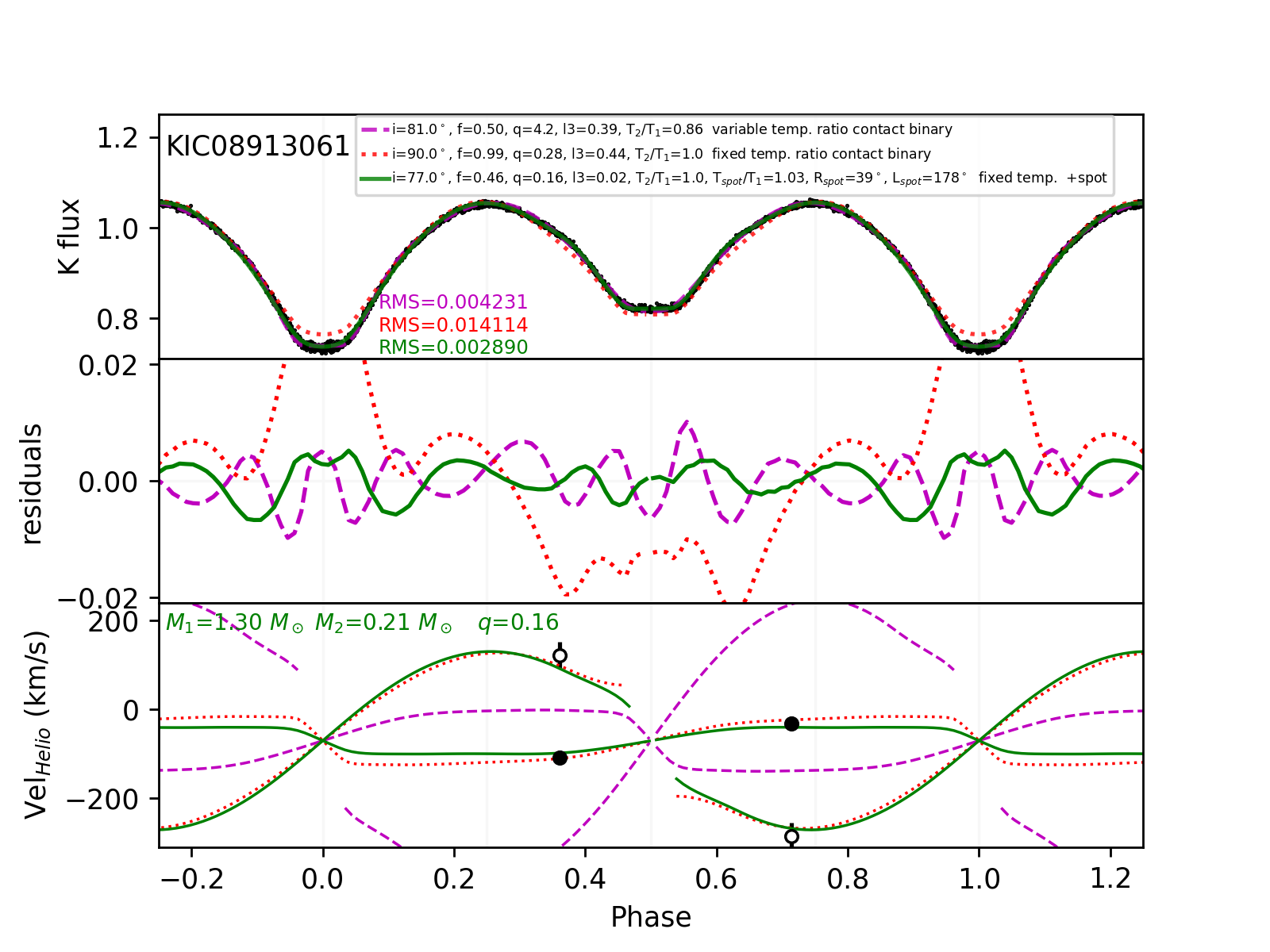}{6in}{}
\caption{Three competing PHOEBE models for KIC08913061 in comparison to the light curve and velocity data.  The nominal variable-temperature-ratio model (magenta) indicates $q$$\approx$4, in marked disagreement with the velocity curve, while either of the \Tratio=1 models yields approximately the correct mass ratio. \label{fig:061LCVC}}
\end{figure}

We performed a PHOEBE Monte Carlo analysis of the KIC08913061 system, finding that the most probable system parameters involve large inclinations $i>$70\degr, rather extreme mass ratios, and \Tratio$\ll$1.  The best-fitting variable-temperature-ratio model shown in Figure~\ref{fig:061LCVC} (upper panel, magenta curve) provides a good fit (RMS=0.0042) to the light curve with  $i$=81.0\degr, $f$=0.50, $q$=4.2 (secondary more massive than primary), $l3$=0.39, and \Tratio=0.86 regardless of whether irradiation effects are included.  Such a large temperature difference between the stars seems inconsistent with a contact system.  {\it Furthermore, this mass ratio is grossly inconsistent with the kinematic data which require $q<$1!} This indicates a limitation of a goodness of fit parameter that weights all orbital phases equally. The small amplitude, but statistically significant, details of the shape of the edge of the secondary eclipse contains important information that is lost in a balance with differences affecting a wider range of orbital phase elsewhere. As an alternative we considered a fixed-temperature-ratio model---\Tratio=1.  The best solution (red curve) is $i$=90\degr, $f$=0.99, $q$=0.28 (roughly consistent with the kinematic data), and $l3$=0.44, but this model produces a less satisfactory fit to the light curve (RMS=0.0141, about 3.5 times larger than the variable-temperature-ratio model), with large residuals (middle panel) near phases 0.0 and 0.5.  Enabling irradiation effects does not improve model.  Detached models fit the data less well than the variable-temperature-ratio contact models and require both components to have \Ronemax$>$1 and \Rtwomax$>$1, indicating that a contact configuration is preferred.  As a third scenario we tried a \Tratio=1 contact model with a hot spot on the primary star.\footnote{In all subsequent spotted models we assume for symmetry reasons that the spot is centered on the star's equator, spot co-latitude of 90\degr\ in the PHOEBE models.}  This model (green curve) produces a superior fit to the light curve (RMS=0.0029) while also matching the kinematic data for parameters $i$=77.0\degr, $f$=0.46, $q$=0.16, $l3$=0.025, $T_{\rm spot}$/$T_{\rm 1}$=1.03, $R_{\rm spot}$=39\degr, and spot co-longitude$_{\rm spot}$=178\degr (on the side of the primary away from the secondary). The lower panel in Figure~\ref{fig:061LCVC} plots the radial velocity curves for each model, illustrating that either of the \Tratio=1 models having $q$$\approx$0.2 produce acceptable fits to the kinematic data while the nominal best-fitting variable-temperature-ratio model yields a mass ratio that is approximately the inverse of the correct one.  

At the inclination and mass ratio of the spotted model, the velocity semi-amplitudes require $M_{\rm 1}$=1.30 \msun\ and $M_{\rm 2}$=0.23 \msun.   This best-fitting spotted model produces a line profile (green crosses in Figure~\ref{fig:061BF}) that matches the primary well in amplitude and radial velocity but slightly underpredicts the velocity amplitude of the secondary at both phases.  The peak of the secondary's BF is also slightly smaller than that predicted by the model line profile function.  Overall, the spotted model produces a satisfactory match between the model velocity profile and the BF data.  By contrast, both of the non-spotted models require substantial third light, $l$=0.39 and $l3$=0.44, respectively, which should show up as deficits in the model line profile function compared to the BF data. High-resolution echelle spectra of KIC08913061 \citet{Cook2022} show no evidence for third light, effectively ruling out the solutions from the first two models.   

The lack of a physical explanation for a hot spot on the primary led us to explore the possibility of a cool spot on the secondary.  Using two symmetrically placed cool spots with $T_{\rm spot}$/$T_{\rm 1}$=0.69 on the secondary near co-long$_{\rm spot}$=30\degr\ also produces a satisfactory fit to the light curve, however, the model line profile function still fails to match the secondary peak in the BF in detail and the light curve fit is not as good as when using the single hot spot.   

\begin{figure}[ht!]
%\fig{Spectra/KIC08913061/triangle_plot_6k_6th.pdf}{7in}{}
\fig{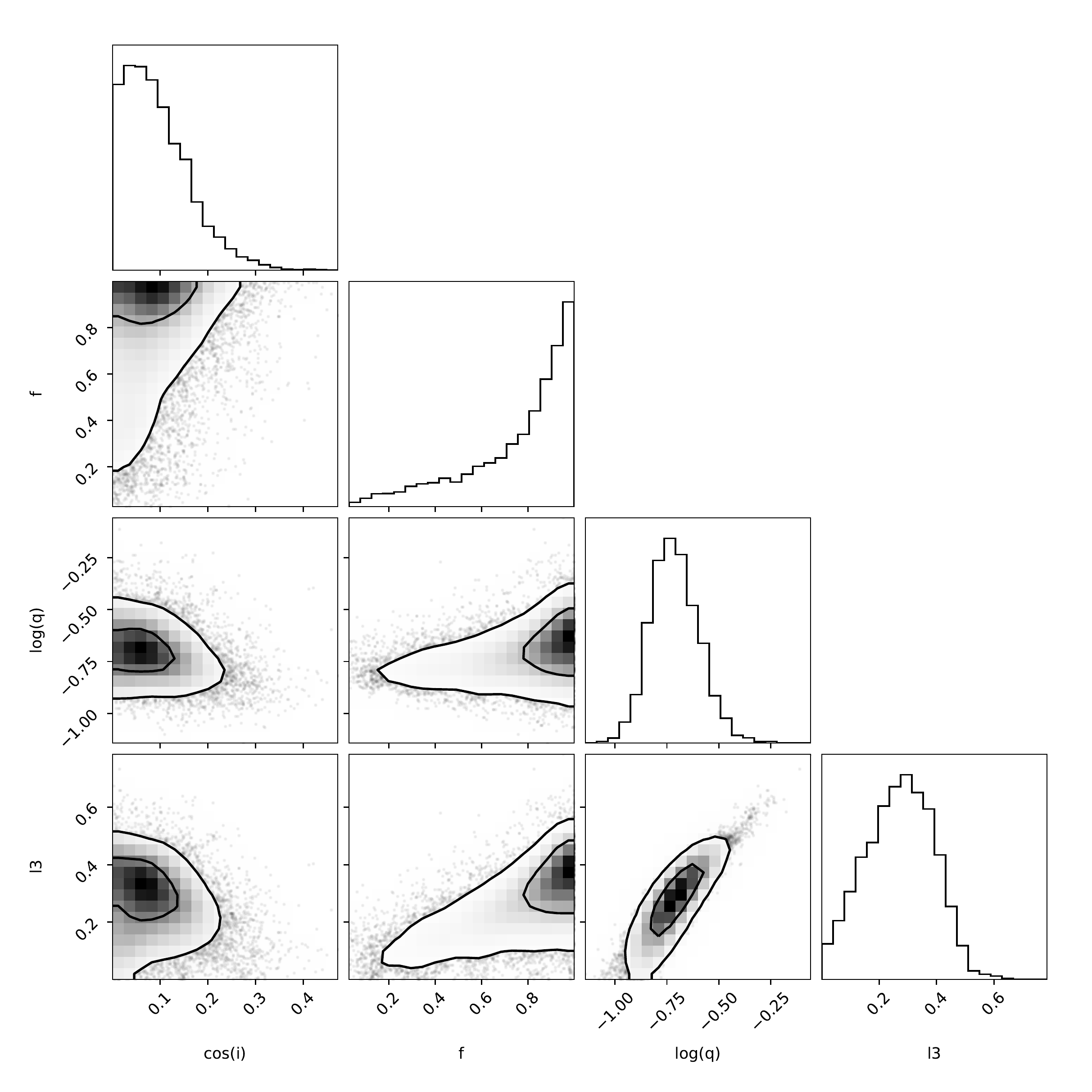}{5in}{}
\caption{Posterior probability distributions for combinations of four free system parameters for the KIC08913061 system when \Tratio=1.    \label{fig:061MCMCfix}}
\end{figure}

We ran MCMC simulations for four free parameters to understand the constraints afforded by the light curve alone.  Figure~\ref{fig:061MCMCfix} presents the distribution of parameters, showing that the key system parameters are reasonably well constrained and that the correct $q$ is identified.   The 16th/50th/84th percentile parameters are \cosi=0.027/0.083/0.160, $f$=0.51/0.84/0.96, \logq=$-$0.824/$-$0.718/$-$0.595, and $l3$=0.14/0.28/0.39.  The most probable $q$=0.19, very similar to the $q$$\approx$0.17 implied by the velocity data and by the best spotted model.  The inclination is high and the mass ratio is extreme.  The fillout factor is large, and there is a modest third light contribution.   

\begin{figure}[ht!]
%\fig{Spectra/KIC08913061/triangle_plot_6k_6th.pdf}{7in}{}
\fig{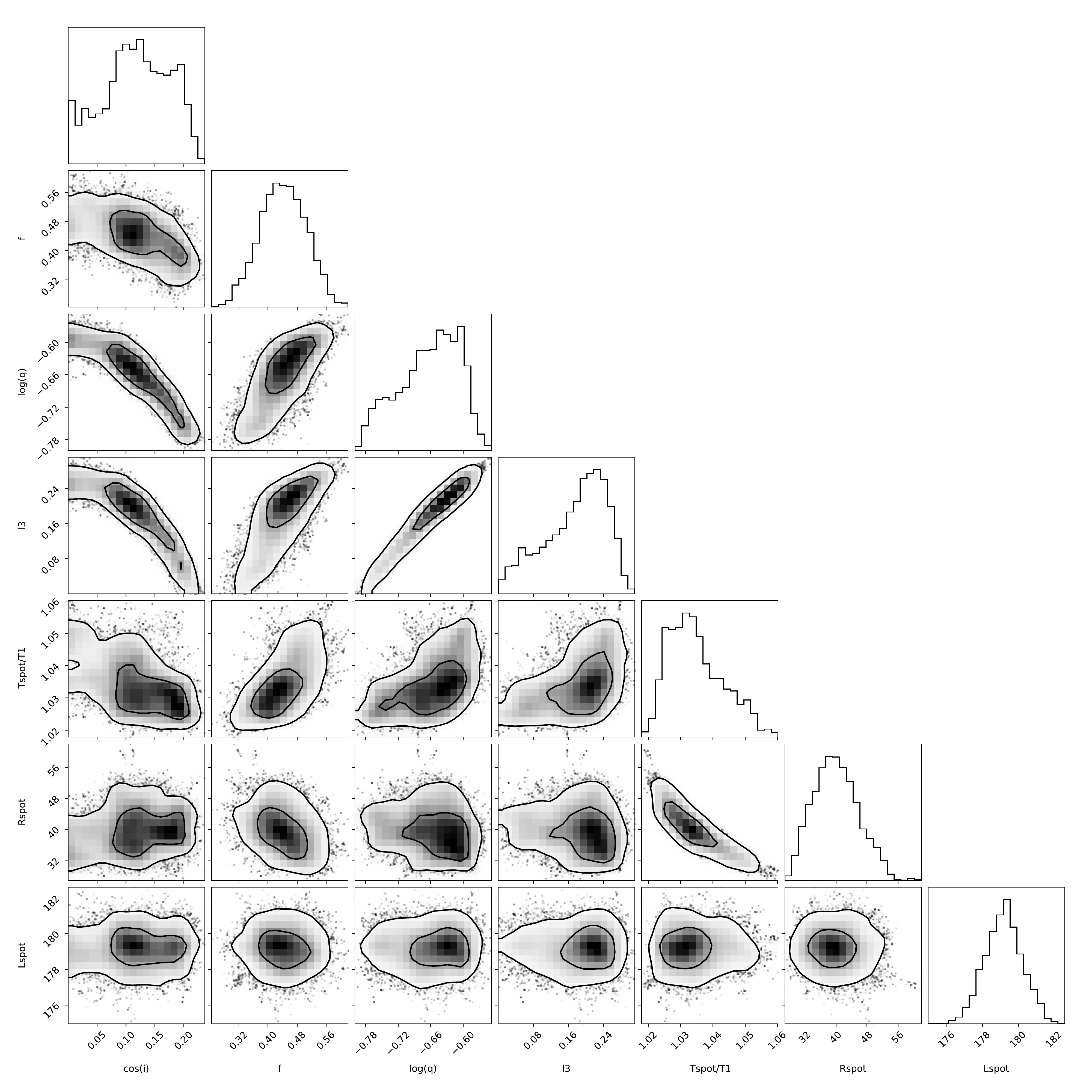}{7in}{}
\caption{Posterior probability distributions for combinations of seven free system parameters for the KIC08913061 system, including a hotspot on the primary.  There is considerable degeneracy among $i$, $q$, and $l3$, but the models still place strong constraints on the system parameters.   \label{fig:061MCMC}}
\end{figure}

We also ran MCMC simulations for a seven-parameter spotted system with \cosi, $f$, $q$, $l3$, $T_{\rm spot}/R_{\rm 1}$, $R_{\rm spot}$, and $L_{\rm spot}$ as free parameters.   Figure~\ref{fig:061MCMC} plots the posterior probabilities.  
The degeneracy between $T_{\rm spot}$/$T_{\rm 1}$ and $R_{\rm spot}$ is evident as a long banana-shaped shaded locus.  There is also considerable degeneracy between $i$ and $l3$ and between $q$ and $l3$---smaller third-light fractions demand smaller mass ratios. Mass ratios between 0.17 and 0.25 are probable.   However, the light curve alone is enough to show that extreme mass ratios near $q$=0.20 are most probable, consistent with the velocities of components measured in the broadening function, albeit with less precision.    The 50th-percentile values and 1$\sigma$ uncertainties are \cosi=0.128$^{+0.049}_{-0.127}$, $f$=0.578$^{+0.109}_{-0.089}$, \logq=$-$0.651$^{+0.063}_{-0.063}$, $l3$=0.139$^{+0.077}_{-0.089}$, $T_{\rm spot}$/$T_{\rm 1}$=1.046$^{+0.053}_{-0.018}$, $R_{\rm spot}$=36.7\degr$^{+16.9\degr}_{-13.9\degr}$, and co-long$_{\rm spot}$=178.6\degr$^{+1.3\degr}_{-1.8\degr}$.  Even with the extra free parameters for the spot, the MCMC simulations are still able to place strong constraints on nearly all of the key parameters.  In particular, the spot longitude is well-constrained. 

KIC08913061 is another long-period, extreme-$q$ system and a probable contact binary.  Despite having a light curve symmetric about $\phi$=0.5, the nominal fixed-temperature-ratio contact binary model was not able to produce a good match to the data without recourse to an additional component, modeled here either as a variable temperature ratio \Tratio=0.86 or using an {\it ad hoc} hot spot on the primary.  Hence, the real geometrical configuration in this system is ambiguous owing to the need for an extra physical component in the model. The physical origin for a hot spot on the primary opposite the secondary is not obvious, as it would be an unusual location for a mass transfer stream to impact.  The lack of agreement between the model line profile function and the BF at the secondary's velocity at {\it both} orbital phases indicates a remaining deficiency in the model; a larger fillout factor would improve agreement with the BF but at the cost of larger \redchi\ in the model light curve.  We are unable to entirely reconcile the best-fitting model light curve and the line profile function produced by this model with the BF.    

KIC08913061 serves as a warning that blind variable-temperature-ratio models fitted to single-band light curves can yield {\it grossly erroneous system parameters.}  In the case of KIC08913061 ($morph$=0.76), $q$ and \Tratio\ turn out to be degenerate in ways that that lead to an incorrect set of system parameters.  The situation becomes more ambiguous when additional physical components are present in the system (e.g., spots).   The correct $q$$\approx$0.2 solution, in fact, does constitute a local minimum in parameter space, but the global minimum with $q$$\approx$4 and \Tratio=0.86 is deeper.  Without {\it a priori} kinematic knowledge of the mass ratio, the correct mass ratio in KIC08913061 is only obtained by constraining the component temperature ratio to \Tratio$\approx$1.  However, fixing \Tratio=1 is likely to be overly constraining in the presence of real component temperature differences, resulting in (inappropriate) adjustments of other parameters as compensation for insufficient flexibility in \Tratio. 

\clearpage

\subsection{KIC09164694}

\begin{figure}[ht!]
%\fig{Spectra/KIC09164694/KIC09164694_phased_all_smaller.pdf}{4in}{}
\fig{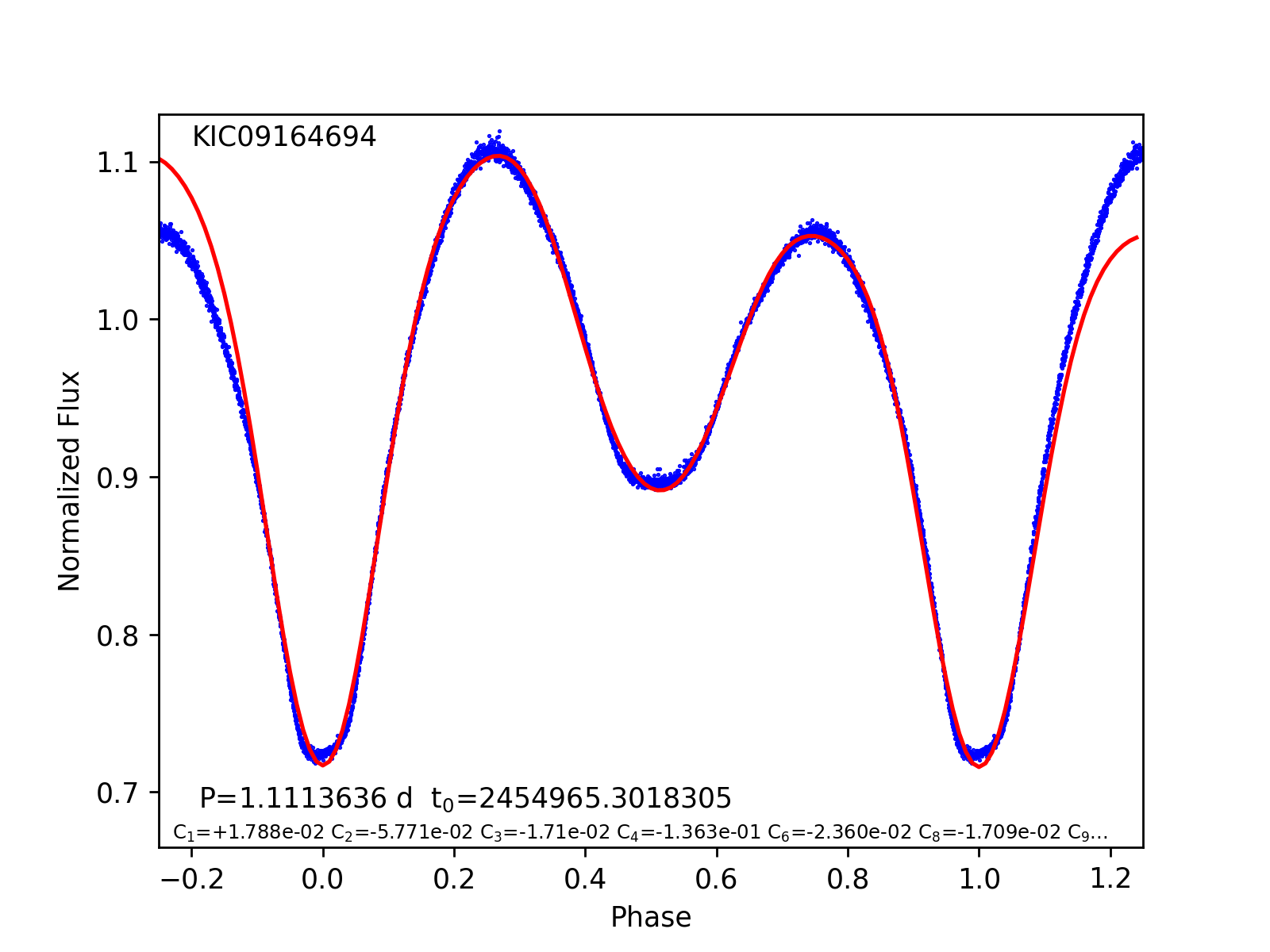}{5in}{}
\caption{Folded {\it Kepler} light curve of KIC09164694. \label{fig:694LC}}
\end{figure}

Figure~\ref{fig:694LC} displays the {\it Kepler} light curve of the $P$=1.11~d system KIC09164694, which shows  greater complexity than any previous examples.  Primary eclipse that is much deeper than secondary eclipse.  The maximum near $\phi$=0.25 is significantly brighter than that at $\phi$=0.75.  The red curve is constructed using the first nine Fourier components, as labeled in the figure.  Some dispersion about the mean light curve is apparent in a small fraction of the time baseline.  The minima have flattened bottoms, suggesting an eclipse of both components.

\begin{figure}[ht!]
%\fig{Spectra/KIC09164694/KIC09164694_O-C.pdf}{4in}{}
\fig{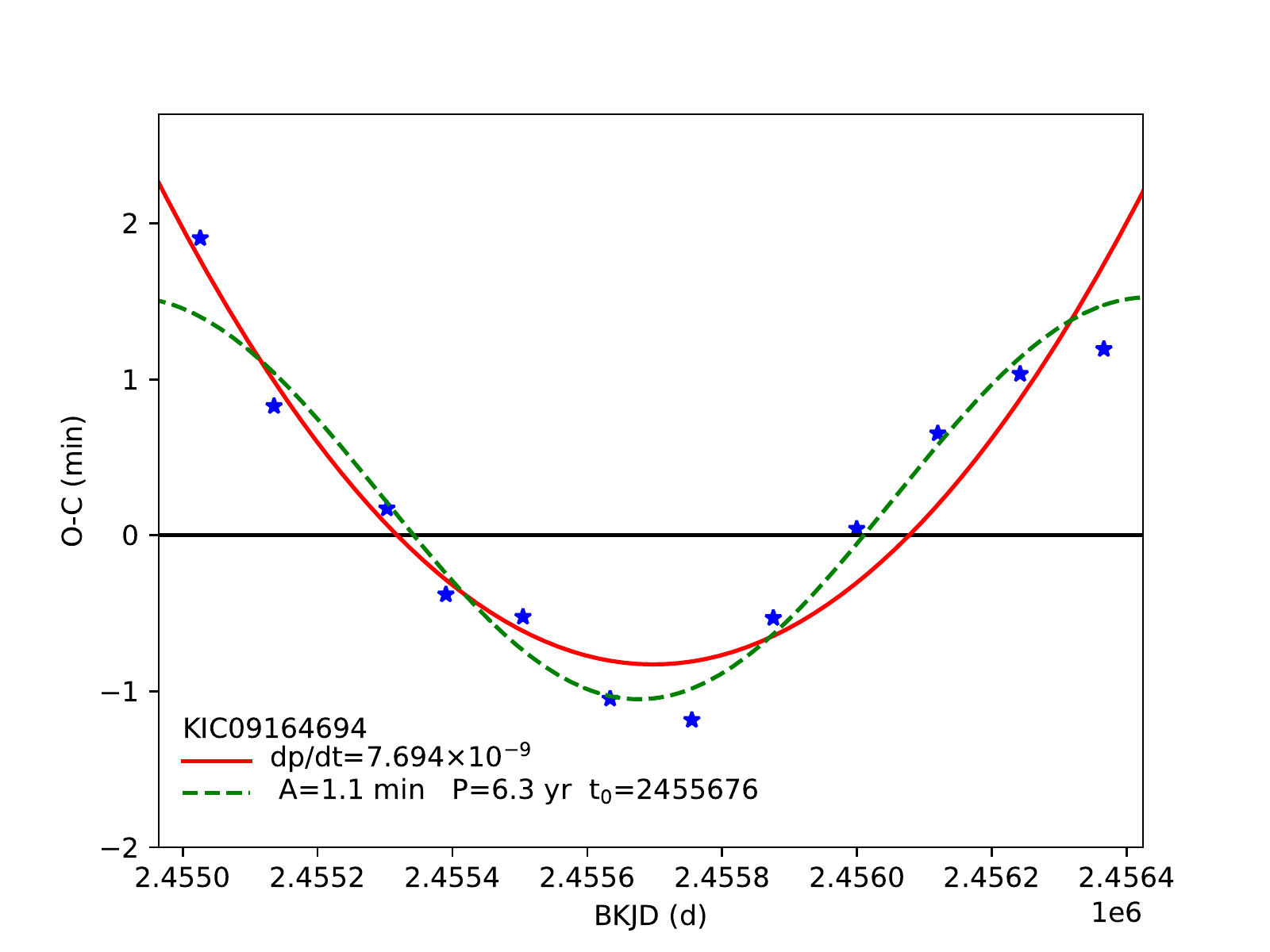}{4in}{}
\caption{\OC\ eclipse timing residuals of KIC09164694. The red solid and green dashed curves depict a parabolic and a sine fit to the data, respectively, either of which provide an approximate match.  \label{fig:694OC}}
\end{figure}

Figure~\ref{fig:694OC} presents the eclipse timing residuals, showing a systematic trend that is represented either by a parabolic function implying a period decrease $dP/dt$=7.7$\times10^{-9}$ or a periodic function (implying an orbit about a third body) with period $P$=6.3 yr, amplitude 1.1 minutes, and $t_{\rm 0}$=2455676 (time when the contact binary is at superior conjunction).  Given the limited time baseline of the {\it Kepler} dataset, it is not possible to distinguish between these possibilities, or a third scenario involving quasi-random orbital period modulations arising from magnetic cycles \citep[e.g.,][]{Applegate1992}.  

\begin{figure}[ht!]
%\fig{Spectra/KIC09164694/KIC09164694_twocomp.pdf}{4in}{}
\fig{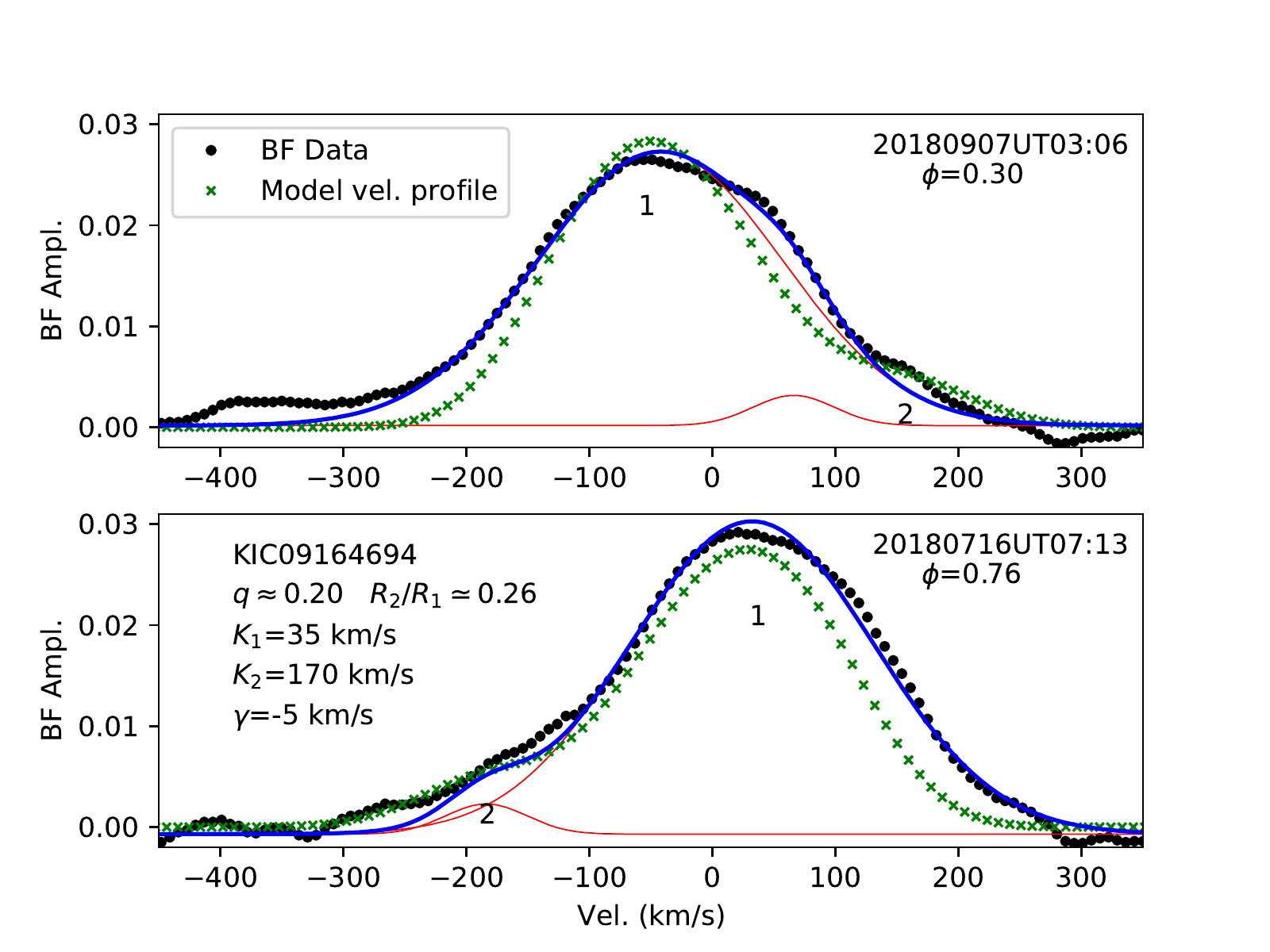}{4in}{}
\caption{Broadening Function of KIC09164694 showing at least two components. \label{fig:694BF}}
\end{figure}

Figure~\ref{fig:694BF} displays the broadening function of KIC09164694.  At $\phi$=0.30 it is asymmetric toward positive velocities and at $\phi$=0.76 the asymmetry lies on the negative side, suggesting a possible third component blended with the dominant component near the systemic velocity.  A secondary component---much fainter---is evident as well at $\pm$150 \kms.  We fit the BF with a two-component Gaussian function to measure velocities and relative luminosities, finding it necessary to constrain the velocity width of the fainter component to be 1/3 that of the brighter (140 \kms)\footnote{This is motivated by the constraint that, for a binary synchronous rotation, the rotational velocity scales with the radius.}, but allowing the centers and normalizations to vary independently.  This leads to a radius ratio of $R_{\rm 2}$/$R_{\rm 1}$$\simeq$0.26, assuming that the temperatures are similar---an assumption that will be relaxed in the analysis that follows.   The ratio of velocity semi-amplitudes indicates a mass ratio $q$$\simeq$0.20.  Because of the BF component blending, the velocities and relative areas are more uncertain than in previous examples where the components are better-separated. The velocity of the fainter component is considerably uncertain at both quadrature phases.

\begin{figure}[ht!]
%\fig{KB/IndivSpectra/KIC09164694/KIC09164694LC+VCnew.png}{5in}{}
\fig{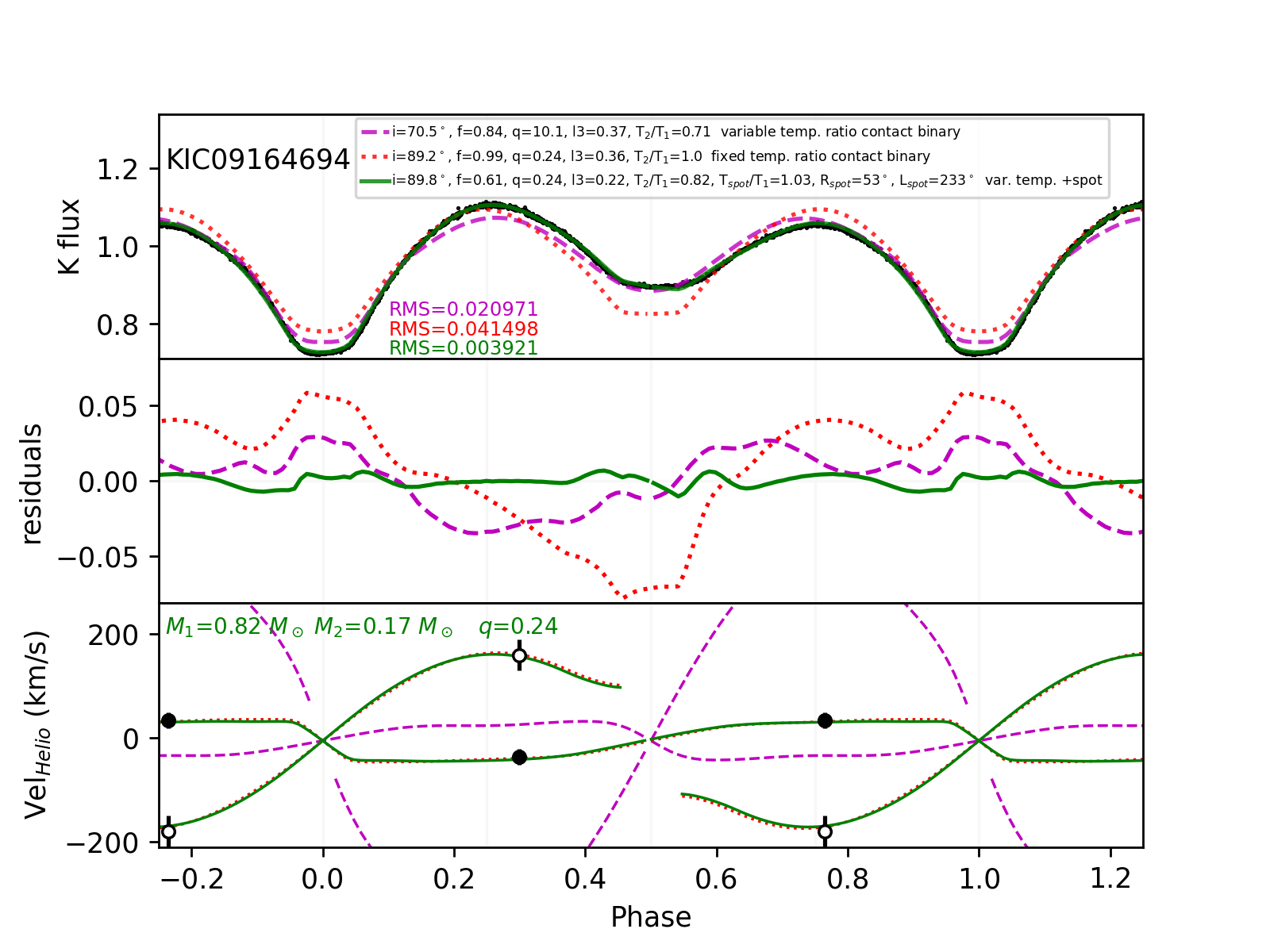}{5in}{}
\caption{Light curve and velocity curve of KIC09164694 along with three competing model light curves and velocity curves.  Both the \Tratio=1 model and the spotted model reproduce the observed component velocities, while the best-fitting variable-temperature-ratio model yields a mass ratios that is inconsistent with the kinematic data.  \label{fig:694LCVC} }
\end{figure}

Figure~\ref{fig:694LCVC} shows light curves, residuals, and velocity curves of KIC09164694 for three competing models.  The light curve is complex, requiring a great degree of fine-tuning to reproduce in detail. The flattish minima require high inclinations.  However, such large inclinations lead to very deep minima---much deeper than observed---a problem that can be remedied by including a modest third-light contribution.  The brighter maximum just after $\phi$=0.25 (when the primary component is approaching) compared to just before $\phi$=0.75 requires the introduction of an asymmetry in the system.  This can be achieved by placing hot spot on the leading face of the primary or the trailing face of the secondary.  We adopt the former, admittedly without physical motivation.  Placing the spot on the larger star allows a less extreme T$_{\rm spot}$/$T_{\rm 1}$ ratio and a smaller spot size than if the spot were placed on the secondary.   

Figure~\ref{fig:694LCVC} shows that the nominal variable-temperature ratio model (magenta dashed curve) yields a reasonable fit (RMS=0.0209) for $i$=70.5, $f$=0.84, $q$=10.1, $l3$=0.37, and an extreme ratio \Tratio=0.71.  Best-fitting detached models have a much larger RMS and require that both components overfill their Roche lobes, so they are not considered further.  Fixed-temperature-ratio models (red dotted curve) produce an RMS that is twice as large, but for very different system parameters: $i$=89.2, $f$=0.99, $q$=0.24, $l3$=0.36.  Only the latter model is consistent with the mass ratio obtained from the kinematic data ($q$$\approx$0.2).  A much better fit (RMS=0.0039) is obtained with the spotted variable-temperature-ratio model (green curve) for $i$=89.6, $f$=0.61, $q$=0.24, $l3$=0.22, \Tratio=0.82, $T_{\rm spot}$/$T_{\rm 1}$=1.03, $R_{\rm spot}$=53\degr,  and $L_{\rm spot}$=233\degr\ (roughly on the leading face of the primary).  At the cost of three additional free parameters, the reduction in RMS for this model is dramatic.  Even so, the best-fitting inclination and mass ratio are nearly identical to the fixed-temperature-ratio model. Reassuringly, the mass ratios of both of these latter models agree with the BF---the spotted model and the \Tratio=1 model both reproduce the observed velocities of the components at quadrature phases.  Including irradiation effects changes the best fits negligibly.

\begin{figure}[ht!]
%\fig{KB/phoebe/KB/Tfixfudgechi2/09164694/09164694_triangle_plot_2k_9th_Tfixfudgechi2.pdf}{7in}{}
\fig{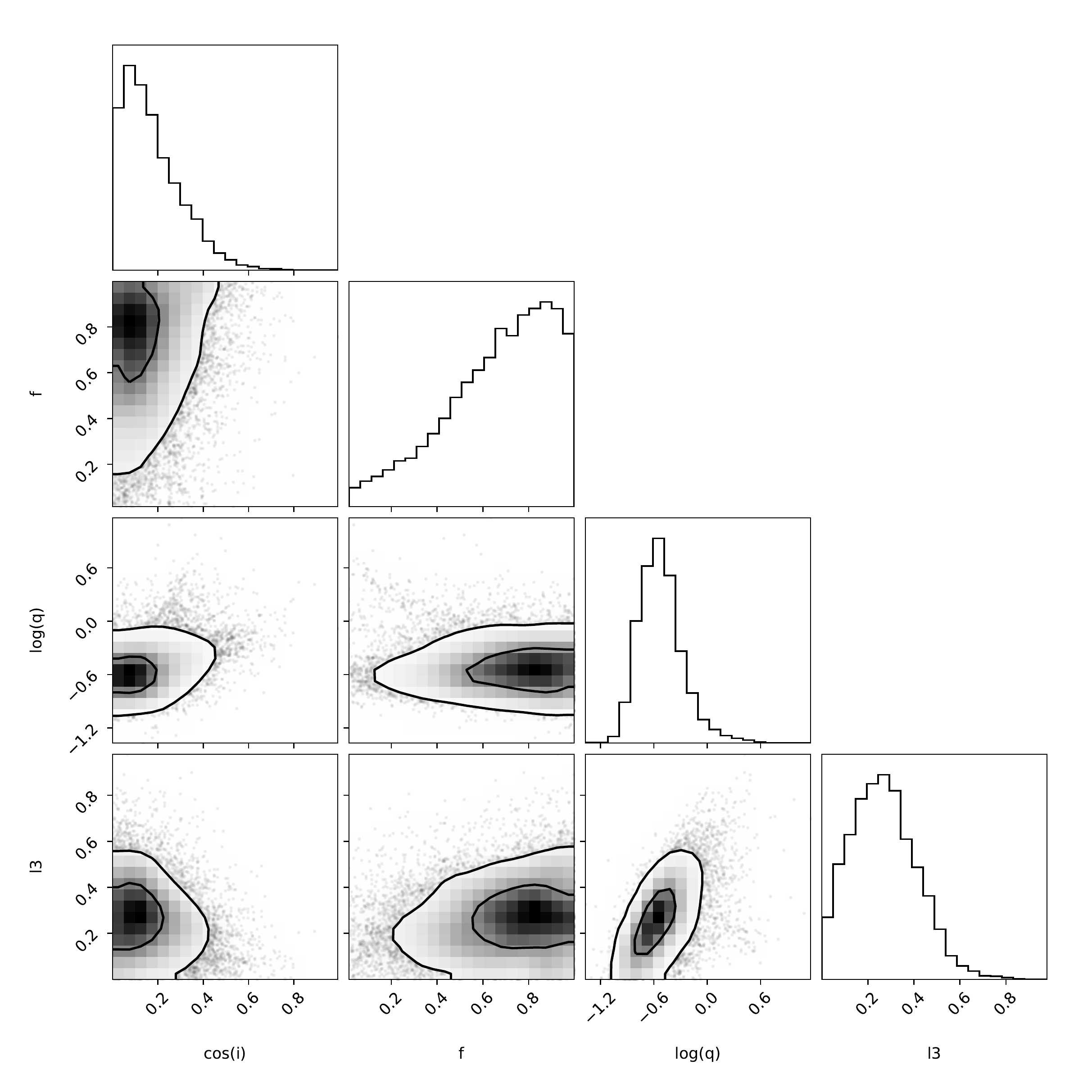}{5in}{}
\caption{Probability distribution for combinations of four free system parameters for the KIC09164694 system.  Despite the asymmetric light curve modeled here as a spot with three free parameters, the constraints on mass ratio and inclination are strong.   \label{fig:694MCMC}}
\end{figure}

Figure~\ref{fig:694MCMC} displays the PHOEBE+MCMC posterior probabilities of a four-parameter fixed-temperature-ratio model. The percentile parameters are $i$=0.053/0.147/0.304, $f$=0.42/0.71/0.90, \logq=$-$0.755/$-$0.543/$-$0.301, and $l3$=0.12/0.26/0.41.  Despite the asymmetric light curve requiring an additional phenomenological feature (a spot) to reproduce, the inclination and mass ratio are well constrained at values consistent with the kinematic data.  A large range of fillout factor and third light are possible.  

\begin{figure}[ht!]
%\fig{KB/phoebe/KB/Tvarfudgechi2/09164694/09164694_triangle_plot_2k_9th_Tvarfudgechi2spot.pdf}{7in}{}
\fig{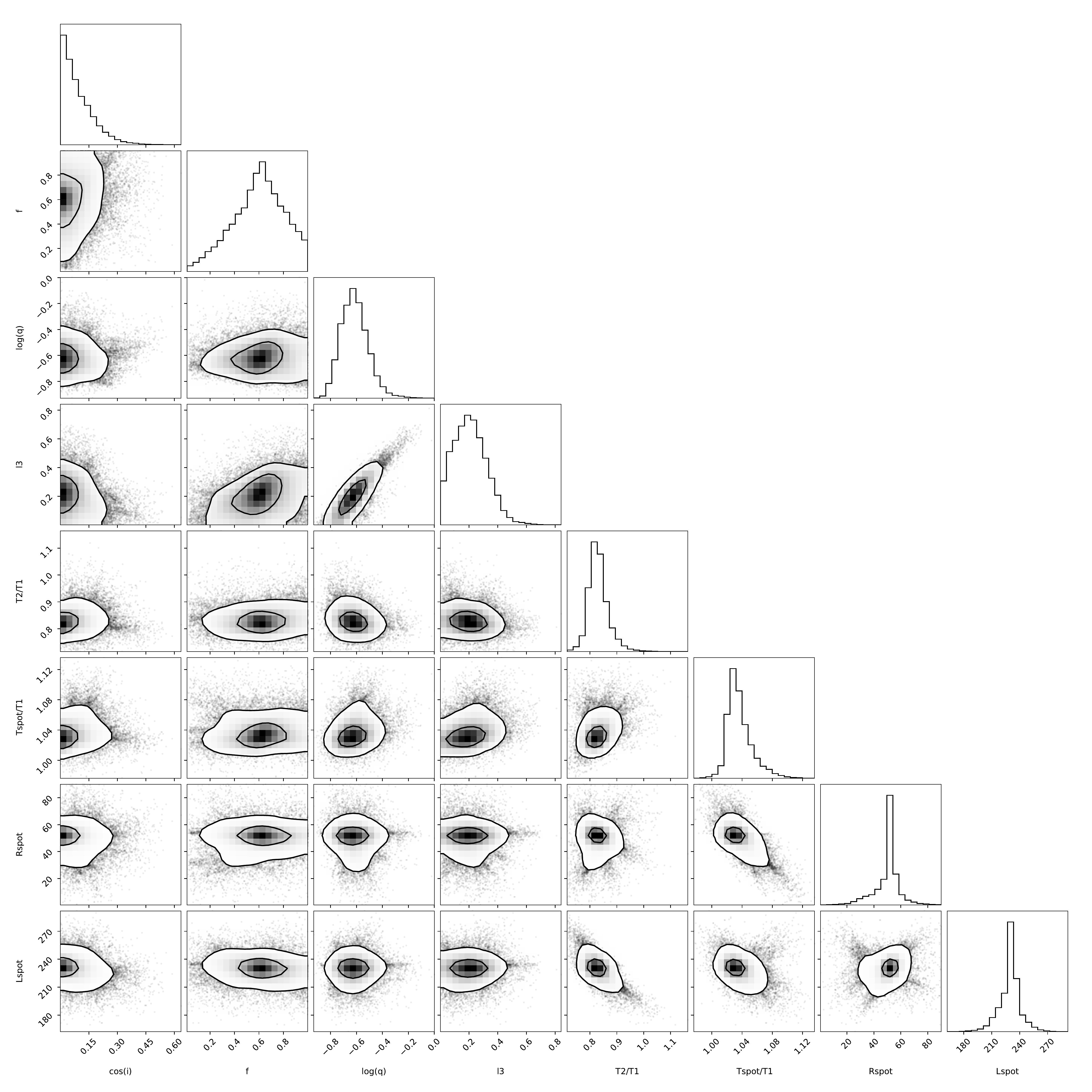}{7in}{}
\caption{Probability distribution for combinations of eight free system parameters for the KIC09164694 system. Even with the additional free parameters entailed by a spot and $T_{\rm 2}<T_{\rm 1}$, the Monte Carlo simulations provide a well-constrained measure of the $q$ that is consistent with the spectroscopic data. \label{fig:694MCMCspot}}
\end{figure}

We also performed MCMC simulations of the full eight-parameter model (variable temperature ratio plus a spot) to understand the extent to which system parameters can be constrained in the presence of additional physical features in the system.  Figure~\ref{fig:694MCMCspot} presents the Bayesian distribution.  As above, the most probable inclinations and mass ratios are consistent with the kinematic data and the allowed range of other parameters remain consistent with the non-spotted model.  There is considerable degeneracy between the spot radius and temperature, but other parameters are well-constrained. Bayesian analysis of KIC09164694 illustrates that the addition of a single physical feature (a spot with three free parameters) greatly improves the model fit and simultaneously provides tight constraints on the principal system parameters and spot parameters. 

KIC09164694, like many systems in the parent sample, has a complex asymmetric light curve ($morph$=0.75) that is not well-fit using a standard four-parameter contact binary model with \Tratio=1.  The five-parameter variable-temperature-ratio model yields a best fit with a {\it wrong} mass ratio ($q$=10 versus the correct $q$$\approx$0.24 measured from the BF), further illustrating the degeneracy between $q$ and \Tratio\ revealed above in the case of KIC08913061. Although there is strong evidence for being a contact binary, the large disparity in retrieved component temperatures \Tratio=0.82 (even with the inclusion of a hot spot on the primary!) seems implausible for a true contact system.  We regard this system as an ambiguous geometry based on the present data.  The presence of some third light indicated by all models is consistent with the \OC\ eclipse timing residuals.  Although a third component comprising $\sim$0.2 of the system luminosity is consistent with the BF, but such a stationary component may easily be masked by the brighter primary component.  At $q$$\approx$0.24, KIC09164694 is another extreme-mass-ratio (near-)contact binary on the long-$P$ tail of the period distribution.    

\clearpage

\subsection{KIC09345838}

\begin{figure}[ht!]
%\fig{Spectra/KIC09345838/KIC09345838_phased_all_smaller.pdf}{4in}{}
\fig{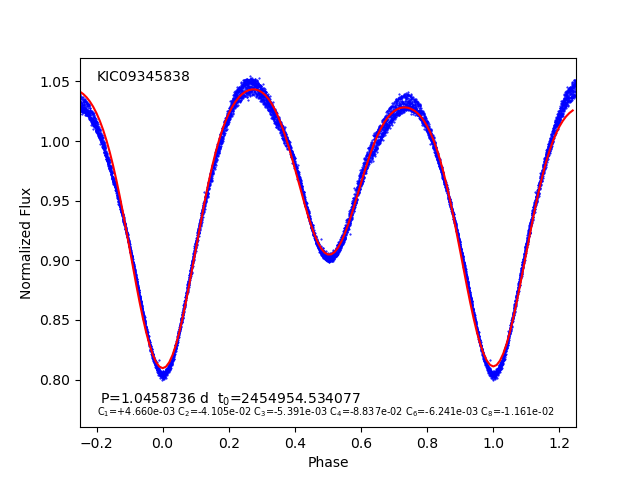}{5in}{}
\caption{Folded {\it Kepler} light curve of KIC09345838 and mean light curve.  \label{fig:838LC}}
\end{figure}

Figure~\ref{fig:838LC} displays the {\it Kepler} light curve of the $P$=1.04~d contact system KIC09345838.  Like KIC09164694, primary minimum is much deeper than secondary minimum and the $\phi$=0.25 maximum is brighter than the $\phi$=0.75 maximum.  The eclipse timing residuals show no systematic variation over the baseline of the {\it Kepler} mission.   

The BF of KIC09345838 is asymmetric with a positive wing at $\phi$=0.32 and a negative wing at $\phi$=0.73, indicating one dominant component and one much less luminous component.  The radial velocity of the fainter component is particularly uncertain at both quadrature phases owing to its low amplitude.  The component velocities lead to semi-amplitudes of $K_{\rm 1}$=32 \kms\ and  $K_{\rm 2}$=163 \kms.  These imply a mass ratio near $q$$\approx$0.20, with considerable uncertainty given the difficulty in measuring the fainter component.  
\begin{figure}[ht!]
%\fig{Spectra/KIC09345838/KIC09345838_twocomp.pdf}{4in}{}
\fig{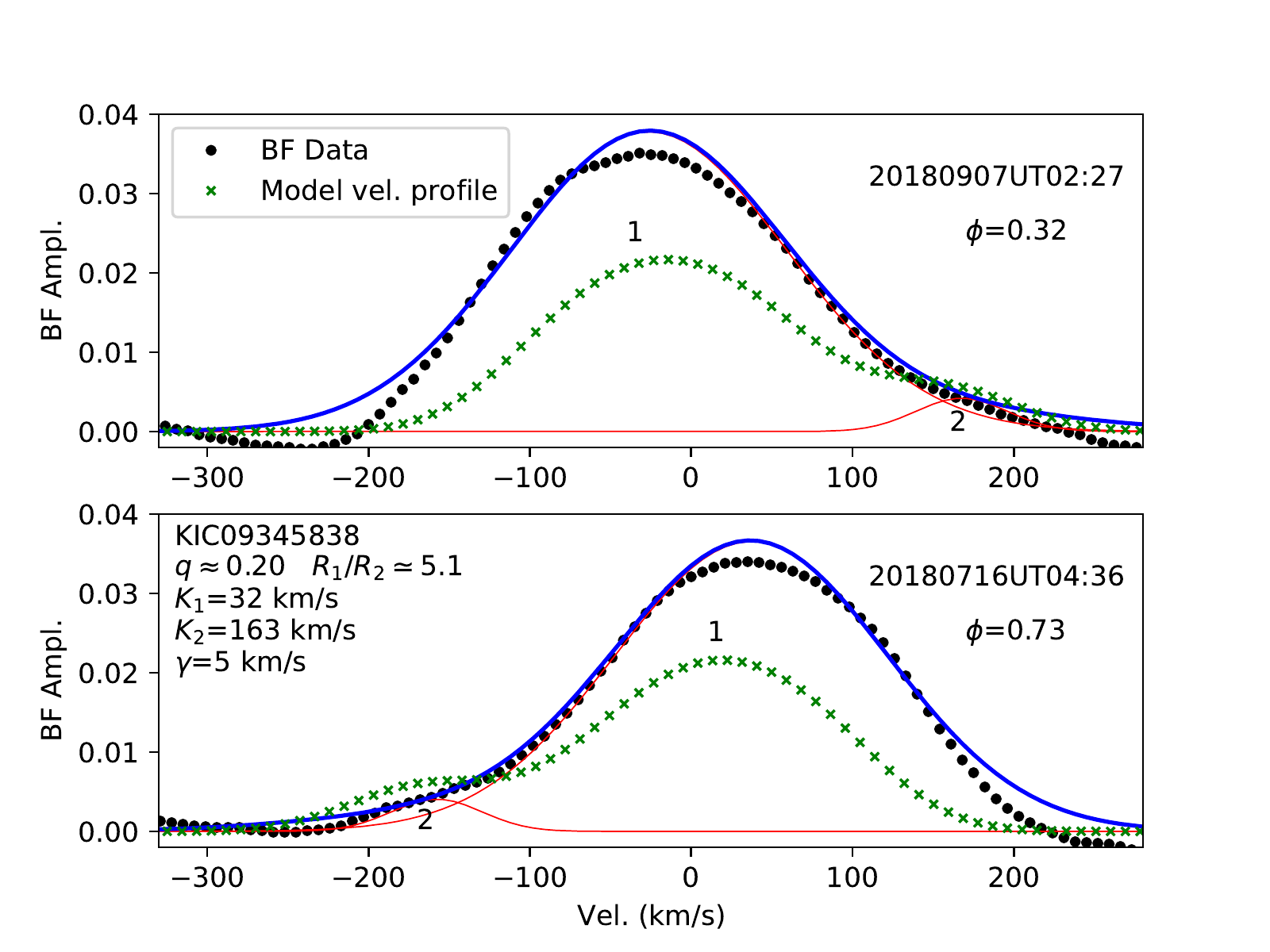}{4in}{}
\caption{Broadening Function and model line profile function of KIC09345838. \label{fig:838BF}}
\end{figure}

\begin{figure}[ht!]
%\fig{Spectra/KIC09345838/KIC09345838LC+VC_smaller.pdf}{4in}{}
\fig{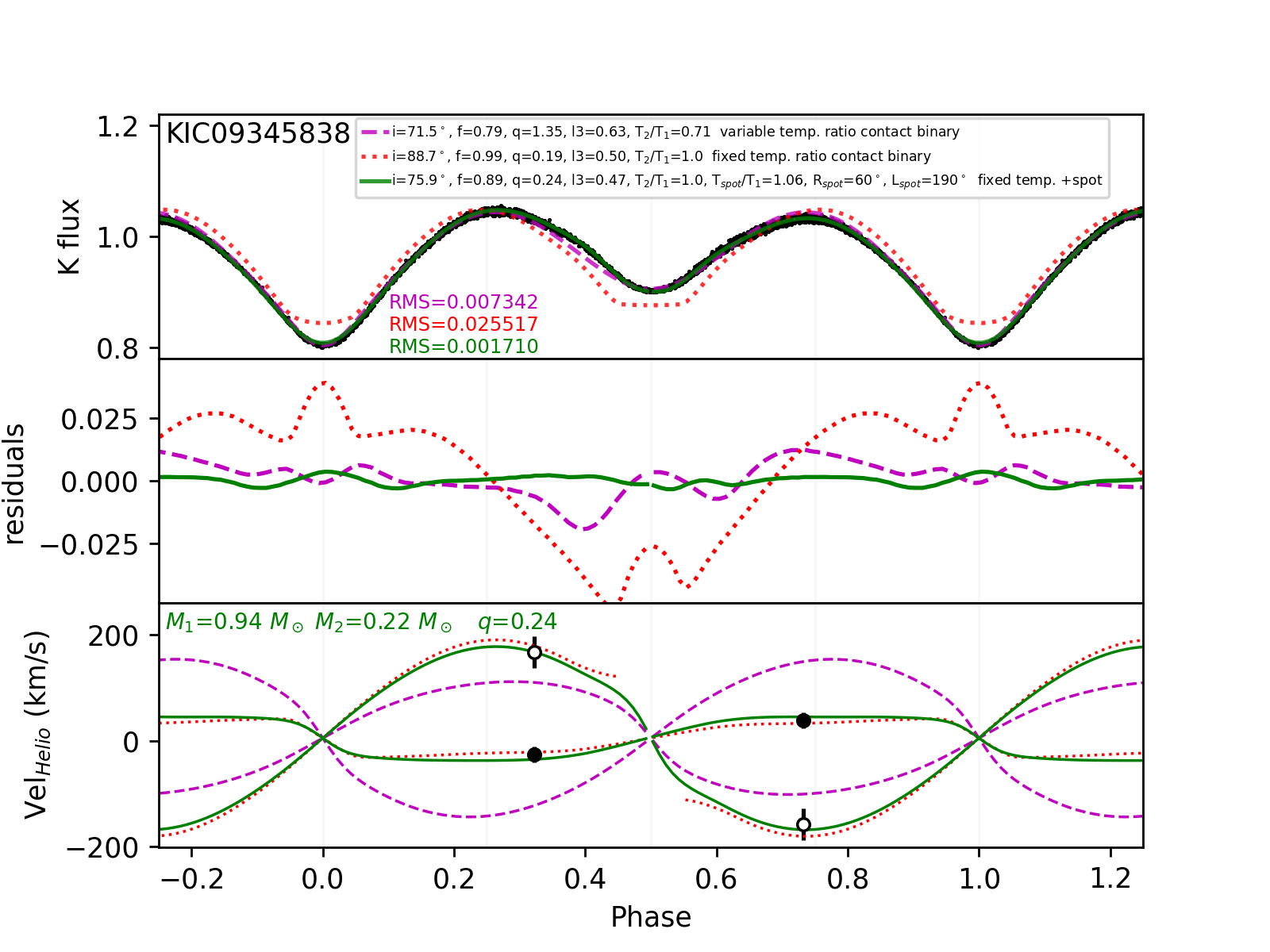}{5in}{}
\caption{Three competing contact binary models for KIC09345838 in comparison to the data.  As in KIC09164694, the variable-temperature-ratio model yields an incorrect mass ratio, while the fixed-temperature-ratio models yield $q$$\approx$0.24, in agreement with the kinematic data.  \label{fig:838LCVC} }
\end{figure}

Figure~\ref{fig:838LCVC} displays the light curves and velocity curves of three competing contact models in comparison to the data.  The results mirror those seen previously with KIC09164694.  Detached models are ruled out, as they all require maximum radii for both components that exceed their respective Roche limits.   A variable-temperature-ratio model (magenta dashed curve) provides a reasonable fit (RMS=0.0073) for parameters $i$=71.5\degr, $f$=0.79, $q$=1.35 (inconsistent with the kinematic data), $l3$=0.63, and \Tratio=0.71.  The great disparity between primary and secondary minumum drives this large temperature difference. A fixed-temperature-ratio mode yields a significantly worse fit (RMS=0.0255) for a larger inclination ($i$=89\degr), $f$=0.99, a vastly lower mass ratio ($q$=0.19), and $l3$=0.50. Only the latter model is consistent with the kinematic data in Figure~\ref{fig:838BF} ($q$$\approx$0.2. A \Tratio=1 spotted model provides a superior match to the data (RMS=0.017) with $i$=75.9\degr, $f$=0.89, $q$=0.24, $l3$=0.47, $T_{\rm spot}$/$T_{\rm 1}$=1.06, $R_{\rm spot}$=60\degr, and $L_{\rm spot}$=190\degr (on the primary opposite the secondary). Either of the models with equal-temperature components correctly predicts the mass ratio.  The modest $l3$ in these models is also consistent with the deficit seen in the model BF (green x's) relative to the data, which indicates substantial third light centered near the primary's velocity.               

\begin{figure}[ht!]
%\fig{/d/zem1/hak/chip/research/Larry/KB/phoebe/KB/Tfixfudgechi2/09345838/09345838_triangle_plot_2k_9th_Tfixfudgechi2.pdf}{6in}{}
\fig{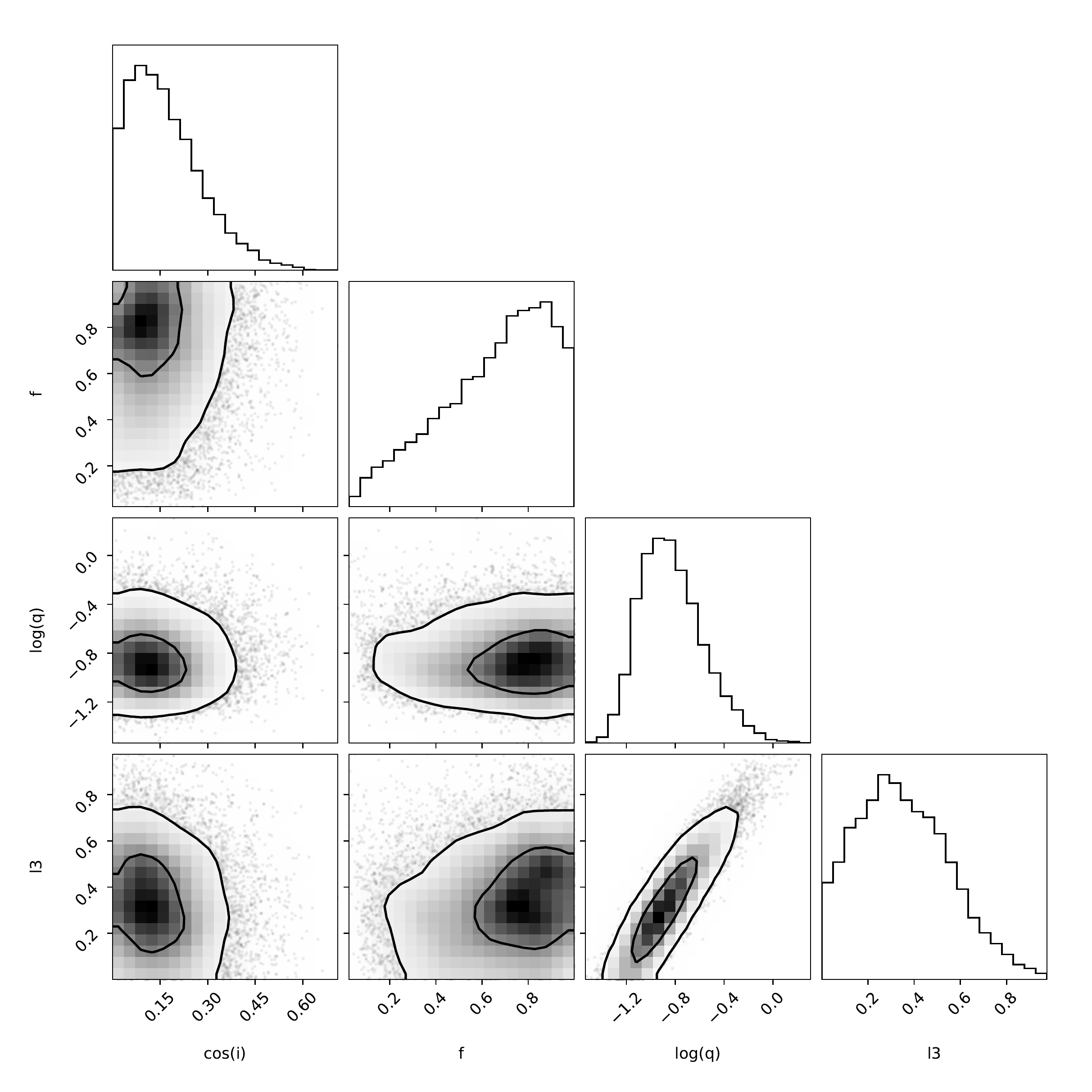}{5in}{}
\caption{Posterior probability distributions for combinations of four free parameters in the KIC09345838 system.  The most probable $q$ of 0.14 provides a good match to the kinematic data. Like KIC09164694 there is some degeneracy between $q$ and $l3$.   \label{fig:838MCMC}}
\end{figure}

Figure~\ref{fig:838MCMC} shows the posterior parameter probabilities resulting from Monte Carlo retrievals for four free parameters: \cosi=0.054/0.149/0.281, $f$=0.39/0.70/89, \logq=$-$01.08/$-$0.858/$-$0.569, and $l3$=0.14/0.33/0.56.  The most probable third-light fractions near $l3$=0.33 are consistent with the (needed but unmodeled) contribution to the overall line profile that would reconcile the theoretical profile with the BF in Figure~\ref{fig:838BF}.  Despite some degeneracy between \logq\ and $l3$, the mass ratio and inclination are well-constrained and the fillout factor is large.   

Figure~\ref{fig:838MCMC} shows the posterior parameter probabilities for a seven-parameter \Tratio=1 spotted model.  All parameters except $f$ are well-constrained.  The allowed ranges overlap with those from the simpler but poor-fitting four-parameter models.  
\begin{figure}[ht!]
%\fig{/d/zem1/hak/chip/research/Larry/KB/phoebe/KB/Tfixfudgechi2/09345838/09345838_triangle_plot_2k_9th_Tfixfudgechi2spot.pdf}{6in}{}
\fig{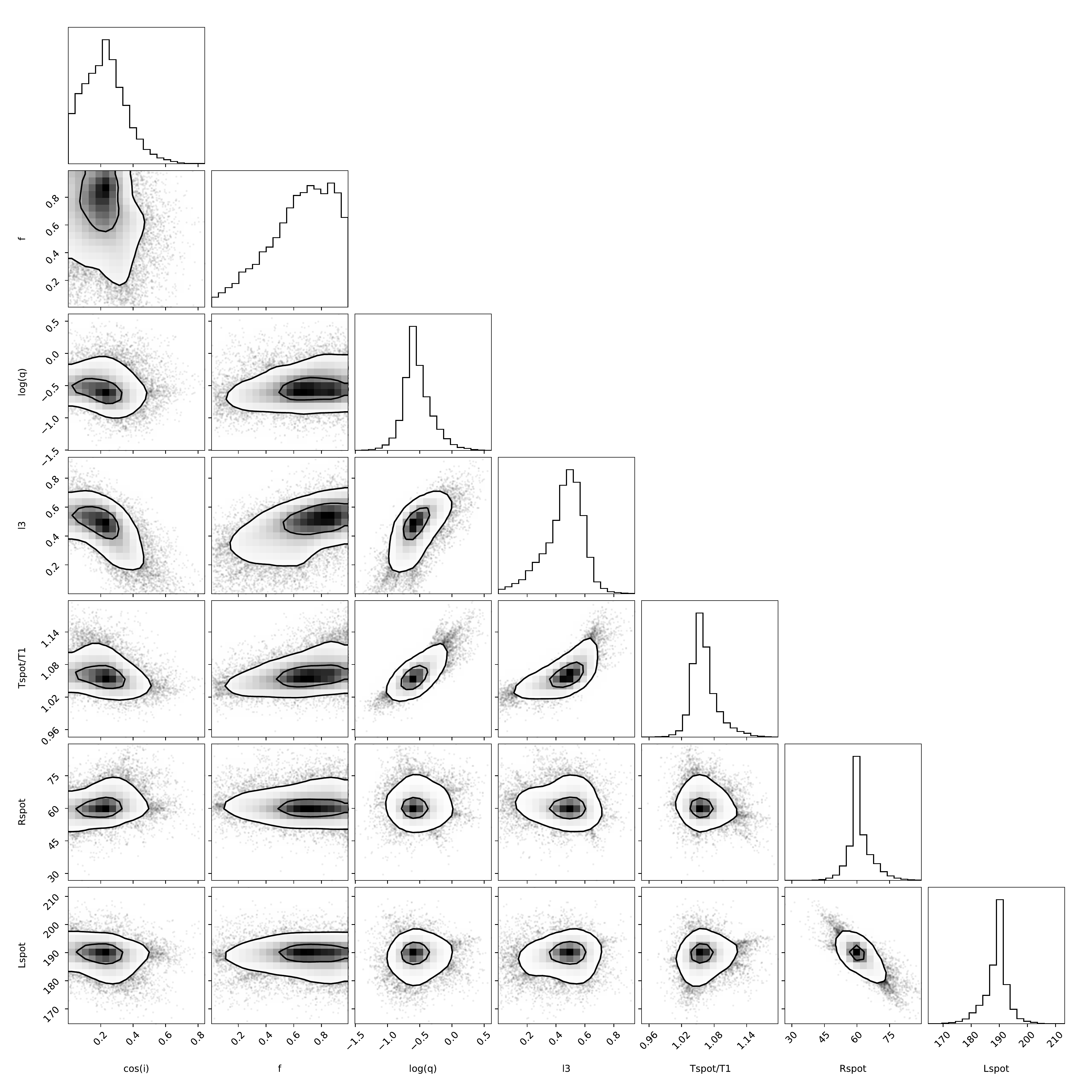}{5in}{}
\caption{Posterior probability distributions for combinations of  seven parameters in the KIC09345838 system when a hot spot on the primary is included.   \label{fig:838MCMCspot}}
\end{figure}

KIC09345838 is another long-period extreme-$q$ system ($morph$=0.75) where the rather extreme mass ratio is constrained on the basis of the light curve alone, even with the additional model complexity of a stationary spot.  However, like KIC08913061 and KIC09164694, the flexibility of the variable-temperature-ratio model leads to an incorrect mass ratio, while a \Tratio$\approx$1 model recovers a mass ratio $q$$\approx$0.2, consistent with the kinematic data.

\clearpage

\subsection{KIC09840412}
\begin{figure}[ht!]
%\fig{Spectra/KIC09840412/KIC09840412_phased_all_smaller.pdf}{4in}{}
\fig{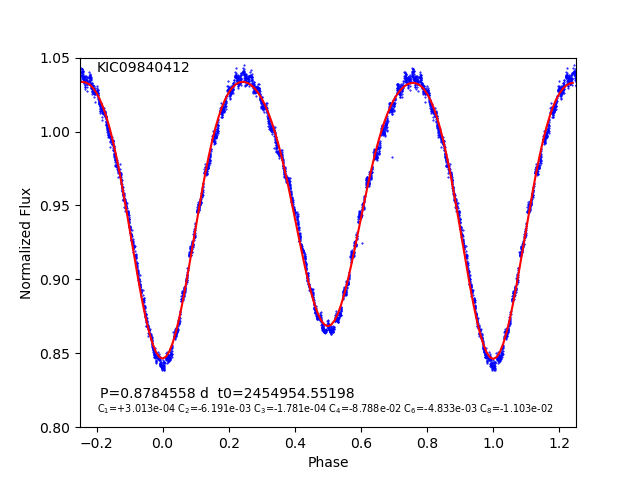}{5in}{}
\caption{Folded {\it Kepler} light curve of KIC09840412. \label{fig:412LC}}
\end{figure}

Figure~\ref{fig:412LC} displays the {\it Kepler} light curve of the $P$=0.88~d contact system KIC09840412.  Like some previous examples, the primary minimum is deeper than secondary minimum.  The two maxima are similar in brightness.  Unlike recent examples, there is relatively little dispersion about the mean light curve.

\begin{figure}[ht!]
%\fig{Spectra/KIC09840412/KIC09840412_O-C+ztf.pdf}{4in}{}
\fig{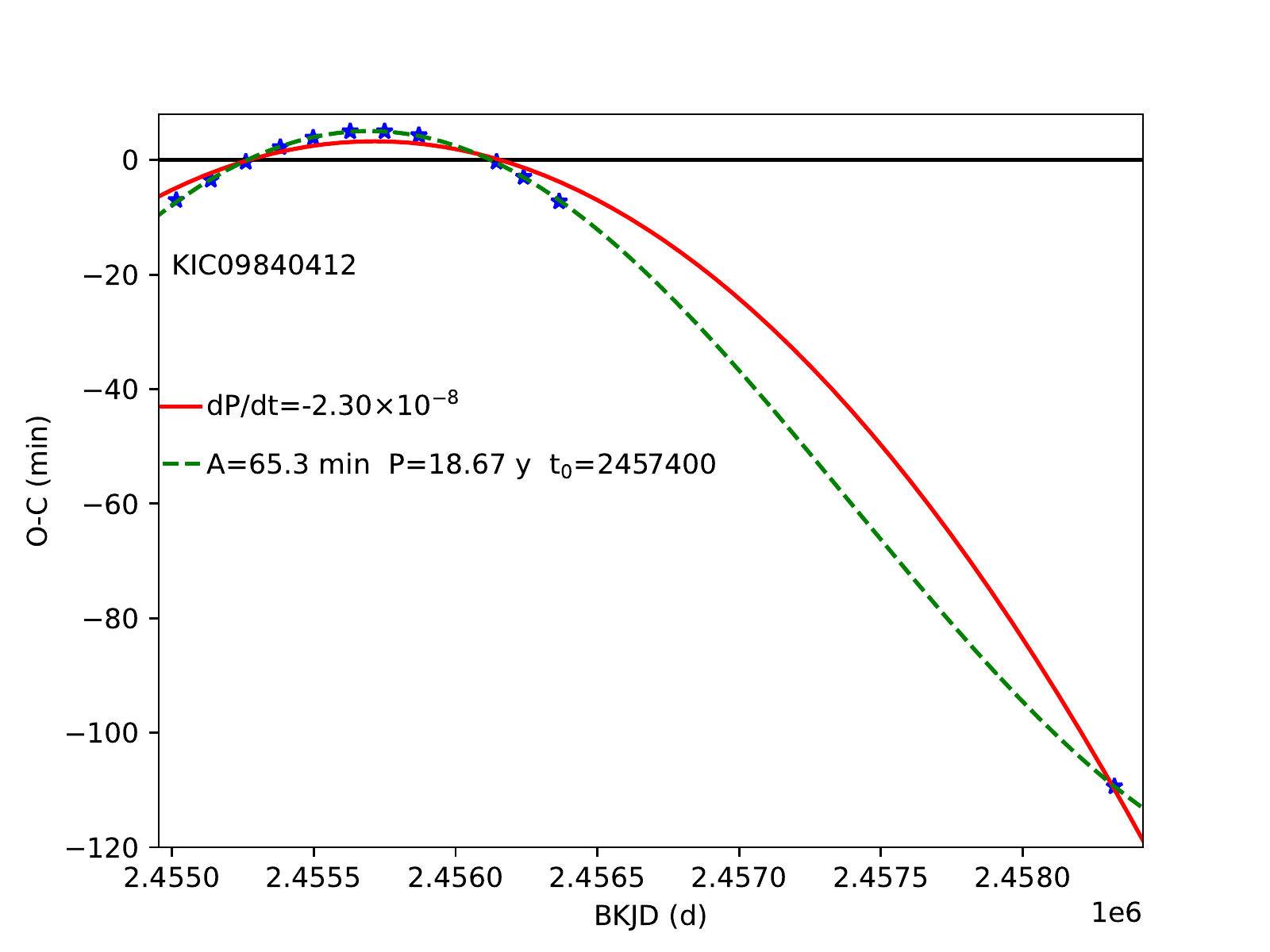}{4in}{}
\caption{O-C eclipse timing residuals of KIC09840412, including one epoch of recent ZTF photometry. \label{fig:412OCztf}}
\end{figure}

Figure~\ref{fig:412OCztf} presents the \OC\ diagram including {\it Kepler} epochs and recent 2018--2019 photometry from the Zwicky Transient Factory \citep{ZTF2014}.  The parabolic fit under the assumption of a constant period derivative (red curve) fits the data less well than the sinusoidal fit (dashed green curve), which provides an excellent match.  The best-fitting sine curve parameters indicate a semi-amplitude of 65.3 min, equivalent to a projected semi-major axis $a_1$sin($i$) for the contact binary about the system barycenter of 7.8 AU.  The best-fitting time of superior conjunction (when the contact binary is farthest away and the third component is nearest) is $t_{\rm 0C}$=2457400.  The best-fitting period is 18.67 years.  For an assumed total mass of the inner contact binary of 1.6 \msun, the minimum implied mass for the unseen third component in a circular orbit would be  3.0 \msun---sufficiently large and bright for any main-sequence star that it should be the dominant spectral signal in the broadening function (which may be the case, as discussed below).  This lower mass limit also rules out white dwarfs and ordinary stellar mass neutron stars as companions.   

\begin{figure}[ht!]
%\fig{Spectra/KIC09840412/KIC09840412_twocomp.pdf}{4in}{}
\fig{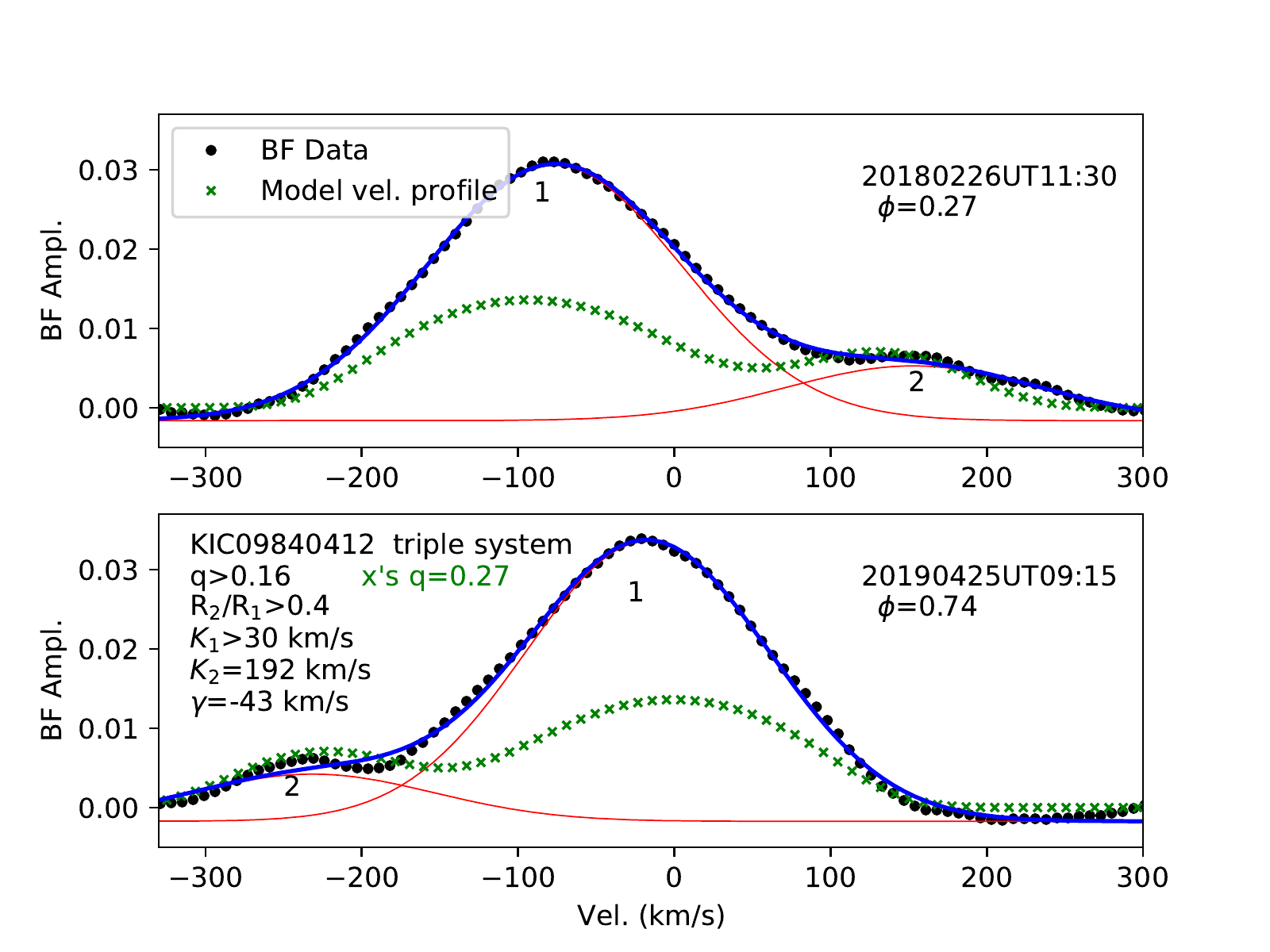}{4in}{}
\caption{Broadening Function and model line profile function of KIC09840412.  The discrepancy between the BF and the model line profile (green x's) suggests substantial third-light contribution, an inference also supported by the light curve modeling.   \label{fig:412BF}}
\end{figure}

Figure~\ref{fig:412BF} shows the broadening function for two epochs of spectroscopy obtained near phases $\phi$=0.27 and $\phi$=0.74, assuming a quadratic ephemeris with $P$=0.8784799~d and $dP/dt$=$-$3.268$\times10^{-8}$ and $t_{\rm 0}$=2454954.556810.\footnote{A linear ephemeris with the average period of $P$=0.87845608~d would lead to phases $\phi$=0.14 and $\phi$=0.56, which are inconsistent with the two clearly separated components in the broadening function.} Two components are visible, one much more luminous than the other.   The estimated velocity semi-amplitudes are $K_{\rm 1}$=30 \kms\ and $K_{\rm 2}$=192 \kms\ at a systemic velocity of $\gamma$=$-$43 \kms, yielding $q$$\approx$0.16.  The ratio of component areas in BF implies $R_{\rm 2}$/$R_{\rm 1}\approx$0.4, but this is too large if third light is present.  The dominant peak in the BF likely contains a luminous third star, meaning that the velocity amplitude of the primary star in the contact binary is underestimated, making the mass ratio a lower limit.  The plotted green x's from the best-fitting light curve model are for $q$=0.27.  

\begin{figure}[ht!]
%\fig{Spectra/KIC09840412/KIC09840412LC+VC_smaller.pdf}{4in}{}
\fig{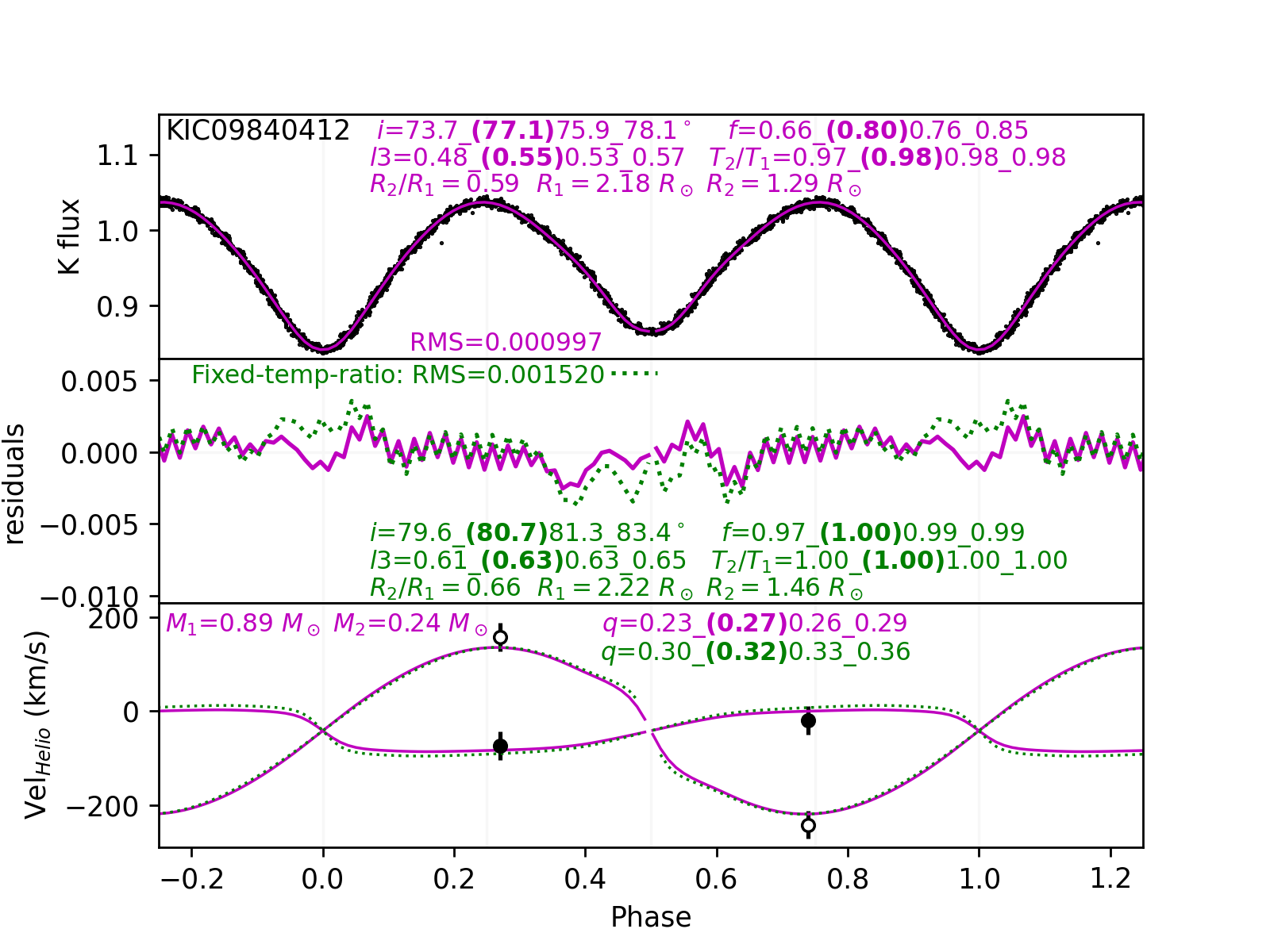}{5in}{}
\caption{Light curves and velocity curves of KIC09840412, along with best-fitting contact binary models. \label{fig:412LCVC} }
\end{figure}

The best-fitting variable-temperature-ratio contact model without irradiation (magenta dashed curve) shown in Figure~\ref{fig:412LCVC} has RMS=0.0010 for $i$=77.1\degr, $f$=0.80, $q$=0.27,  $l3$=0.55, and \Tratio=0.98.  The corresponding fixed-temperature-ratio model (RMS=0.0015) is very similar at  $i$=80.7\degr, $f$=0.99, $q$=0.32,  $l3$=0.63, leading to $R_{\rm 2}$/$R_{\rm 1}$=0.66, $R_{\rm 1}$=2.22 \rsun,  $R_{\rm 2}$=1.46 \rsun.  Both are consistent with the BF results for $q$ and for $l3$.  Models implementing irradiation effects have a similar slightly worse RMS but similar best-fitting parameters.  Best detached models have a much larger RMS and are not considered further. Adopting the best model $i$ and $q$ and the observed $K_{\rm 2}$=192 \kms\ yields $M_{\rm 1}$=0.89 \msun\ and $M_{\rm 2}$=0.24 \msun.   The model velocity curve in the lower panel of Figure~\ref{fig:412LCVC} shows the the BF data underestimate the amplitude of the primary, ostensibly because of the bright tertiary that comprises $\approx$63\% of the light and is blended near zero velocity with $M_{\rm 1}$.  The presence of the third component is unsurprising, given the eclipse timing variations.    

\begin{figure}[ht!]
%\fig{KB/phoebe/KB/Tvarlimfudgechi2/09840412/09840412_triangle_plot_1k_9th_Tvarlimfudgechi2.pdf }{5in}{}
\fig{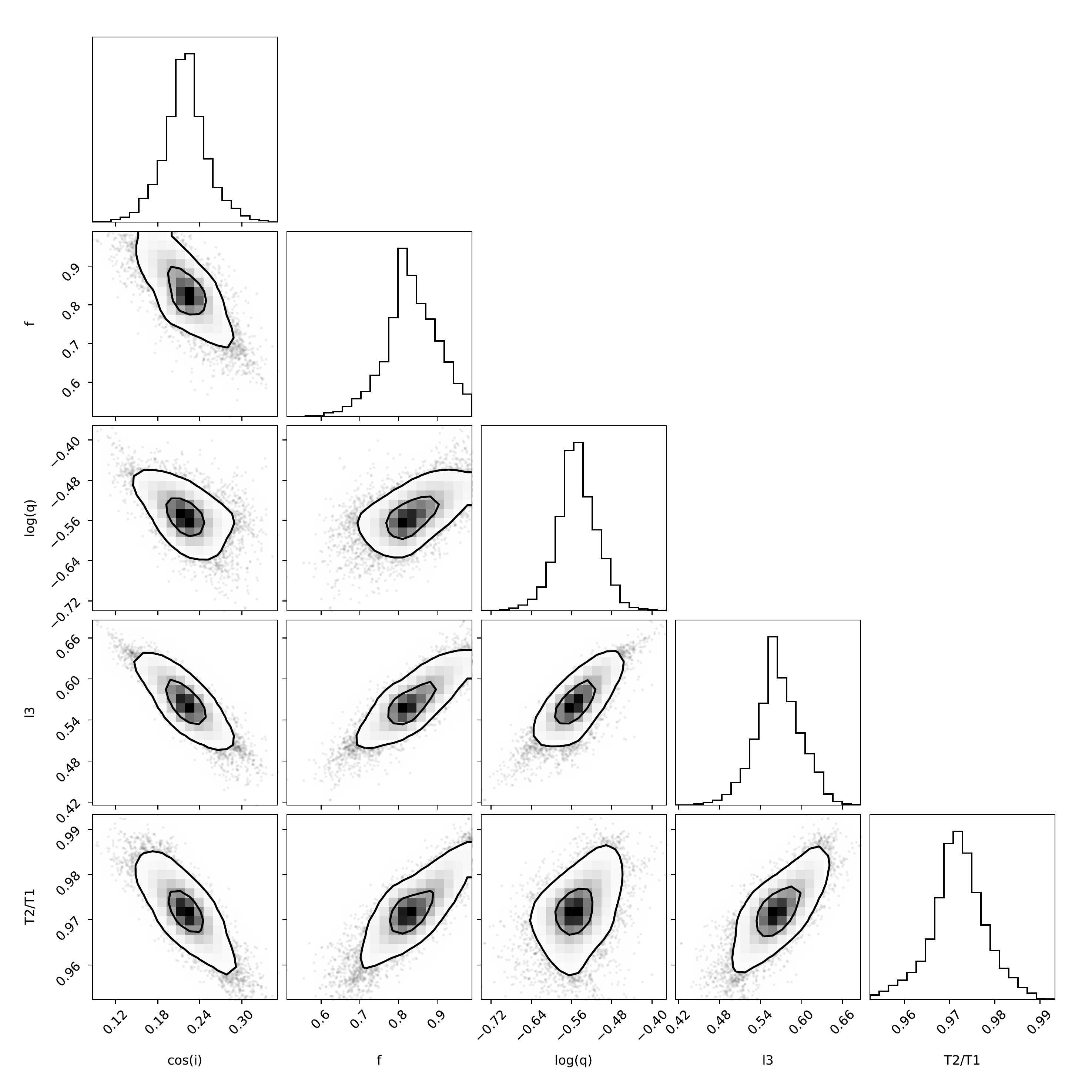}{5in}{}
\caption{Posterior probability distribution for combinations of five parameters in the KIC09840412 system.  \label{fig:412MCMC}}
\end{figure}

Figure~\ref{fig:412MCMC} (left) shows the relative probabilities resulting from Monte Carlo simulations using the \Tapprox\ model for KIC09840412.  The 16th/50th/84th percentile parameters are \cosi=0.190/0.219/0.247, $f$=0.77/0.83/0.90, \logq=$-$0.584/$-$0.549/$-$0.509,  $l3$=0.53/0.56/0.60 and \Tratio=0.967/0.972/0.977.  KIC09840412 is a good example of how third light can be degenerate with both \logq\ and $i$.  The necessity of third-light is corroborated by the strength of the primary peak in the BF in Figure~\ref{fig:412BF} relative to the model line profile curve.  The radial velocity of the primary is also offset from the peak of the BF, in the direction consistent with the third star having a velocity near the system's center of mass.  The discrepancy is consistent with the third star providing about 60\% of the system light.   Nevertheless, on the basis of the light curve alone the principal system parameters are well-constrained, as illustrated by Figure~\ref{fig:412MCMC}.  Compared to the other long-period contact binaries discussed previously, the most probable mass ratio of $q$=0.31 appears rather modest.  

KIC09840412 is a clear case of a contact binary ($morph$=0.79) with nearly equal-temperature components.  Variable-temperature-ratio models provide a somewhat better fit and produce a slightly shifted locus of system parameters.  Bayesian retrieval of system parameters are consistent with the kinematic data.  

\clearpage

\subsection{KIC09953894}
\begin{figure}[ht!]
%\fig{/d/zem1/hak/chip/research/Larry/KB/IndivSpectra/KIC09840412/KIC09953894_phased_all.png}{4in}{}
\fig{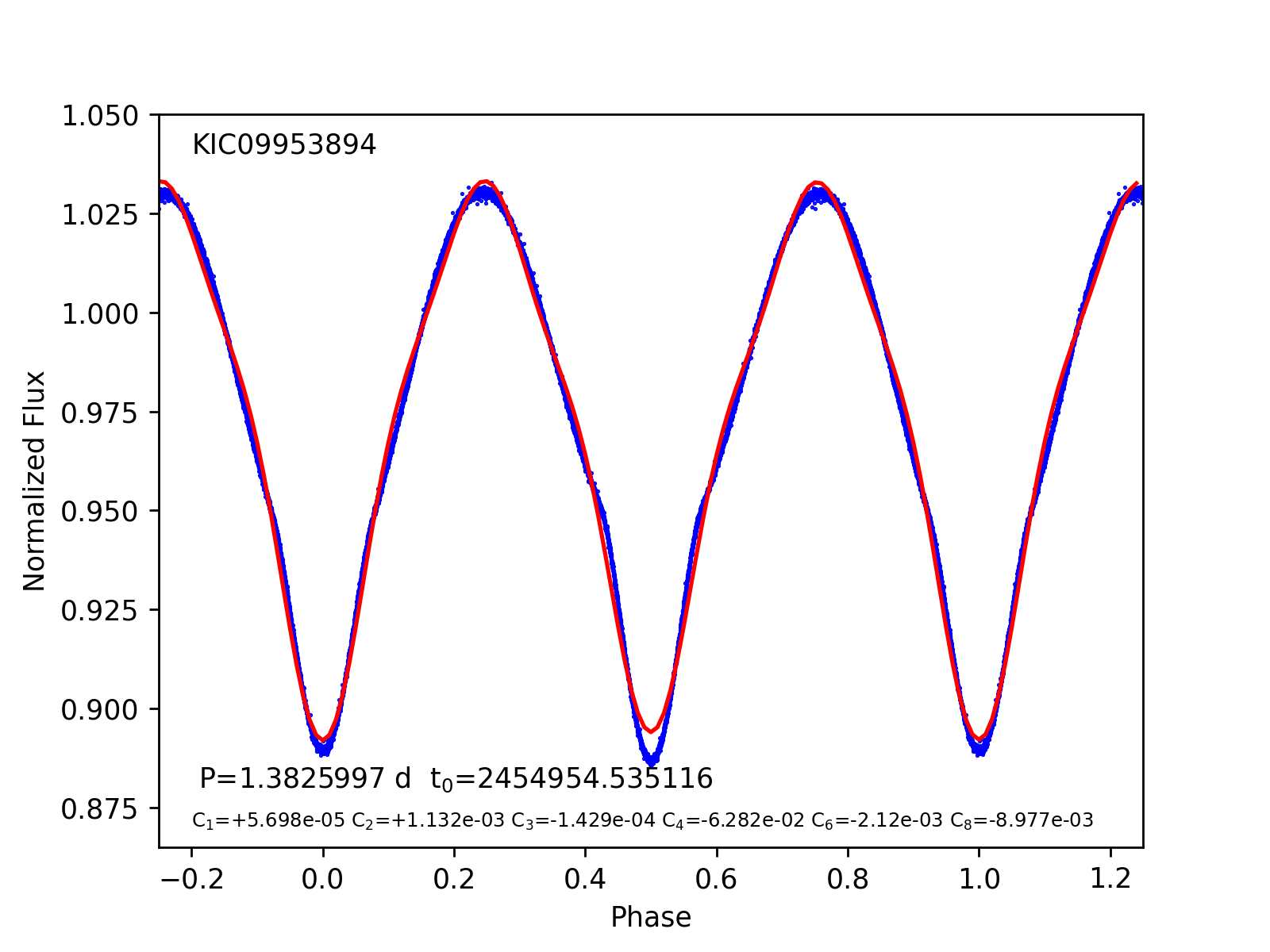}{4in}{}
\caption{Folded {\it Kepler} light curve of the detached system KIC09953894. \label{fig:894LC}}
\end{figure}

Figure~\ref{fig:894LC} displays the {\it Kepler} light curve of the $P$=1.38~d system KIC09953894, displaying both both strong ellipsoidal modulation and eclipses.   Eclipse timing residuals are $<$0.2 minutes and show no systematic pattern, indicating that a linear ephemeris is adequate.   The light curve morphology parameter of 0.74 is the lowest in our pilot sample and consistent with the possibility of a detached system, indicated also by the V-shaped inflection near primary and secondary minimum.

\begin{figure}[ht!]
%\fig{/d/zem1/hak/chip/research/Larry/KB/IndivSpectra/KIC09953894/KIC09953894_twocomp.pdf}{4in}{}
\fig{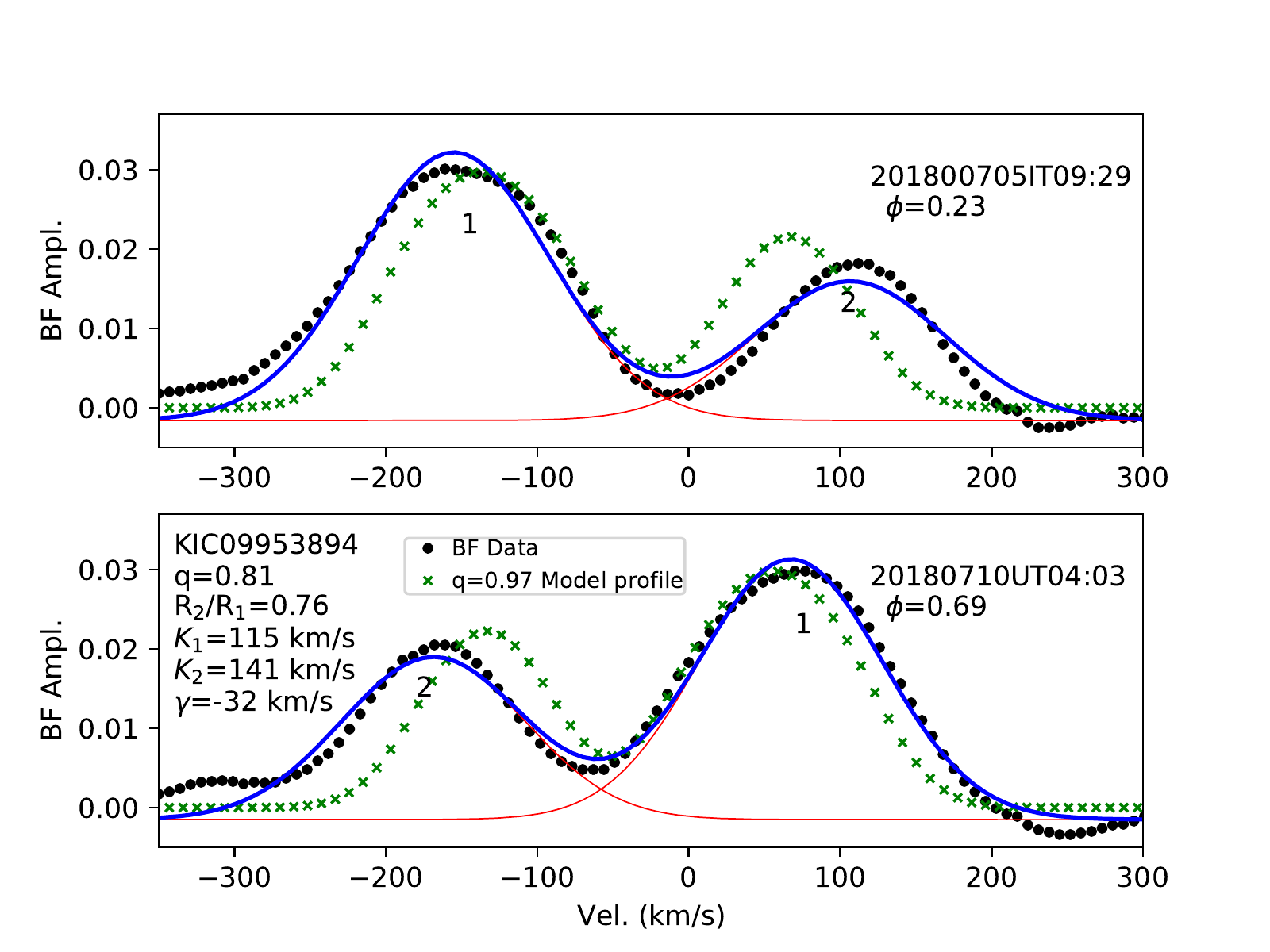}{4in}{}
\caption{Broadening Function and model line profile of KIC09953894. The green x's are for the best-fitting detached model including irradiation which significantly underpredicts the velocity semi-amplitudes unless the spectroscopically determined $M_{\rm 1}$=1.3 \msun\ is used in place of the default $M_{\rm 1}$=1 \msun. \label{fig:894BF}}
\end{figure}

Figure~\ref{fig:894BF} shows the broadening function of KIC09953894 for two epochs of spectroscopy obtained at phases $\phi$=0.23 and $\phi$=0.69.  Two well-separated components indicate a primary that is more luminous than the secondary.  Component velocities are well-measured and yield  $K_{\rm 1}$=115 \kms, $K_{\rm 2}$=141 \kms, $\gamma$=$-$32 \kms, and $q$=0.81.  The ratio of component radii is near $R_{\rm 2}$/$R_{\rm 1}$=0.77 if the temperatures are similar. The best model line profile, however, provides a poor match to the data and underpredicts the mass ratio and the velocity semi-amplitude. 

\begin{figure}[ht!]
%\fig{Spectra/KIC09953894/KIC09953894LC+VC_smaller.pdf}{4in}{}
\fig{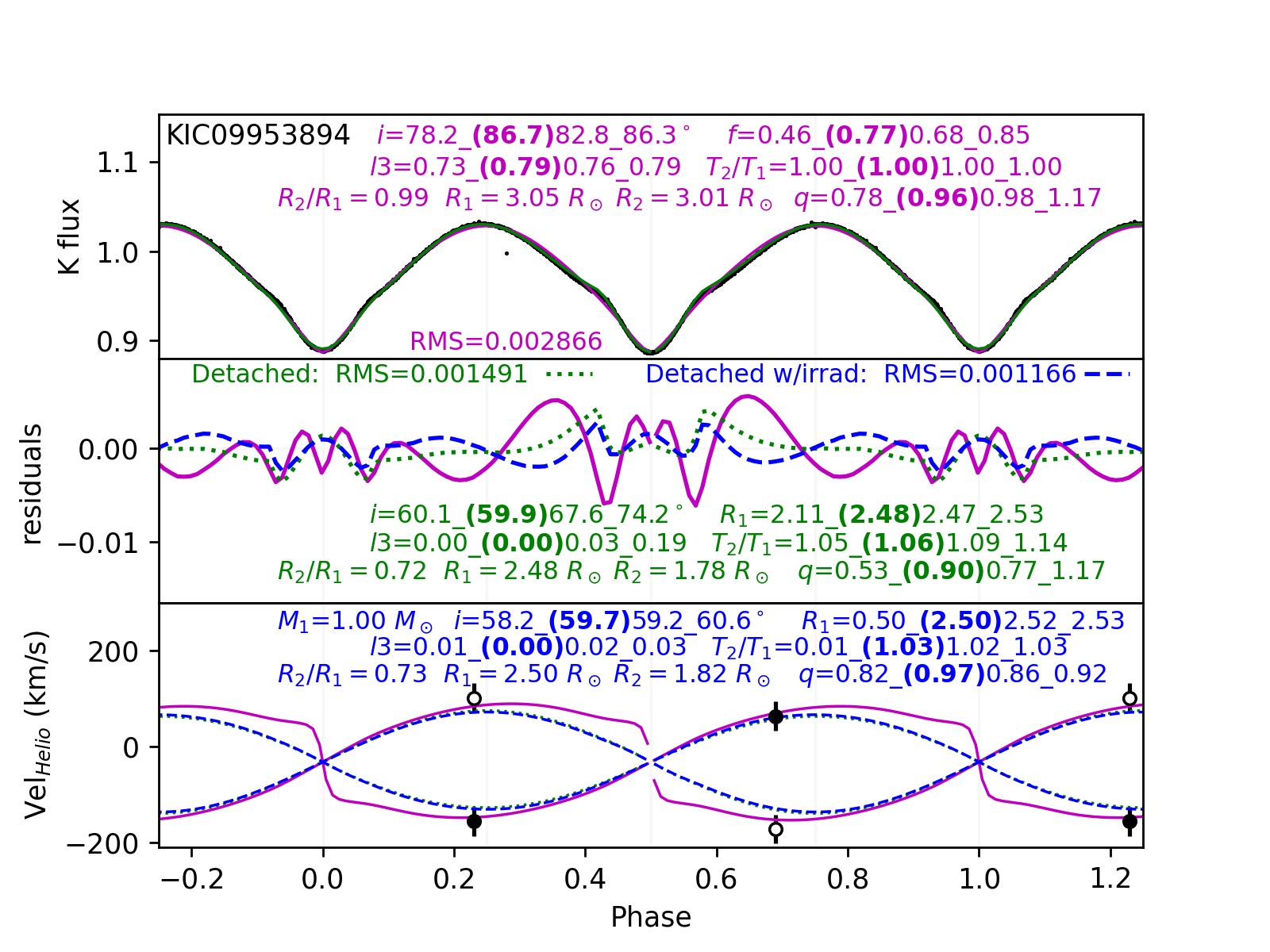}{4in}{}
\caption{Three competing model light curves and velocity curves in comparison to the data of the detached system KIC09953894. \label{fig:894LCVC}}
\end{figure}

Figure~\ref{fig:894LCVC} presents the model light curves and velocity curves for three competing models in comparison to the KIC09953894 data.  The nominal fixed-temperature-ratio model (magenta curve) provides a reasonable fit (RMS=0.0028) for $i$=86.7\degr, $f$=0.77, $q$=0.96, and $l3$=0.79, but the residuals are significant and systematic particularly surrounding $\phi$=0.5, as the contact model is unable to reproduce the inflection indicative of eclipses. The 16th/50th/84th percentile ranges are especially broad ($q$=0.78--1.17). The model does, however, approximately reproduce the velocity semi-amplitudes of the components in the lower panel.  A detached model (green curves) provides a much better fit (RMS=0.0014) for a much lower inclination $i$=60\degr, similar $q$=0.90, much lower $l3$=0.00, and \Tratio=1.06 (a hotter secondary).  The middle panel shows that the residuals for this model are dominated by imperfect agreement near $\phi$=0.5.  Including irradiation and reflected light (blue curve) reduces the residuals near $\phi$=0.5 (RMS=0.0011) but does not eliminate them, suggesting either additional physical effects in the system or an imperfect implementation of irradiation in the PHOEBE models.  The resulting parameters are similar to the detached model except that $q$=0.97, closer to unity than allowed by the kinematic data ($q$=0.81).  However, $q$=0.81 is within the $\pm$1$\sigma$ uncertainties in all three models.  In both of the detached models the best-fitting parameters yield \Ronemax$>$0.99, indicating a primary close to overflowing.  Both the detached models underpredict the velocity semi-amplitudes; this is a consequence of fixing $M_{\rm 1}$=1 \msun\ in the models, while the kinematic data dictate $M_{\rm 1}$=1.3 \msun\ for $i$=60\degr. With an indicated primary temperature of 7295~K, KIC09953894 appears to be  a relatively hot system probably containing two late-F stars, which would be consistent with the masses derived from the spectra, assuming a main-sequence temperature-mass relation.   The blue curve shows the best-fitting detached model with irradiation. This model provides a slightly better fit than the detached model without irradiation effects, but the residuals near phases $\phi$=[0.0, 0.5] are still systematic.   KIC09953894 serves as a warning that for detached systems light curves alone are insufficient to obtain an unique solution; kinematic data are required to fix either $q$ or $M_{\rm 1}$.  Even so, the Bayesian confidence interval limits $q$ to values that include the spectroscopically measured mass ratio.   

\begin{figure}[ht!]
%\fig{/d/zem1/hak/chip/research/Larry/KB/phoebe/KB/Detfudgechi2/09953894/09953894_triangle_plot_2k_9th_Detfudgechi2.pdf }{6in}{}
\fig{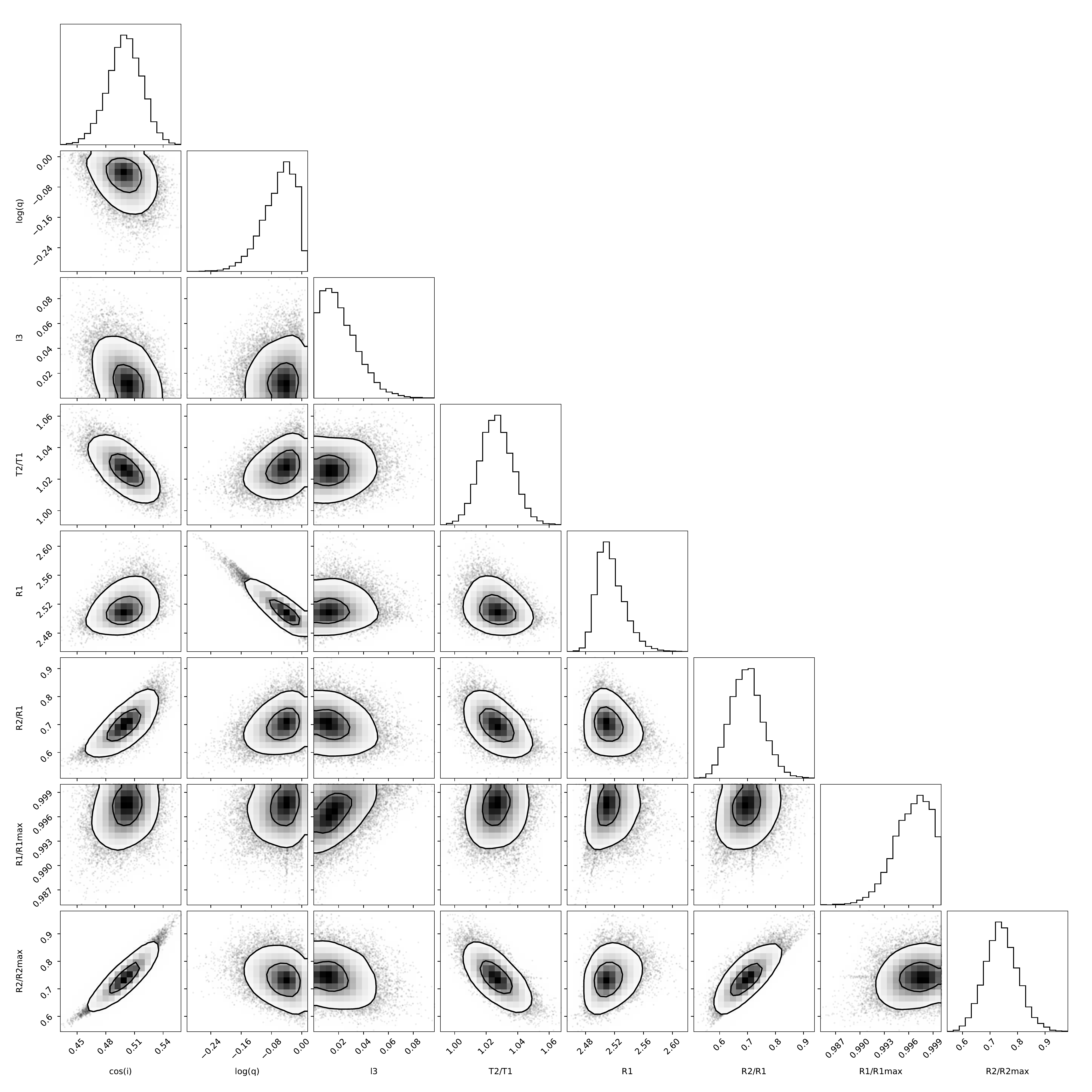}{5in}{}
\caption{Bayesian probability distribution of KIC09953894 system parameters when modeled as a detached configuration and a primary mass of $M_{\rm 1}$=1 \msun is assumed. \label{fig:894MCMC}}
\end{figure}

Figure~\ref{fig:894MCMC} shows the MCMC distribution of system parameters for the KIC09953894 system for the detached model.  Percentile ranges are \cosi=0.482/0.500/0.517, \logq=$-$0.105/$-$0.054/$-$0.018, $l3$=0.00/0.02/0.04, and \Tratio=1.02/1.03/1.04.   By comparison the most probable parameters for a worse-fitting contact model are \cosi=0.065/0.125/0.205 (much larger than the detached model), $f$=0.46/0.67/0.85 (not applicable in the detached model)  \logq=$-$0.106/$-$0.008/0.067 (similar to the detached model), and $l3$=0.72/0.76/0.79 (much larger than the detached model).  While the mass ratio distribution encompasses the correct $q$$\sim$0.81, the ranges for the preferred detached model are very different than the contact model despite a similar RMS.  The large third light fraction retrieved from the contact model is inconsistent with the $l3$$\approx$0 indicated by the detached models and by the BF.  The secondary is slightly hotter than the primary (\Tratio=1.03) and the primary is close to overflowing (\Ronemax=0.99) but the secondary is not (\Rtwomax=0.8), yielding large ellipsoidal modulations of the light curve and shallow eclipses at both quadrature phases.   

The data and models indicate that KIC09953894 is a detached system with two inflated components and a slightly hotter but smaller secondary.  Contact models can approximate the light curve and yield a mass ratio broadly consistent with detached models and spectral data, but the contact models erroneously predict large third light and larger inclinations.  Including irradiation and reflected light in the models improves the agreement with the light curve for this \Tratio=1.03 system.  Without a spectrosopcially determined primary mass or mass ratio, both detached models (with and without irradiation effects) underpredict the velocity semi-amplitudes but do an excellent job of reproducing the light curve.

\subsection{KIC10292413}

\begin{figure}[ht!]
%\fig{/d/zem1/hak/chip/research/Larry/KB/IndivSpectra/KIC10292413/KIC10292413_phased_all.png}{4in}{}
\fig{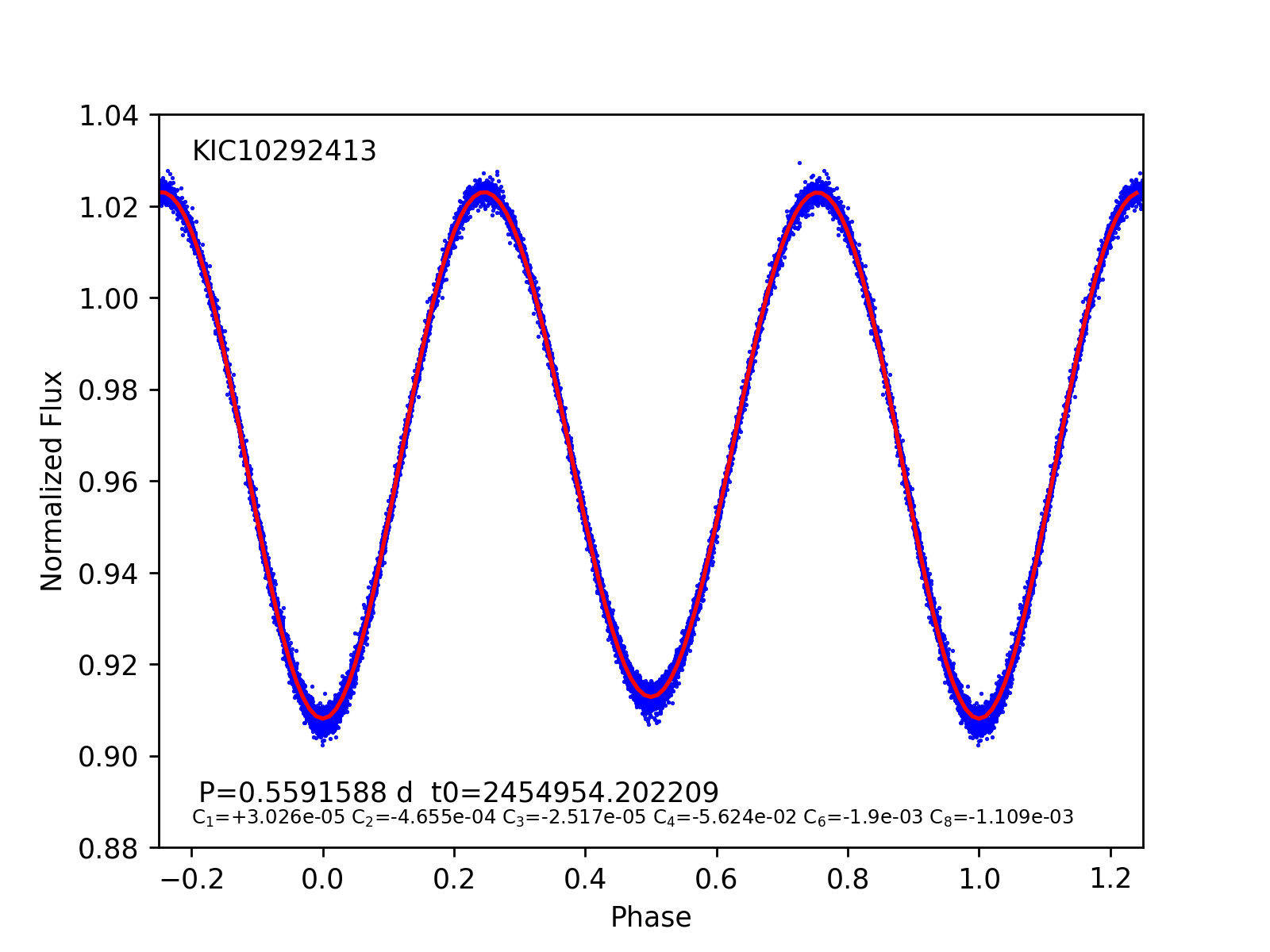}{4in}{}
\caption{Folded {\it Kepler} light curve of KIC10292413. \label{fig:413LC}}
\end{figure}

Figure~\ref{fig:413LC} displays the {\it Kepler} light curve of the $P$=0.56~d contact system KIC10292413, the shortest period among our pilot sample. The primary minimum is very slightly deeper than secondary minimum.  This target also has a close visual companion within 2\arcsec\ which may affect the measured flux of the {\it Kepler} data since it lies within one detector pixel.  Dilution of a light curve by a constant third light makes the maxima fainter and the minima brighter, reducing the overall amplitude of modulation, in agreement with the fairly small amplitude in Figure~\ref{fig:413LC}.      

\begin{figure}[ht!]
%\fig{/d/zem1/hak/chip/research/Larry/KB/IndivSpectra/KIC10292413/PanStars_KIC10292413.pdf}{4in}{}
\fig{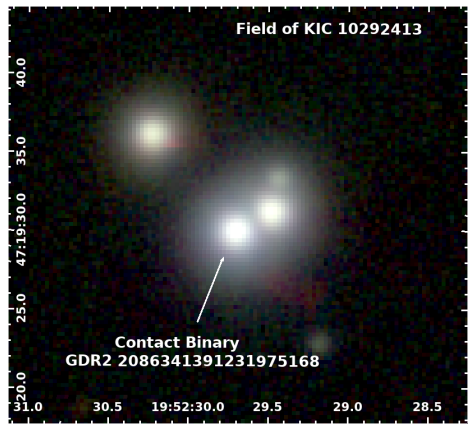}{3in}{}
\caption{Pan-STARRS g'/r'/i' image of the KIC10292413 field with the contact binary source labeled. \label{fig:413IMG} }
\end{figure}

The Pan-STARRS \citep{Kaiser2010}  images of this target in Figure~\ref{fig:413IMG} show it to be a visual double consisting of  approximately equal brightness components at a position angle of 120\degr\ at a 2\farcs{3} separation.  We spectroscopically determined that the indicated star to the southeast is the contact binary, identified as $Gaia$ DR2 ID 2086341391231975168 (G=14.6 mag). We aligned the spectrograph slit angle to cover only this star and minimize contributions from the neighbor to the northwest ($Gaia$ DR2 ID 2086341391231974656; G=15.1 mag).  

\begin{figure}[ht!]
%\fig{/d/zem1/hak/chip/research/Larry/KB/IndivSpectra/KIC10292413/KIC10292413_O-C.pdf}{4in}{}
\fig{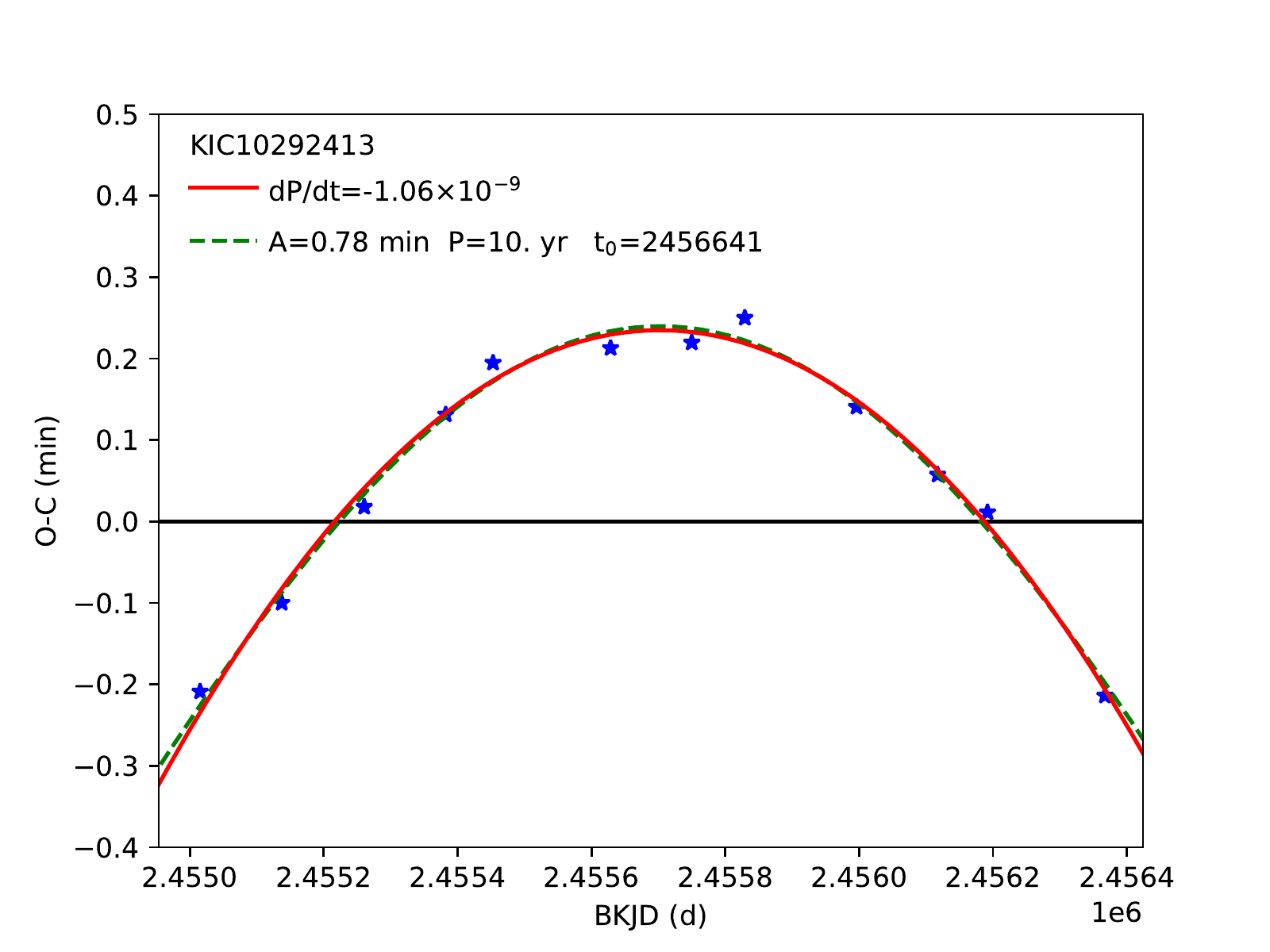}{4in}{}
\caption{\OC\ eclipse timing residuals of KIC10292413. \label{fig:413OC}}
\end{figure}

Figure~\ref{fig:413OC} presents the eclipse timing residuals of KIC10292413, which show some evidence of a decreasing period with $dP/dt$=$-$1.06$\times10^{-9}$ (red parabola).  This is much smaller than in previous examples and is at the limit of what can be detected in the data.   A sine function with a period of 10 yr is also a good fit to these data.  The amplitude of 0.78 minutes corresponds to a minimum projected light crossing distance of 0.1 AU for the orbit of the contact binary about a common center of mass with a hypothetical third body.  A very low-$i$ orbit about a $M_{\rm 3}$$\approx$0.5 \msun\ star could plausibly explain these \OC\ variations.  We conclude that the evidence for eclipse timing variations is strong in this object, but we are unable to distinguish between a constant period derivative and a periodic \OC\ variations that would indicate a third body in the system.  In any case, the visual companion at 2\farcs{3} separation cannot constitute the third body responsible for the eclipse timing variations, as it has a very different $Gaia$ parallax and must be physically unrelated.   

\begin{figure}[ht!]
%\fig{Spectra/KIC10292413/KIC10292413_twocomp.pdf}{4in}{}
\fig{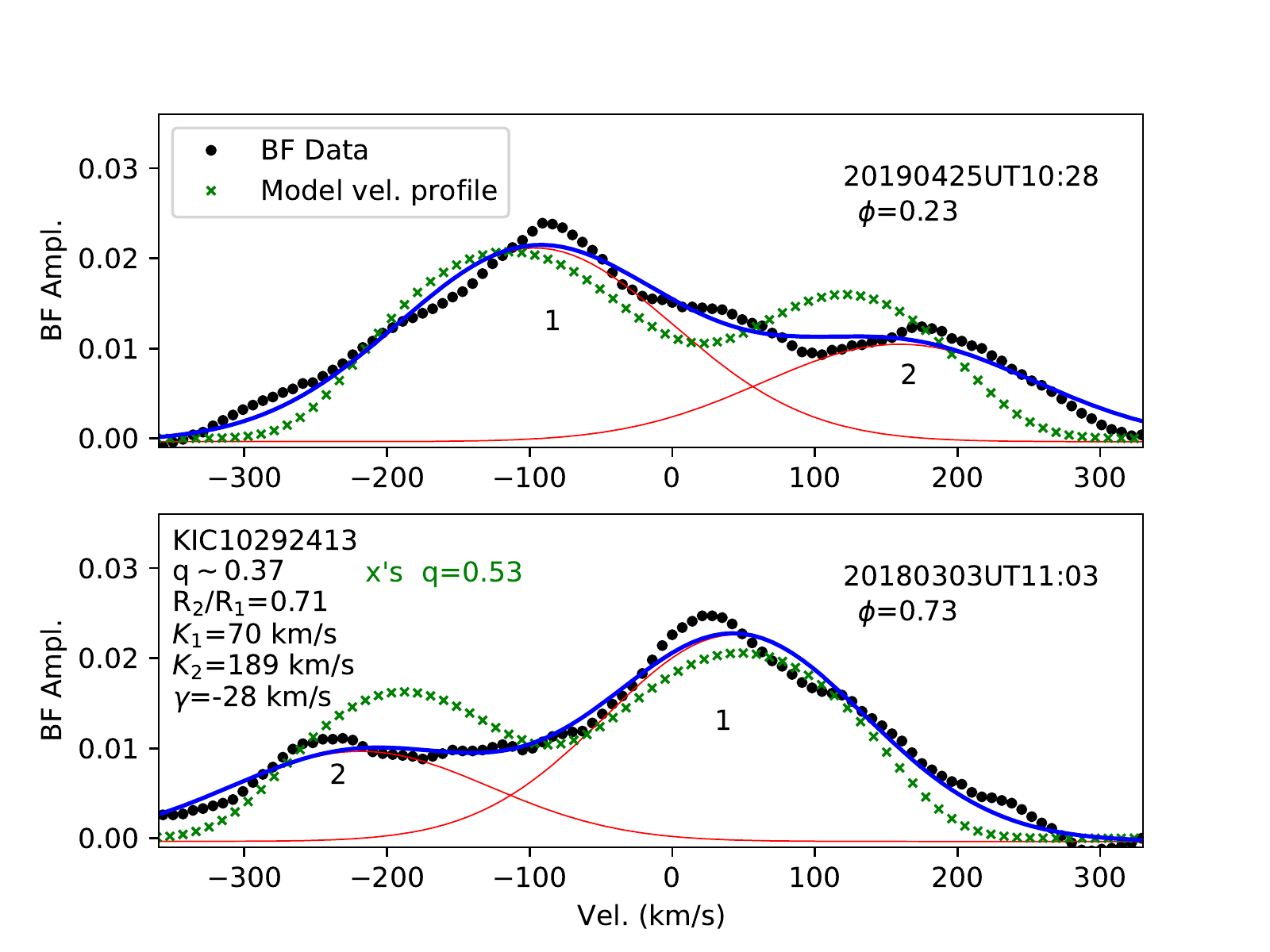}{4in}{}
\caption{Broadening Function of KIC10292413.  The BF does not include light from the third star.  The computed BF is noisier than previous examples owing to the faintness of the source. The model line profile function is generated with $q$=0.53 (green x's) while the two-component Gaussian fit to the BF (red curves) suggests $q$=0.37.     \label{fig:413BF}}
\end{figure}

Figure~\ref{fig:413BF} shows the broadening function of KIC10292413 for two epochs of spectroscopy obtained at phases $\phi$=0.23 and $\phi$=0.73 (adopting the best linear ephemeris).  Judicious slit placement allowed us to exclude light from the close visual companion.  Owing to the faintness of this source (G=14.6 mag) the signal-to-noise ratio is lower than for other targets.  Most of the information in the broadening function comes from the H$\alpha$ line which is the strongest in the spectrum.  There is no reported $T_{\rm eff}$ in the Kepler Input Catalog, so we adopt 6200~K, a value consistent with the strong H$\alpha$ line.  There are two peaks in the (rather noisy) BF, allowing us to estimate component velocities and a mass ratio near $q$$\approx$0.37, with considerable uncertainty.  The estimated velocity semi-amplitudes are $K_{\rm 1}$=70 \kms\ and $K_{\rm 2}$=189 \kms\ at a systemic velocity of $\gamma$=$-$28 \kms.  The ratio of component areas in BF implies $R_{\rm 2}$/$R_{\rm 1}\approx$0.71.  However, the presence of third light likely skews the velocities of one or both components, rendering the kinematic measurements significantly uncertain.  

\begin{figure}[ht!]
%\fig{/d/zem1/hak/chip/research/Larry/KB/IndivSpectra/KIC10292413/KIC10292413LC+VCnew.png}{4in}{}
\fig{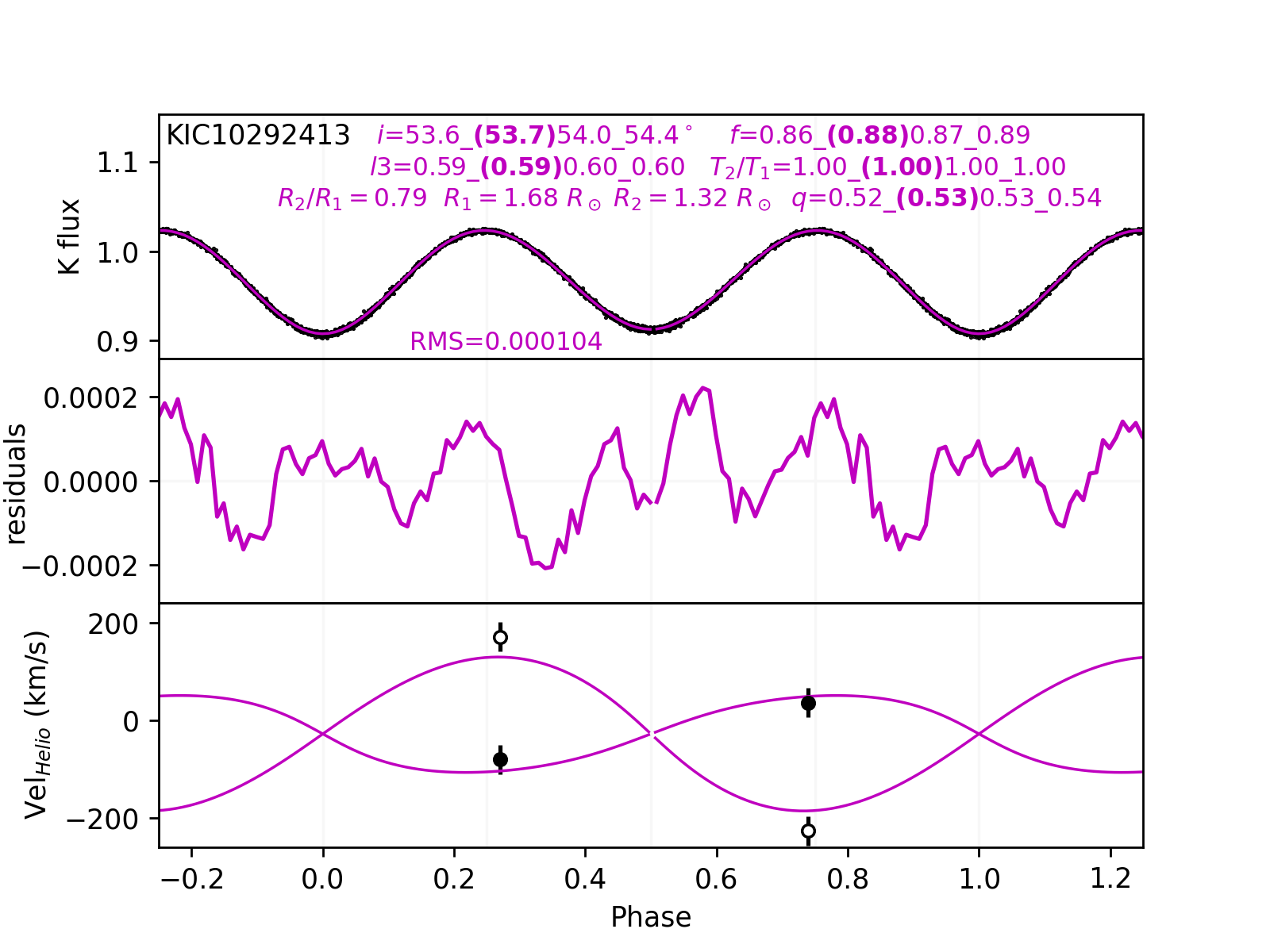}{3.5in}{}
\caption{Light curve and velocity curve of KIC10292413, along with a best-fitting contact model. \label{fig:413LCVC}}
\end{figure}

Figure~\ref{fig:413LCVC} presents the best-fitting fixed-temperature-ratio contact binary light curve and velocity curve for KIC10292413.   A excellent fit (RMS=0.00010) with no systematic residuals  is achieved with a contact binary system using  $i$=53.7\degr, $f$=0.88, $q$=0.53 (somewhat larger than the value inferred from the BF), and $l3$=0.59, leading to $R_{\rm 2}$/$R_{\rm 1}$=0.79, $R_{\rm 1}$=1.68 \rsun,  and $R_{\rm 2}$=1.32 \rsun.  At this inclination the velocity semi-amplitudes imply component masses $M_{\rm 1}$=1.34 \msun and $M_{\rm 2}$=0.71 \msun.  Detached configurations produce a much larger RMS and are not considered further. Models employing a variable temperature ratio provide an essentially identical solution, supporting equal-temperature components.   The model line profile (green x's in Figure~\ref{fig:413BF}) matches the BF reasonably, despite the lower SNR of the spectra.  Yet, the deficit between the BF and the model line profile suggests a third component at a different radial velocity in the two spectra obtained more than a year apart.  Two sources of third light are possible.  The visual 2\arcsec\ companion to the northwest constitutes 38\% of the combined system light, as it falls within one {\it Kepler} pixel.  The \OC\ variations also suggest third light within the KIC10292413 system.  Together, these could easily account for the $l3$$\approx$0.59 suggested by the best model. 
\clearpage

\begin{figure}[ht!]
%\fig{/d/zem1/hak/chip/research/Larry/KB/phoebe/KB/Tfixfudgechi2/10292413/10292413_triangle_plot_2k_9th_Tfixfudgechi2alt.pdf}{5in}{}
\fig{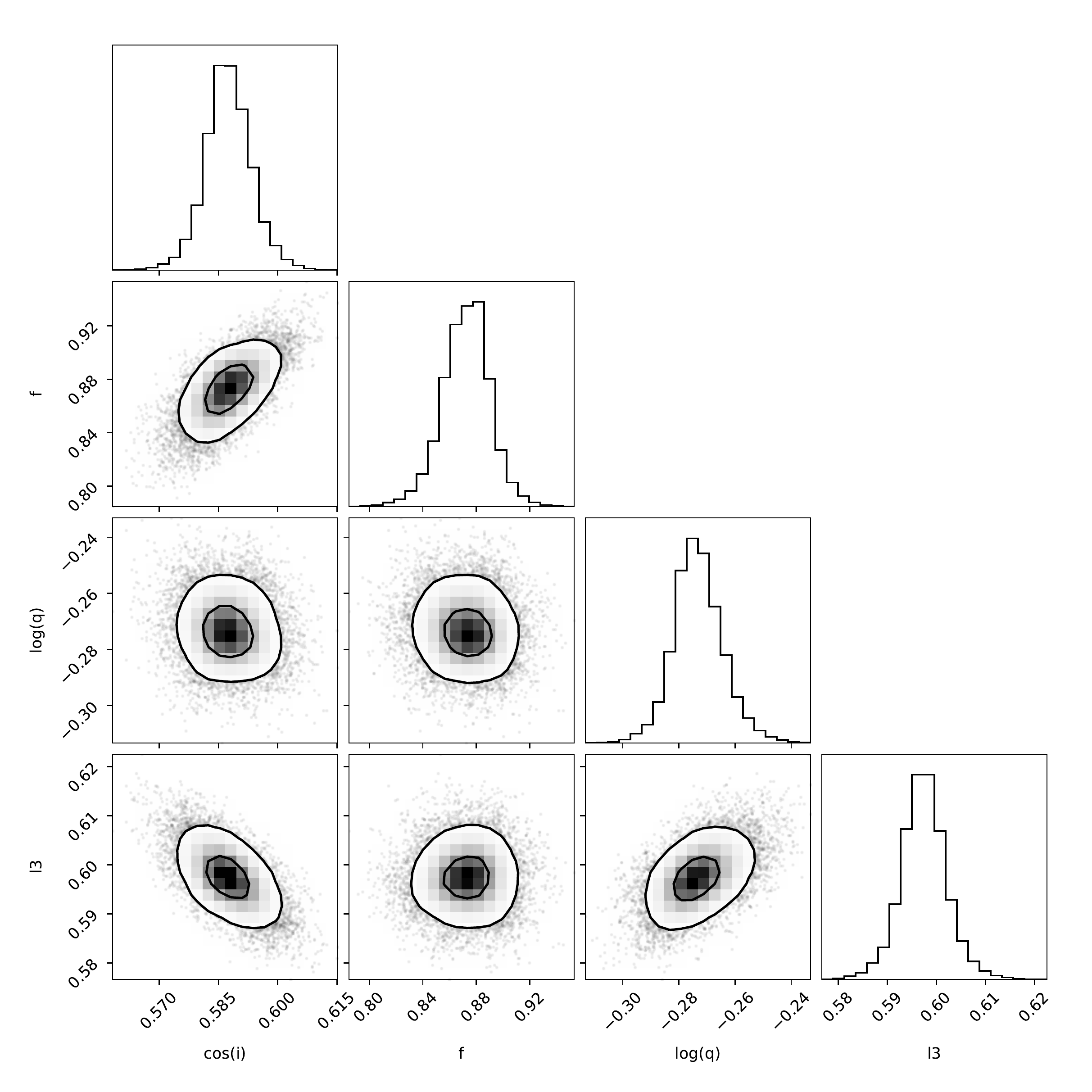}{4in}{}
\caption{Posterior probability distributions for combinations of four free light curve parameters in the KIC10292413 system, indicating very tight constraints on all system parameters.  \label{fig:413MCMC}}
\end{figure}

Figure~\ref{fig:413MCMC} shows the posterior probabilities from Monte Carlo simulations for the fixed-temperature-ratio contact model. Percentile ranges are \cosi=0.582/0.587/0.593, $f$=0.86/0.87/0.89, \logq=$-$0.281/$-$0.276/$-$0.269, and $l3$=0.59/0.60/0.60.  These are the tightest constraints on any system in the pilot sample, enabled by the very small dispersion in the {\it Kepler} light curve. The RMS of the best-fitting model (0.00010) is actually smaller than typical photometric uncertainty of 0.00018, indicating very little intrinsic variability in the system over the four-year baseline and no significant physical features such as spots.     

Despite the large light curve morphology parameter of 0.95 that might suggest a detached ellipsoidal variable, KIC10292413 appears to be a genuine contact system with a substantial third light contribution.  KIC10292413 may be a tight triple system with a velocity-variable third component helping to generate the irregular shape of the BF in Figure~\ref{fig:413BF}.      

\clearpage

\subsection{KIC11097678}

\begin{figure}[ht!]
%\fig{/d/zem1/hak/chip/research/Larry/KB/IndivSpectra/KIC11097678/KIC11097678_phased_all.png}{4in}{}
\fig{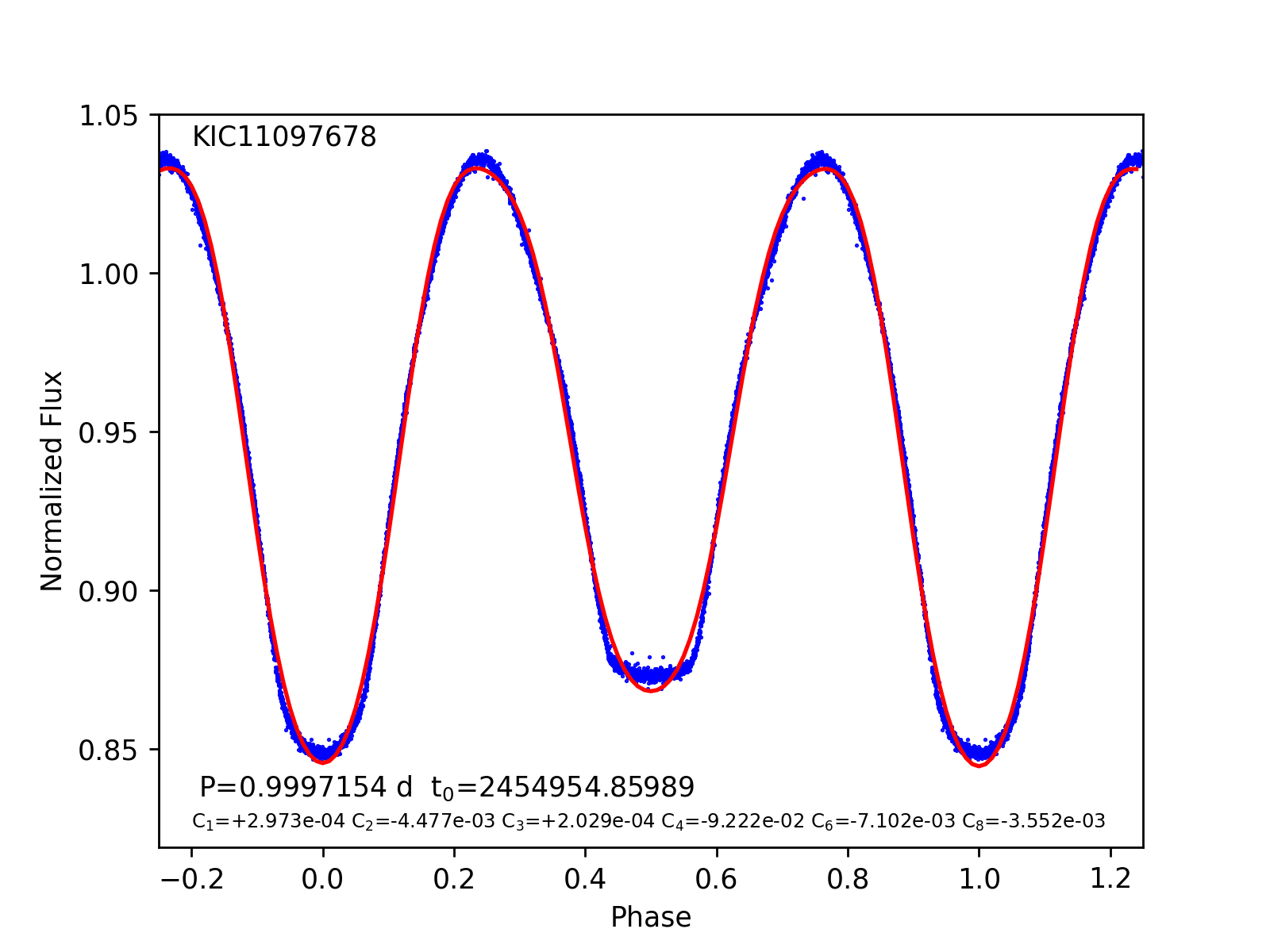}{5in}{}
\caption{Folded {\it Kepler} light curve of KIC11097678. \label{fig:678LC}}
\end{figure}

Figure~\ref{fig:678LC} displays the folded {\it Kepler} light curve of the $P$=0.999~d contact system KIC11097678. The secondary minimum is not as deep as the primary minimum, and the minima are flat-bottomed, indicating an eclipsing system seen at high inclination angle. 

\begin{figure}[ht!]
%\fig{/d/zem1/hak/chip/research/Larry/KB/IndivSpectra/KIC11097678/KIC11097678_O-C.pdf}{5in}{}
\fig{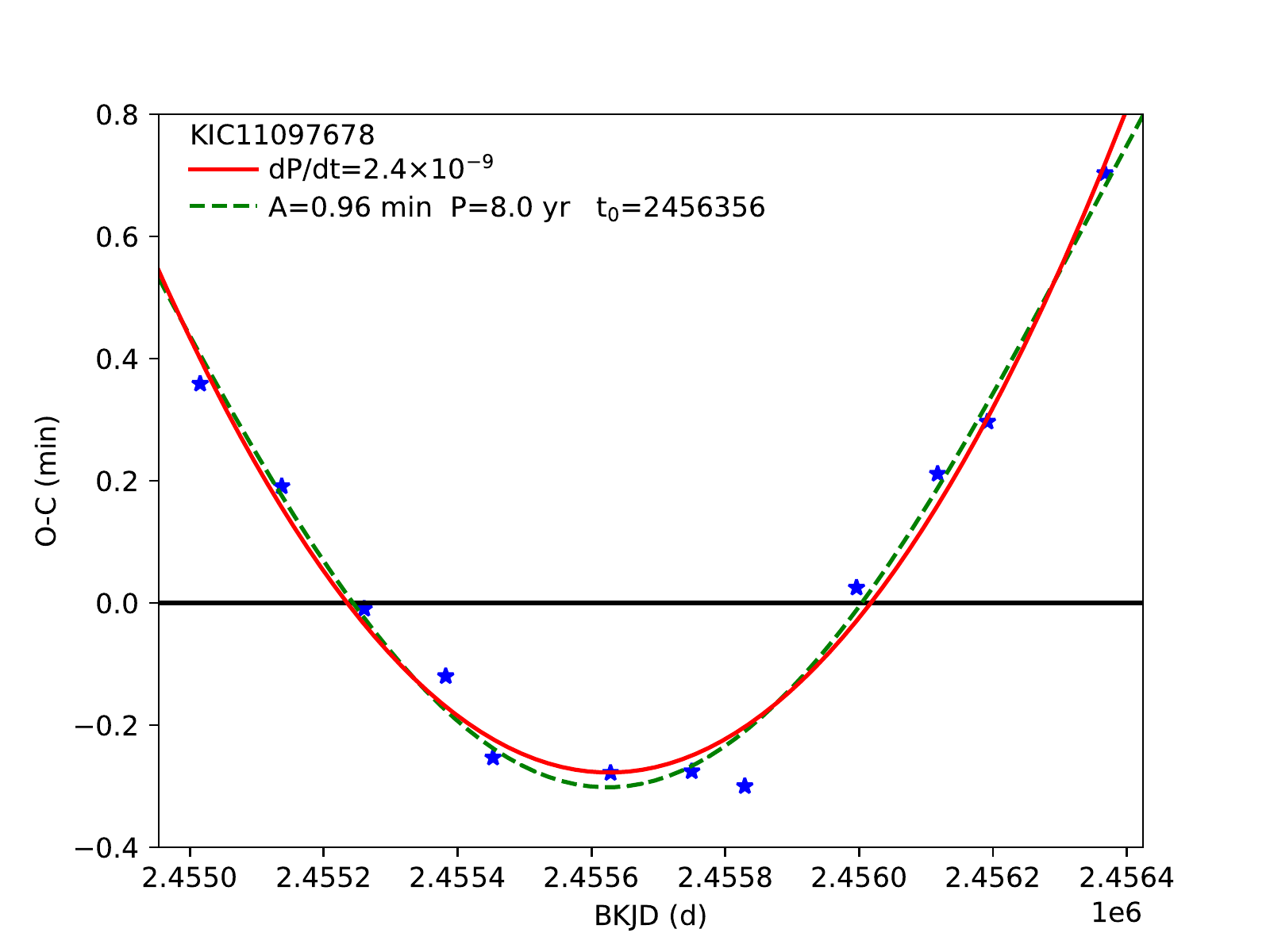}{4in}{}
\caption{\OC\ residuals of KIC11097678. \label{fig:678OC}}
\end{figure}

Figure~\ref{fig:678OC} presents the eclipse timing residuals of KIC11097678, which show some evidence of an increasing period with $dP/dt$=2.4$\times10^{-9}$.  Alternatively, the \OC\ data can be fit with a sine function having amplitude $A$=0.96 min, $P$=8 yr, $T_{\rm 0}$=2476356.    The short duration of the {\it Kepler} dataset precludes more robust conclusions regarding the eclipse timing variations.  Unfortunately, the $ZTF$ data are not helpful because of the limited phase range sampled on this $P$$\approx$1.00 d system.      

\begin{figure}[ht!]
%\fig{/d/zem1/hak/chip/research/Larry/KB/IndivSpectra/KIC11097678/KIC11097678_twocomp.pdf}{4in}{}
\fig{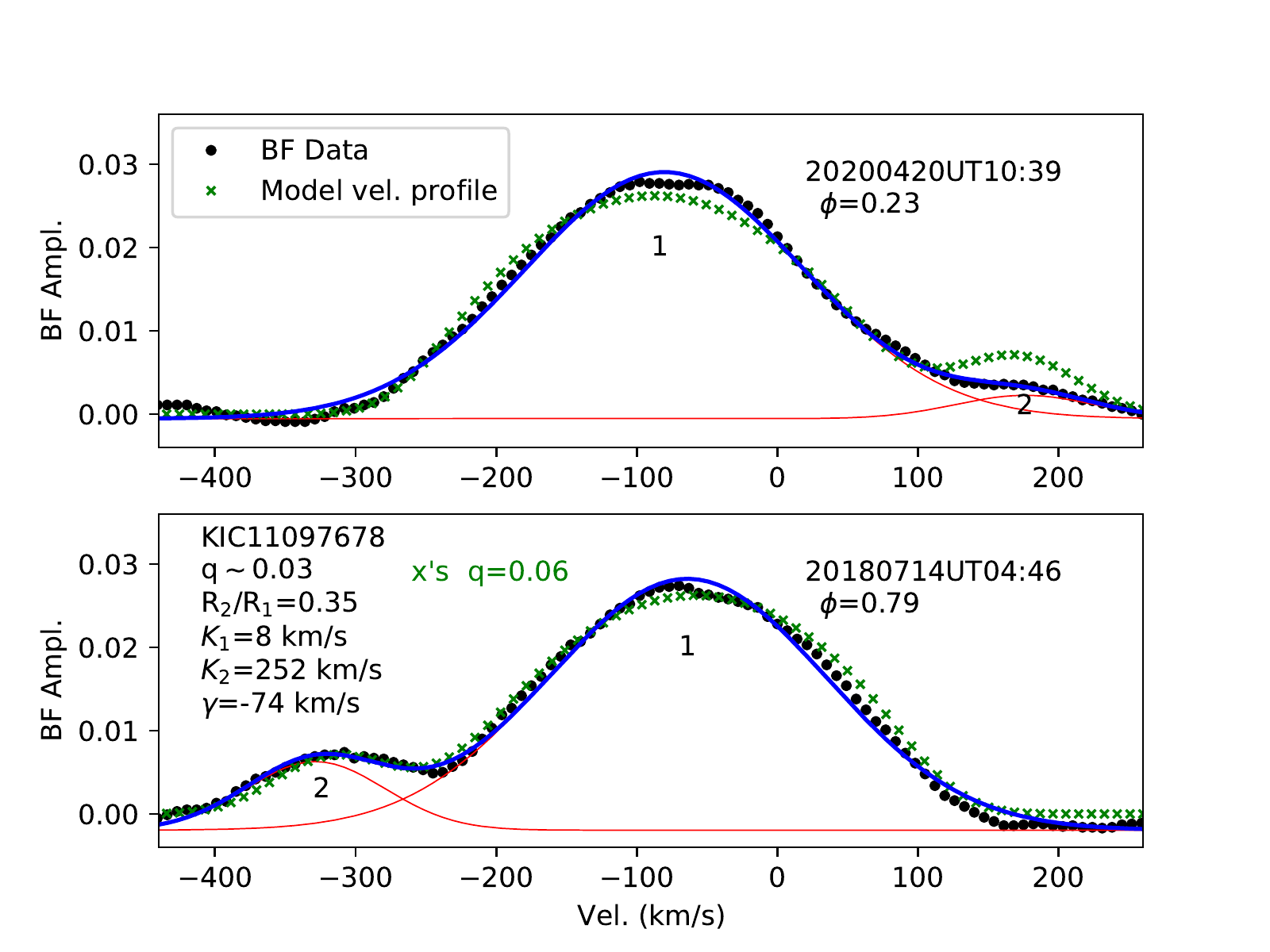}{5in}{}
\caption{Broadening Function and model line profile of KIC11097678 showing two spectral components.   \label{fig:678BF}}
\end{figure}

Figure~\ref{fig:678BF} shows the broadening function of KIC11097678 for two epochs of spectroscopy obtained at phases $\phi$=0.23 and $\phi$=0.79.  There are two clear peaks in the BF.  The velocity of the fainter component at $\phi$=0.23 is not well-constrained on account of the smaller signal of the secondary in the BF.   The ratio of areas of the two Gaussian components implies a radius ratio $R_{\rm 2}$/$R_{\rm 1}$=0.35. The radial velocities of the contact components give velocity semi-amplitudes $K_{\rm 1}$=8 \kms\ and $K_{\rm 2}$=252 \kms\ with $\gamma$=-74 \kms,  implying a mass ratio near $q$$\sim$0.03.   The green x's depict the model line profile function of the $q$=0.10 model that provides the best fit to the light curve, but we find that a model with an even lower $q$ would improve the fit to the BF. The velocity of the secondary is not well measured at the first quadrature phase, but an extreme mass ratio is clearly indicated by the velocities.  At this phase the BF shows a substantial deficit relative to the model, consistent with a real physical effect that decreases the flux from the secondary.  If there were a third component blended with the primary near the systemic velocity, our measurement of $K_{\rm 1}$ and $q$ would be  lower limits.   The Pan-STARRS image data show a visual companion 1\farcs{5} to the west, 3.3 mag fainter in the $Gaia$ photometric system, implying a third-light contribution of about 5\% within the {\it Kepler} pixel.    Such a faint companion does not contribute measurably to the spectra or show up in the BF.  It has a substantially larger parallax and cannot be physically associated with KIC11097678.  

\begin{figure}[ht!]
%\fig{/d/zem1/hak/chip/research/Larry/KB/IndivSpectra/KIC11097678/KIC11097678LC+VCnew.png}{5in}{}
\fig{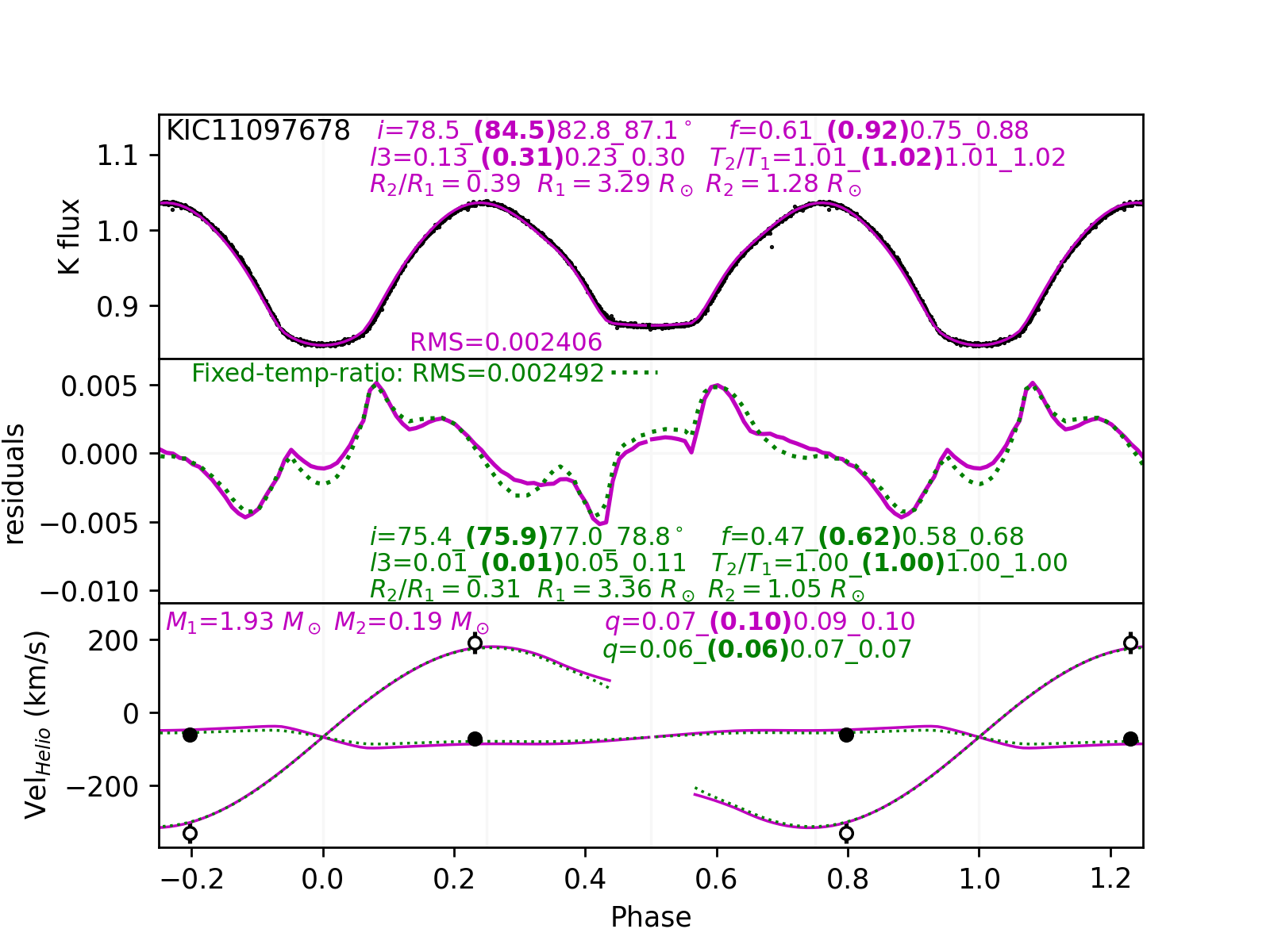}{5in}{}
\caption{Model light curves and velocity curves of KIC11097678 in comparison to the data with the data.  \label{fig:678LCVC}}
\end{figure}

Figure~\ref{fig:678LCVC} shows a light curve and velocity curves for KIC11097678 along with a best-fitting (RMS=0.0024) variable-temperature-ratio contact model (magenta curve) having $i$=84.5\degr, $f$=0.92, $q$=0.10 (somewhat larger than the $q$$\approx$0.03 measured from the BF), $l3$=0.31, and \Tratio=1.02 (a slightly hotter secondary).  Enabling irradiation effects does not improve the fit. However, the fixed-temperature-ratio contact model provides a similarly good fit (RMS=0.0024) for $i$=65.9\degr, $f$=0.62, $q$=0.06, and $l3$=0.01.  Competing detached models imply \Ronemax\ and \Rtwomax$\gg$0.99 and are not viable, having RMS at least three times larger.  The best model component radii are $R_{\rm 1}$=3.36 \rsun\ and $R_{\rm 2}$=1.05 \rsun, yielding $R_{\rm 2}$/$R_{\rm 1}$=0.31, very similar to that inferred above from the BF.   Component masses approximately $M_{\rm 1}\sim$1.93 and $M_{\rm 2}\sim$0.19 using the velocity data and best model $i$. This PHOEBE model produces a line profile function (green crosses in the BF plot of Figure~\ref{fig:678BF}) that is in excellent agreement with the BF. (in fact, the significant third light implied by the variable-temperature-ratio model (with or without irradiation) cannot be accommodated in the BF, lending credence to the \Tratio=1 model!) The agreement with the BF is good in the lower panel at $\phi$=0.79 except that the secondary peak has a larger amplitude than the data.  In the upper panel at $\phi$=0.23, the primary's BF peak is shifted to positive velocities relative to the model and the secondary appears too faint in the BF compared to the model profile, echoing the deficit noted previously for KIC04999357 and KIC10292413.   This difference exists regardless of the range of wavelengths used to construct the broadening function. We conclude that it is not a consequence of H$\alpha$ emission or some other portion of the spectrum that varies with orbital phase.\footnote{This difference is also seen in a $\phi$=0.21 spectrum obtained 2020 April 20UT:10:39.}  We find that the model line profile at $\phi$=0.23 can better reproduce the BF if we add a large cool spot ($T_{\rm spot}$/$T_{\rm 2}$=0.8; or indeed any phenomenon that suppresses the signal from the secondary) on the trailing face of the secondary (co-longitude $\sim$90\degr).  Such a spot configuration would not alter the good agreement between the model and BF at $\phi$=0.79. Large transient spots are known to exist on contact binaries \cite[e.g.,][]{Binnendijk1970, Maceroni1994, Kouzuma2019}. Although such large spots would be visible in the light cure, it is possible they were not present (or present only for a brief duration) during the 2009--2013 {\it Kepler} epochs.      

\begin{figure}[ht!]
%\fig{/d/zem1/hak/chip/research/Larry/KB/phoebe/KB/Tfixfudgechi2/11097678/11097678_triangle_plot_2k_9th_Tfixfudgechi2.pdf}{5in}{}
\fig{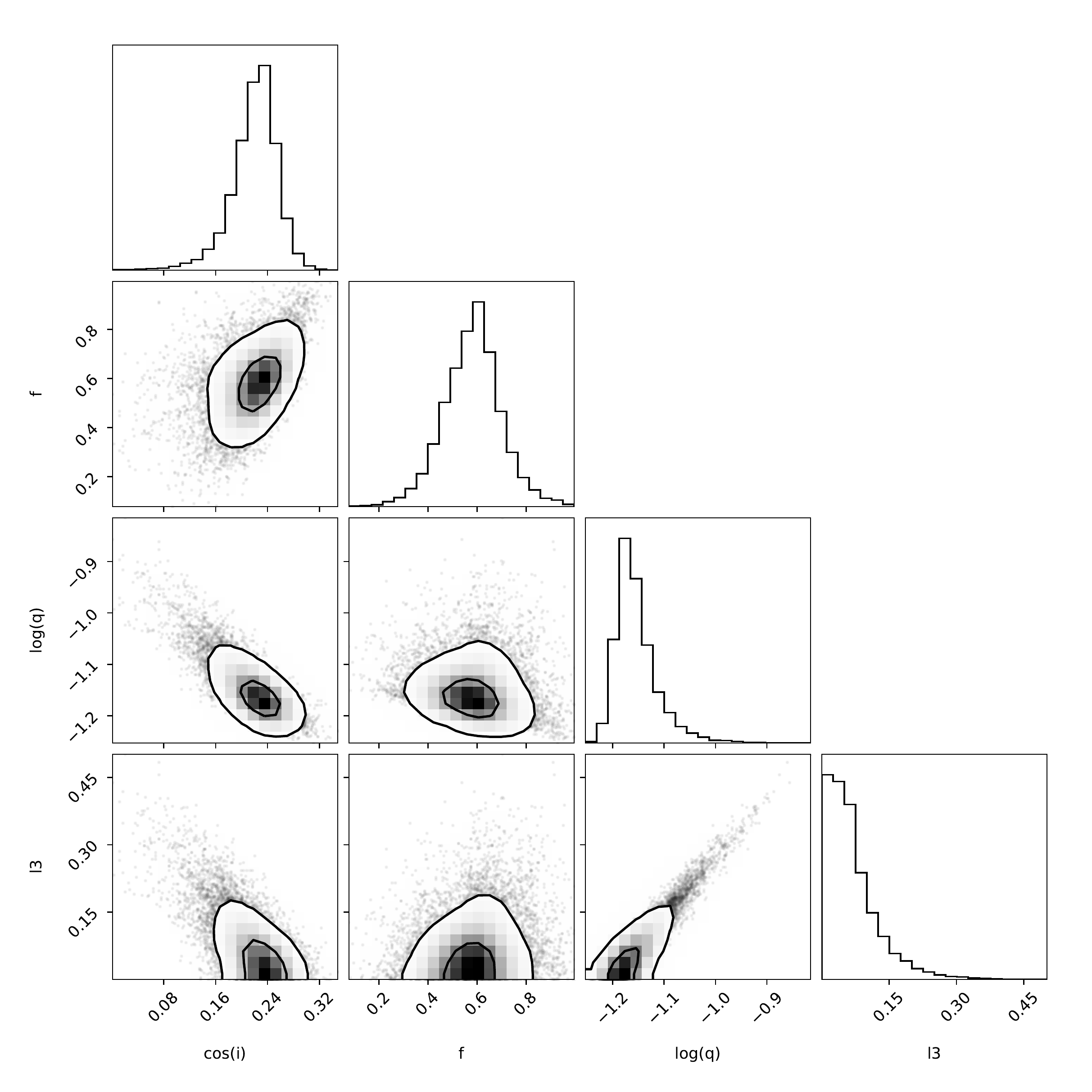}{5in}{}
\caption{Posterior probability distributions for combinations of four parameters for KIC11097678. \label{fig:678MCMC}}
\end{figure}

Figure~\ref{fig:678MCMC} shows the relative probabilities resulting from Monte Carlo simulations using the four-parameter fixed-temperature-ratio models. The percentile parameters are  \cosi=0.195/0.225/0.251, $f$=0.47/0.58/0.67, \logq=$-$1.19/$-$1.16/$-$1.12, and  $l3$=0.01/0.05/0.10, implying strong constraints on system parameters. These parameters are consistent with those of the best fits shown in Figure~\ref{fig:678LCVC}, indicating very small $q$ and small third-light fraction.  The most probable $l3$ is consistent with the estimated 5\% light contribution from the identified 1\farcs{5} visual companion (GDR2 2086752574218680704). Contours in Figure~\ref{fig:678MCMC} are helpful in illustrating the degeneracy between \logq\ and $l3$, here in the sense that larger third light would lead to less extreme mass ratios.

\begin{figure}[ht!]
%\fig{/d/zem1/hak/chip/research/Larry/KB/phoebe/KB/Tfixfudgechi2/11097678/11097678_triangle_plot_2k_9th_Tfixfudgechi2horvat.pdf}{5in}{}
\fig{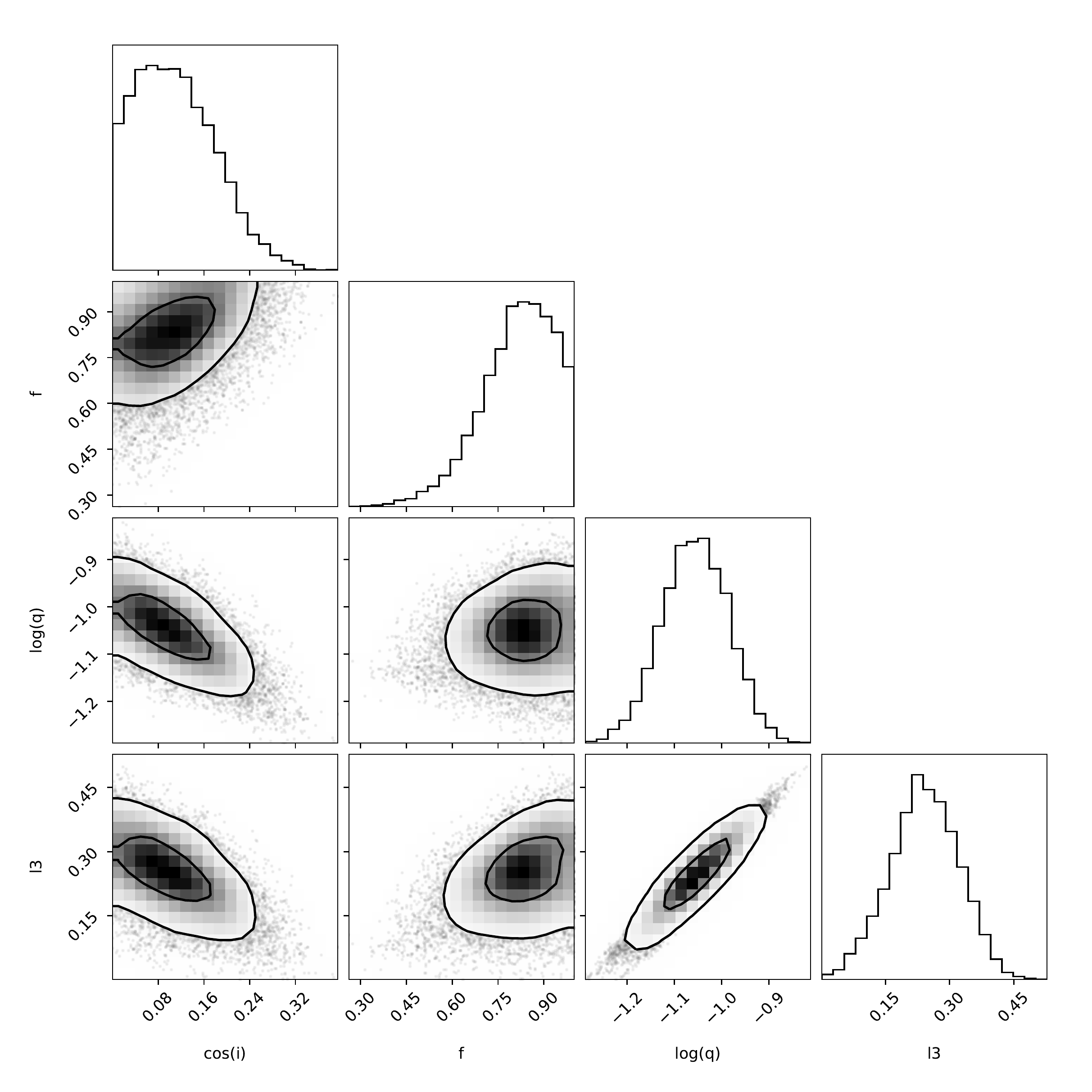}{5in}{}
\caption{Posterior probability distributions for combinations of four free parameters for KIC11097678 when irradiation effects are included in the PHOEBE models. The distributions are shifted markedly from those in Figure~\ref{fig:678MCMC}, particularly regarding third light which is now $l3$$\simeq$0.24 instead of zero. \label{fig:678MCMCirrad}}
\end{figure}

Including irradiation effects in the models changes the RMS negligibly but substantially shifts the locus of most probable parameters toward larger inclinations, larger fillout factors, and larger third-light fractions.   Figure~\ref{fig:678MCMCirrad} shows the posterior distribution of parameters from MCMC simulations where irradiation effects are included.  The most probable parameters are significantly different than in Figure~\ref{fig:678MCMC} where irradiation is disabled:  \cosi=0.039/0.105/0.183, $f$=0.69/0.83/0.93, \logq=$-$1.12/$-$1.05/$-$0.99, and  $l3$=0.16/0.24/0.32.  The non-negligible $l3$ found when irradiation is included is inconsistent with the BF data in Figure~\ref{fig:678BF} and with our BF analysis of high-resolution echelle spectroscopy \citep[to be presented in ][]{Cook2022} which limits the third-light fraction to $<$0.01. The \OC\ variations may, then, result from a very faint low-mass companion or from magnetic activity.  The substantial third-light fractions implied by irradiation-enabled models cast doubt upon the blind application of this particular implementation, at least in this particular system.  Our analysis of KIC11097678 suggests that caution is appropriate when invoking irradiation physics, but we are not able to speculate as to the reason for this inconsistency.  Nevertheless, the most probable mass ratios are still extreme for any model-based Bayesian analysis---near nominal limits for onset of the Darwin instability.   Despite the large $morph$=0.90, KIC11097678 is a clear case of a contact system.         

\subsection{Summary of the Pilot Study}

In the pilot sample, spectroscopic data provided radial velocities of (near-)contact binary stars that permitted the measurement of component masses when combined with orbital inclinations inferred from the light curves.  The systems' mass ratios were already well-constrained by the light curves alone, even in the presence of additional features modeled as spots on the stars (KIC08913061, KIC09164694, KIC09345838).   The spectroscopic data, nevertheless, provided confirmation of the mass ratios and the third-light fractions retrieved from the light curve analyses.  Sometimes the third component could be seen directly as a distinct peak in the BF.  Most often the third component was blended in velocity space with the brighter and more massive primary component, adversely affecting the measurement of the primary star's velocity and systematically decreasing the mass ratio inferred from the BF.   Notably, all but one of the ten pilot study systems show some evidence for a third component either from the Monte Carlo light curve analysis, the \OC\ variations, the BF, or some combination of the three.   In all but one of the systems, the evidence for a third-light contribution is compelling.

Light curve shapes are insensitive to the mass of the primary, so kinematic data are required if component masses are to be measured.  We performed PHOEBE+MCMC simulations of the KIC09345838 system (which required a hot spot on the primary star to achieve good fits to the mean light curve) allowing the mass of the primary, $M_{\rm 1}$ to vary in addition to the seven parameters modeled in Figure~\ref{fig:838MCMC}.  The posterior probability distributions of all parameters are nearly identical to when $M_{\rm 1}$ is fixed, and the uncertainties are as well.  This is expected and merely serves to confirm that the light curve morphology is sensitive to the mass ratio and not to the overall system mass.  

Table~\ref{tab:tensample} provides a short synopsis of the best-fitting model parameters and spectroscopically determined parameters for the (near-)contact systems in our pilot study.  Column 1 is the KIC ID, columns 2--6 give the $i$, $f$, $q$, $l3$, and \Tratio\ from the best-fitting contact binary model, columns 7--10 give the BJD, observed orbital phase, and component Heliocentric radial velocities for the first spectroscopic observation near $\phi$=0.25, columns 11--14 give the BJD, orbital phase, and component Heliocentric radial velocities for the second spectroscopic observation near $\phi$=0.75, column 15 lists the computed systemic velocity, and column 16 gives the spectroscopic mass ratio, $q_{\rm spec}$,  if measurable from the BF.  Table footnotes provide additional insights on each system gleaned from the spectroscopic analyses.    

Five systems from the pilot study (KIC04999357, KIC06844489, KIC09840412,  KIC10292413, and KIC11097678) have contact models that provide an excellent match to the light curves and have RMS considerably less than the best detached models.   Variable-temperature-ratio models provide slightly better fits and have 0.98$<$\Tratio$<$1.02 but yield system parameters very similar to the \Tratio=1 models.  Both sets of contact models produce radial velocity profiles consistent with the broadening function and with the mass ratio measured from the kinematic data.    

KIC04853067 is equally well modeled as either a contact or a detached system with an extreme mass ratio seen at low inclination.  It serves as a cautionary example that for low-$i$ extreme-$q$ systems, contact and detached models are indistinguishable and the true geometry is uncertain.  Such ambiguity can be expected as a matter of course when fitting large samples of (near)-contact binary light curves even with exquisite photometric precision.       

KIC08913061 provides a striking example of a system with a low-dispersion symmetric light curve where the  variable-temperature-ratio models produce a superior RMS compared to fixed-temperature-ratio models yet select the {\it wrong} mass ratio.  \Tratio=1 models select the correct mass ratio.  Addition of a single physical component (a hot spot on the primary) to the fixed-temperature-ratio models reduces the RMS by a factor of almost two compared to variable-temperature-ratio models.  This is an indication that real physical features present in the star systems but not in the nominal contact binary models are both detectable and large compared to the {\it Kepler} photometric precision, dominating the RMS in many systems.  

KIC09164694 and KIC09345838 both display asymmetric light curves and require the addition of a model component (e.g., a hot spot on the primary) to produce a satisfactory fit to the light curve.  Like KIC08913061, the variable-temperature-ratio model select the {\it wrong} mass ratio, further illustrating a dangerous degeneracy between \Tratio\ and $q$ when only single-band light curves are modeled.  Models using \Tratio=1 and a hot spot provide superior fits and identify a mass ratio consistent with the kinematic data.  

KIC09953894 is a detached system, yet with large ellipsoidal modulations that resemble a contact binary.  The detached models provide a superior fit and indicate significantly smaller inclinations and third-light fractions (yet similar mass ratios) than the best contact models.  Models including irradiation effects further improve the fit and require less extreme temperature differences (\Tratio=1.03 versus 1.06) compared to the non-irradiation models.  A Bayesian analysis shows that all the key parameters in the detached model are well constrained, underscoring the utility of this type of analysis for detached systems having large ellipsoidal modulations.  The component velocity semi-amplitudes (the BF), however, are not well reproduced without independent knowledge that allows one of the component masses to be fixed {\it a priori}.         

\clearpage

\section{Bayesian Retrieval of System Parameters for \nummodeled\ {\it Kepler} Close Binaries  \label{sec:MCMC}}

In the general case of a contact binary system where component velocities are not available, the Monte Carlo simulations presented above demonstrate that single-band high-quality light curves alone are sufficient to retrieve inclinations, fillout factors, mass ratios, third-light fractions and crude temperature ratios from \Tapprox\ contact models and, sometimes, from detached models if the ellipsoidal variations are large.  Given the success of the Bayesian modeling techniques applied above, we undertook a statistical study of \underbar{all} candidate contact binaries in the {\it Kepler} field using the same methods.  We conducted Bayesian modeling for \numksample\ out of 2878 total {\it Kepler} eclipsing binaries tabulated by \citet{Kirk2016}.  The selection criteria included systems having  periods $P$$<$2.0~d and morphology parameter $morph$$>$0.70 indicative of contact configuration or ellipsoidal variations from at least one of the aspherical components.   For 45 systems we assigned a different reference time $T_{\rm 0}$ shifted by half an orbital period compared to the one tabulated in \citet{Kirk2016} in order to place the deeper eclipse at phase zero.   We removed from the sample KIC02831097, as it is a pulsating RR Lyrae type star described by \citet{Sodor2017}. We also removed \numremoved\ systems owing to their extremely low level of photometric variability where the amplitude of modulation is less than about 10 times the RMS in each phase bin.  Such systems are likely to be low-$i$ and/or high-$l3$ systems, and the data do not yield useful constraints on system parameters.  We further removed KIC07733540 which showed large \OC\ variations and abnormally large photometric variability.  This left \nummodeled\ systems to model.  

%\footnote{These are KIC03966491, KIC05386810, KIC05428668, KIC05975712, KIC05991936, KIC07620664, KIC07630690, KIC07749504, KIC07976783, KIC08177958, KIC08285970, KIC08937663, KIC09030447, KIC09085955, KIC09470054, KIC09480977, KIC10014536, KIC10417135, KIC10919564, KIC10972830, and KIC11038312 for low-amplitude variability, plus KIC06436038 and KIC07733540.} 

We performed MCMC simulations of all \nummodeled\ systems, modeling  each as both a \Tapprox\ (constrained temperature ratio) contact binary using five free parameters and as a detached system using six free parameters.  Both sets of models include irradiation effects.  Although contact binaries can exhibit  small temperature differences between the components, the degeneracies documented with the pilot sample using the variable-temperature-ratio contact model motivated us to abandon this unconstrained temperature ratio option as dangerously unreliable.  Photometric data from a second bandpass could, presumably, break this degeneracy and retrieve all of the system parameters with better fidelity.

No single criterion can reliably discriminate between contact and detached geometries.  We employ three model-based metrics that provide approximate classification as probable contact binaries (C), probable detached systems (D), or ambiguous cases (A). Our classification scheme (based on lessons learned from the pilot sample) utilizes the RMS from the best fitting contact versus detached models (i.e., RMS$_{\rm con}$ versus RMS$_{\rm det}$), the temperature ratio \Tratio from the best-fitting contact model, and the fraction of the components' Roche lobes filled from the best-fitting detached model (i.e., \Ronemax\ and \Rtwomax). 
\begin{itemize}
    \item (C) Contact---\numcontacts\ systems: RMS$_{\rm con}$$<$RMS$_{\rm det}$ \underbar{AND} RMS$_{\rm con}$$<$0.005 \underbar{AND} 0.96$<$\Tratio$\leq$1.04 \underbar{AND} ( \Ronemax$\geq$0.95 \underbar{OR} \Rtwomax$>$0.95). This last criterion requires at least one of the components to fill its Roche lobe to at least 95\% of the maximum value in the best-fitting detached model.  
    \item (D) Detached---\numdetached\ systems: RMS$_{\rm det}$$<$RMS$_{\rm con}$ \underbar{AND} RMS$_{\rm det}$$<$0.005 \underbar{AND} ( \Ronemax$<$0.95 \underbar{OR} \Rtwomax$<$0.95).  This last criterion requires at least one of the components to have a Roche lobe volume less than 95\% of the maximum possible  value in the best-fitting detached model.  
    \item (A) Ambiguous---\numambiguous\ systems:  everything else. The bulk of these consist of the union of the set of \numbadfit\ systems having large RMS values with the set of  \numtextreme\ systems with \Tratio\ more than 4\% from unity.  
\end{itemize}

Table~\ref{tab:contactsbest} provides the parameters of the best-fitting contact binary model for all the \nummodeled\ systems, regardless of classification.  Column 1 lists the KIC identification number, column 2 is the adopted orbital period in days, column 3 is the adopted reference Julian Date of deeper eclipse, column 4 is the stellar $T_{\rm eff}$, column 5 is the light curve morphology parameter from \citet{Kirk2016}, column 6 is the inclination in degrees, column 7 is the fillout factor, column 8 is the mass ratio\footnote{$q$$\equiv M_{\rm 2}/M_{\rm 1}$.  Values $<$1 indicate that the star $M_{\rm 1}$ producing the deeper eclipse at superior conjunction is the more massive. Values $>$1 indicate that the star $M_{\rm 2}$ eclipsed at secondary minimum is more massive.}, column 9 is the third-light fraction,  column 10 is the temperature ratio \Tratio, column 11 is the RMS of the best-fitting model in units of the normalized light curve, and column 12 is a flag designating whether the system is best regarded as a contact binary (C), a detached system (D), or an ambiguous case (A).   Table~\ref{tab:contactsprob} provides the values of the 16th, 50th, and 84th percentile for each of the five free parameters resulting from contact models of systems in Table~\ref{tab:contactsbest}.   Column 1 lists the KIC identification number, columns 2--4 are the 16th/50th/84th percentiles of \cosi, columns 5--7 are the percentiles of $f$, columns 8--10 are the percentiles of \logq, columns 11--13 are the percentiles of $l3$, columns 14-16 are the percentiles of \Tratio, and column 17 is the Contact/Detached/Ambiguous identification flag.  Tables~\ref{tab:contactsbest} and \ref{tab:contactsprob} contain the first ten rows to provide an example of the form and content.  The full tables are available in electronic form as machine-readable files.   

Table~\ref{tab:detachedbest} provides the parameters of the best-fitting detached binary model for the \numdetached\ probable detached systems.  Column 1 the KIC identification number, column 2 is the orbital period in days, column 3 is the adopted reference Julian Date of deeper eclipse, column 4 is the stellar $T_{\rm eff}$, column 5 is the light curve morphology parameter from \citet{Kirk2016}, column 6 is the inclination in degrees, column 7 is $q$, column 8 is $l3$, column 9 is \Tratio, column 10 is the primary's stellar radius $R_{\rm 1}$ in solar units, column 11 is the ratio of stellar radii \Rratio, column 12 is \Ronemax, column 13 is \Rtwomax, column 14 is the RMS of the best-fitting detached modelin units of the normalized light curve, and column 14 is the flag designating the system as a probable detached binary (D).  Table~\ref{tab:detachedprob} provides the  16th, 50th, and 84th percentiles for each of the six free parameters in the detached models, following the format of  Table~\ref{tab:contactsprob}.

\subsection{System Parameters for (Near)Contact Binaries {\label{sec:contacts}} }

\begin{figure}[ht!]
%\fig{KB/phoebe/KB/Pversusmorph.pdf}{5in}{}
\fig{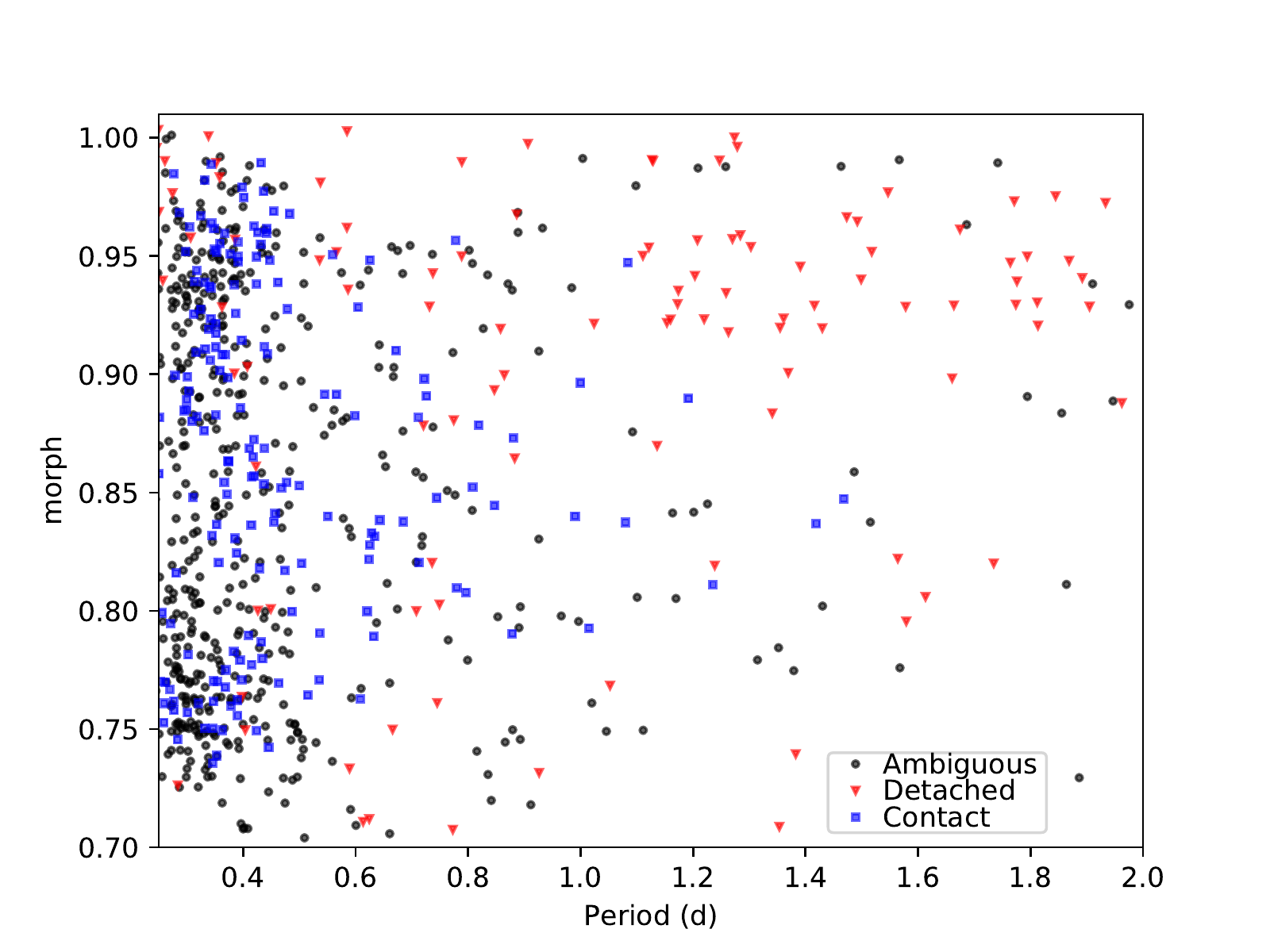}{5in}{}
\caption{Light curve morphology versus orbital period for the probable contact binaries, detached binaries, and ambiguous cases, as denoted by symbol type. \label{fig:Pversusmorph}}
\end{figure}

Figure~\ref{fig:Pversusmorph} plots the light curve morphology parameter versus orbital period, with blue squares/red triangles/black circles representing the contact/detached/ambiguous systems, respectively. A small dispersion has been added to the data to reduce marker pileup.  Sample selection criteria ($morph$$>$0.7) already select against most eclipsing detached systems but do allow ellipsoidal variables which tend to have $morph$$>$0.9 \citep{Kirk2016}.  Probable contact systems congregate preferentially at short periods.  Ambiguous systems are found at all morphology parameters but cluster strongly at the shortest periods.  Long-period systems $P$$>$0.8 d are preferentially detached systems which populate the upper right corner at $morph$$>$0.9  as expected for ellipsoidal variables.  Very few contact systems lie at $morph$$<$0.75.  Ambiguous systems concentrate at short periods and span the full range of $morph$ but show a concentration near $morph$$\sim$0.75.  Many ambiguous systems receive their designation for having poor fits to either model, generally because the light curve is asymmetric. We suspect that asymmetric light curves end up with $morph$$\approx$0.75 designations as a consequence of the information loss in downprojecting a higher dimensional manifold to a single dimension \citep[see][and descriptions therein]{Kirk2016}. 

\begin{figure}[ht!]
%\fig{/d/zem1/hak/chip/research/Larry/KB/phoebe/KB/Tvarfudgechi2/histerrs.pdf}{6in}{}
\fig{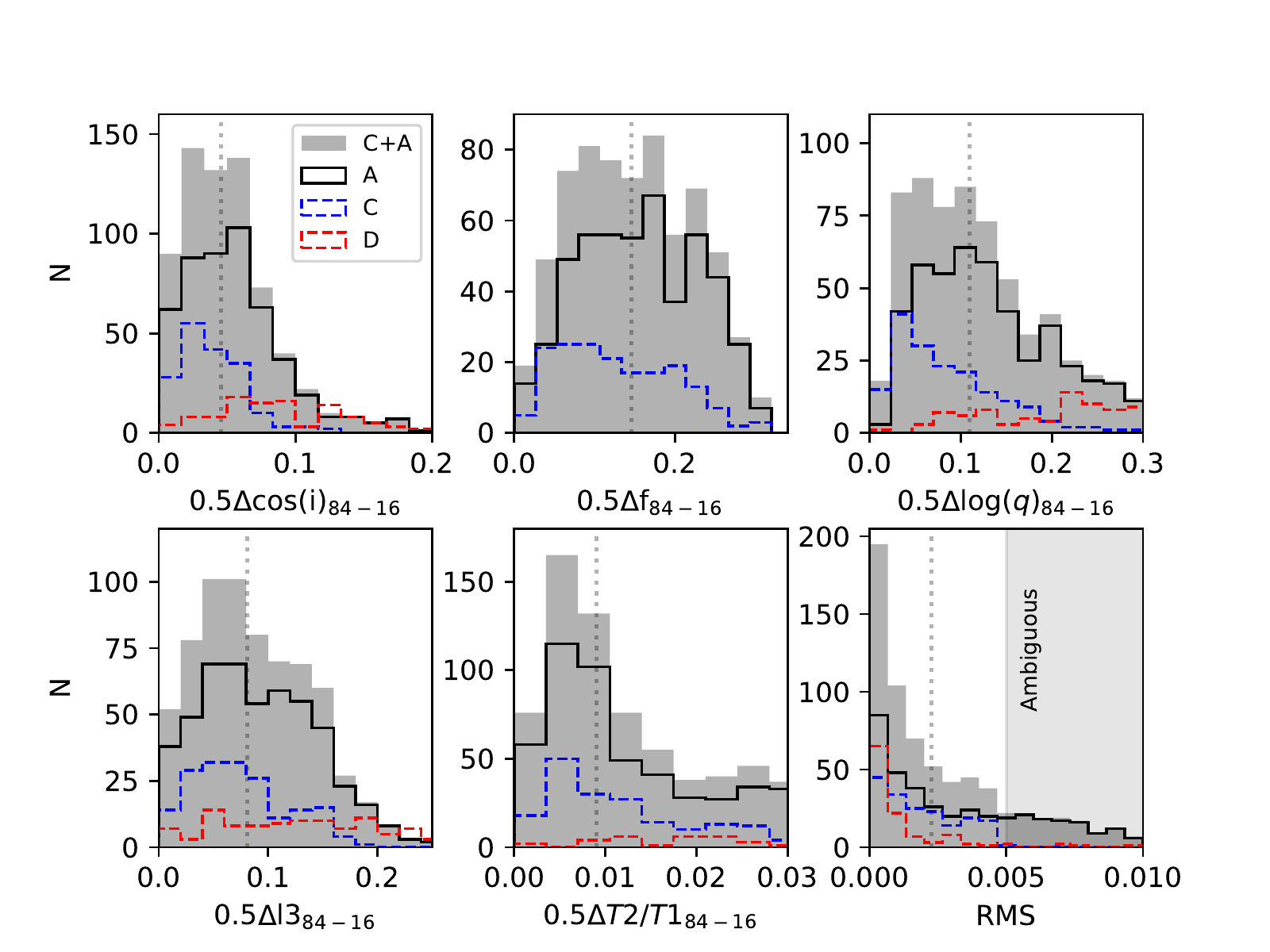}{6in}{}
\caption{Histograms of the uncertainties on systems classified as a contact binary (C), detached (D), or ambiguous (A), characterized as one-half the difference between the 84th and 16th percentile values of each parameter.  Vertical dotted lines mark the median of each C+A distribution. \label{fig:contactserrs_hist}}
\end{figure}

Figure~\ref{fig:contactserrs_hist} presents histograms of the  uncertainties on each fitted parameter, characterized as 0.5 times the difference between the 84th and 16th percentile value, i.e., 0.5$\Delta$$\cos$($i$)$_{\rm 84-16}$, for systems modeled as contact binaries.  These values constitute approximate 1$\sigma$ uncertainties as long as the distributions are nearly Gaussian (they are often not). The gray shaded histogram includes all \numcontactsplusambiguous\ C+A systems.  Systems classified as contact {\it (blue)}, detached {\it(red)}, or ambiguous {\it(black)} are plotted separately by color. The median 0.5$\Delta$$\cos$($i$)$_{\rm 84-16}$ is 0.045 (upper-left panel), and all values are $<$0.22, indicating that the MCMC simulations generally provide strong constraints on inclination.  This is especially true for contact (C) systems which have smaller median   uncertainties than the other subsets.  The median 0.5$\Delta{f}_{\rm 84-16}$ is 0.15 (upper middle panel) but smaller for the contact subset.  The median 0.5$\Delta$$\log$($q$)$_{\rm 84-16}$ is 0.11 (upper right panel), with a few cases as large as 0.38, indicating the the probabilistic constraints on \logq\ are strong in most instances---again, much smaller for the contact subset.  Detached systems have larger uncertainties, as expected, given that the detached geometry light curve is not sensitive to $q$.  The median third light uncertainties (lower left panel) are 0.5$\Delta{l3}_{84-16}$=0.08.  The median \Tratio\ (lower middle panel) is 0.009, indicating that the temperature ratio is constrained to better than 1\% in most cases, although the artificial hard bounds imposed on \Tratio\ at [0.95, 1.05] mask the true extent of possible variations.  The lower right panel presents the distribution of the RMS between the mean of the data and the best-fitting contact binary model, where the median RMS is 0.0022.  This is seven times larger than the nominal {\it Kepler} photometric uncertainty of 0.0003, affirming that the data contain the signatures of real physical features not present in the nominal PHOEBE models.  However, the distribution peaks sharply at small RMS for all subsets, meaning that a large fraction of the sample exhibit reasonable agreement with the models.  The shaded region at right marks the upper limit of RMS=0.005 defining one criterion for the ambiguous designation.

\begin{figure}[ht!]
%\fig{KB/phoebe/KB/errsVScosi.pdf}{5in}{}
\fig{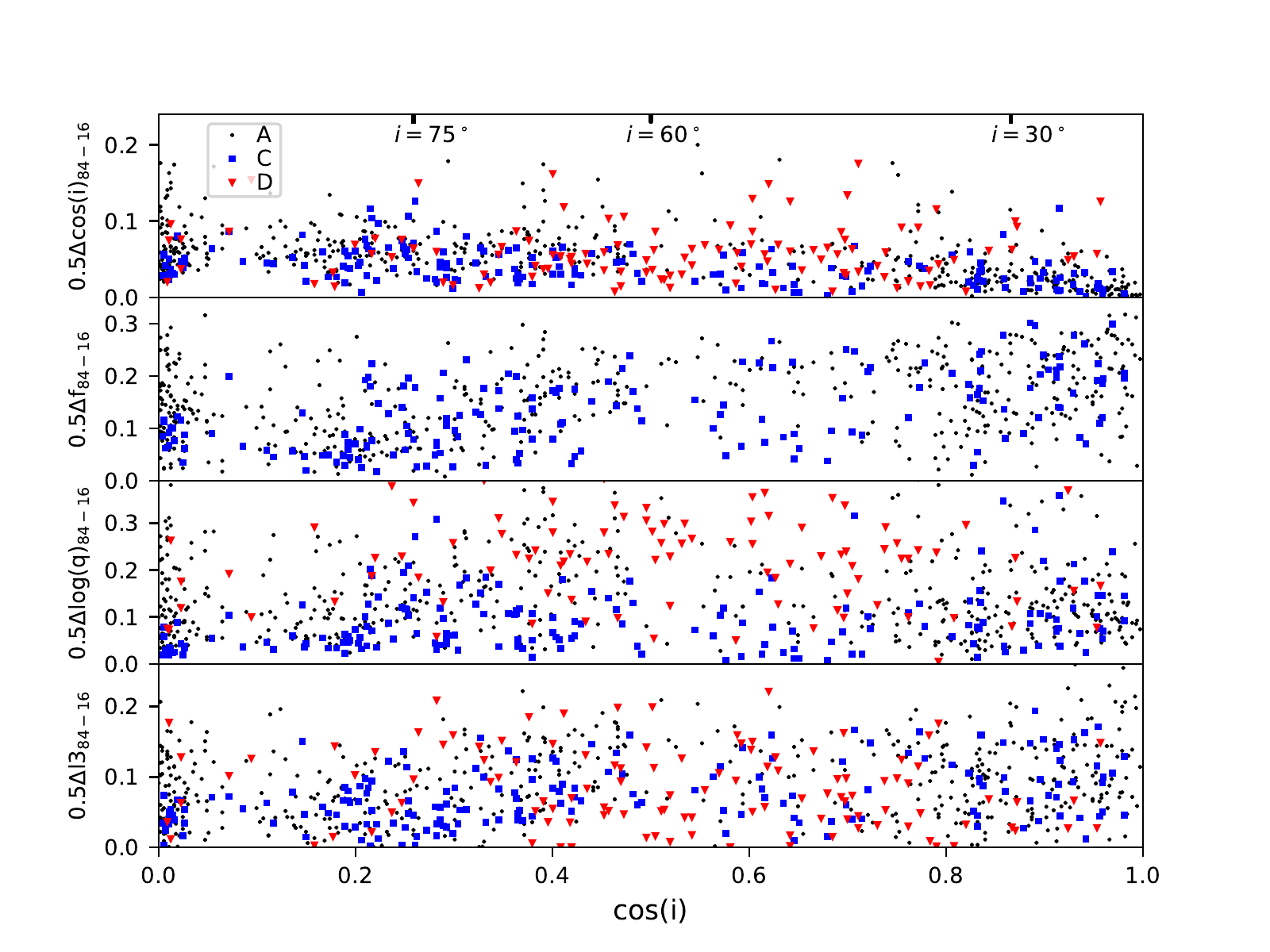}{5in}{}
\caption{Bayesian uncertainty versus \cosi\ for probable contact binaries, detached binaries, and ambiguous cases. \label{fig:errsVScosi}}
\end{figure}

Figure~\ref{fig:errsVScosi} plots the Bayesian uncertainties versus \cosi, with contact/detached/ambiguous systems represented as blue squares/red triangles/grey dots, respectively.  The top panel show that the typical uncertainty on \cosi\ is smallest among contact systems and that the uncertainty decreases toward large \cosi\ (small $i$; face-on systems).  This may be understood as a consequence of all the projected quantities in a binary system (e.g., linear size, velocity) scaling as $\sin$($i$), which changes most rapidly at $i$$\approx$0\degr, allowing for better discrimination between models with similar inclination.  The uncertainties on $f$ are similar between C and A systems, and there is a modest trend with \cosi\ in the sense that $f$ is better constrained at large inclinations (small \cosi). This is where the eclipsing geometry creates light curves laden with information about the system parameters.  Constraints on the mass ratio (third panel) are much stronger for C than for A or D systems, unsurprisingly, regardless of inclination.  The mass ratio in contact systems becomes less well constrained at large \cosi.  A similar pattern is evident in the lower panel where the typical uncertainty on third-light fraction grows with \cosi.  Constraints are similar for all subsets.  Taken together, these panels indicate that the best-constrained systems will be those at large $i$ where the (near-)eclipsing geometry provides the mostly highly structured light curves containing information on the principal system parameters.  

\begin{figure}[ht!]
%\fig{phoebe/KB/histvals.pdf}{6in}{}
\fig{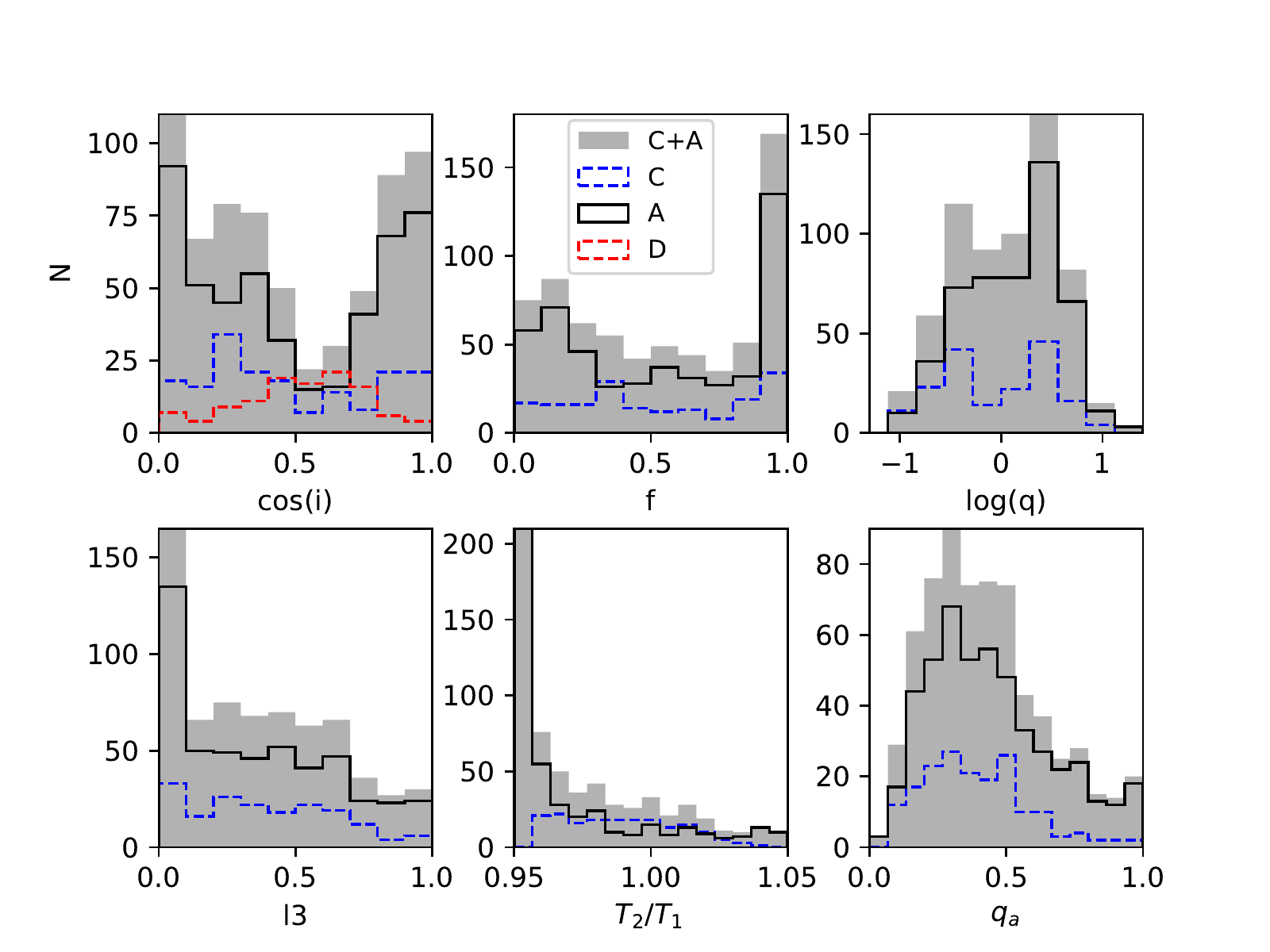}{6in}{}
\caption{Histogram of best-fitting model contact binary parameters for contact plus ambiguous systems.  \label{fig:contacts_hist}}
\end{figure}

Figure~\ref{fig:contacts_hist} displays histograms of best-fitting model parameters for all \numcontactsplusambiguous\ systems modeled as contact binaries or ambiguous cases.  The upper left panel shows that the distribution of \cosi\ is strongly peaked at small and large values when all C+A systems (gray shaded histogram) are considered.  The majority of these are ambiguous systems which have ill-fitting models (RMS$>$0.005) or large temperature differences between the stars (\Tratio$<$0.96 or $>$1.04).  The distribution becomes nearly flat, as expected for random viewing angles, when only probable contact systems are considered (dashed blue histogram) but there is still a statistically significant excess at inclinations  \cosi$\approx$0.3 ($i$$\approx$70\degr). Investigation shows that some of these systems have light curve asymmetries and RMS values near the 0.005 cutoff.  A more stringent RMS$<$0.004 cutoff significantly reduces this excess, which we tentatively attribute to the effects of starspots.  Furthermore, spots at longitudes 0\degr\ or 180\degr\ in moderate-inclination (\cosi$\approx$0.6) systems will change the ratio of primary:secondary minima, displacing them toward more edge-on orientations (smaller), further contributing to the apparent excess of \cosi$\approx$0.3 systems.  Spots appear to be abundant in contact systems, but a full exploration of spot properties is beyond the scope of this work.   

It is also noteworthy in Figure~\ref{fig:contacts_hist}  that the distribution of probable detached systems---which have poorly measured inclinations, in general---peaks in the \cosi$\approx$0.5 regime.  At these viewing angles detached but nearly overflowing stars undergo grazing eclipses are most likely to be identified as they produce distinctive V-shaped inflections in the light curve.  Detached systems with very low inclinations will exhibit ellipsoidal modulations and appear much like contact systems and may be incorrectly identified as such.  As a case in point, KIC09953894 from the pilot study has an indicated $i$$\approx$87\degr\ when modeled as a contact system but $i$$\approx$59\degr\ when correctly modeled as a detached system.  At the other extreme, detached systems with very large (eclipsing) inclinations have already been removed from the sample by virtue of the initial selection criterion $morph$$>$0.7, thereby producing the apparent deficit of \cosi$\approx$0 detached systems.

The left middle panel of Figure~\ref{fig:contacts_hist} shows that the distribution of fillout factor is slowly declining with $f$  contact systems. There is a small excess near $f$=1 for contact systems that would indicate a preference for Roche lobes filled near the maximum volume.   The upper right panel ($\log q_{50}$) shows that mass ratio distribution for both the C+A and C subsets is similar, evincing two peaks on either side of zero.  As discussed in connection with Figure~\ref{fig:fourbyfourmodels}, the \logq$<$0 peak contains systems at inclinations $i$$\gtrsim$50\degr\ while the systems having \logq$>$0 are those with $i$$\lesssim$50\degr.   The lower left panel shows that contact binaries have appreciable third-light contributions.  The distribution is flat or slowly falling out to $l3$=0.8 where few  systems are found.   All subsets show a preference for considerable third light.  The distribution demonstrates that 77\% of {\it Kepler} probable contact binaries have $l3$$>$0.15, indicating that luminous triple systems are common in close binaries---much more common than among the general population where this fraction is about $\gtrsim$40\% among short-period binaries but much lower for longer-period systems \citep{Tokovinin1997}.  Our finding for contact binaries is generally consistent with the 59$\pm$8\% tertiary fraction discovered among a spectroscopically studied sample of close binaries \citep{Pribulla2006}.  Our result may be slightly biased by our initial removal of the \numremoved\ systems having very low levels of photometric variability, some of which are expected to have large $l3$, but some of these \numremoved\ will also be low-$i$ (high-\cosi) systems as well.  

The lower middle panel of Figure~\ref{fig:contacts_hist} shows that a large fraction of the C+A subset is best fit by models having significant temperature differences, \Tratio$<$0.96.  This is not surprising as this subset contains unidentified detached systems plus a large portion of ill-fitting light curves with strong asymmetries.  The probable contact systems, by our adopted criteria, have 0.96$\leq$\Tratio$\leq$1.04, and the distribution is nearly flat across the allowed range, with a preference for \Tratio$<$1.  The distribution of absolute mass ratio  in the lower right panel peaks near  $q_{\rm a}$$\approx$0.4, indicating that the majority of close binaries have mass ratios far from unity, both for the C+A sample and the more restrictive C subset.   There are very few contact systems having $q_{\rm a}$$>$0.8.  

\begin{figure}[ht!]
%\fig{phoebe/KB/cornerall-contactTlimqabs.pdf}{7in}{}
\fig{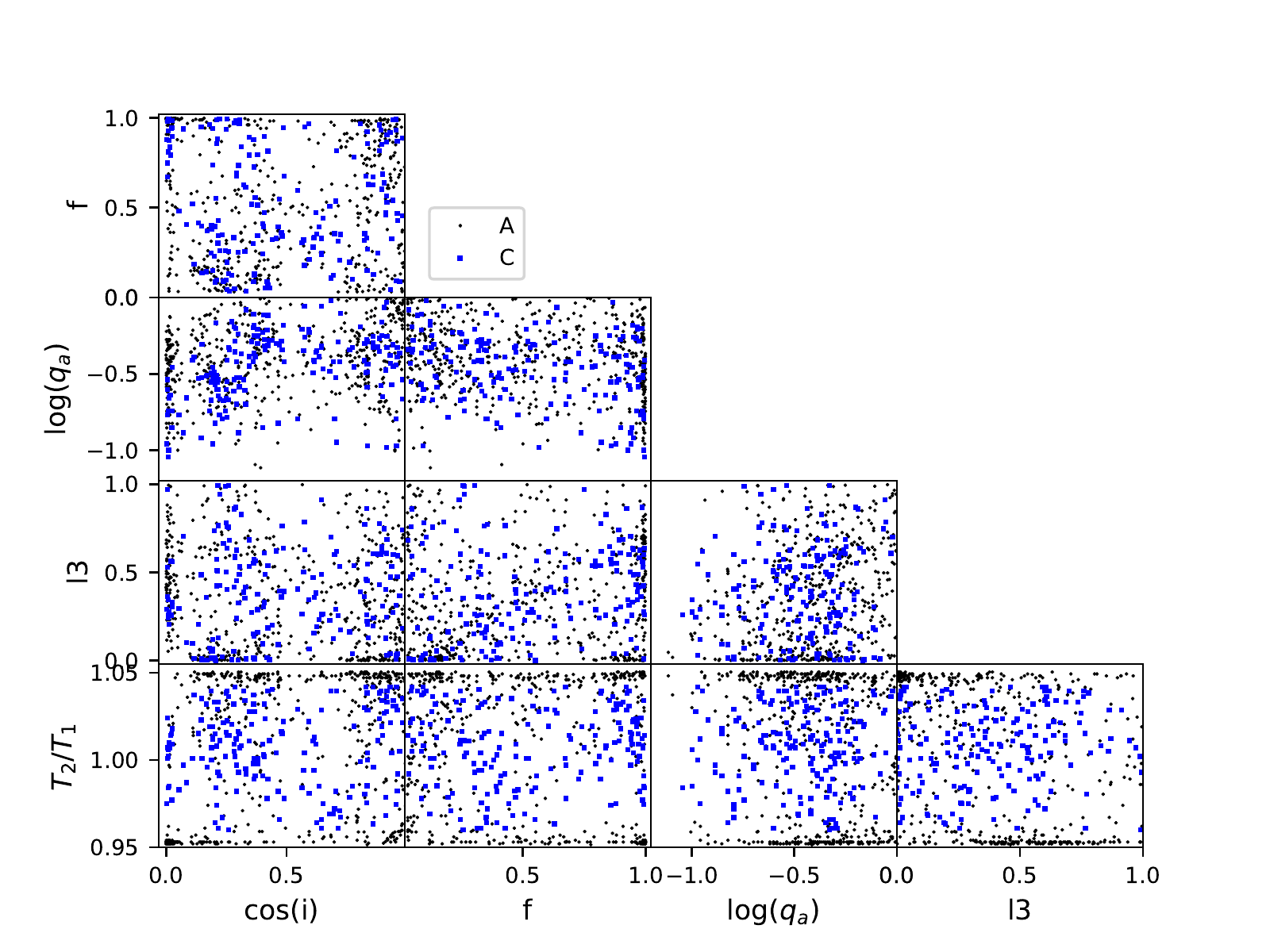}{6in}{}
\caption{Best-fitting values of \cosi, $f$, $\log(q)$, $l3$, and \Tratio\ for all \numcontactsplusambiguous\ systems designated as contact binaries (blue squares) or ambiguous (grey points). \label{fig:corner}}
\end{figure}

Figure~\ref{fig:corner} plots each of the principle contact binary parameters against one another, with grey points  representing the A subset and blue squares marking the probable contact subset.  In cases where the mass ratios has been inverted to obtain $q_{\rm a}$ we have also inverted \Tratio. The pileup of points at \Tratio$=$0.95 or 1.05 along the bottom/top in the lower row of panels illustrates the abundance of either detached systems or those having ill-fitting light curves.  The distribution of contact binaries shows no strong correlations or patterns across each parameter pair, with a few exceptions.  In the $q_{\rm a}$ versus \cosi\ panel there is a paucity of contact systems having moderate mass ratios $-$0.4$<$$\log(q_a)$$<$$-$0.1 at small \cosi\ ($i$$>$60 edge-on systems). A substantial population of ambiguous systems do populate this regime.  At these large inclinations the constraints on \logq\ are quite strong (Figure~\ref{fig:errsVScosi}), so we speculate that spots or other physical phenomena  not present in the models force the best-fitting solutions toward larger \cosi\ and smaller $\log(q_a)$ or cause true contacts to receive an ambiguous designation.  The distribution of $\log(q_a)$ versus $l3$ shows that extreme-$q_{\rm a}$ systems are found across the full range of $l3$, but less commonly when third light is large.  This further underscores the warning that system parameters become less certain in the presence of substantial third light.        

\clearpage

\section{The Dependence of Mass Ratio on Period for Contact Binaries from a New Set of Model Predictions\label{sec:qp}}

\subsection{Theoretical Background} 
The dependence of mass ratio on orbital period provides the most direct test of evolutionary theories of contact binary systems. The distribution of mass ratio in short-period $P<$100~d detached systems with $\approx$1 \msun\ primaries (progenitors of contact systems) is observed to be nearly uniform \citep{Niu2020, Moe2017}, so deviations from a uniform distribution in contact systems must be a consequence of evolution. The most commonly cited modification that is expected is a lower limit on mass ratio due to the Darwin instability \citep{Darwin1893}, generally taken to be a fixed value around 0.09 \citep{Rasio1995} or  $\approx$0.07 \citep{Arbutina2009,Li2006}. The evolutionary models of \citet{Molnar2019} and \citet{Molnar2022} yield orbital period dependence of this minimum, along with period dependences of a maximum $q_{\rm a}$ and of typical $q_{\rm a}$ values.

The central premise of the contact binary star evolution theory of \citet{Molnar2022} is that it is driven by nuclear evolution of the primary (more massive) star and proceeds in a steady fashion. The MESA \citep{Paxton2011} package is used to follow in detail the evolution of the primary star including gradual change in mass and angular velocity due to interaction with the companion star. The calculations yield mass, radius, luminosity and orbital period as a function of time, properties that can be compared in detail with observations of individual contact binary systems and the statistics of ensembles. They also yield the evolving moment of inertia which determines the mass ratio at the onset of tidal instability, the immediate cause of stellar merger and subsequent red nova explosion. Finally, they make specific predictions about two regions of mass ratio-orbital period space that should be devoid of contact binaries: low mass ratios unaccessible due to the tidal instability and high mass ratios quickly depleted due to a secular instability.

The binary evolution computations follow a procedure analogous to the binary evolution code of MESAbinary (\citep{Paxton2015}). In that code the mass receiving star is treated as a point mass assumed capable of receiving whatever mass is transferred to it while the structure of the mass donor star is followed in detail. The mass transfer rate is set by the requirement that the donor star continues to fill its Roche lobe and is driven by evolution of that star (and optional orbital angular momentum loss). \citet{Paxton2015} say the code cannot be used for contact binary systems. The innovation of \citet{Molnar2022} is to realize that in contact systems the roles played by the stars are reversed but the computation is analogous. The mass donor star is assumed capable of transferring mass at whatever rate is required and need not be followed in detail while the mass transfer rate is set by the requirement that the mass-receiving star continues to fill its Roche lobe and is driven by evolution of that star. MESA is used to compute the evolution of the mass-receiving star in a series of time steps small enough to follow the changing stellar rotation rate and mass accretion rate (which are computed to conserve system mass and angular momentum in tandem in a separate program much as MESAbinary does for the cases to which it applies).

\begin{figure}[]
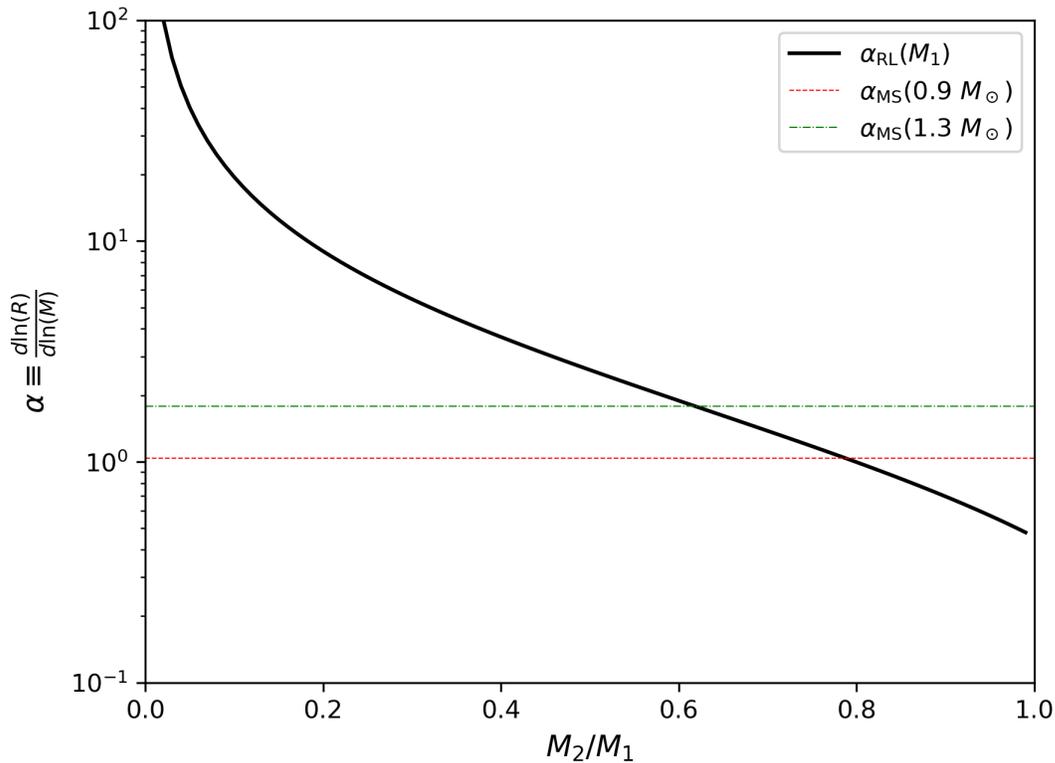

\fig{Fig62.png}{6in}{}
\caption{The power law index $\alpha$ relating the change in size of the primary star to its changing mass as a function of mass ratio. The solid black line is  $\alpha_{\rm RL}$, the change in the equivalent Roche lobe radius. The red dashed and green dot-dashed lines mark $\alpha_{\rm MS}$, the indices for equilibrium main sequence stars at 1 Gy age and mass 0.9 and 1.3 $M_\odot$, respectively. Stable accretion onto the primary can occur where $\alpha_{\rm RL}>\alpha_{\rm MS}$.  \label{fig:stability}}
\end{figure}

It is necessary to show that steady state evolution is possible. The typical viewpoint in the literature goes back to \citet{Flannery1976} and subsequent works that found no steady solution. Instead their computations showed an oscillation between contact and detached stages (and between mass transfer of alternating signs) on a thermal time scale. They concluded this behavior is general to all mass ratios. It was based on computation of one case with near unity mass ratio which they expected to be the most stable case. However, stable accretion by the primary can occur when an increment of mass transferred increases the effective Roche lobe radius more than the stellar radius. Figure~\ref{fig:stability} compares the power law index $\alpha\equiv\frac{d{\rm ln}R}{d{\rm ln}m}$ versus mass ratio for the Roche lobe of the primary to that of an equilibrium main sequence star. Steady evolution can occur when $\alpha_{\rm RL}>\alpha_{\rm MS}$. Hence steady state evolution is possible for $q<$0.6--0.8 (for primary masses in the observed range of 0.9--1.3~$M_\odot$). Nuclear evolution sets the mass transfer rate in that range with values low enough to justify the assumption of equilibrium main sequence structure. Systems that have initial mass ratios $q$$\gtrsim$0.8 will experience a secular instability that likely results in significant mass transfer on a thermal timescale until the mass ratio reaches the stable regime (a result consistent with \citet{Flannery1976}). This would be followed by lower, steady transfer rates thereafter. As thermal timescales are much shorter than than nuclear timescales, this leads to an effective maximum mass ratio that should be observed---a distinctive prediction of this model.

Given that a steady state solution is possible it is necessary next to identify mechanisms that drive the secondary to the stable mass transfer rate. Rates above the steady rate are self-limiting as they drive the system out of contact. The response to rates below the steady rate (or even negative rates) depend on the nature of the secondary star. Companion stars are observed to fill their Roche lobes and to have surface temperatures close to that of the primary, which requires radii and luminosities significantly greater than expected for main sequence stars. The consensus in the literature is that the luminosity is powered by the primary star which shares its output by advection of the surface layers. \citet{Molnar2019} suggest the steady state structure of the secondary star could be modeled by extending this idea: compute the structure using a surface boundary condition fixed at the temperature and pressure of the primary star at the inner Lagrangian point. The result would likely approximate a composite of a core much like a main sequence star (as the nuclear burning is insensitive to surface boundary conditions) surrounded by a nearly isothermal envelope (as the nuclear burning contributes relatively little to the surface luminosity). Densities would drop exponentially with radius in the core but more slowly in the envelope. This very approximate description of the secondary is sufficient to show how it would respond to mass transfer at less than the steady state rate. The inner Lagrangian point would move deeper into the primary star increasing the pressure and temperature of the boundary condition. Energy from the primary would then drive expansion of the secondary which in turn would increase the mass transfer rate. 

In summary, steady mass transfer driven by the nuclear evolution of the primary is possible in contact systems with mass ratio less than 0.6 to 0.8, and mechanisms exist that maintain the transfer from the secondary at the steady rate.

\subsection{Observational Tests} 

The present data set is well suited to testing both historical and new predictions regarding the upper and lower limits on $q$ for contact systems.  It has the largest unbiased sample analyzed in a uniform fashion, no period dependence in the selection of systems, and no period dependence in any systematic biases in mass ratio determination.  We first present the data and then discuss them in context of the \citet{Molnar2022} evolutionary models.

\begin{figure}[ht!]
%\fig{phoebe/qversusP.pdf}{7in}{}
\fig{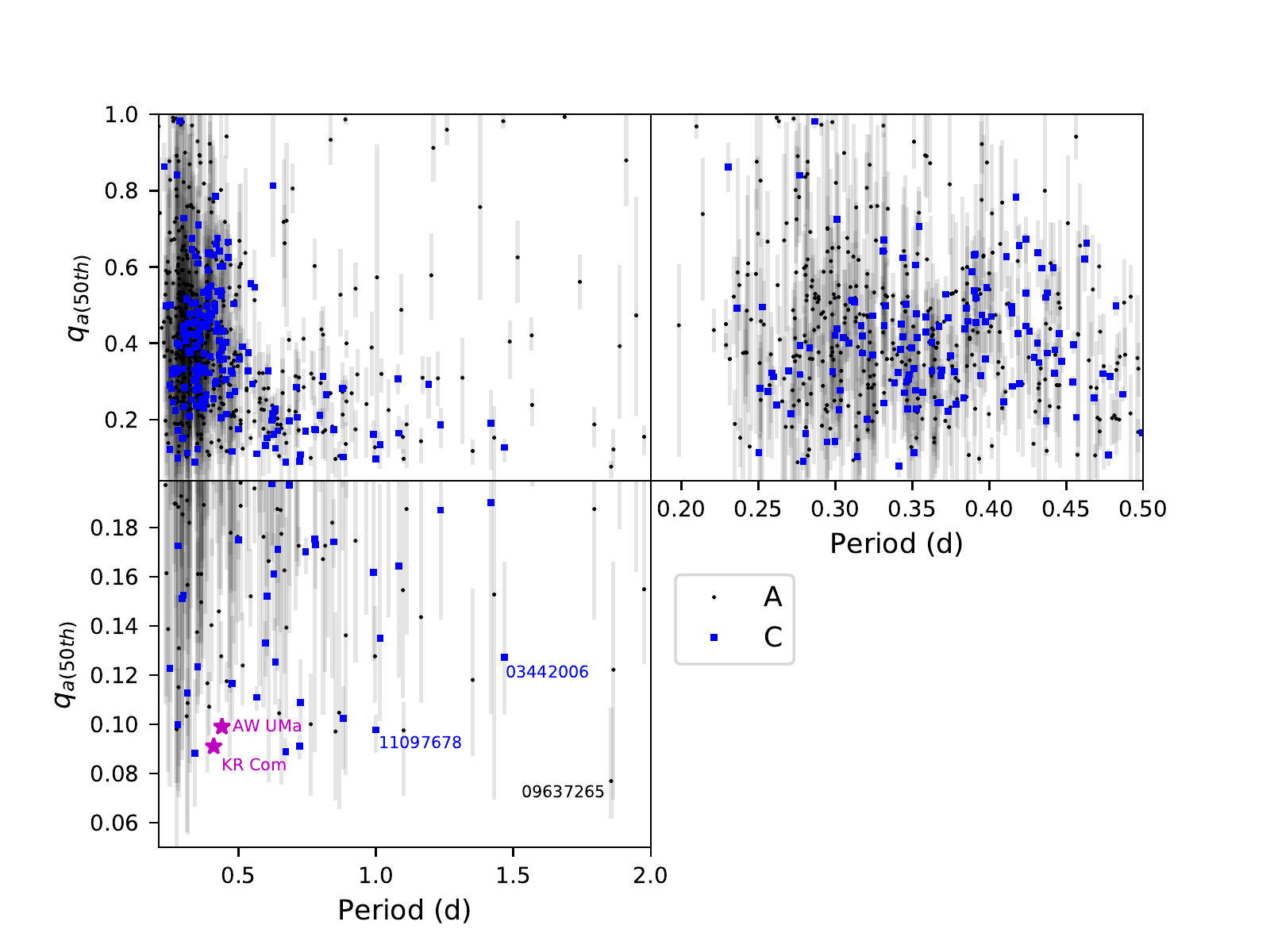}{6in}{}
\caption{50th-percentile absolute mass ratios $q_{\rm a}$ versus $P$ for probable contact binaries and ambiguous systems.  \label{fig:qversusP}}
\end{figure}

Figure~\ref{fig:qversusP} plots the absolute mass ratio $q_{\rm a}$ versus period for the probable contact systems and ambiguous systems modeled as contact binaries. We focus our analysis on probable contact systems represented by the blue squares. The majority of the ambiguous systems, represented by small black points, are probable contact systems as well and can be considered for confirmation of trends, although with more spread in values. Symbols with vertical error bars depict the 50th-percentile and 84th--16th percentile ranges.  Each of the three panels shows a different window of mass ratios and periods, allowing the full dynamic range to be conveyed.  The upper left panel displays the full range in both parameters, confirming the well-known concentration of contact binary periods in the range 0.25--0.5 days. The mass ratio in this period range is revealed to concentrate in the range 0.2--0.8.  There are very few systems with $q_{\rm a}>0.8$ at any period.  At periods longer than $P\approx0.5$~d the median mass ratio of blue squares shifts toward more extreme values.  At $P>0.5$~d nearly all of the systems have $q_{\rm a}<0.3$. The lower left panel better shows the distribution of mass ratios at longer periods.  No systems are found at $q_{\rm a}<$0.08. This lower bound becomes more clear when some additional extreme systems from the literature are added.  AW UMa, for instance, is a well-studied short-period ($P$=0.43~d) system with a spectroscopically determined mass ratio of $q=$0.099 \citep{Rucinski2015}. A magenta star marks its position in Figure~\ref{fig:qversusP}, lying near the lower limit in the $q_{\rm a}$ versus $P$ plane.  An additional star marks KR Com \citep{Rucinski2002b}, another spectroscopically confirmed extreme-$q$ system that lies on this boundary. While a few other extreme $q_{\rm a}$=0.05--0.10 candidates exist in the literature, they are based on photometric measurements only, which carry larger uncertainties on account of the (often unknown) third-light contributions. The lower left panel also shows that long-period contact systems are rare and have $q_{\rm a}>$0.1 when $P\geq$1 d, avoiding the most extreme mass ratio regime.  The upper right panel shows the short-period portion of the range at higher resolution. It shows the few systems with $q_{\rm a}>0.8$ are among the shortest period systems ($P\lesssim$0.3~d). 

% possible violators listed in figure 63
 % 03442006 i=17, f=0.99, l3~0, RMS_contact is << RMS_det; appears to be a contact system; looks reliable
 % 03966491 i=12, very low amplitude LC;  f=0.37; RMS_con/RMS_det~0.86; best detached models suggest R/Rmax 0.436,0.738, so some arguments here for detached
 % 11097678 see pilot sample  ; seems robust;
   % Evan's high res spectra show no third light, consistent with T1=T2 fitting
 % 08430210 i=30, l3=86%, very low amplitude; f=0.04, l3=0.7 and RMS_con/RMS_det~0.98, so a good case can be made that this one is detached.  
 % 09637265  RMScon/RMSdet=0.971, R2/R2max=0.971 f=0.48  l3=0.36 so perhaps detached

\clearpage 

\begin{figure}[ht!]
%\fig{KB/qversusPspecial.pdf}{6in}{}
\fig{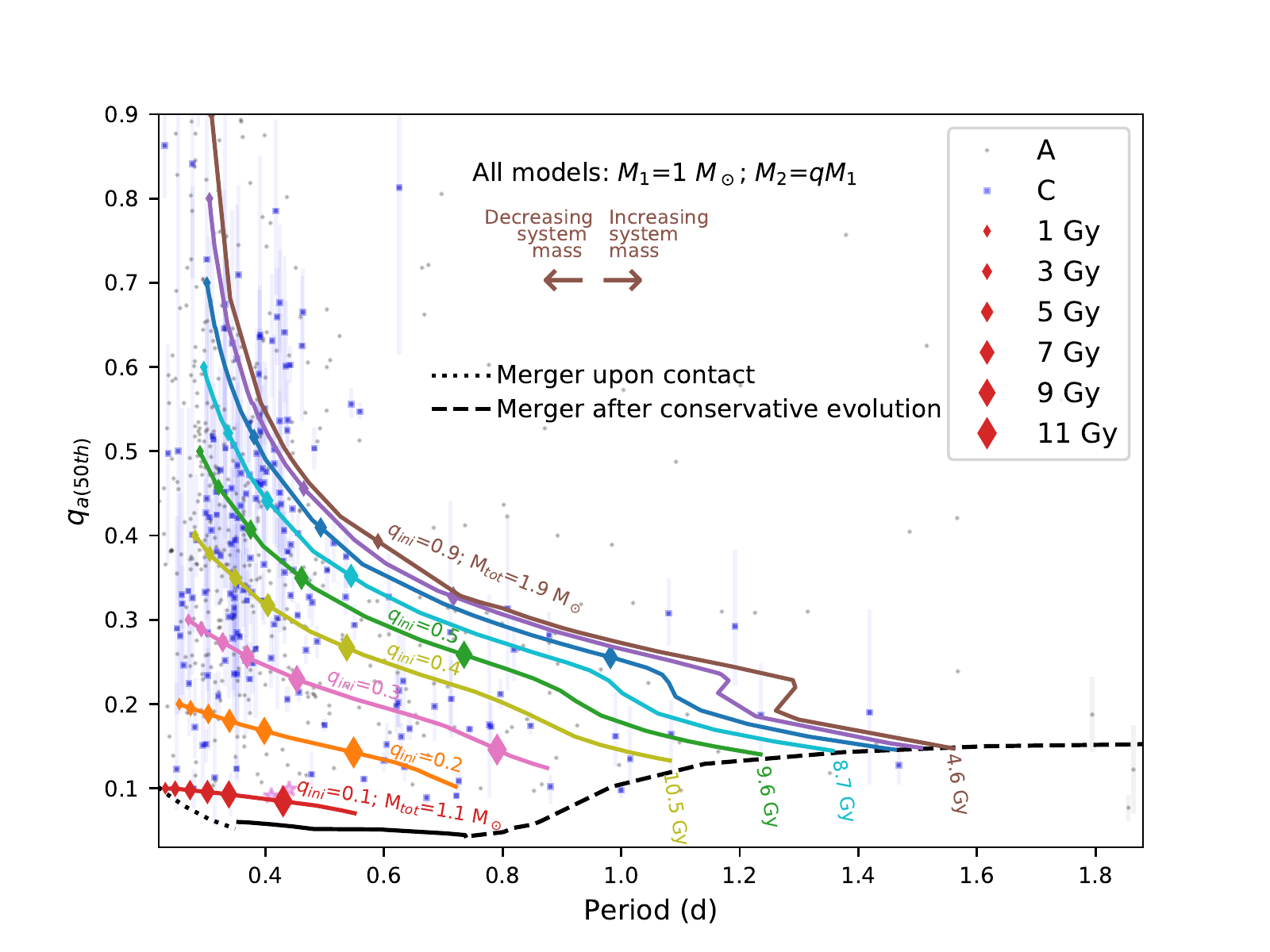}{6in}{}
\caption{Mass ratio versus orbital period as in Figure~\ref{fig:qversusP}.  The colored tracks show the evolution of contact binaries---from upper left toward lower right---of contact systems with initial mass ratios $q_{ini}$=0.1--0.9, as labeled.  Diamonds mark equal time intervals every 2 Gyr.  Decreasing/Increasing the total system mass shifts models slightly to the left/right, as indicated by arrows. \label{fig:qpfigure}}
\end{figure}

Figure~\ref{fig:qpfigure} plots mass ratio versus orbital period for contact and ambiguous systems, as in Figure~\ref{fig:qversusP}.  Colored curves depict evolutionary tracks for contact systems with initial primary mass $M_{
\rm 1}$=1 \msun\ and initial mass ratios $q_{ini}$=0.1--0.9, color coded and labeled \citep[based on the mass and angular momentum conserving models of][]{Molnar2019,Molnar2022}. The stellar contact components begin as evolved single stars at an age of 1 Gyr, based on the  $>$0.6--3 Gyr  required for Lidov-Kozai cycles and magnetic braking to bring close detached systems into contact \cite{Stepien1995,Chen2016,Hwang2020}. Increasing or decreasing the primary mass shifts the model tracks to the right or left, respectively, by a small amount.  Diamonds of increasing size along each track mark the progression of 2 Gyr intervals from 1 Gyr to 11 Gyr, from upper left toward lower right, as the system transfers mass to the more massive component.  The dashed curve defines a lower limit on $q$ and an upper limit on $P$ at which the suite of all model systems (including those of larger and smaller $M_{\rm 1}$ not depicted) reach the critical threshold for the onset of the Darwin instability when the rotational angular momentum of the components exceeds one-third the orbital angular momentum.   The short section of dotted curve in the lower left designates a limit for a small subset of initially very extreme mass ratio systems where the conditions for merger are satisfied upon initial contact.

The distribution of systems in Figure~\ref{fig:qpfigure}, both contact and ambiguous, stands in good agreement with the model predictions.  Short-period systems with $q\approx$0.8--1 at the upper left are rare, consistent with the expectation of rapid mass transfer during the first 1 Gyr in such systems, the result of a previously unnamed instability discussed further in \citet{Molnar2022}.  The short-period bound along the left edge seen in the data is reflected in the models.  The small number of points that lie to the left of model tracks can be reproduced by systems having a slightly smaller total mass or by a slightly younger system age.  The upper right portion of the plot is devoid of points, except for a few ambiguous systems that are likely to be detached systems.  A small concentration of contact systems near 0.4 d and $q_{\rm a}$=0.6 lying to the right of the $M_{\rm tot}$=1.9 \msun\ model track may be more slightly more massive systems.  No contact systems lie below the dashed line demarcating the locus of stellar mergers; KIC03442006 at $P$=1.47 d comes the closest, but its 1$\sigma$ uncertainties straddle the limit.  One ambiguous system, KIC09637265, lies below this limit at $P$=1.85 d, a location where detached systems are permitted. Overall, the bulk of the data points lie at short periods and modest mass ratios---locations where the models predict contact binaries spend the majority of their lifetime.      

Figure~\ref{fig:qpfigure} implies several predictions that can, in principle, be tested by detailed studies of individual systems or groups of systems.  Systems near the top of the Figure at $q\approx$1 are expected to be young and rapidly evolving in $q$, perhaps showing signatures of vigorous mass transfer.  Long-period contact systems are expected to exhibit extreme mass ratios.  Any long-period systems found below the predicted limit are expected to be detached systems.  Short-period contact binaries near the left edge of the model tracks are expected to be young, achieving contact only recently.  The lower left corner of the $q$ versus $P$ should be populated by systems with small total mass and the upper right populated by large-$M_{\rm tot}$ systems.   A better knowledge of outliers and systems at the periphery of the model tracks will help inform the boundary conditions for additional modeling efforts and perhaps identify different contact binary formation/evolution channels than the one envisioned in this set of models.

\clearpage

\subsection{A Comparison of Best Detached Model Solutions with Contact Model Solutions {\label{sec:betternoncontacts}} }

\begin{figure}[ht!]
%\fig{phoebe/DCratiovsP.pdf}{7in}{}
\fig{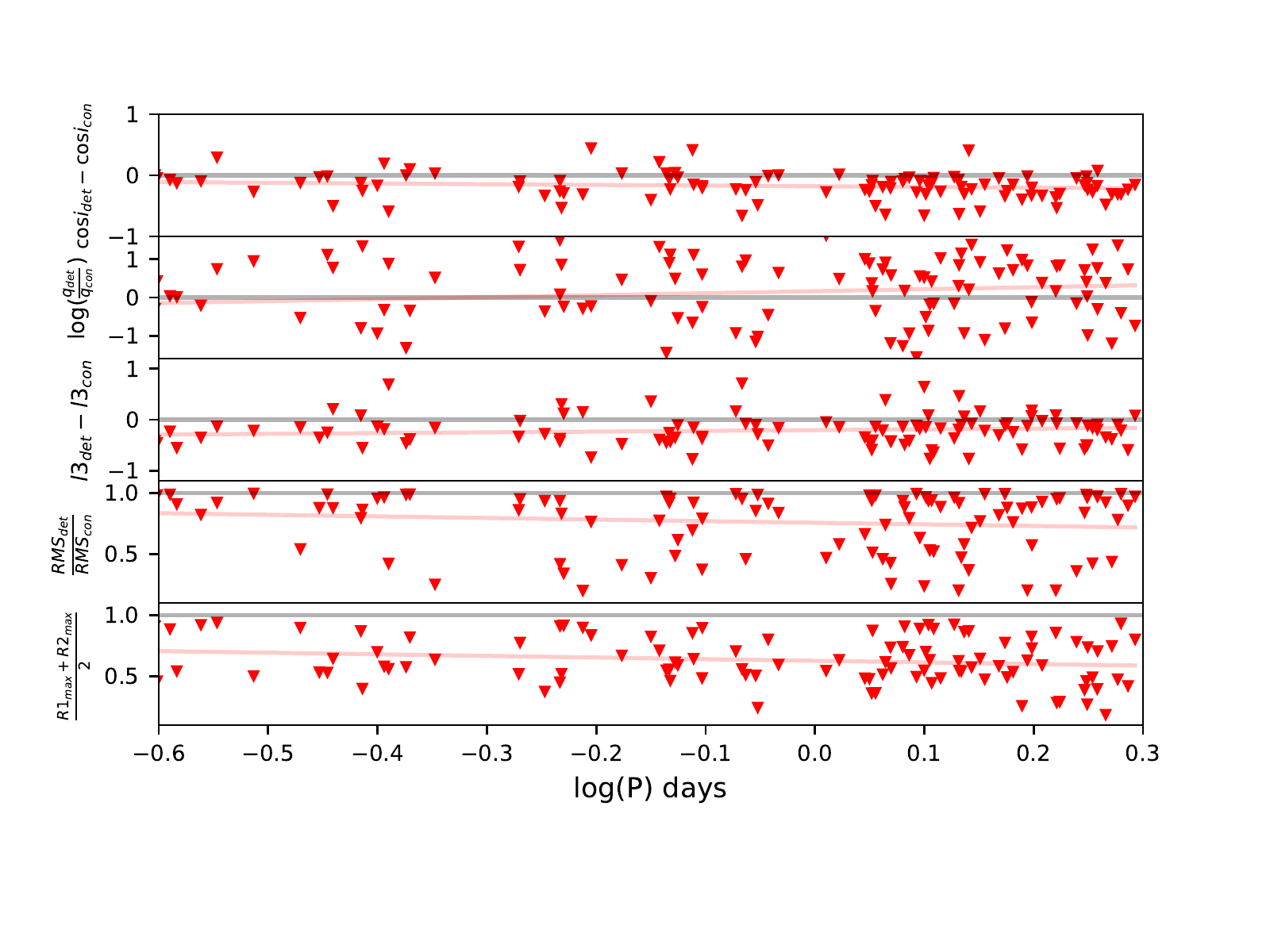}{5in}{}
\caption{Differences or ratios of 50th-percentile parameters comparing  detached models and the \Tapprox\ contact models for the \numdetached\ probable detached binaries.  Parameters from best detached model can be very different from the best contact model.  \label{fig:DCratio}}
\end{figure}

Figure~\ref{fig:DCratio} compares 50th-percentile parameters as inferred from the detached model with the 50th-percentile parameters from the \Tapprox\ contact model versus \logp\ for the \numdetached\ probable detached systems.  Unsurprisingly, there are more detached systems at long periods than short periods. The red line in each panel shows the linear fit to the data.   The upper panels shows the difference between most probable \cosi\ values, where the RMS dispersion is 0.25. At short periods the differences are symmetric about zero, but at long periods they becomes systematically negative; contact models yield systematically larger \cosi\ (smaller inclinations) than detached models in long-period systems that have detached configurations.  This means that detached systems mistakenly modeled as contact systems will have \cosi\ values that are too large ($i$ too small). The second panel plots the difference in \logq\ between detached and contact models.  The dispersion is large (0.78) but not systematic, primarily confirming an earlier conclusion that mass ratio is not well constrained for detached systems.  The third panel shows the difference in third light fractions, which is consistently negative at all periods with a mean of $-$0.22. Contact models yield systematically larger third light fractions than detached models; detached systems mistakenly modeled as contact systems will yield $l3$ values that are too large.   The fourth panel plots the ratio of the RMS of the best-fitting detached/contact models;  this ratios ranges from near unity to 0.1 across the period range, indicating that sometimes the models are comparable and sometimes the detached model provides a much-superior fit to the light curve.  The RMS alone is not a reliable metric to distinguish between contact and detached geometries.  The bottom panel shows the average fraction of the components' Roche lobes that are filled in the best-fitting detached model.  This derived parameter is akin to a fillout factor for detached systems.  At the shortest periods this fraction is near unity, indicating that both stars nearly overflow.  At longer periods this ratio increasingly departs from unity, indicating that one or both stars are not close to overflowing.  Such long-period systems produce the classical ellipsoidal light curve modulations.   Figure~\ref{fig:Pversusmorph} has already shown that most of the long-period systems are detached and have $morph>$0.9.  Figure~\ref{fig:DCratio} illustrates that the best  detached solutions can be very different from the best contact solutions, highlighting the necessity of making a correct geometrical identification that dictates the light curve modeling strategy. 

\clearpage

\section{Conclusions \label{sec:conclusions} }

We have presented an analysis of \nummodeled\ $P<$2 d close binaries using MCMC methods in conjunction with the state-of-the-art PHOEBE software to model {\it Kepler} light curves.  The reliability of the Bayesian retrievals of system parameters is predicated on the success of the models demonstrated using a pilot sample of ten systems for which phase-resolved spectroscopy provides independent measurements of the systems' mass ratios and estimates of third-light fractions.  The level of intrinsic variability in all of these systems far exceeds the exquisite photometric precision of the {\it Kepler} data, indicating the ubiquity of intrinsic variability from phenomena that may include pulsations, spots, and active regions that vary both in time and location upon the stellar surfaces.  Accordingly, the best-fitting and 16th/50th/84th-percentile parameters presented reflect the mean properties as averaged over the three-year baseline of the {\it Kepler} dataset.  Our detailed photometric and spectroscopic analysis of the ten-object pilot sample provides a representative tour of the types of systems and physical phenomena comprising the {\it Kepler} close binaries: contact and detached systems with a variety of fillout factors, spots, temperature differences, \OC\ variations, and third components.      

\begin{enumerate}

\item{PHOEBE models plus Markov-Chain Monte Carlo analyses of single-band light curves reliably recover the mass ratios for contact binary systems determined independently using spectroscopy as long as the ratio of stellar temperatures is constrained to near unity (our \Tapprox\ model).  Temperature ratio \Tratio\ and mass ratio $q$ can be degenerate in ways that yield incorrect mass ratios if \Tratio is unconstrained.  Light curve modeling combined with phase- and velocity-resolved spectroscopy---in particular, the broadening function analysis---provides strong constraints on system parameters for the ten near-(contact)systems having both datasets.   Adopted practices include two rounds of MCMC model fitting, the first to locate the global minimum within the multi-modal parameter space and the second to characterize the shape and extent of the minimum after marginalizing over the (unknown and unmodeled) parameters by re-normalizing the reduced chi-squared of the best-fitting model to one.  After experimenting with a small suite of competing models, we adopt either a five-free-parameter contact model or a six-free-parameter detached model.  Even when additional model components such as a spot are required (KIC08913061, KIC09164694, KIC09345838) the mass ratio is recovered with good fidelity.  The MCMC 1$\sigma$ uncertainties are generally small for \logq\ and \cosi\, while $f$ and $l3$ are still constrained but with less precision.  }

\item{The PHOEBE model line profiles agree reasonably well with the broadening functions determined from the $R\approx$4000 optical spectra obtained near quadrature phases.  The differences between the model line profiles and the broadening functions provide measures of the third-light fractions that are usually in good agreement with those independently obtained from the best-fitting models. The broadening functions in four of the ten spectroscopically studied systems (KIC04999357, KIC06844489, KIC10292413, and KIC11097678) show a deficit relative to the model  at velocities associated with the secondary (less luminous) component.  Such a deficit could be caused by cool spot or line-emitting material on the secondary.  We favor the former possibility because the deficit persists when portions of the spectrum such as H$\alpha$ are excluded from the BF analysis. These BF deficits in extreme-$q$ may also be explained in terms of accretion flows between components, as in AW UMa \citep{Rucinski2015}. The characteristics of persistent surface inhomogeneities on one or both components could be further constrained using time-resolved multi-band photometry and/or spectroscopy.  } 

\item{At least eight of the ten spectroscopically studied contact binaries show evidence for a third component, either in the \OC\ residuals or the broadening function, consistent with prior evidence that most contact binaries exist in triple systems \citep{Pribulla2006}.   }

\item{Our exploration of the \citet{Horvat2019} implementation of irradiation effects yields mixed results on the ten pilot study systems.  Sometimes including irradiation in the models leads to a significantly lower RMS (KIC04999357), and sometimes the RMS is the same or larger (most of the pilot sample).  Often the impact of spots and/or intrinsic variability swamps the effects of irradiation (KIC08913061, KIC09164694, KIC09345838).  Generally, inclusion of irradiation effects changes the best-fitting parameters very little.  Among the probable contact systems \Tratio\ becomes slightly less extreme and mass ratios become slightly less extreme when irradiation effects are included in the models.   Given the ubiquity of photometric variability arising from spots and other stellar activity, these effects may mask or be degenerate with those caused by temperature differences and irradiation.   We conclude that implementing irradiation effects makes a minimal or negligible difference in the retrieval of parameters for the majority of contact binary systems. Detached systems having large temperature differences may show otherwise.  }

\item{For \underbar{detached} binaries evincing large ellipsoidal modulations like KIC09953894, we show that it is possible to retrieve approximate system parameters, including $q$, {\it without} spectroscopic knowledge of the mass ratio.  The allowed range of parameters is larger, however, than when $q$ is known {\it a priori}. }

\item{Contact binary models of \nummodeled\ {\it Kepler} systems with orbital periods $P<$2 d yield \numcontacts\ (23\%) that are best reproduced using the \Tapprox\ contact configuration.  Cos~$i$ and \logq\ are generally well constrained (median 1$\sigma$ ranges 0.045 and 0.10, respectively) while fillout factors and third-light fractions are less certain (1$\sigma$ ranges 0.15 and 0.08, respectively) and often have some degree of degeneracy with other parameters.  Constraints are less strong among the \numdetached\ (15\%) of systems deemed detached binaries.    Over half (62\%) of modeled systems were classified as ambiguous geometries; most of these have best-fitting models with large RMS resulting from asymmetric light curves or best-fitting models requiring unequal temperature components.  From single-band light curves alone it is not possible to distinguish a low-fillout-factor contact system from a detached configuration. Especially for systems having $P<$0.5 d the RMS for the best-fitting contact and detached models are often similar.   At these short periods the best-fitting solutions from contact and detached models tend to yield very similar parameters, but at longer periods the solutions can differ considerably. }

\item{The distribution of third-light fractions presented in Figure~\ref{fig:contacts_hist}} may be the most robust characterization of this parameter yet achieved, given the high precision of the {\it Kepler} photometry coupled with the unbiased and complete nature of the {\it Kepler} contact binary sample.  We find that 77\% of contact binaries have third light fractions $l3>$0.15, meaning that bright tertiaries are common among contact binary systems and that the true fraction of triple systems among this population is even larger.      

\item{The vast majority of systems at long periods $P>$0.5 d have either detached or ambiguous classifications, indicating that true contact binaries become rare at these longer periods. The \citet{Kirk2016} light curve morphology parameter serves as a rough indicator of the geometrical configuration of the close binaries studied (0.70$<$$morph$$<$1.0), but genuine contact and detached system can be found across this entire range.  Our modeling process identifies a preponderance of detached configurations at $morph>$0.9 and long periods, affirming the consistency of these independent approaches to characterizing physical geometry through analysis of the light curves. } 

\item{Among probable contact systems, the distribution of fillout factors covers the full range from 0--1 but shows a small excess near 1, indicating a measurable subpopulation with Roche lobes  near the maximum permitted volumes.}

\item{The observed distribution of $q$ with $P$ is consistent with the new set of model predictions of \citet{Molnar2022} in several ways. The models compute binary evolution with mass transfer to the primary star at a rate driven by the nuclear evolution of that star resulting in increasingly long orbital periods and extreme mass ratios with time (under the assumption of conservation of total mass and angular momentum).  Mass ratios $q_{\rm a}>$0.8 are rare among the probable contact binary subset and the ambiguous subset as shown in Figure~\ref{fig:qpfigure} (Section~\ref{sec:qp}). \citet{Molnar2022} show Roche geometry precludes steady mass transfer in this high-mass-ratio regime. Systems beginning in this mass range evolve rapidly (on a thermal timescale) towards lower mass ratios. The frequency of contact systems decreases with $P$ as does typical values of $q$. The models show more rapid evolution for greater mass ratios and as the primaries leave the main sequence (systems with larger $P$).  Finally, mass ratios $q_{\rm a}<$0.1 are rare among the whole sample and appear to be absent among contact binaries. The lower limit is expected when the post-main sequence primary rapidly increases its moment of inertia and the system reaches the Darwin instability. The models show the mass ratio at instability depends modestly on initial mass ratio and total mass, with the limit being reached for $P>$0.75 d at mass ratios increasing from 0.045 to 0.15 with increasing $P$.}

\end{enumerate}

These conclusions affirm the power of modern binary modeling codes combined with MCMC techniques to retrieve system parameters for large populations of close binaries from single-band light curves.  Although the computational effort required is considerable ($\gtrsim$80,000 CPU days for this project involving \nummodeled\ systems) suitable resources are becoming common even outside of major supercomputer centers.  The addition of photometric data at a second bandpass would provide further constraints on third-light parameters and reduce degeneracies owing to \Tratio, thereby narrowing the dispersion of  posterior parameter distributions.  Multi-band high-cadence long-baseline datasets are imminent, if not already available.  When coupled with planned time-domain photometric and spectroscopic surveys, the methods demonstrated here are a potent tool  when applied in a statistical manner  to reveal the evolutionary pathways and role of distinct physical processes (Lidov-Kozai cycles, tidal friction, mass transfer, magnetic cycles, thermal relaxation oscillations, common envelope evolution, etc.) among large populations of close binaries.

\acknowledgments

We thank undergraduate students Elle Buser and Alexander Schultz for helping obtain data in support of this program. We are grateful to Adam Myers for consultation on the use of MCMC methods and the {\tt emcee} code.  We thank Natalie Batalha for communications regarding the selection of the {\it Kepler} target sample.  We thank Andre Pr\'sa and the PHEOBE 2 team for making a powerful tool available to the community.   Suggestions from an efficient and dedicated anonymous reviewer improved this work.  This work made use of the Advanced Research Computing Center (2018) Teton Computing Environment, Intel x86\_64 cluster, University of Wyoming, Laramie, WY https://doi.org/10.15786/M2FY47.  This paper was made possible, in part, by the extended isolation enforced by the 2020--2021 SARS-CoV2 pandemic.  It is dedicated to Eugene M. Hermansen (U.S. Army, 1951--1953), who became one of the  casualties during the 2020--2022 pandemic.  This paper includes data collected by the Kepler mission and obtained from the MAST data archive at the Space Telescope Science Institute (STScI). Funding for the Kepler mission is provided by the NASA Science Mission Directorate. STScI is operated by the Association of Universities for Research in Astronomy, Inc., under NASA contract NAS 5-26555. This research was supported by the National Science Foundation through grant AST-1716718.

\vspace{5mm}

\facilities{WIRO, APO, KEPLER}

\software{IRAF \citep{Tody1986}, PHOEBE \citep{Prsa2016}, emcee \citep{Foreman-Mackey2013}}

\clearpage

\begin{deluxetable*}{lrrrrr}
\tablecaption{Pilot Study Systems with Spectroscopic Data\label{tab:sample}}
\tablewidth{0pt}
\tablehead{
\colhead{Name}&\colhead{$K$}   &\colhead{$T_{\rm eff}$}  &\colhead{$morph$}  &\colhead{$P$}      &\colhead{$t_{\rm 0}$}  \\
\colhead{}    &\colhead{(mag.)}&\colhead{(K)}            & \colhead{}      & \colhead{(d)}  & \colhead{(BJD)}     \\   
\colhead{(1)} &\colhead{(2)}   & \colhead{(3)}           & \colhead{(4)}  & \colhead{(5)}        & \colhead{(6)}  
}
\startdata
KIC04853067 & 13.1 & 5838  & 0.88 & 1.3409129  & 2454954.8641   \\
KIC04999357 & 12.0 & 6662  & 0.84 & 0.9901294  & 2454954.8807   \\ 
KIC06844489 & 14.1 & 6077  & 0.84 & 1.0797607  & 2454954.5790    \\ %     1.10 ztf is okay
KIC08913061 & 12.6 & 5414  & 0.76 & 1.0199709  & 2454954.1088   \\  %    0.99  ztf not needed
KIC09164694 & 14.2 & 6213  & 0.75 & 1.1113642  & 2454954.1862   \\  %    1.03 ztf data too fint
KIC09345838 & 12.4 & 6761  & 0.75 & 1.0458734  & 2454954.5341    \\  %    0.96  ztf not needed
KIC09840412 & 12.8 & 6507  & 0.79 & 0.8784699  & 2454954.5568    \\  %    1.05 ztf data good
KIC09953894 & 11.1 & 7295  & 0.74 & 1.3825997  & 2454954.5352    \\  %   1.02
KIC10292413 & 15.1 & 6200  & 0.95 & 0.5591594 & 2454954.2022   \\  %    1.02 zrtf not helpful!
KIC11097678 & 13.2 & 6493  & 0.90 & 0.9997155  & 2454954.8597    \\   % no third light    0.90  ztf data no good because P~1d
\enddata
\tablecomments{(1)--{\it Kepler Input Catalog} identifier, (2)---mean {\it Kepler} band magnitude, (3)--stellar effective temperature from the {\it Kepler Input Catalog}, (4)--light curve morphology parameter from \citet{Kirk2016}, (5)--orbital period determined as described in this work,  (6)--reference time of superior conjunction, as determined in this work, }
\end{deluxetable*}

%\begin{rotatetable}
\begin{deluxetable*}{lrrrrrrrrrrrrrcc}
\tablecaption{Best-fitting Contact Binary Model Parameters and Measured Spectroscopic Parameters for the Pilot Sample \label{tab:tensample} }
\tabletypesize{\scriptsize}
\tablewidth{0pt}
\tablehead{
\colhead{Name} &\colhead{$i$} & \colhead{$f$}& \colhead{$q_a$} & \colhead{$l3$}& \colhead{\Tratio} &\colhead{Date$_1$} & \colhead{$\phi_1$}& \colhead{$v_{\rm 1}$}& \colhead{$v_{\rm 2}$}&\colhead{Date$_2$}& \colhead{$\phi_2$}& \colhead{$v_{\rm 1}$}& \colhead{$v_{\rm 2}$}& \colhead{$\gamma_{\odot}$}& \colhead{$q_{spec}$} \\
\colhead{}     &\colhead{(\degr)}& \colhead{}& \colhead{}    & \colhead{}   & \colhead{}     &\colhead{BJD (d)}  & \colhead{}        & \colhead{(\kms)}     & \colhead{(\kms)}      & \colhead{BJD (d)}& \colhead{}	   & \colhead{(\kms)}	  & \colhead{(\kms)}	 & \colhead{(\kms)}  & \colhead{}  \\
\colhead{(1)}  &\colhead{(2)}    & \colhead{(3)} & \colhead{(4)} & \colhead{(5)} & \colhead{(6)}  & \colhead{(7)}     & \colhead{(8)}	     & \colhead{(9)}	    & \colhead{(10)}   & \colhead{(11)}    & \colhead{(12)}	  & \colhead{(13)}	 & \colhead{(14)}    & \colhead{(15)} & \colhead{(16)}
}
\startdata
KIC04853067 & 21.4 & 0.03 & 0.11 & 0.73 & 0.95 & 2458484.95970 & 0.18 &\nodata& 8  & 2458173.91641 & 0.64 & \nodata &   -5  & 2   & \nodata \tablenotemark{abcd} \\ 
KIC04999357 & 67.9 & 0.70 & 0.17 & 0.35 & 1.02 & 2459109.69654 & 0.24 & -87  & 137 & 2458365.65278 & 0.77 & -5 & -236  & -48 & 0.22  \tablenotemark{b}   \\
KIC06844489 & 73.0 & 0.98 & 0.31 & 0.63 & 0.98 & 2458365.81667 & 0.25 & -79  & 185 & 2458317.77222 & 0.76 & 23 & -232  & -26 & 0.25   \tablenotemark{bc}  \\
KIC08913061 & 77.0 & 0.46 & 0.16 & 0.02 & 1.00 & 2458959.90284 & 0.36 & -100 & 123 & 2458365.62014 & 0.71 & -29 & -281 & -72 & 0.16  \tablenotemark{e}   \\
KIC09164694 & 89.8 & 0.61 & 0.24 & 0.22 & 0.82 & 2458368.62917 & 0.30 & -36 & 160 & 2458315.80069 & 0.76 &  34 & -179 & -5  & 0.20 \tablenotemark{ce}    \\
KIC09345838 & 75.9 & 0.89 & 0.24 & 0.47 & 1.00 & 2458368.60208 & 0.32 & -26 & 167 & 2458315.69167 & 0.73 &  38 & -158 & 5  & 0.20 \tablenotemark{be}    \\
KIC09840412 & 87.1 & 0.80 & 0.27 & 0.55 & 0.98 & 2458175.97917 & 0.27 & -77 & 153 & 2458598.88542 & 0.74 & -17 & -231 & -43 & $>$0.16 \tablenotemark{bc}    \\
KIC09953894 & 59.7 & \nodata  & 0.97 & 0.00 & 1.03 & 2458304.89514 & 0.24 & -150 & 113 & 2458309.66875 & 0.69 & 79 & -168 & -32 & 0.81   \tablenotemark{d} \\
KIC10292413 & 53.6 & 0.88 & 0.53 & 0.59 & 1.00 & 2458598.92917 & 0.23 & -96 & 159 & 2458180.96042 & 0.73 &  44 & -219 & -28 & 0.37\tablenotemark{bcf}    \\
KIC11097678 & 75.9 & 0.62 & 0.06 & 0.01 & 1.00 & 2458970.94550 & 0.23 & -80 & 176 & 2458313.69861 & 0.79 & -64 & -328 & -74 & $>$0.03\tablenotemark{cg}    \\ 
\enddata
\tablenotetext{a}{The broadening function shows a single peak at the listed velocities and the components are not resolved.  }
\tablenotetext{b}{Significant third light is indicated; systemic velocity of primary may be uncertain. }
\tablenotetext{c}{Systematic \OC\ variations suggest a triple system. }
\tablenotetext{d}{Probable/Possible detached system. }
\tablenotetext{e}{Best-fitting model involves a hot spot on the primary. }
\tablenotetext{f}{A visual companion at 2\arcsec\ separations and $\Delta$G=0.5 mag may contribute light within the {\it Kepler} pixel. }
\tablenotetext{g}{The secondary is very faint and its radial velocity is significantly uncertain. }
\end{deluxetable*}
%\end{rotatetable}

\begin{deluxetable*}{lrrrrrrrrrrr}
\tabletypesize{\scriptsize}
\tablecaption{Best Parameters for \nummodeled\ Contact, Detached, and Ambiguous Systems when Modeled Using a Contact Configuration \label{tab:contactsbest}}
\tablewidth{0pt}
\tablehead{
\colhead{KIC ID} &\colhead{Period} & \colhead{$t_{\rm 0}$} & \colhead{KIC T$_{\rm eff}$} & \colhead{$morph$} &\colhead{$i$} & \colhead{$f$} & \colhead{$q$}& \colhead{$l3$} & \colhead{\Tratio} & \colhead{rms} & \colhead{Flag}  \\
\colhead{}     &\colhead{(d)}    & \colhead{(BJD)}        & \colhead{(K)}       & \colhead{}       &\colhead{(deg.)}    & \colhead{}       & \colhead{}      & \colhead{} & \colhead{} & \colhead{} & \colhead{}   \\
\colhead{(1)}  &\colhead{(2)}      & \colhead{(3)}      & \colhead{(4)}          & \colhead{(5)}          & \colhead{(6)}     & \colhead{(7)}    & \colhead{(8)}     & \colhead{(9)} & \colhead{(10)} & \colhead{(11)} & \colhead{(12)} 
}
\startdata
09612468 & 0.1334715 & 2454953.60380 &  7202 & 0.99 & 10.9 & 0.529 &  1.08 & 0.924 & 1.010 & 0.000006 & A   \\ 
06613627 & 0.1507996 & 2455115.34485 &  7090 & 0.98 & 41.5 & 0.080 &  0.95 & 0.936 & 0.997 & 0.000119 & A   \\ 
05302006 & 0.1511712 & 2454980.50904 &  6536 & 0.97 & 10.6 & 0.091 &  0.96 & 0.714 & 0.983 & 0.000109 & A   \\ 
09898401 & 0.1527742 & 2454964.73397 &  7376 & 0.99 & 33.7 & 0.162 &  1.03 & 0.966 & 1.004 & 0.000013 & A   \\ 
07375612 & 0.1600729 & 2454953.63870 &  6682 & 0.97 & 39.1 & 0.036 &  1.10 & 0.711 & 1.003 & 0.001247 & D   \\ 
05872696 & 0.1726132 & 2454953.79871 &  6000 & 0.99 & 38.9 & 0.037 &  1.04 & 0.929 & 1.004 & 0.000045 & D   \\ 
07767774 & 0.1733883 & 2455007.74026 &  6000 & 0.99 & 14.1 & 0.685 &  1.16 & 0.855 & 1.017 & 0.000039 & D   \\ 
12350008 & 0.1845296 & 2455002.01444 &  6108 & 0.71 & 88.7 & 0.031 &  1.36 & 0.997 & 0.979 & 0.000066 & D   \\ 
10684673 & 0.1925563 & 2455026.59881 &  7106 & 1.00 & 24.4 & 0.053 &  0.99 & 0.718 & 0.991 & 0.000071 & D   \\ 
09532219 & 0.1981551 & 2455001.94539 &  5031 & 0.77 & 75.0 & 0.141 &  1.48 & 0.850 & 0.978 & 0.002218 & A   \\ 
\enddata
\tablecomments{Table~\ref{tab:contactsbest} is published in its entirety in the machine-readable format.  A portion is shown here for guidance regarding its form and content. A (1) in the final column designates systems having superior model fits using detached configuration parameters summarized in Table~\ref{tab:detachedbest}.}
\end{deluxetable*}

\clearpage
%\begin{rotatetable}
\begin{deluxetable*}{lrrrrrrrrrrrrrrrr}
\movetabledown=1cm
\setlength{\tabcolsep}{2pt}
\tabletypesize{\scriptsize}
\tablecaption{Bayesian Percentile Parameters for \nummodeled\ Contact, Detached, and Ambiguous Systems Modeled Using a Contact Configuration \label{tab:contactsprob}}
\tablewidth{0pt}
\tablehead{
\colhead{Name}  & \colhead{$\cos$($i$)$_{\rm 16}$} &\colhead{$\cos$($i$)$_{\rm 50}$} & \colhead{$\cos$($i$)$_{\rm 84}$} & \colhead{$f_{\rm 16}$}& \colhead{$f_{\rm 50}$} & \colhead{$f_{\rm 84}$} & \colhead{$\log$($q$)$_{\rm 16}$}& \colhead{$\log$($q$)$_{\rm 50}$} & \colhead{$\log$($q$)$_{\rm 84}$} & \colhead{$l3_{\rm 16}$} & \colhead{$l3_{\rm 50}$} & \colhead{$l3_{\rm 84}$}  & \colhead{$T_{\rm 2}/T_{{\rm 1}_{\rm 16}}$ } & \colhead{$T_{\rm 2}/T_{{\rm 1}_{\rm 50}}$} & \colhead{$T_{\rm 2}/T_{{\rm 1}_{\rm 84}}$ }& \colhead{Flag}\\
\colhead{}         & \colhead{}       &\colhead{}    & \colhead{}       & \colhead{}      & \colhead{}       & \colhead{}    & \colhead{}    & \colhead{}    & \colhead{} & \colhead{}    & \colhead{} & \colhead{} & \colhead{}    & \colhead{} & \colhead{} & \colhead{}     \\
\colhead{(1)}  &\colhead{(2)}      & \colhead{(3)}      & \colhead{(4)}          & \colhead{(5)}          & \colhead{(6)}     & \colhead{(7)}    & \colhead{(8)}         & \colhead{(9)}          & \colhead{(10)}     & \colhead{(11)}      & \colhead{(12)}     & \colhead{(13)} & \colhead{(14)}      & \colhead{(15)}     & \colhead{(16)} & \colhead{(17)}     
}
\startdata
09612468 & 0.966 & 0.976 & 0.983 & 0.589 & 0.811 & 0.944 & -0.002 &  0.044 &  0.089 & 0.932 & 0.951 & 0.963 & 0.999 & 1.009 & 1.018 & A \\ 
06613627 & 0.750 & 0.800 & 0.849 & 0.096 & 0.171 & 0.247 & -0.073 & -0.011 &  0.047 & 0.914 & 0.929 & 0.943 & 0.983 & 0.998 & 1.014 & A \\ 
05302006 & 0.960 & 0.976 & 0.986 & 0.101 & 0.174 & 0.308 & -0.056 & -0.020 &  0.030 & 0.694 & 0.799 & 0.891 & 0.968 & 0.991 & 1.021 & A \\ 
09898401 & 0.819 & 0.835 & 0.856 & 0.110 & 0.163 & 0.219 & -0.027 &  0.009 &  0.050 & 0.961 & 0.965 & 0.968 & 0.989 & 1.003 & 1.016 & A \\ 
07375612 & 0.728 & 0.779 & 0.840 & 0.063 & 0.120 & 0.223 & -0.179 & -0.034 &  0.088 & 0.659 & 0.736 & 0.799 & 0.967 & 0.994 & 1.022 & D \\ 
05872696 & 0.756 & 0.781 & 0.805 & 0.043 & 0.069 & 0.111 & -0.043 & -0.002 &  0.035 & 0.925 & 0.933 & 0.941 & 0.988 & 1.000 & 1.012 & D \\ 
07767774 & 0.953 & 0.965 & 0.974 & 0.440 & 0.658 & 0.796 &  0.011 &  0.079 &  0.149 & 0.828 & 0.873 & 0.901 & 1.004 & 1.022 & 1.040 & D \\ 
12350008 & 0.024 & 0.062 & 0.122 & 0.039 & 0.055 & 0.089 & -0.194 & -0.057 &  0.146 & 0.997 & 0.997 & 0.997 & 0.968 & 0.986 & 1.004 & D \\ 
10684673 & 0.911 & 0.923 & 0.934 & 0.078 & 0.136 & 0.208 & -0.044 & -0.005 &  0.035 & 0.680 & 0.718 & 0.745 & 0.968 & 0.993 & 1.020 & D \\ 
09532219 & 0.077 & 0.185 & 0.278 & 0.110 & 0.224 & 0.353 &  0.100 &  0.344 &  0.532 & 0.828 & 0.855 & 0.877 & 0.960 & 0.973 & 0.990 & A \\ 
\enddata
\tablecomments{Table~\ref{tab:contactsprob} is published in its entirety in the machine-readable format.  A portion is shown here for guidance regarding its form and content.}
\end{deluxetable*}
%\end{rotatetable}

\clearpage

\clearpage
\begin{deluxetable*}{lrrrrrrrrrrrrrr}
%\tablenum{5}
\movetabledown=1cm
\setlength{\tabcolsep}{2pt}
\tabletypesize{\scriptsize}
\tablecaption{Best Parameters for \numdetached\ Probable Detached Systems Modeled Using a Detached Configuration \label{tab:detachedbest}}
\tablewidth{0pt}
\tablehead{
\colhead{KIC ID} &\colhead{Period} & \colhead{$t_{\rm 0}$}& \colhead{T$_{\rm eff}$} & \colhead{$morph$} &\colhead{$i$}  & \colhead{$q$}& \colhead{$l3$} & \colhead{$T_{\rm 2}/T_{\rm 1}$} & \colhead{$R_{\rm 1}$} &\colhead{$R_{\rm 2}/R_{\rm 1}$}  & \colhead{$R_{\rm 1}/R_{\rm 1max}$}& \colhead{$R_{\rm 2}/R_{\rm 2max}$} &\colhead{RMS} &\colhead{Flag} \\
\colhead{}     &\colhead{(d)}    & \colhead{(BJD)}        & \colhead{(K)}       & \colhead{}       &\colhead{(deg.)}    & \colhead{}       & \colhead{}      & \colhead{}   & \colhead{(\rsun)}       &\colhead{}    & \colhead{}       & \colhead{}   & \colhead{}   & \colhead{}    \\
\colhead{(1)}  &\colhead{(2)}      & \colhead{(3)}      & \colhead{(4)}          & \colhead{(5)}          & \colhead{(6)}     & \colhead{(7)}    & \colhead{(8)}         & \colhead{(9)}   & \colhead{(10)}     & \colhead{(11)}    & \colhead{(12)}         & \colhead{(13)} & \colhead{(14)}  & \colhead{(15)}        
}
\startdata
07375612 & 0.1600729 & 2454953.63870 &  6682 & 0.97 & 46.8 &  0.755 & 0.185 & 0.994 & 0.46 & 0.92 & 0.767 & 0.799 & 0.001235 & D  \\ 
05872696 & 0.1726132 & 2454953.79871 &  6000 & 0.99 & 60.3 &  0.058 & 0.116 & 0.962 & 0.43 & 0.37 & 0.518 & 0.678 & 0.000044 & D  \\ 
07767774 & 0.1733883 & 2455007.74026 &  6000 & 0.99 & 41.4 &  0.044 & 0.174 & 0.962 & 0.44 & 0.34 & 0.521 & 0.704 & 0.000039 & D  \\ 
12350008 & 0.1845296 & 2455002.01444 &  6108 & 0.71 & 80.9 &  0.439 & 0.996 & 0.974 & 0.65 & 0.69 & 0.934 & 0.943 & 0.000049 & D  \\ 
10684673 & 0.1925563 & 2455026.59881 &  7106 & 1.00 & 29.5 &  0.959 & 0.107 & 0.999 & 0.51 & 0.98 & 0.749 & 0.752 & 0.000062 & D  \\ 
06699679 & 0.2012033 & 2454953.61283 &  6000 & 1.00 & 46.4 &  0.049 & 0.357 & 0.946 & 0.57 & 0.35 & 0.619 & 0.831 & 0.000016 & D  \\ 
06287172 & 0.2038732 & 2454953.75384 &  6646 & 0.95 & 71.0 &  0.087 & 0.170 & 0.988 & 0.34 & 0.44 & 0.380 & 0.497 & 0.000260 & D  \\ 
10030943 & 0.2357801 & 2454953.67463 &  6704 & 1.00 & 54.1 &  1.104 & 0.632 & 1.002 & 0.49 & 1.04 & 0.670 & 0.664 & 0.000018 & D  \\ 
12216817 & 0.2462731 & 2455002.06796 &  6681 & 1.00 & 43.1 &  1.242 & 0.087 & 1.006 & 0.66 & 1.08 & 0.903 & 0.880 & 0.000454 & D  \\ 
08122124 & 0.2492776 & 2454964.61283 &  6250 & 1.00 & 34.9 &  0.557 & 0.520 & 0.976 & 0.74 & 0.82 & 0.882 & 0.945 & 0.000152 & D  \\ 
\enddata
\tablecomments{Table~\ref{tab:detachedbest} is published in its entirety in the machine-readable format.  A portion is shown here for guidance regarding its form and content. $R_{\rm 1}/R_{\rm 1max}$ and $R_{\rm 2}/R_{\rm 2max}$ are approximate values predicated on the assumption of $M$=1~\msun\ for the more massive component. $R_{\rm 1max}$ and $R_{\rm 2max}$ scale weakly with the adopted $M_{\rm 1}$.  $R_{\rm 1}$ is the radius of a spherical star having equivalent surface area to the tidally distorted star. $R_{\rm 1max}$ is the equivalent radius of a tidally distorted star of maximum possible size without overflow.}
\end{deluxetable*}

\clearpage
\begin{deluxetable*}{lrrrrrrrrrrrrrrrrrrr}
%\tablenum{6}
\movetabledown=1cm
\setlength{\tabcolsep}{2pt}
\tabletypesize{\scriptsize}
\tablecaption{Bayesian Percentile Parameters for \numdetached\ Probable Detached Systems Modeled Using a Detached Configuration \label{tab:detachedprob}}
\tablewidth{0pt}
\tablehead{
\colhead{KIC ID} &  \colhead{$\cos$($i$)$_{\rm 16}$} &\colhead{$\cos$($i$)$_{\rm 50}$} & \colhead{$\cos$($i$)$_{\rm 84}$}  & \colhead{$\log$($q$)$_{\rm 16}$}& \colhead{$\log$($q$)$_{\rm 50}$} & \colhead{$\log$($q$)$_{\rm 84}$} & \colhead{$l3_{\rm 16}$} & \colhead{$l3_{\rm 50}$} & \colhead{$l3_{\rm 84}$}  & \colhead{$T_{\rm 2}/T_{{\rm 1}_{\rm 16}}$}& \colhead{$T_{\rm 2}/T_{{\rm 1}_{\rm 50}}$} & \colhead{$T_{\rm 2}/T_{{\rm 1}_{\rm 84}}$} & \colhead{$R_{{\rm 1}_{\rm 16}}$}& \colhead{$R_{{\rm 1}_{\rm 50}}$} & \colhead{$R_{{\rm 1}_{\rm 84}}$} & \colhead{$R_{\rm 2}/R_{{\rm 1}_{\rm 16}}$}& \colhead{$R_{\rm 2}/R_{{\rm 1}_{\rm 50}}$} & \colhead{$R_{\rm 2}/R_{{\rm 1}_{\rm 84}}$}  &\colhead{Flag} \\
\colhead{}          & \colhead{}       &\colhead{}    & \colhead{}       & \colhead{}      & \colhead{}   & \colhead{}       &\colhead{}    & \colhead{}       & \colhead{}   & \colhead{}  & \colhead{}   & \colhead{}   & \colhead{(\rsun)}   & \colhead{(\rsun)}   & \colhead{(\rsun)}   & \colhead{}   & \colhead{}   & \colhead{}   & \colhead{}      \\
\colhead{(1)}  &\colhead{(2)}      & \colhead{(3)}      & \colhead{(4)}          & \colhead{(5)}          & \colhead{(6)}     & \colhead{(7)}    & \colhead{(8)}         & \colhead{(9)}   & \colhead{(10)}     & \colhead{(11)}    & \colhead{(12)}         & \colhead{(13)} & \colhead{(14)} & \colhead{(15)} & \colhead{(16)} & \colhead{(17)} & \colhead{(18)} & \colhead{(19)} & \colhead{(20)}         
}
\startdata
07375612 & 0.507 & 0.599 & 0.722 & -0.948 & -0.477 & -0.239 & 0.053 & 0.179 & 0.368 & 0.954 & 0.989 & 1.038 & 0.344 & 0.433 & 0.500 & 0.610 & 0.728 & 0.892 & D \\ 
05872696 & 0.433 & 0.493 & 0.590 & -1.292 & -1.037 & -0.679 & 0.146 & 0.282 & 0.436 & 0.962 & 0.973 & 0.985 & 0.354 & 0.416 & 0.465 & 0.355 & 0.432 & 0.590 & D \\ 
07767774 & 0.482 & 0.607 & 0.745 & -1.470 & -1.254 & -0.953 & 0.209 & 0.416 & 0.592 & 0.943 & 0.964 & 0.981 & 0.336 & 0.425 & 0.504 & 0.314 & 0.366 & 0.472 & D \\ 
12350008 & 0.155 & 0.255 & 0.345 & -0.669 & -0.395 & -0.085 & 0.992 & 0.994 & 0.996 & 0.948 & 0.969 & 0.990 & 0.579 & 0.648 & 0.682 & 0.483 & 0.644 & 0.852 & D \\ 
10684673 & 0.512 & 0.740 & 0.843 & -0.349 & -0.013 &  0.105 & 0.024 & 0.080 & 0.216 & 0.914 & 0.977 & 1.007 & 0.389 & 0.435 & 0.490 & 0.417 & 0.651 & 1.054 & D \\ 
06699679 & 0.512 & 0.579 & 0.649 & -1.528 & -1.425 & -1.297 & 0.184 & 0.274 & 0.384 & 0.936 & 0.946 & 0.964 & 0.514 & 0.557 & 0.585 & 0.296 & 0.320 & 0.358 & D \\ 
06287172 & 0.347 & 0.545 & 0.769 & -1.334 & -0.892 & -0.296 & 0.184 & 0.500 & 0.780 & 0.939 & 0.985 & 1.038 & 0.250 & 0.355 & 0.503 & 0.354 & 0.507 & 0.824 & D \\ 
10030943 & 0.517 & 0.577 & 0.618 & -0.029 &  0.018 &  0.074 & 0.543 & 0.617 & 0.659 & 0.998 & 1.001 & 1.003 & 0.454 & 0.496 & 0.517 & 0.971 & 1.015 & 1.067 & D \\ 
12216817 & 0.657 & 0.724 & 0.788 &  0.006 &  0.113 &  0.259 & 0.038 & 0.137 & 0.261 & 0.980 & 1.004 & 1.024 & 0.588 & 0.654 & 0.719 & 1.001 & 1.084 & 1.200 & D \\ 
08122124 & 0.700 & 0.779 & 0.828 & -0.642 & -0.283 & -0.048 & 0.243 & 0.408 & 0.540 & 0.970 & 0.992 & 1.024 & 0.611 & 0.674 & 0.727 & 0.642 & 0.799 & 0.964 & D \\ 
\enddata
\tablecomments{Table~\ref{tab:detachedprob} is published in its entirety in the machine-readable format.  A portion is shown here for guidance regarding its form and content. $R_{\rm 1}/R_{\rm 1max}$ and $R_{\rm 2}/R_{\rm 2max}$ are approximate values predicated on the assumption of $M$=1~\msun\ for the more massive component. $R_{\rm 1max}$ and $R_{\rm 2max}$ scale weakly with the adopted $M_{\rm 1}$.  $R_{\rm 1}$ is the radius of a spherical star having equivalent surface area to the tidally distorted star. $R_{\rm 1max}$ is the equivalent radius of a tidally distorted star of maximum possible size without overflow.}
\end{deluxetable*}

%% For this sample we use BibTeX plus aasjournals.bst to generate the
%% the bibliography. The sample63.bib file was populated from ADS. To
%% get the citations to show in the compiled file do the following:
%%
%% pdflatex sample63.tex
%% bibtex sample63
%% pdflatex sample63.tex
%% pdflatex sample63.tex

\bibliography{ms}{}
\bibliographystyle{aasjournal}

\end{document}